\documentclass[pdflatex,sn-mathphys-num]{sn-jnl}

\usepackage{graphicx}%
\usepackage{multirow}%
\usepackage{amsmath,amssymb,amsfonts}%
\usepackage{amsthm}%
\usepackage{mathrsfs}%
\usepackage[title]{appendix}%
\usepackage{xcolor}%
\usepackage{textcomp}%
\usepackage{manyfoot}%
\usepackage{booktabs}%
\usepackage{algorithm}%
\usepackage{algorithmicx}%
\usepackage{algpseudocode}%
\usepackage{listings}%

\usepackage{threeparttable}
\theoremstyle{thmstyleone}%
\theoremstyle{thmstyletwo}%

\theoremstyle{thmstylethree}%

\usepackage{color}

\usepackage{soul}

\newcommand{\Msunh}{\>h^{-1} {\rm M_\odot}}
\newcommand{\Msunhh}{\>h^{-2} {\rm M_\odot}}

\newcommand{\mpch}{\>h^{-1}{\rm {Mpc}}}

\begin{document}


\title[Blue dwarf galaxies]{Blue dwarf galaxies reveal the mass and environmental dependence of galaxy quenching}


\author[1,3]{\fnm{Qingyang} \sur{Li}}
\equalcont{These authors contributed equally to this work.}

\author[1,2,3]{\fnm{Yirong} \sur{Wang}}
\equalcont{These authors contributed equally to this work.}

\author[1,2,3]{\fnm{Yizhou} \sur{Gu}}
\equalcont{These authors contributed equally to this work.}

\author*[1,2,3]{\fnm{Xiaohu} \sur{Yang}}\email{xyang@sjtu.edu.cn}

\author[1,3]{Xiaoju Xu}
\author[1,3]{Wenting Wang}
\author*[1,3]{Jiaxin Han}\email{jiaxin.han@sjtu.edu.cn}
\author[1,3]{Wensheng Hong}
\author[1,2,3]{Jiaqi Wang}
\author[1,3]{Yunliang Zheng}
\author[1,3]{Yiyang Guo}
\author[1,2,3]{Junde Li}
\author[1,3]{Zhongxu Zhai}
\author[1,2,3]{Yipeng Jing}
\author[1,2,3]{Ying Zu}
\author[1,4]{Zhejie Ding}
\author[5]{Jessica Nicole Aguilar}
\author[6]{Steven Ahlen}
\author[7,8]{Davide Bianchi}
\author[9]{David Brooks}
\author[5]{Andrei Cuceu}
\author[10]{Axel de la Macorra}
\author[9]{Peter Doel}
\author[11,12]{Jaime E. Forero-Romero}
\author[13,14,15]{Enrique Gaztañaga}
\author[5]{Satya Gontcho A Gontcho}
\author[16]{Gaston Gutierrez}
\author[17]{ChangHoon Hahn}
\author[18,19]{Klaus Honscheid}
\author[20]{Dick Joyce}
\author[21]{Robert Kehoe}
\author[5]{Theodore Kisner}
\author[5]{Anthony Kremin}
\author[9]{Ofer Lahav}
\author[5]{Martin Landriau}
\author[22]{Laurent Le Guillou}
\author[23,24]{Marc Manera}
\author[20]{Aaron Meisner}
\author[25,24]{Ramon Miquel}
\author[26]{John Moustakas}
\author[14]{Seshadri Nadathur}
\author[27,28,29]{Will Percival}
\author[30]{Francisco Prada}
\author[31]{Ignasi Pérez-Ràfols}
\author[32]{Graziano Rossi}
\author[33]{Eusebio Sanchez}
\author[34]{Edward Schlafly}
\author[5]{David Schlegel}
\author[35]{Hee-Jong Seo}
\author[5]{Joseph Harry Silber}
\author[20]{David Sprayberry}
\author[36]{Gregory Tarlé}
\author[20]{Benjamin Alan Weaver}
\author[37]{Hu Zou}



\affil*[1]{Department of Astronomy, School of Physics and Astronomy, Shanghai Jiao Tong University, Shanghai 200240, China}
\affil[2]{State Key Laboratory of Dark Matter Physics, Tsung-Dao Lee Institute, Shanghai Jiao Tong University, Shanghai 200240, China}
\affil[3]{Shanghai Key Laboratory for Particle Physics and Cosmology, and Key Laboratory for Particle Physics, Astrophysics and Cosmology, Ministry of Education, Shanghai Jiao Tong University, Shanghai 200240, China}
\affil[4]{Nanjing Institute of Astronomical Optics \& Technology, Chinese Academy of Sciences, Nanjing 210042, China}
\affil[5]{Lawrence Berkeley National Laboratory, 1 Cyclotron Road, Berkeley, CA 94720, USA}
\affil[6]{Physics Department, Boston University, 590 Commonwealth Avenue, Boston, MA 02215, USA}
\affil[7]{Dipartimento di Fisica ``Aldo Pontremoli'', Università degli Studi di Milano, Via Celoria 16, I-20133 Milano, Italy}
\affil[8]{INAF-Osservatorio Astronomico di Brera, Via Brera 28, 20122 Milano, Italy}
\affil[9]{Department of Physics and Astronomy, University College London, Gower Street, London WC1E 6BT, United Kingdom}
\affil[10]{Instituto de Física, Universidad Nacional Autónoma de México, Circuito de la Investigación Científica, Ciudad Universitaria, Cd. de México C.P. 04510, México}
\affil[11]{Departamento de Física, Universidad de los Andes, Cra. 1 No. 18A-10, Edificio Ip, CP 111711, Bogotá, Colombia}
\affil[12]{Observatorio Astronómico, Universidad de los Andes, Cra. 1 No. 18A-10, Edificio H, CP 111711 Bogotá, Colombia}
\affil[13]{Institut d’Estudis Espacials de Catalunya (IEEC), c/ Esteve Terradas 1, Edifici RDIT, Campus PMT-UPC, 08860 Castelldefels, Spain}
\affil[14]{Institute of Cosmology and Gravitation, University of Portsmouth, Dennis Sciama Building, Portsmouth PO1 3FX, United Kingdom}
\affil[15]{Institute of Space Sciences (ICE-CSIC), Campus UAB, Carrer de Can Magrans s/n, 08913 Bellaterra, Barcelona, Spain}
\affil[16]{Fermi National Accelerator Laboratory, PO Box 500, Batavia, Illinois 60510, USA}
\affil[17]{Steward Observatory, University of Arizona, 933 N. Cherry Avenue, Tucson, Arizona 85721, USA}
\affil[18]{Center for Cosmology and AstroParticle Physics, The Ohio State University, 191 West Woodruff Avenue, Columbus, Ohio 43210, USA}
\affil[19]{Department of Physics, The Ohio State University, 191 West Woodruff Avenue, Columbus, Ohio 43210, USA}
\affil[20]{NSF NOIRLab, 950 N. Cherry Ave., Tucson, AZ 85719, USA}
\affil[21]{Department of Physics, Southern Methodist University, 3215 Daniel Avenue, Dallas, Texas 75275, USA}
\affil[22]{Sorbonne Université, CNRS/IN2P3, Laboratoire de Physique Nucléaire et de Hautes Énergies (LPNHE), FR-75005 Paris, France}
\affil[23]{Departament de Física, Serra Húnter, Universitat Autònoma de Barcelona, 08193 Bellaterra (Barcelona), Spain}
\affil[24]{Institut de Física d’Altes Energies (IFAE), The Barcelona Institute of Science and Technology, Edifici Cn, Campus UAB, 08193 Bellaterra (Barcelona), Spain}
\affil[25]{Institució Catalana de Recerca i Estudis Avançats (ICREA), Passeig de Lluís Companys 23, 08010 Barcelona, Spain}
\affil[26]{Department of Physics and Astronomy, Siena College, 515 Loudon Road, Loudonville, New York 12211, USA}
\affil[27]{Department of Physics and Astronomy, University of Waterloo, 200 University Ave W, Waterloo, Ontario N2L 3G1, Canada}
\affil[28]{Perimeter Institute for Theoretical Physics, 31 Caroline St. North, Waterloo, Ontario N2L 2Y5, Canada}
\affil[29]{Waterloo Centre for Astrophysics, University of Waterloo, 200 University Ave W, Waterloo, Ontario N2L 3G1, Canada}
\affil[30]{Instituto de Astrofísica de Andalucía (CSIC), Glorieta de la Astronomía s/n, E-18008 Granada, Spain}
\affil[31]{Departament de Física, EEBE, Universitat Politècnica de Catalunya, c/Eduard Maristany 10, 08930 Barcelona, Spain}
\affil[32]{Department of Physics and Astronomy, Sejong University, 209 Neungdong-ro, Gwangjin-gu, Seoul 05006, Republic of Korea}
\affil[33]{CIEMAT, Avenida Complutense 40, E-28040 Madrid, Spain}
\affil[34]{Space Telescope Science Institute, 3700 San Martin Drive, Baltimore, MD 21218, USA}
\affil[35]{Department of Physics and Astronomy, Ohio University, 139 University Terrace, Athens, Ohio 45701, USA}
\affil[36]{Department of Physics, University of Michigan, 450 Church Street, Ann Arbor, Michigan 48109, USA}
\affil[37]{National Astronomical Observatories, Chinese Academy of Sciences, A20 Datun Road, Chaoyang District, Beijing 100101, People’s Republic of China}


\abstract{Low-mass dark matter haloes provide a critical test of models of galaxy formation, where stellar feedback and environmental processes are expected to regulate the growth and star-formation activity of dwarf galaxies. However, the quenching of central and satellite dwarf galaxies, and its dependence on host halo mass, remain poorly constrained observationally. Here we develop an \(N_{\max}\)–\(V_{\max}\) framework to measure the red-galaxy fractions (\(f_{\rm red}\)), used as our primary observational proxy for quenching, and the complementary quiescent-galaxy fractions of central and satellite galaxies. Applying this framework to the Dark Energy Spectroscopic Instrument (DESI) Data Release 2, we obtain measurements down to stellar masses of \(M_*\sim10^7\,h^{-2}M_\odot\). We find a substantial population of blue dwarf galaxies among both central and satellite systems, while the red fraction of satellite galaxies increases systematically with host halo mass. These measurements reveal distinct stellar-mass and halo-mass dependences of dwarf-galaxy quenching. Comparisons with the IllustrisTNG simulations show systematic discrepancies in both central and satellite populations: TNG predicts stronger suppression of star formation in low-mass central galaxies and a stronger host-halo-mass dependence of satellite quenching than inferred from DESI. We further quantify these trends with a five-parameter model and validate the measurement procedure using end-to-end mock surveys that incorporate DESI selection effects and our group-finding pipeline. Finally, abundance matching between the central-galaxy SMF and the ELUCID halo mass function suggests a nearly flat stellar-to-halo mass ratio over the low-mass halo range probed. Together, these results provide new observational constraints on how stellar mass and halo environment regulate the evolution of dwarf galaxies.

}

\maketitle

\section{Introduction}\label{sec1}

Low-mass dark matter haloes hosting dwarf galaxies provide important laboratories for testing models of galaxy formation, where the effects of baryonic processes on galaxy growth can be examined in relatively low-mass systems. A central prediction of the prevailing paradigm is that the conversion of baryons into stars is strongly suppressed in these haloes. Processes including cosmic reionization, stellar winds, and supernova feedback are invoked to expel or heat gas, thereby alleviating the ``overcooling problem'' and inhibiting efficient star formation within low-mass haloes. This framework predicts that dwarf galaxies residing in such haloes are predominantly red and quiescent \cite{Mo2010, Vogelsberger2014}, and is incorporated into modern cosmological simulations such as IllustrisTNG \cite{Nelson2019}, EAGLE \cite{Schaye2015}, and SIMBA \cite{Dave2019}. Previous studies have revealed important discrepancies between the observed quiescent fractions of low-mass galaxies and predictions from state-of-the-art hydrodynamical simulations. In particular, Dickey et al. \cite{Dickey2021} showed that several major simulations overproduce quiescent isolated low-mass galaxies, even after accounting for observational selection effects. However, their analysis primarily focused on isolated systems, leaving the environmental dependence of dwarf-galaxy quenching, particularly among satellite galaxies residing in haloes of different masses, less well constrained. A comprehensive understanding of dwarf-galaxy evolution therefore requires a joint investigation of central and satellite populations, with particular attention to how their quenching properties depend on both stellar mass and host halo mass. Meanwhile, observations with the James Webb Space Telescope (JWST) have revealed unexpectedly massive galaxies at high redshift, raising further questions about the efficiency and timing of early galaxy formation. These findings motivate more stringent tests of galaxy formation models across a broad range of halo masses. In the local Universe, although legacy surveys such as the Sloan Digital Sky Survey (SDSS) have established the observational foundation for studying galaxy populations, their statistical power becomes limited for dwarf galaxies below $M_\ast \sim 10^{8.5} \Msunhh$, leaving the galaxy--halo connection in this regime poorly constrained \cite{yang2009galaxy, Lan2016}.

The fraction of red galaxies, as identified from their optical colours, provides a widely used empirical diagnostic of the suppression of star formation and the evolutionary pathways of galaxies \cite{Blanton2005,Baldry2006,Conselice2014,Masters2010}. Since the discovery of the bimodal distribution of galaxy colours \cite{Strateva2001, Baldry2004}, numerous studies have investigated how the red-galaxy fraction depends on stellar mass and environment. In particular, Baldry et al. \cite{Baldry2006} used SDSS data to measure red fractions across different stellar-mass and environmental bins, describing their joint dependence through an empirical relation, \(f_{\rm red}=F(\Sigma,M_{\ast})\). Building on this framework, Peng et al. \cite{Peng2010} interpreted the observed trends in terms of two distinct quenching channels: mass quenching and environmental quenching. More recently, Wang et al. \cite{Wang2018}, using galaxy groups identified from the SDSS \cite{yang2007galaxy}, described these trends in terms of stellar-mass-dependent quenching and halo-mass-dependent quenching, highlighting the respective roles of stellar mass and host halo mass. Although the red fractions of relatively massive galaxies (\(M_{\ast}\gtrsim10^{8.5}\,h^{-2}M_\odot\)) have been extensively studied using both observations and numerical simulations \cite{Dickey2021,Donnari2021,Geha2012,Schaefer2019}, observational constraints remain limited in the dwarf-galaxy regime, particularly below \(M_{\ast}\sim10^{7.5}\,h^{-2}M_\odot\). Consequently, it remains unclear how the relative importance of stellar-mass-dependent and halo-mass-dependent quenching evolves towards the lowest galaxy masses.

The Dark Energy Spectroscopic Instrument (DESI) survey provides an unprecedented opportunity to extend these studies to the dwarf-galaxy regime. Here we use the DESI Data Release 2 (DR2) Bright Galaxy Survey (BGS), which combines extensive sky coverage with high spectroscopic completeness, to conduct a statistical study of dwarf galaxies. We identify galaxy groups over the redshift range \(0\leq z\leq0.6\) using an extended halo-based group finder \cite{yang2021extended}, which enables us to distinguish central and satellite galaxies and estimate their host halo masses. To make optimal use of the flux-limited DESI sample, we calculate the maximum redshift, \(z_{\max}\), to which each galaxy could be observed and account for the associated selection effects. By linking galaxies to their host dark matter haloes, we go beyond the conventional \(V_{\max}\) weighting scheme and introduce an additional \(N_{\max}(M_{\rm h})\) weighting scheme, which accounts for the maximum number of groups within a given halo-mass range over the redshift interval in which a galaxy could be detected. This framework enables us to measure the red and quiescent fractions of central and satellite galaxies using, respectively, the stellar mass functions (SMFs) of central galaxies and the conditional stellar mass functions (CSMFs) of satellite galaxies, without requiring separate stellar-mass-complete and halo-mass-complete samples \cite{van2008, Wang2018}. These measurements allow us to investigate the dependence of dwarf-galaxy quenching on both stellar mass and host halo mass.

With this framework, we first measure the SMFs of central galaxies and the CSMFs of satellite galaxies, together with their red and quiescent fractions, down to the dwarf-galaxy regime. We then investigate how the quenching properties of central and satellite galaxies depend on stellar mass and host halo mass, and compare our measurements with predictions from the IllustrisTNG simulations to assess the performance of current galaxy formation models in the low-mass regime. Finally, we derive the stellar-to-halo mass relation (SHMR) down to halo masses of \(\sim10^{10}\,h^{-1}M_\odot\), using the ELUCID simulation \cite{Tully2019} to account for the impact of the local underdensity \cite{Wang2016}. Together, these analyses provide complementary observational constraints on the abundance, quenching, and stellar-to-halo mass connection of dwarf galaxies, enabling us to examine how star formation is regulated in low-mass dark matter haloes and their surrounding environments.

\section{The stellar mass functions of central and satellite dwarf galaxies}
\label{sec_cen_smf}

We start our investigation by measuring the SMFs for both the total and central-galaxy population, covering all, red/blue, and quiescent/star-forming systems using the standard $V_{\rm max}$ approach. Before moving forward, we note that, for a fixed apparent-magnitude cut of $r \le 19.5$, the blue, star-forming and red, quiescent galaxy populations become incomplete in stellar mass at different redshift thresholds. To visually and quantitatively evaluate whether these selection effects introduce systematic biases for dwarf galaxies, we construct a 2D map of galaxy number density. By inspecting this map directly, we search for any galaxy populations that exhibit systematic shortfalls, or disappear entirely, in the number-density distribution at the lowest stellar masses. In general, we find that red/quiescent galaxies can be robustly measured for $M_{\ast} > 10^7\,\Msunhh$, whereas blue/star-forming galaxies can be tracked down to $M_{\ast} > 10^6\,\Msunhh$. These checks enable us to robustly estimate the red/blue and star-forming/quiescent fractions in the Bright Galaxy Survey (BGS) bright sample down to $M_{\ast} \sim 10^7\,\Msunhh$.

\begin{figure}
    \centering
    \includegraphics[width=0.905\textwidth]{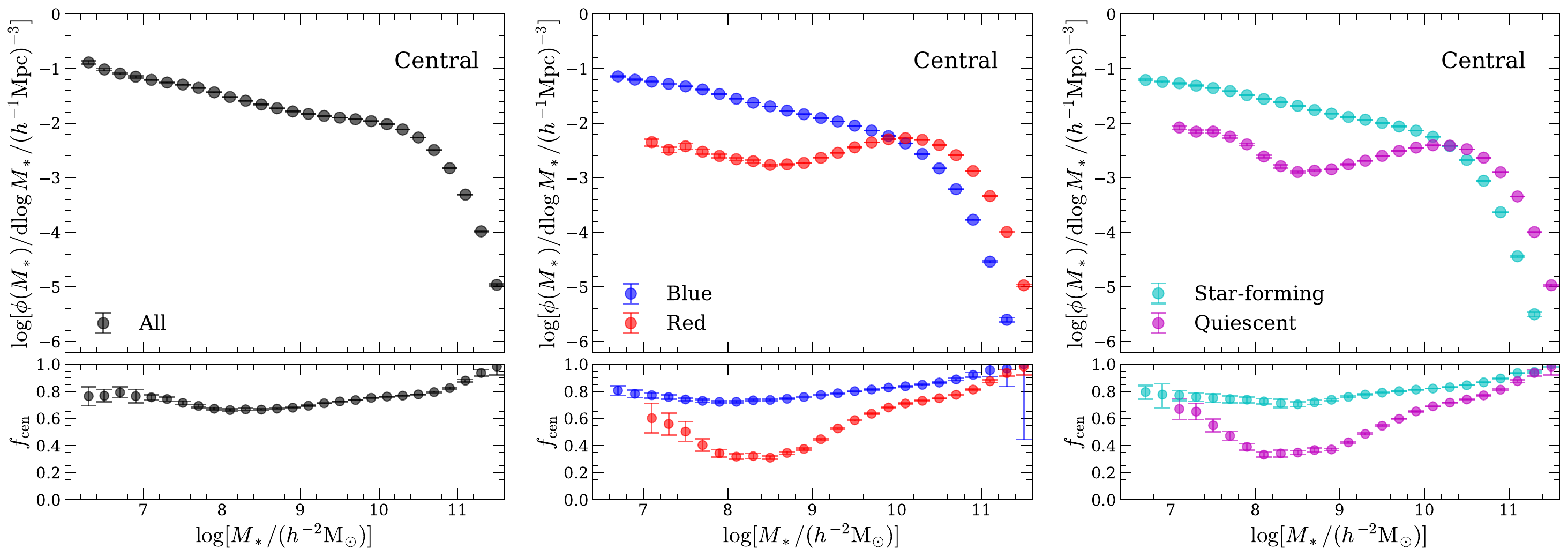}
    \includegraphics[width=0.925\textwidth,height=0.4\textwidth]{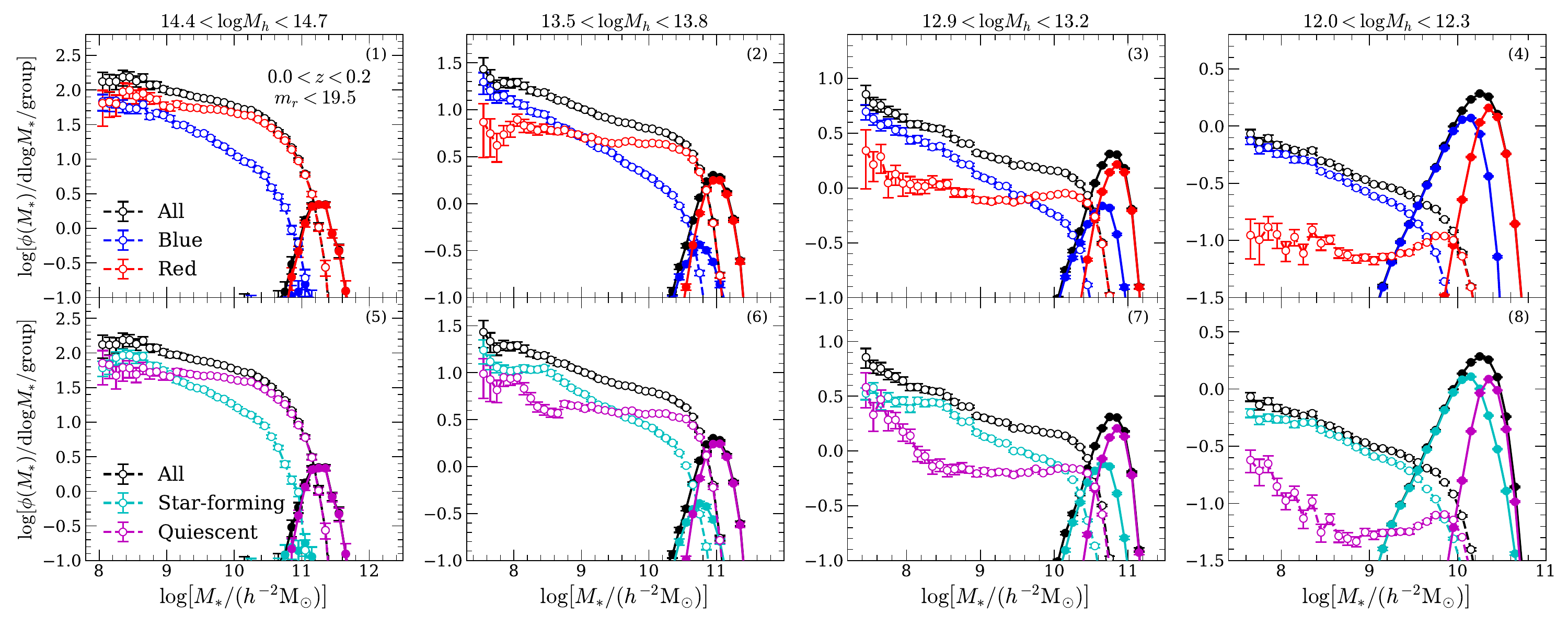}
\caption{
Upper panels: Stellar mass functions for all, red/blue, star-forming/quiescent central galaxies in the left, middle and right panels, respectively. The error bars are obtained from the standard deviation of 100 bootstrap resamplings of galaxies. Below each panel, the fraction of central galaxies, $f_{\rm cen}$, is shown in the corresponding subpanel. Lower panels: Conditional stellar mass functions for all, red and blue galaxy populations (upper row panels), for all, quiescent and star-forming populations (lower panels) based on DESI BGS galaxy samples with an apparent magnitude cut of 19.5 in $r$-band. Each column corresponds to a different halo mass bin as indicated on top of each panel. The filled and open points represent the central and satellite galaxy populations, respectively. The error bars are obtained from the standard deviation of 100 bootstrap resamplings of galaxy groups. 
}
    \label{fig:CSMF_measure}
\end{figure}

We present the resulting SMFs for the central galaxies as open circles in the upper panels of Fig. \ref{fig:CSMF_measure}, with the all, red/blue, and quiescent/star-forming populations shown in the left, middle, and right panels, respectively, where the SMFs for red and quiescent galaxies are reported only for $M_{\ast}>10^{7}\Msunhh$. Our results show that the massive regime ($M_{\ast}>10^{10}\Msunhh$) is dominated largely by red and quiescent systems. By comparison, at lower stellar masses ($10^{7.0}\Msunhh<M_{\ast}<10^{10}\Msunhh$), the SMFs are mainly shaped by blue and star-forming galaxies, whereas the red and quiescent component becomes increasingly minor toward lower stellar masses.  We also measure the SMFs of the total galaxy population and present the central-to-total ratios, \(f_{\rm cen}\), in the corresponding subpanels. For the full galaxy population, \(f_{\rm cen}\) is close to unity at the high-mass end and decreases to approximately 0.8 at the low-mass end, reaching a minimum around \(M_{\ast}\sim10^{8.5}\,\Msunhh\). The trends are more pronounced for red and quiescent galaxies, whose \(f_{\rm cen}\) values increase again towards lower stellar masses. Our MGRS tests indicate that this upturn is not reproduced by the group-finding procedure alone, suggesting that it is unlikely to arise solely from central/satellite misidentification. Instead, it is likely associated with the local-void (LV) effect, whereby the underrepresentation of low-mass red satellite galaxies in the local volume leads to an artificially elevated central fraction of red galaxies at the low-mass end. This effect should therefore be taken into account when interpreting the abundance and environmental distribution of low-mass red galaxies.

In addition to measuring the SMFs of galaxies with different properties, we compute the CSMFs, \(\Phi(M_{\ast}\mid M_{\rm h})\), which describe the average number of galaxies per halo as a function of stellar mass, \(M_{\ast}\), for haloes of a given mass, \(M_{\rm h}\). The CSMFs, computed using the \(N_{\max}(M_{\rm h})\) method described in Section \ref{sec_CSMF}, for the total, red, and blue galaxy populations across different halo-mass bins are shown in the first row of the lower panels of Fig. \ref{fig:CSMF_measure}. The corresponding CSMFs for the total, quiescent, and star-forming populations are presented in the second row. The large volume and depth of DESI DR2, combined with our halo-based group finder, enable us to constrain the CSMFs of different galaxy populations well into the dwarf-galaxy regime. In particular, we find substantial populations of blue and star-forming satellite galaxies at low stellar masses. Together with the central-galaxy SMFs presented above, these measurements demonstrate that blue dwarf galaxies are abundant in both central and satellite systems.

We compare our measurements with previous CSMF studies based on both spectroscopic and photometric surveys. At \(M_{\ast}>10^9\,h^{-2}M_\odot\), our measurements agree well with previous SDSS results \cite{yang2005galoccup, yang2009galaxy}. At lower stellar masses, however, we find a substantially steeper faint-end slope for the satellite-galaxy CSMFs, indicating a larger abundance of low-mass satellites than inferred from earlier spectroscopic measurements. This trend is consistent with the results from DESI DR1 \cite{Wang2024}. Similar steep faint-end slopes have also been reported in studies using photometric data \cite{Lan2016, Wang2025}. Our measurements therefore extend previous constraints on the CSMFs of different galaxy populations to the dwarf-galaxy regime, providing a more detailed view of the abundance of low-mass galaxies in haloes of different masses.

The CSMFs further allow us to investigate the dependence of galaxy populations on stellar mass and host halo mass. In particular, the red fractions of both central and satellite galaxies decrease towards lower stellar masses, while the red fraction of satellite galaxies also decreases towards lower host halo masses. The quiescent fractions exhibit broadly similar trends over most of the mass range. However, below a characteristic stellar mass of \(M_{\ast}\sim10^{8.5}\,h^{-2}M_\odot\), the quiescent fraction of satellite galaxies shows a different behaviour, increasing towards lower stellar masses. This difference between the red and quiescent fractions highlights the importance of distinguishing color-based and sSFR-based classifications when characterizing the quenching of dwarf galaxies. We investigate these trends quantitatively in the following section.

\section{The stellar-mass and halo-mass dependence of dwarf-galaxy quenching}

\begin{figure}
    \centering
    \includegraphics[width=0.45\linewidth, height=0.35\linewidth]{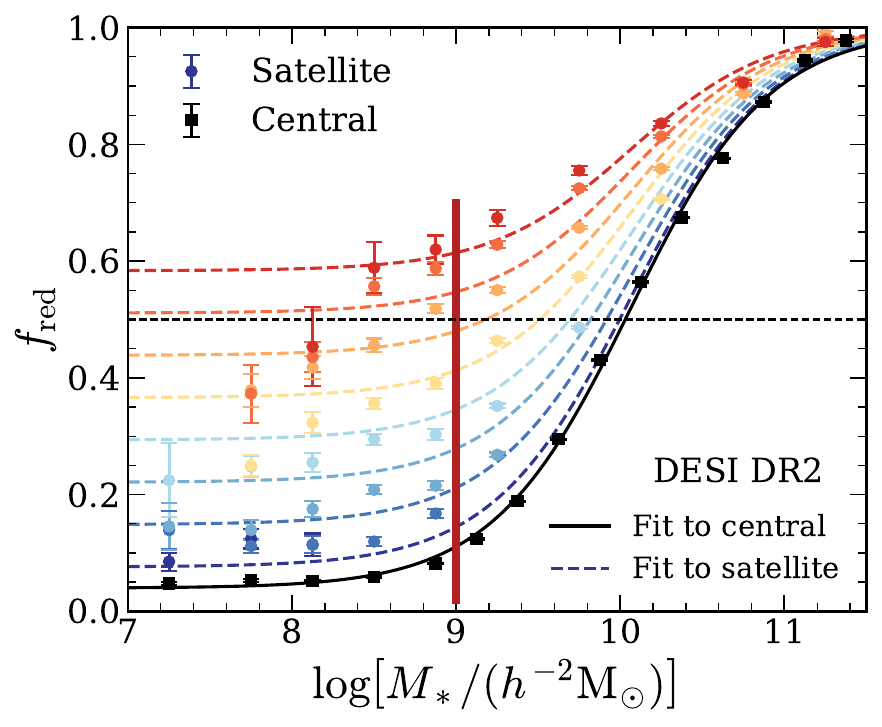}\hspace{0.5cm}
    \includegraphics[width=0.45\linewidth,height=0.35\linewidth]{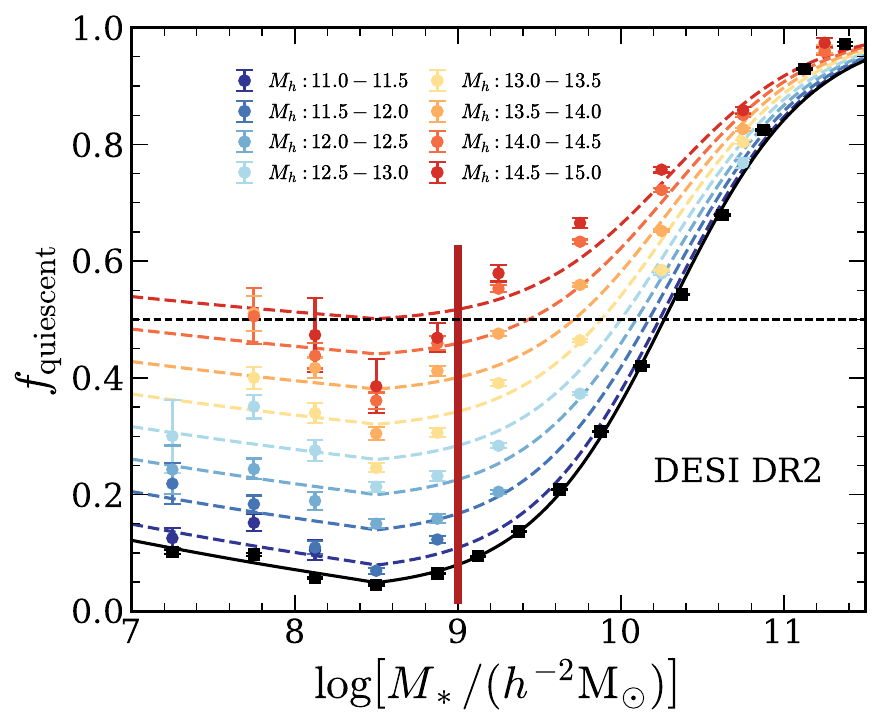}
    \includegraphics[width=0.45\linewidth, height=0.35\linewidth]{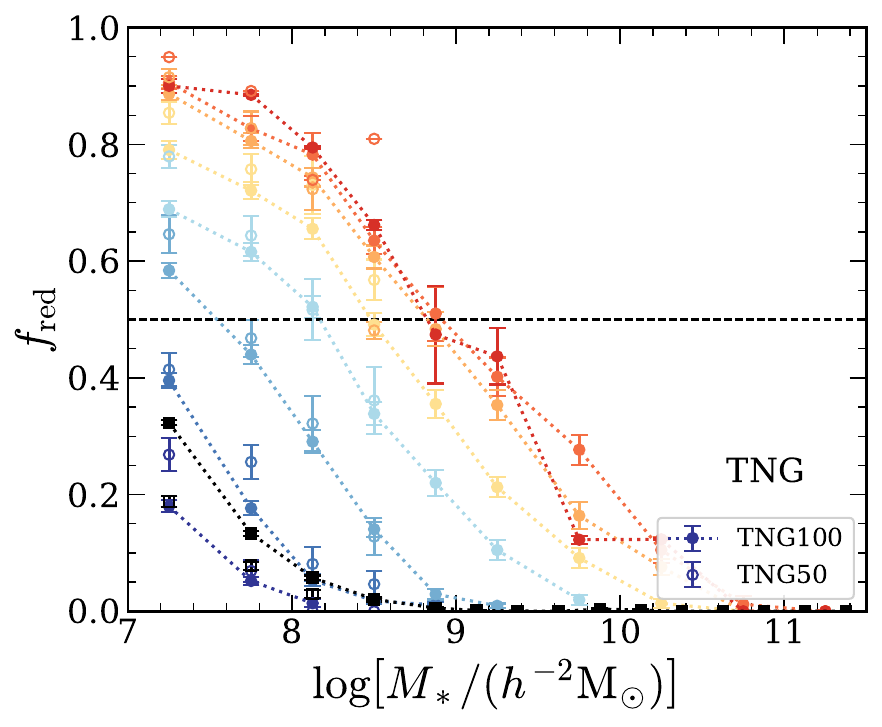}\hspace{0.5cm}
    \includegraphics[width=0.45\linewidth,height=0.35\linewidth]{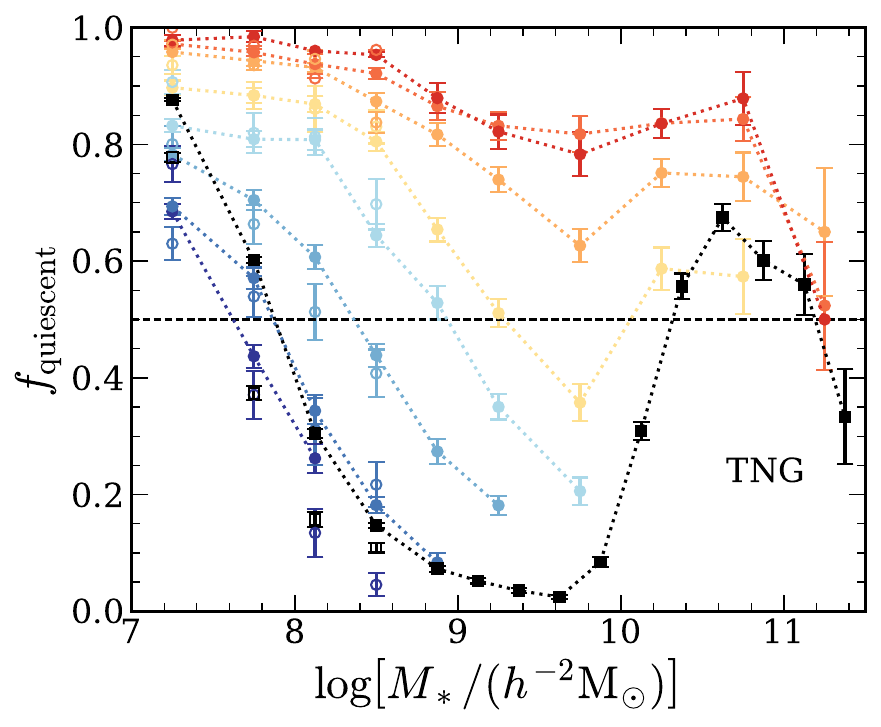}
    \caption{Upper panels: Red fraction, \(f_{\rm red}\) (left), and quiescent fraction, \(f_{\rm quiescent}\) (right), for central and satellite galaxies as a function of stellar mass. Different colours distinguish the halo-mass bins, and the measurements for the full halo-mass range are also shown for the DESI DR2 sample. Error bars are estimated from the standard deviations of 100 bootstrap resamplings, using galaxy groups for satellite galaxies and individual galaxies for central galaxies. The fractions of central and satellite galaxies are derived from the central-galaxy SMFs measured with the \(V_{\max}\) method and the satellite-galaxy CSMFs measured with the \(N_{\max}\) method, respectively. We use rest-frame optical colours as our primary classification, with \(f_{\rm red}\) as the main quenching indicator; the sSFR-based quiescent fraction, \(f_{\rm quiescent}\), is shown for comparison. Our measurements extend robust constraints on these fractions into the dwarf-galaxy regime, down to stellar masses of \(10^7\,\Msunhh\), approximately two orders of magnitude lower than in previous studies, as indicated by the vertical red line. The best-fitting phenomenological model is shown by the curves, with dashed and solid lines representing satellite and central galaxies, respectively. The model is based on the functional form \(f_{\rm active}(M_{\ast},M_{\rm h})=[1-Q_{\rm stellar}(M_{\ast})][1-Q_{\rm halo}(M_{\rm h})]\). Lower panels: The corresponding red and quiescent fractions measured from the TNG-100 and TNG-50 simulations. The same colour and sSFR selection criteria as those adopted for the DESI observations are applied to the simulated galaxies to facilitate a consistent comparison of their quenching statistics.
    }
    \label{fig:fred_Mh} 
\end{figure}

With the above SMF and CSMF measurements, we proceed to investigate the quenching of star formation in the dwarf-galaxy regime. We first examine the fraction of red galaxies (\(f_{\mathrm{red}}\)) as a function of stellar mass for central galaxies and for satellite galaxies residing in host haloes of different masses. Since both the SMF and CSMF measurements properly account for the maximum redshift, \(z_{\rm max}\), at which a galaxy can be detected, our results are, in principle, not affected by the survey magnitude limit. Nevertheless, as demonstrated in our end-to-end mock evaluation (see Methods, Section \ref{sec_validate}), the abundance of red central galaxies may be mildly overestimated owing to a combination of survey selection effects and the group-finding methodology. We therefore correct the measured \(f_{\mathrm{red}}\) for central galaxies by rescaling it according to the discrepancy between the input and recovered SMFs of red central galaxies in the MGRS mocks. The resulting measurements are shown by the symbols with error bars in the upper-left panel of Fig. \ref{fig:fred_Mh}.

A striking result is that the red fraction of central galaxies decreases sharply towards lower stellar masses, from approximately 95\% at \(M_{\ast}\sim10^{11}\,h^{-2}M_{\odot}\) to only about 5\% at \(M_{\ast}\sim10^{7}\,h^{-2}M_{\odot}\). Thus, blue galaxies dominate the central dwarf population, indicating that star formation remains active in a substantial fraction of low-mass haloes, rather than being universally suppressed as predicted by some galaxy-formation models. Satellite galaxies also show a declining red fraction towards lower stellar masses, but, importantly, their red fractions exhibit a clear dependence on host halo mass: at a given stellar mass, satellites in more massive haloes are more likely to be red. This trend provides direct evidence that the quenching of satellite dwarf galaxies is linked to their large-scale environment. Compared with the early study of Baldry et al. \cite{Baldry2006}, which investigated the environmental dependence of galaxy colors, our measurements show consistent trends at \(M_{\ast}>10^9\,h^{-2}M_{\odot}\). The large volume and depth of DESI, combined with our halo-based group finder, now allow us to extend these measurements roughly two orders of magnitude lower in stellar mass, and to quantify the dependence of satellite quenching on host halo mass well into the dwarf-galaxy regime.

In addition to examining the red fraction, we also computed the quiescent fraction, $f_{\rm quiescent}$, of galaxies as a joint function of stellar mass and halo mass for centrals and satellites, respectively. Here, we also correct the measured \(f_{\rm quiescent}\) for central galaxies using the discrepancy between the input and recovered SMFs of quiescent central galaxies in the MGRS mocks. The corrected measurements are shown in the upper-right panel of Fig. \ref{fig:fred_Mh}.  Relative to the red fraction displayed in the left panel, we identify a characteristic stellar mass of $10^{8.5}\Msunhh$, corresponding to typical  peak subhalo masses of $\sim 10^{11}\Msunh$, at which $f_{\mathrm{quiescent}}$ reaches a minimum. At stellar masses above this characteristic scale, $f_{\rm quiescent}$ follows trends that closely match those of $f_{\rm red}$. At lower masses, however, the quiescent fraction shows a mild rise toward decreasing stellar mass. We emphasize that this behavior is sensitive to our definition of $f_{\rm quiescent}$, because in the dwarf-galaxy regime the sSFR does not display as clear a bimodality as galaxy colors do (see \hyperref[subsec:stmass_sfr]{Methods}); this result is therefore presented mainly for reference. Any quantitative comparison should be carried out within a consistent set of definitions.

\begin{table}[h]
\rmfamily 
    \caption{Best-fit parameters for the red and quiescent fractions of central and satellite galaxies derived from Eq. \ref{eq:f_act}. The parameters $f_{\rm min}$, $k_c$, $\log M_q$ and $b$ characterize stellar-mass-dependent quenching, whereas $a$ characterizes halo-mass-dependent quenching.}
    \label{tab:fit_fraction}
\centering

\begin{tabular}{cccccc}
\hline 
Quenching indicator   &  $f_{\rm min}$ & $k_c$  & $\log M_q$ & $a$ & $b$\\
\hline 
$f_{\rm red}$ & 0.040 & 2.369 & 10.065 & 0.151 & - \\
$f_{\rm quiescent}$ & 0.034 & 2.321 & 10.297  & 0.127 & $-$1.660 \\
\hline
\end{tabular}
\end{table}

Our measurements of \(f_{\rm red}\) and \(f_{\rm quiescent}\) span four orders of magnitude in stellar mass and five orders of magnitude in halo mass, revealing clear dependencies on both stellar mass and host halo mass. To quantify these trends and provide a benchmark for theoretical models of galaxy formation, we introduce a simple empirical phenomenological model that separates the observed quenching trends into stellar-mass-dependent and halo-mass-dependent components. The model is intended to provide a compact parameterization of the observed galaxy populations, rather than a first-principles description of the physical processes responsible for quenching.

We model the red and quiescent fractions as the complements of the corresponding active fractions. Specifically, the active fraction is defined as the blue fraction when fitting \(f_{\rm red}\), and as the star-forming fraction when fitting \(f_{\rm quiescent}\). Thus, we write
\begin{equation}
    f_{\rm red}=1-f_{\rm blue},
    \qquad
    f_{\rm quiescent}=1-f_{\rm star-forming}.
\end{equation}
For both classifications, we parameterize the active fraction at a given stellar mass and halo mass as
\begin{equation}
\label{eq:f_act}
f_{\rm active}(M_{\ast}, M_{\rm h}) =
[1-Q_{\rm stellar}(M_{\ast})]
[1-Q_{\rm halo}(M_{\rm h})]\,.
\end{equation}
Here, \(Q_{\rm stellar}\) describes the stellar-mass-dependent component of quenching, while \(Q_{\rm halo}\) represents an additional contribution associated with the host halo mass and applies only to satellite galaxies. We parameterize the stellar-mass-dependent term as
\begin{equation}
\label{eq:fqch_c}
Q_{\rm stellar}(M_{\ast}) =
\left[
f_{\rm min}+
\frac{1-f_{\rm min}}
{1+\exp[-k_{\rm c}(\log M_{\ast}-\log M_{\rm q})]}
\right]
\left\{1+b\min[0,(\log M_{\ast}-8.5)]\right\},
\end{equation}
where \(f_{\rm min}\) sets the low-mass asymptotic value of the baseline sigmoid component, \(k_{\rm c}\) controls the steepness of the transition, and \(M_{\rm q}\) characterizes the stellar mass at which the quenching fraction rises. The additional parameter \(b\) allows for a modification of the low-mass behaviour, motivated by the distinct feature observed in \(f_{\rm quiescent}\) below \(M_{\ast}\sim10^{8.5}\,h^{-2}M_{\odot}\). We set \(b=0\) when fitting \(f_{\rm red}\).

For satellite galaxies, we further introduce a halo-mass-dependent term,
\begin{equation}
\label{eq:haloquench}
Q_{\rm halo}(M_{\rm h}) =
\max\left\{
\min[1,a(\log M_{\rm h}-11.0)],\,0
\right\},
\end{equation}
where \(a\) determines how rapidly this additional quenching contribution increases with host halo mass above the characteristic scale of \(10^{11}\,h^{-1}M_{\odot}\). For central galaxies, we set \(Q_{\rm halo}=0\), so that their quenching is described entirely by the stellar-mass-dependent term.

We fit this phenomenological model to the measured red and quiescent fractions shown in Fig. \ref{fig:fred_Mh}. The best-fitting models are shown by the solid lines for central galaxies and the dotted lines for satellite galaxies, with the corresponding parameter values listed in Table \ref{tab:fit_fraction}. Overall, the model provides a good description of the observed fractions across the stellar-mass and halo-mass ranges considered. By separating the stellar-mass and host-halo-mass dependencies, this parameterization provides a quantitative benchmark for testing theoretical predictions of dwarf-galaxy quenching. We emphasize, however, that the model is an empirical description of the observed trends: its parameters should not be interpreted as direct measurements of individual physical processes, nor does the model specify the underlying mechanisms responsible for quenching. Finally, red-galaxy selection based on optical color identifies quiescent-like systems statistically, but red galaxies can still host residual star formation. Thus, \(f_{\rm red}\) serves as a practical proxy for quenching rather than an absolute indicator of the absence of star formation.

We compare our measurements with predictions from the IllustrisTNG hydrodynamical simulations, using both the TNG-100 and TNG-50 runs. To ensure a consistent comparison, we apply the same color and sSFR selection criteria to the simulated galaxies as those adopted for the DESI observations. The resulting red and quiescent fractions are shown in the lower panels of Fig. 2. As discussed by Donnari et al. \cite{Donnari2021}, comparisons of quenched fractions at \(M_\ast>10^9\,h^{-2}M_\odot\) can be sensitive to the adopted quenching definition, SFR tracer, aperture, and sample selection. We therefore focus on the dwarf-galaxy regime, particularly at \(M_\ast<10^9\,h^{-2}M_\odot\), where our measurements extend the observational constraints to substantially lower stellar masses and allow us to examine the quenching properties of both central and satellite galaxies.

The comparison with IllustrisTNG reveals systematic differences between the simulated and observed quenching statistics in the dwarf-galaxy regime. These discrepancies are evident for both central and satellite galaxies, but manifest in different ways. For low-mass central galaxies, TNG predicts substantially higher red and quiescent fractions than those inferred from DESI, indicating that star formation is more strongly suppressed in the simulation than implied by the observations. In particular, the large population of blue central dwarf galaxies identified in DESI suggests that star formation remains active in a substantial fraction of low-mass haloes. The discrepancy therefore points to potential limitations in the prescriptions governing star formation in low-mass central galaxies, including the effects of stellar feedback and cosmic reionization, as well as the subsequent replenishment and reaccretion of gas.

For satellite galaxies, the most prominent difference is the dependence of the red fraction on host halo mass. Although both DESI and TNG predict higher red fractions in more massive host haloes, this dependence is considerably stronger in TNG than observed. This suggests that the environmental processes incorporated in the simulation may suppress star formation in satellites too efficiently, or may produce an overly strong dependence of quenching on host halo mass. Processes such as ram-pressure stripping, starvation, tidal interactions and preprocessing can all contribute to satellite evolution, but our measurements do not identify their individual contributions. Instead, they provide a quantitative constraint on the combined effects of these processes and on how their dependence on host halo mass is represented in galaxy-formation models.

Taken together, these comparisons reveal two complementary challenges for current galaxy-formation models: reproducing the high fraction of star-forming central dwarf galaxies in low-mass haloes, and capturing the more moderate dependence of satellite quenching on host halo mass observed by DESI. Our empirical phenomenological model provides a quantitative framework for characterizing these two trends separately. The observed discrepancies do not uniquely identify the physical processes responsible, but highlight the need to examine how stellar feedback, reionization, gas removal, and gas replenishment and reaccretion jointly regulate star formation in central and satellite dwarf galaxies. Further comparisons of the gas content and star-formation histories of simulated galaxies with the DESI measurements will be important for identifying the origin of these discrepancies.

\section{The galaxy-halo connection at the low-mass end} 
\label{sec_SHMR}

The stellar-to-halo mass ratio (SHMR) is a key quantity linking galaxies to their host dark matter haloes. It provides an important constraint on the galaxy--halo connection and on models of galaxy formation. Since central galaxies are associated with their host haloes, the SHMR can be inferred from the central-galaxy stellar mass function (SMF) using halo abundance matching \cite{Vale2004, Wechsler2018}. However, the local Universe is known to reside within a substantial local underdensity, or local void (LV), which can affect the measured abundance of low-mass galaxies \cite{Tully2019}. Using the ELUCID reconstruction simulation, Chen et al. \cite{Chen2019} quantified the impact of this effect and provided correction factors for recovering the intrinsic SMFs. In this work, we instead use the ELUCID halo catalogues to construct a halo mass function matched to the SDSS footprint and derive the SHMR through abundance matching, thereby accounting for the impact of the local environment on the halo abundance.

\begin{figure}
    \centering

    \begin{minipage}[c]{0.34\textwidth}
        \centering
        \includegraphics[width=\linewidth]{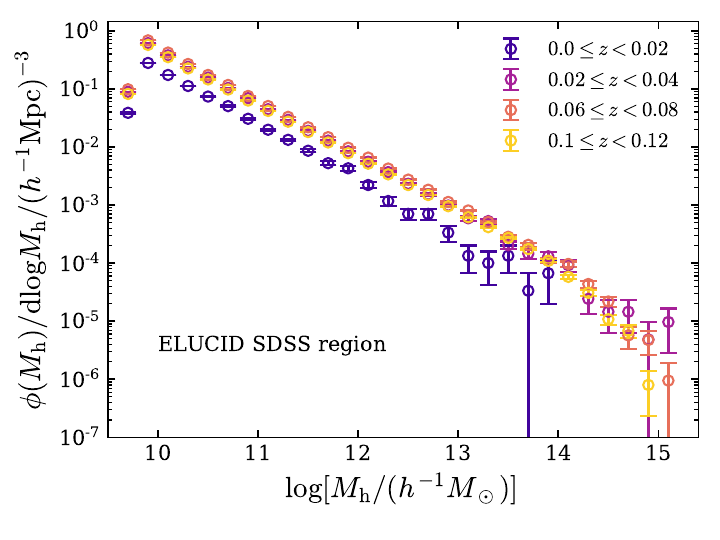}
        \vspace{0.17cm}
        \includegraphics[width=\linewidth]{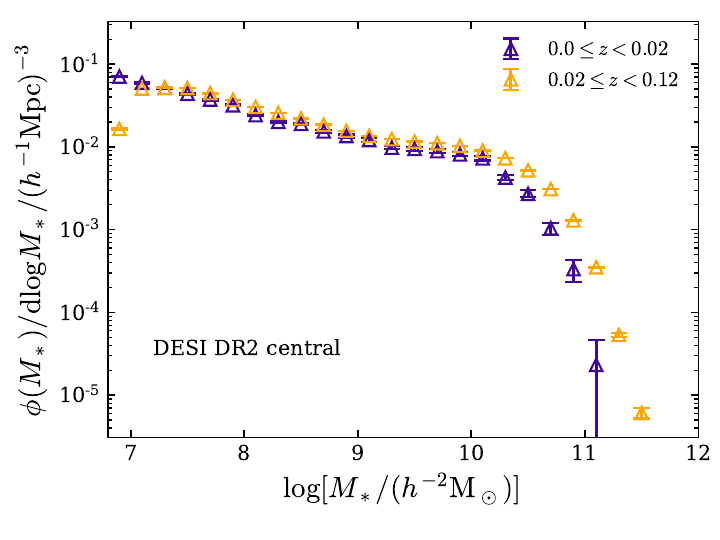}
    \end{minipage}
    \hfill
    \begin{minipage}[c]{0.63\textwidth}
        \centering
        \includegraphics[width=\linewidth, height=0.82\linewidth]{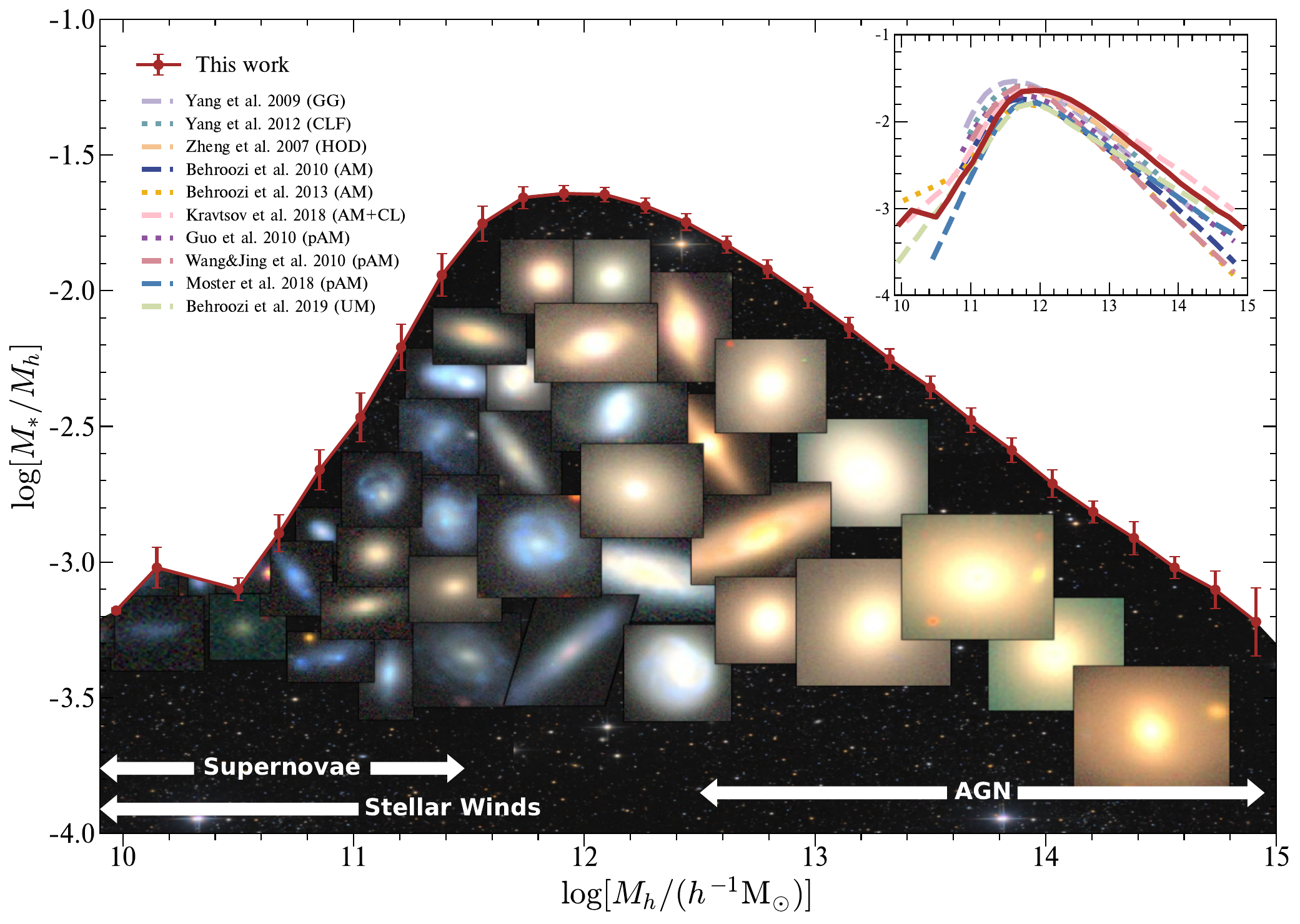}
    \end{minipage}    
    \caption{The stellar-to-halo mass ratio (SHMR) of central galaxies inferred from the DESI DR2 BGS sample and the ELUCID simulation using halo abundance matching. The red points show the median SHMR inferred in this work, with red error bars indicating the uncertainties propagated from the central-galaxy SMF measurements. Previous SHMR measurements and model constraints compiled by Wechsler et al. \cite{Wechsler2018} are shown in the inset panel for comparison. Representative images of central galaxies as a function of halo mass are displayed below the relation, with images obtained from the DESI Legacy Imaging Surveys. The upper-left panel shows the halo mass function measured within the SDSS footprint of the ELUCID simulation, while the lower-left panel presents the central-galaxy SMFs measured from the DESI DR2 BGS sample. The inferred SHMR is approximately flat over the low-mass halo range, with a transition around \(M_{\rm h}\sim10^{10.5}\,h^{-1}M_\odot\). The interpretation of this low-mass behaviour is subject to the resolution limit of the ELUCID halo catalogue and the uncertainties in the central-galaxy SMF used for abundance matching.
    }
    \label{fig:MlMhratio}
\end{figure}

The ELUCID simulation, developed to reconstruct the structure of the local Universe using SDSS observations and assuming a WMAP5 cosmology, was carried out with \(3072^3\) particles in a cubic volume of side length \(500\,h^{-1}\mathrm{Mpc}\) \cite{Wang2016}. ELUCID resolves haloes down to a minimum mass of approximately \(10^{10}\,h^{-1}M_\odot\). To provide a consistent framework for connecting central galaxies to their host haloes, we measure the halo mass function (HMF) within the SDSS footprint of the ELUCID simulation. We compute the HMF in several redshift intervals, as shown in the upper-left panel of Fig. \ref{fig:MlMhratio}. In the lowest-redshift bin, the HMF is substantially reduced and lacks the most massive haloes compared with measurements from larger volumes, illustrating the impact of the local underdensity. This environmental variation may affect the abundance and distribution of galaxies in the local volume. In particular, the underrepresentation of low-mass red satellite galaxies can lead to an elevated central fraction of red galaxies at low stellar masses, as discussed in the Results section on the central-galaxy SMFs. We therefore account for the local-volume effects when interpreting the low-mass red-galaxy statistics.  We measure the central-galaxy SMFs in two redshift bins, \(0\leq z<0.02\) and \(0.02\leq z<0.12\), shown in the lower-left panel of Fig. \ref{fig:MlMhratio}. We then apply abundance matching between the central-galaxy SMFs and the corresponding ELUCID HMFs to infer the SHMR, accounting for the redshift-dependent stellar-mass completeness.

The resulting SHMR is shown in Fig. \ref{fig:MlMhratio}, with the error bars representing uncertainties propagated from the central-galaxy SMF measurements. We also show representative images of central galaxies as a function of halo mass, using halo masses from our group catalogue and images from the DESI Legacy Imaging Surveys\footnote{\url{https://www.legacysurvey.org/viewer}}. The images illustrate the diversity of galaxy colours and morphologies across the halo-mass range, from predominantly red, elliptical systems in massive haloes to blue and diffuse galaxies in lower-mass haloes. For comparison, we show previous SHMR measurements and model constraints compiled by Wechsler et al. \cite{Wechsler2018} in the inset panel.

Our inferred SHMR suggests that the stellar-to-halo mass ratio becomes approximately flat over the low-mass halo range, with a transition around \(M_{\rm h}\sim10^{10.5}\,h^{-1}M_\odot\). This behaviour differs from the more rapidly declining SHMR predicted by some galaxy-formation models in which stellar feedback strongly suppresses the stellar content of low-mass haloes \cite{Wechsler2018}. A nearly flat SHMR implies that the stellar masses of central galaxies scale approximately in proportion to their halo masses over this range. It therefore provides a useful constraint on models of galaxy formation and on the processes regulating the stellar content of low-mass haloes. However, the low-mass end of the relation should be interpreted in light of the resolution limit of the ELUCID halo catalogue and the uncertainties in the central-galaxy SMF used for abundance matching. Together with the large population of blue central dwarf galaxies found in our analysis, this result provides complementary evidence that the stellar content of low-mass haloes may be less strongly suppressed than predicted by some current models.

\section{Discussion} 

By combining the large population of blue central dwarf galaxies with the halo-mass dependence of satellite quenching, our results provide a more comprehensive picture of dwarf-galaxy evolution. The high fraction of blue central dwarfs indicates that star formation remains active in many low-mass haloes, whereas the increasing red fraction of satellite galaxies with host halo mass points to an additional environmental dependence of quenching. This contrast suggests that the evolution of dwarf galaxies is governed by different combinations of processes depending on whether they reside at the centres of their own haloes or have been accreted into larger systems. In particular, the dependence of satellite quenching on host halo mass provides an important constraint on how environmental effects regulate star formation in dwarf galaxies. We also find that red and quiescent fractions do not always exhibit identical trends at the lowest stellar masses, emphasizing the importance of distinguishing color-based and sSFR-based classifications when interpreting dwarf-galaxy quenching.

Our study provides new observational constraints on dwarf-galaxy formation and evolution by extending measurements of their abundance and quenching properties to stellar masses of approximately \(10^7\,h^{-2}M_\odot\). Beyond reaching this low-mass regime, our halo-based analysis enables us to investigate central and satellite dwarf galaxies within a unified framework and to quantify the dependence of satellite quenching on host halo mass. The comparison with IllustrisTNG reveals systematic discrepancies in both the quenching of low-mass central galaxies and the environmental dependence of satellite quenching, highlighting challenges for current prescriptions of stellar feedback and environmental processes. Together with the nearly flat low-mass end of the inferred stellar-to-halo mass relation, these results provide complementary constraints on the abundance, star formation activity, and galaxy–halo connection of dwarf galaxies. They establish a quantitative observational benchmark for testing and improving theoretical models of galaxy formation in low-mass dark matter haloes.

\section{Methods}\label{sec11}

\renewcommand{\figurename}{\bf Extended Data Fig.}
\setcounter{figure}{0}
\renewcommand*{\theHfigure}{ED.\arabic{figure}}
\renewcommand{\tablename}{\bf Extended Data Table}
\setcounter{table}{0}
\renewcommand*{\theHtable}{ED.\arabic{table}}

\subsection{Galaxy statistics and the galaxy-halo connection}\label{sec_connection}

Over the past two decades, large-scale galaxy surveys, from the Two-degree Field Galaxy Redshift Survey (2dFGRS) \cite{Colless1999} and the Sloan Digital Sky Survey (SDSS) \cite{York2000} to the ongoing Dark Energy Spectroscopic Instrument (DESI) survey, have greatly advanced our understanding of galaxy formation and evolution. These surveys enable precise measurements of statistical properties of galaxies, including luminosity functions (LFs), stellar mass functions (SMFs), and two-point correlation functions (2PCFs), which characterize their abundance and spatial distribution \cite{Norberg2002, blanton2003galaxy, zehavi2005luminosity, wang2007cross, li2009distribution, zehavi2011galaxy, 2013ApJ...767...50M, zhao2000, Wang2021, 2021MNRAS.500.3776W, 2025ApJ...986..218W}. Although these statistics do not directly identify the physical processes governing galaxy evolution, they provide essential constraints on the combined effects of gravitational collapse, gas cooling, star formation, mergers, and feedback \cite{Naab2017,Smercina2018,Katsianis2021,Sales2022}.

A particularly useful approach to interpreting these observables is to establish the connection between galaxies and their host dark matter haloes. Empirical frameworks such as the halo occupation distribution (HOD) \cite{jing1998spatial, peacock2000halo, tinker2005mass, Zheng2005, Zheng2009, brown2008red, 2015Zu, 2016Zu, 2018Zu, 2019Wang, 2020Alam, 2018Yuan, 2022Yuan}, conditional luminosity function (CLF) \cite{yang2003constraining, van2003linking, van2007towards, yang2008galaxy, cacciato2009galaxy, Yang2012}, and subhalo abundance matching (SHAM) \cite{Vale2004, conroy2006modeling, Behroozi2010, neistein2011tale, Guo2016} provide complementary ways of describing the statistical connection between galaxies and dark matter haloes \cite{Wechsler2018,Katsianis2023}. These approaches have substantially improved our understanding of how galaxy properties depend on halo mass and environment.

An important complementary development is the use of halo-based group-finding algorithms to identify galaxies residing in the same dark matter halo and to infer the properties of their host haloes \cite{yang2005halo}. In particular, the extended adaptive halo-based group finder \cite{yang2021extended}, which incorporates both spectroscopic and photometric redshift information, enables the identification of galaxy groups over a wide range of survey conditions, including regions with incomplete spectroscopic coverage. This method has been applied to the SDSS and the DESI Legacy Imaging Surveys, enabling measurements of CLFs and  CSMFs as functions of halo mass \cite{yang2008galaxy,Wang2024}. By identifying central and satellite galaxies and associating them with their host haloes, such group catalogues provide a direct observational basis for studying the galaxy-halo connection. However, previous measurements have primarily focused on relatively luminous galaxies and massive haloes, leaving the dwarf-galaxy regime comparatively poorly constrained \cite{yang2005galoccup, yang2008galaxy, yang2009galaxy, Lan2016, To2020, Meng2023, Golden-Marx2022, Wang2021, Tinker2021}. Extending these measurements to low stellar masses is particularly important for understanding the efficiency of star formation and the role of feedback in low-mass haloes.

Galaxy properties can also be characterized through their optical colors and specific star formation rates (sSFRs), which provide complementary information about their stellar populations and star formation activity. The bimodal color distribution of galaxies, commonly described in terms of the blue cloud and red sequence, has been well established in large surveys such as the SDSS \cite{Strateva2001,Baldry2004,Baldry2006,Blanton2005}. In general, blue galaxies are associated with ongoing star formation, whereas red galaxies tend to have older stellar populations and lower star formation activity. The fractions of red galaxies also vary systematically with stellar mass and environment \cite{Baldry2004,Baldry2006,Blanton2005}. However, optical color and sSFR do not provide identical classifications: red galaxies may retain residual star formation, while recently quenched galaxies may not yet have developed red colors. Measuring both color-based and sSFR-based populations therefore provides complementary constraints on the quenching of galaxies and its dependence on stellar mass and halo environment.

In this study, we use the DESI Data Release 2 (DR2) sample to investigate the abundance and quenching properties of galaxies in the dwarf-galaxy regime. We measure the global LFs and SMFs of all galaxies and central galaxies, as well as the CSMFs of satellite galaxies in different host halo mass bins. We further divide galaxies into red and blue populations according to their \(g-r\) colors, and into quiescent and star-forming populations using their sSFRs derived from spectral energy distribution fitting with the widely used \textsc{cigale} code \cite{Boquien2019,Yang2020,Yang2022,Noll2009}. By combining these measurements with the halo-based group catalogue, we investigate how galaxy abundance and quenching depend on stellar mass, central or satellite status, and host halo mass, providing new constraints on the galaxy–halo connection and the formation of dwarf galaxies.

\subsection{DESI DR2 galaxy and group samples}
\label{sec_data}

\subsubsection{DESI DR2 galaxy sample}

The galaxy sample used in this study is based on the seed galaxy catalogue described in Wang et al. \cite{Wang2024}, which was originally constructed from the DESI Legacy Imaging Surveys Data Release 9 (LS DR9). The catalogue combines photometric measurements from the imaging surveys with spectroscopic redshifts as they become available from DESI \cite{Guy2023, Brodzeller2023, Moustakas2023, Schlafly2023}. Further details of the DESI survey can be found in a series of survey overview and instrument papers \cite{DESI_Collaboration2016a,DESI_Collaboration2016b,DESI_Collaboration2022,Levi2013,Silber2023,Miller2023}. For the majority of objects without spectroscopic redshifts, photometric redshifts are estimated using optical photometry in the three \(grz\) bands from the Beijing-Arizona Sky Survey \cite{Zou2017}, the Mayall \(z\)-band Legacy Survey, and the DECam Legacy Survey \cite{Dey2019}, together with the two mid-infrared bands W1 (3.4 \(\mu\mathrm{m}\)) and W2 (4.6 \(\mu\mathrm{m}\)) from the Wide-field Infrared Survey Explorer (WISE) \cite{Zhou2021}.

We use the version of the seed catalogue updated with DESI Data Release 2 (DR2) spectroscopic observations. The resulting parent catalogue covers approximately \(14,650\,\mathrm{deg}^2\) and contains 23,397,368 galaxies over the redshift range \(0.0\leq z\leq0.6\), with an apparent \(r\)-band magnitude limit of \(m_r=20.175\). It includes targets from both the bright and faint components of the BGS, and has an overall spectroscopic completeness of approximately 55 per cent.

For the construction of our primary BGS sample, we impose a redshift selection of \(0.0\leq z\leq0.2\) and a bright-sample magnitude limit of \(m_r\leq19.5\). We further apply a photometric quality cut based on the \(FRACFLUX\) parameter provided in the DESI Legacy Imaging Surveys catalogues \cite{Darragh-Ford2023, Dirk2024, Pucha2025}. This parameter helps identify sources whose photometry may be affected by contamination from neighbouring objects or by photometric shredding. A recent analysis by Manwadkar et al. (in preparation) indicates that sources with \(FRACFLUX>0.2\) are likely to be significantly affected by shredding. We therefore retain only galaxies with \(FRACFLUX<0.2\) to reduce the impact of such photometric artefacts on our measurements.

\begin{figure}
\centering
\includegraphics[width=0.8\textwidth]{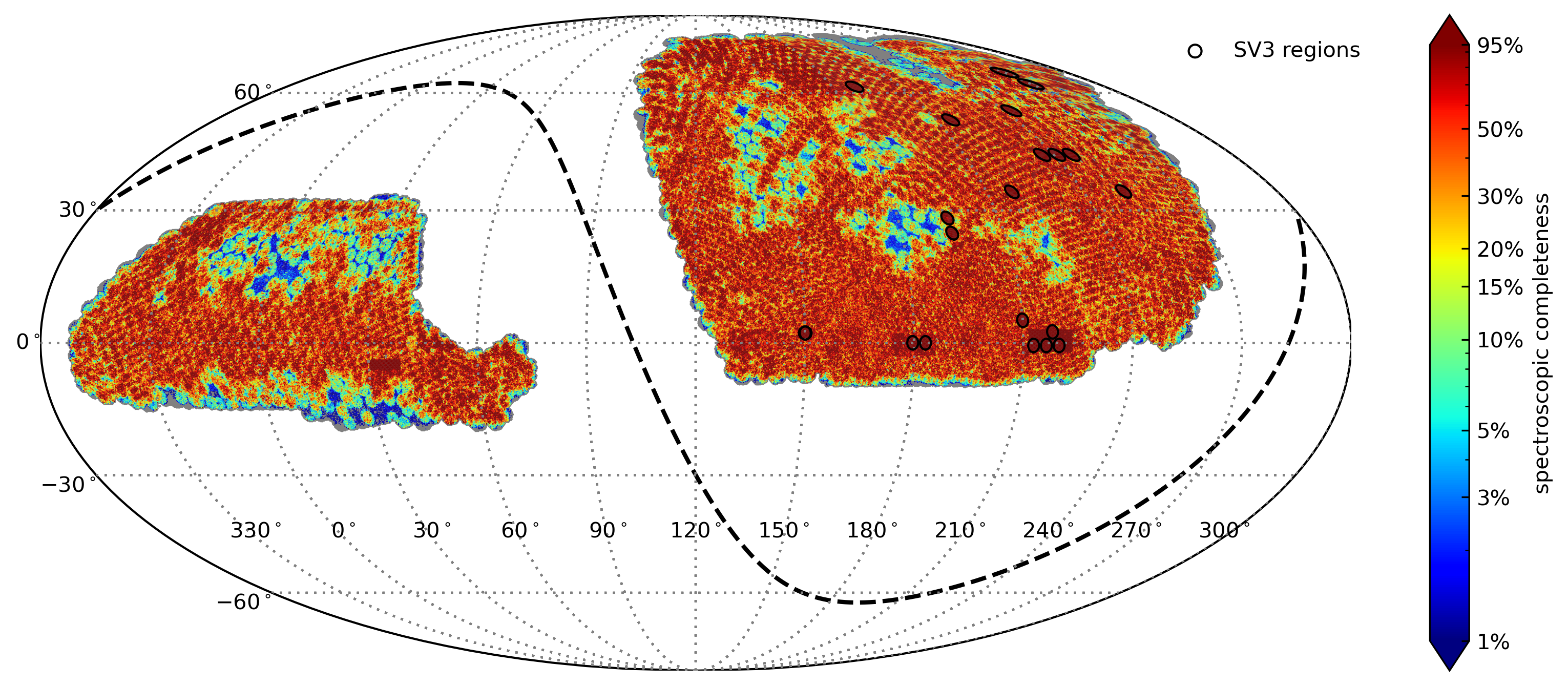}
\caption{Sky coverage of the DESI spectroscopic data DR2 sample used in this study. The spectroscopic completeness indicated by the color bar is calculated with respect to the total number of galaxies with $m_r \leq 19.5$. The black dashed line represents the Galactic plane, and the black circles mark the DESI SV3 regions. We cut Galactic latitude $|b| < 15^\circ$ to avoid the underobserved areas of the south region in the SGC. } \label{sky_cov}
\end{figure}

We determine the angular footprint of the DESI DR2 spectroscopic sample using the HEALPix framework \cite{Gorski2005}, with \(\mathrm{nside}=256\), corresponding to an area of approximately \(5.246\times10^{-2}\,\mathrm{deg}^2\) per pixel. Pixels containing at least one galaxy are initially included in the footprint. As illustrated in Extended Data Fig. \ref{sky_cov}, the resulting footprint covers the regions with DESI spectroscopic observations used in this study. We additionally exclude regions at Galactic latitudes \(|b|<15^\circ\), where the spectroscopic coverage is relatively incomplete, particularly in the Southern Galactic Cap. After these selections, the resulting DESI DR2 BGS sample contains 5,431,946 galaxies and has a spectroscopic completeness of approximately 74 per cent.

\subsubsection{Galaxy group identification and final sample}

Since identifying galaxy groups and their host dark matter haloes is essential to this study, we first apply the extended halo-based group finder \cite{yang2021extended} to the full DESI DR2 parent catalogue, which contains 23,397,368 galaxies over \(0.0\leq z\leq0.6\). The algorithm can incorporate both spectroscopic and photometric redshift information, allowing us to use the full parent catalogue rather than restricting group identification to the spectroscopic BGS sample alone. This approach provides a more comprehensive galaxy distribution for identifying groups and estimating their host halo masses.

For the subsequent analysis, we further impose a spatial completeness mask to mitigate the effects of incomplete spectroscopic sampling, particularly those associated with fibre collisions. Specifically, we exclude sky regions with spectroscopic completeness below \(0.75\). After applying this mask, our final analysis sample contains \(3,126,721\) galaxies with \(m_r\leq19.5\) within \(0.0\leq z\leq0.2\), covering a total sky area of \(9,959\,\mathrm{deg}^2\). This sample is used for the measurements presented in the following sections.

\subsubsection{Absolute magnitudes and colour-based galaxy classification}
\label{subsec:lum_color}

Following the procedure described in Wang et al. \cite{Wang2024}, we calculate the absolute magnitudes in the \(g\), \(r\), and \(z\) bands (\(M_g\), \(M_r\), and \(M_z\)) from the apparent magnitudes and redshifts of individual galaxies. Since our analysis focuses on galaxies at \(z\leq0.2\), we apply \(K\)-corrections to a common reference redshift of \(z=0.1\). The individual \(K\)-corrections are calculated using the \textsc{Kcorrect} model (version 4.3) developed by Blanton et al. \cite{Blanton2007}. For the calculation of the maximum observable redshift, \(z_{\rm max}\), we instead use an analytical fit to the average \(K\)-correction as a function of redshift.

The \(z=0.1\) absolute magnitude in band \(x\) (\(x=g,r,z\)) is given by
\begin{equation}
\label{eq:mag_eachgal}
M^{0.1}_{x}-5\log h =
m_x-\mathrm{DM}(z_{\rm obs})-K^{0.1}_{x}(z_{\rm obs}),
\end{equation}
where \(K^{0.1}_{x}\) is the \(K\)-correction to the reference redshift \(z=0.1\), and \(\mathrm{DM}(z_{\rm obs})\) is the distance modulus at the observed redshift:
\begin{equation}
\label{eq:distmeas}
\mathrm{DM}(z_{\rm obs}) =
5\log\left[\frac{D_L(z_{\rm obs})}{\mathrm{Mpc}}\right]+25,
\end{equation}
with \(D_L\) denoting the luminosity distance. The corresponding luminosity in each band is calculated as
\begin{equation}
\label{eq:lum}
\log\left(\frac{L_x}{h^{-2}L_\odot}\right)
=0.4\left(M_{\odot,x}^{0.1}-M_x^{0.1}\right),
\end{equation}
where \(M_{\odot,x}^{0.1}\) is the solar absolute magnitude in the corresponding band after applying the \(K\)-correction to \(z=0.1\). We adopt \(M_{\odot,g}=5.12\), \(M_{\odot,r}=4.61\), and \(M_{\odot,z}=4.50\), with solar \(K\)-corrections of \(K_{\odot,g}^{0.1}=-0.19\), \(K_{\odot,r}^{0.1}=-0.22\), and \(K_{\odot,z}^{0.1}=-0.10\), respectively.

\begin{figure}
\centering
\includegraphics[width=1\textwidth]{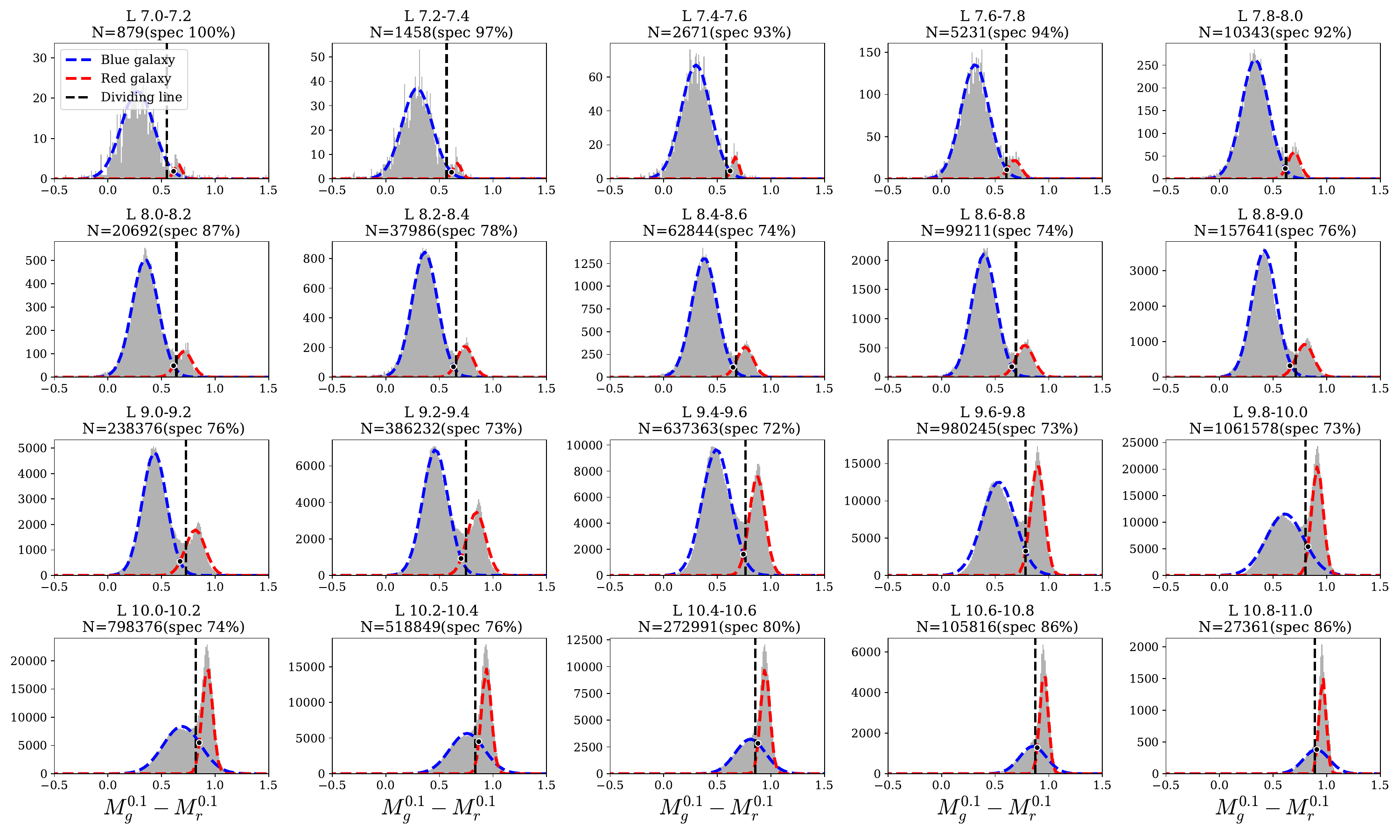}
\caption{Galaxy color bimodal distribution across different luminosity bins (width = 0.2~dex). The red and blue dashed lines represent the red and blue galaxy populations classified via the Gaussian bimodal fitting. The black point in each panel marks the boundary position that separates the two Gaussian distributions with the same probability.  All vertical dividing lines are used to split the galaxies into red and blue populations. The number of galaxies within the luminosity bin is indicated in the title of each panel with the fraction of galaxies having spectroscopic redshift illustrated in parentheses.} \label{fig:colorbimodal}
\end{figure}

We separate galaxies into red and blue populations using their rest-frame colour, defined as
\[
(g-r)^{0.1}=M_g^{0.1}-M_r^{0.1}.
\]
As shown in Extended Data Fig. \ref{fig:colorbimodal}, the colour distributions exhibit a clear bimodal structure over a wide range of luminosities. We divide the sample into luminosity bins of width 0.2 dex, with the luminosity range of each bin indicated at the top of each panel. The number of galaxies and the corresponding spectroscopic completeness are also reported in each panel. In each luminosity bin, we fit the colour distribution with a double-Gaussian function, whose blue and red components describe the blue cloud and red sequence, respectively. We define the colour threshold separating the two populations as the colour at which the two Gaussian probability distributions intersect, corresponding to equal probabilities for a galaxy to belong to either component. These intersection points are indicated by the black symbols in Extended Data Fig. \ref{fig:colorbimodal}.

To obtain a continuous luminosity-dependent colour threshold, we fit a linear relation to the intersection points over the luminosity range \(10^7\)–\(10^{11}\,h^{-2}L_\odot\). The best-fitting relation is
\begin{equation}
\label{eq:color_fit}
(g-r)^{0.1} =
0.0898\log\left(\frac{L_r}{h^{-2}L_\odot}\right)-0.0880.
\end{equation}
This relation, shown by the vertical dashed lines in the corresponding panels of Extended Data Fig. \ref{fig:colorbimodal}, is adopted as the luminosity-dependent colour criterion for classifying galaxies as red or blue throughout this study. Galaxies with colours above the threshold are classified as red, while those below it are classified as blue.

\begin{table}
\caption{Parameter setting for \textsc{cigale} SED fitting. Other parameter settings not provided in this table is the same as Table~1 of Siudek et al.\cite{Siudek2024}.}
 \label{cigale_paras}
\centering
\renewcommand{\arraystretch}{1}
\begin{tabular}{ccc}
    \hline 
         Parameter & Symbol & Values \\ \hline
         $\tau$ of the main stellar population (Myr) & $\tau_{main}$ & 100, 500, 850, 1000, 3000, 5000, 8000 \\ \hline
         \multirow{2}{*}{Age of the main stellar population (Myr)} & \multirow{2}{*}{$t_1$}       & 10,50,100,500,1000,3000, \\
         & & 4500,6000,8000,10000,13000 \\ \hline
         Age of the burst stellar population (Myr) & $t_{burst}$ & 1,5,10 \\
    \hline
    \end{tabular}
\end{table}

\subsubsection{Stellar masses, star formation rates, and the classification of star-forming and quiescent galaxies}
\label{subsec:stmass_sfr}

Stellar mass and star formation rate (SFR) are fundamental quantities for characterizing galaxy formation and evolution. Although they are inferred from observational data through model-dependent methods, they provide important complementary information to luminosities and colours and are widely used in comparisons with theoretical models. In this study, we use two independent sets of stellar mass estimates: one derived from spectral energy distribution (SED) fitting with the Code Investigating GALaxy Emission (\textsc{cigale}; \cite{Boquien2019,Yang2020,Yang2022}), and the other obtained using \textsc{fastspecfit} \cite{2023ascl.soft08005M}. As discussed below, the two estimates are in good agreement, including when compared with the results of our previous study \cite{Wang2024}, which adopted the \textsc{fastspecfit}-based stellar masses. We therefore use the \textsc{cigale}-based stellar masses throughout this work, since the same SED-fitting procedure also provides estimates of SFRs, allowing us to examine the star formation activity of galaxies in a consistent framework.

\textsc{Cigale} is an SED-fitting code that estimates galaxy properties by comparing observed multi-wavelength photometry with synthetic SEDs generated from stellar population models \cite{Noll2009,Boquien2019}. Its modelling framework combines several components, including star formation histories, simple stellar population models, dust attenuation and re-emission, and emission from active galactic nuclei. The physical properties of galaxies, such as stellar mass, SFR, and stellar age, are inferred from the likelihood distributions of the models that best reproduce the observed SEDs. The code has been widely applied to the study of galaxy properties, including systems with possible AGN contributions \cite{Ciesla2015,Yang2023}.

For the DESI sample, we adopt the general parameter configuration described by Siudek et al. \cite{Siudek2024}, with modifications to the ages of the main and burst stellar populations. These modifications were motivated by our tests of the resulting distributions in the sSFR--\(M_\ast\) plane, for which we aimed to avoid an artificial truncation of the low-mass galaxy population. The adopted parameter values are summarized in Extended Data Table \ref{cigale_paras}, while the remaining settings follow Table 1 of Siudek et al. \cite{Siudek2024}. This configuration generates 672,000 SED models at each redshift. We use the resulting stellar masses and SFRs to characterize the galaxy population. Our primary objective is not to obtain highly precise absolute SFR measurements, but rather to use the inferred SFRs to provide a complementary classification of galaxies according to their star formation activity.

\begin{figure}
    \centering
    \includegraphics[width=0.98\linewidth]{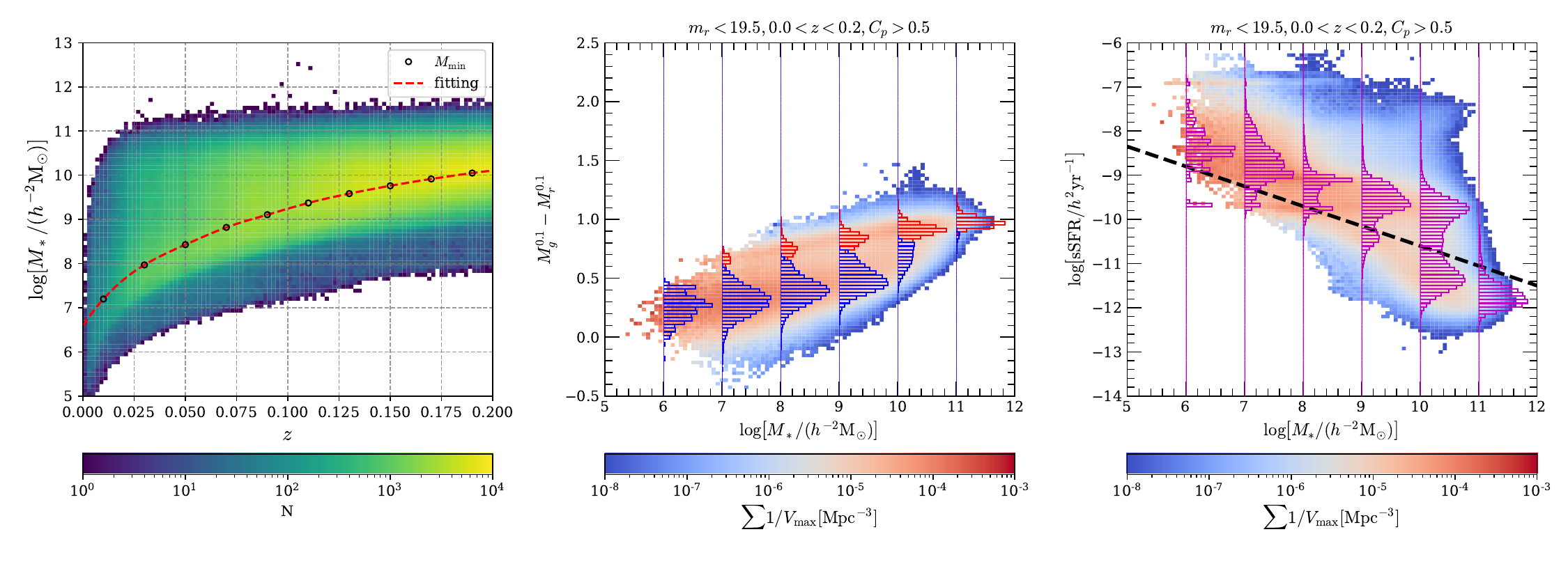}
    \includegraphics[width=0.98\linewidth]{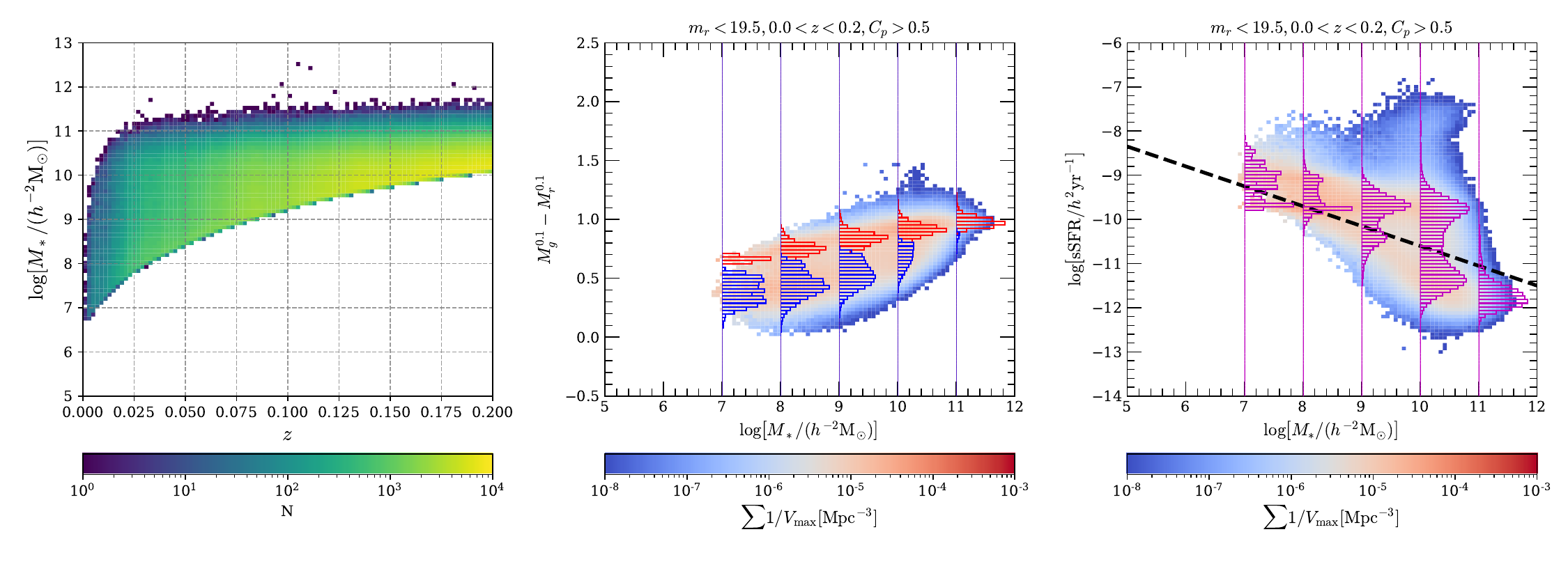}
    \caption{$M_*$–$z$ (left), color–$M_*$ (center), and sSFR–$M_\ast$ (right) diagrams. The upper panels show the galaxies with $r\le 19.5$ in the redshift range $0.0 < z < 0.2$. In the upper-left panel, the black circles mark the minimum stellar mass as a function of redshift for this sample, determined following the procedure of \cite{Pozzetti2010}; the corresponding best-fit relation is shown as a red dashed line. The lower panels present the same diagrams but only for galaxies whose redshifts place them below the stellar mass completeness limit indicated by the dashed curves in the upper-left panel. In the middle and right panels, the color bars represent the sum of $1/V_{\rm max}$ within each cell. In the central panels, the vertical red and blue histograms represent the distributions of red and blue galaxies in stellar mass bins of width 0.2 dex. The two populations are divided using the color threshold $(g-r)^{0.1} = M^{0.1}_{g} - M^{0.1}_{r} = 0.0898 \times \log(L_{r}/h^{-2}\rm L_{\odot}) - 0.0880$, derived from a fit to the color-separation data displayed in Extended Data Fig. \ref{fig:colorbimodal}. In the right panels, the black dashed line indicates the dividing relation for the bimodal distribution, $\log(\mathrm{sSFR}/h^2\mathrm{yr}^{-1}) = -0.45 \times \log (M_\ast/h^{-2}M_{\odot}) - 6.1$, which we use to separate the galaxy sample into star-forming and quiescent populations. Cells in the middle and right panels containing fewer than 5 galaxies are omitted.}
    \label{fig:r19.5}
\end{figure}

We characterize star formation activity using the specific star formation rate, defined as \(\mathrm{sSFR}=\mathrm{SFR}/M_\ast\). The sSFR distribution of galaxies is known to be bimodal, broadly separating actively star-forming systems from quiescent populations \cite{Brinchmann2004,Salim2007,Peng2010,Tamburri2014,Salim2018}. As shown in the upper-right panel of Extended Data Fig. \ref{fig:r19.5}, the galaxy distribution in the sSFR--\(M_\ast\) plane exhibits a clear bimodal structure. We therefore adopt a mass-dependent threshold to separate star-forming and quiescent galaxies:
\begin{equation}
\label{eq:ssfr_cut}
\log(\mathrm{sSFR}/h^2\,\mathrm{yr}^{-1})
=
-0.45\log(M_\ast/h^{-2}M_\odot)-6.1.
\end{equation}
Galaxies above this threshold are classified as star-forming, while those below it are classified as quiescent. This division provides a complementary measure of galaxy quenching based on SED-derived star formation activity.

We note that several diagnostics can be used to distinguish star-forming and quiescent galaxies, including the \(D_n(4000)\) spectral index. Although \(D_n(4000)\) is a useful indicator of the age and star formation history of stellar populations, its interpretation can depend on metallicity and other stellar population properties. This is particularly relevant for dwarf galaxies, whose low metallicities can weaken the 4000-\(\mathring{\mathrm A}\) break even in systems with relatively old stellar populations. In addition, for the low-mass BGS galaxies considered here, the relatively low continuum signal-to-noise around 4000 \(\mathring{\mathrm A}\), possible contamination from emission lines, and the limited spatial coverage of DESI fibres can complicate the interpretation of individual \(D_n(4000)\) measurements. We therefore do not adopt this index as our primary classification criterion. Instead, we use the rest-frame optical colour defined in Section \ref{subsec:lum_color} to separate red and blue galaxies, and employ the SED-based sSFR classification as an independent, complementary diagnostic. The two classifications need not identify exactly the same populations: optical colours reflect the integrated stellar light, whereas sSFR provides an estimate of the more recent star formation activity. This distinction is important when interpreting the quenching properties of dwarf galaxies. Stacked \(D_n(4000)\) measurements may provide additional constraints on their longer-term star formation histories, but such an analysis is beyond the scope of this work.

\subsubsection{Assessing the completeness of red and blue, and star-forming and quiescent galaxies}
\label{sec:assess_comp}

The DESI BGS reaches approximately 1.8 magnitudes deeper than the SDSS Main Galaxy Sample, allowing us to extend measurements of galaxy  LFs and SMFs, as well as the fractions of red/blue and quiescent/star-forming galaxies, into the dwarf-galaxy regime. To make full use of the flux-limited sample while accounting for the associated selection effects, we calculate the maximum observable redshift, \(z_{\rm max}\), for each galaxy. Since our group catalogue provides a host halo mass estimate for each galaxy, we can further calculate both \(V_{\rm max}\), the maximum comoving volume over which a galaxy could be detected, and \(N_{\rm max}(M_{\rm h})\), the maximum number of groups within which it could be identified, as a function of halo mass. These quantities allow us to account for the different selection effects when measuring galaxy abundances and conditional stellar mass functions.

Before presenting our measurements, we first assess the stellar-mass completeness of different galaxy populations. In a flux-limited survey, galaxies with different colours and star formation activities can have substantially different detection limits in stellar mass, even when subject to the same apparent-magnitude cut. In particular, red and quiescent galaxies generally have higher stellar mass-to-light ratios than blue and star-forming galaxies, making them more difficult to detect at a given stellar mass. Consequently, the magnitude limit can affect their measured abundances differently. This effect is especially important at the low-mass end, where a population may become severely underrepresented or fall below the detection limit. Although \(1/V_{\rm max}\) weighting corrects for the accessible volume of detected galaxies, it cannot recover objects that are entirely absent from the observed sample. We therefore examine the selection effects separately for the different populations using their distributions in the colour--stellar mass and sSFR--stellar mass planes.

We first consider the flux-limited sample. As shown in the upper-middle and upper-right panels of Extended Data Fig. \ref{fig:r19.5}, we divide the colour--stellar mass and sSFR--stellar mass planes into \(100\times100\) grids. For each galaxy, we calculate its \(1/V_{\rm max}\) weight and assign it to the corresponding grid cell. The weighted galaxy number density in each cell is then estimated by summing the \(1/V_{\rm max}\) contributions of the galaxies within it. To avoid displaying poorly constrained regions, we exclude cells containing fewer than five galaxies. The resulting distributions are shown by the colour maps. In the upper-middle panel, the red and blue histograms show the stellar mass distributions of the two colour-selected populations, respectively, with a bin width of 0.2 dex. The upper-right panel similarly shows the stellar mass distributions of star-forming and quiescent galaxies, separated by the adopted sSFR threshold.

These distributions indicate that the abundances of red and quiescent galaxies can be reliably constrained down to stellar masses of approximately \(10^7\,h^{-2}M_\odot\). In contrast, the blue and star-forming populations remain detectable to substantially lower stellar masses, allowing their number densities to be constrained down to approximately \(10^6\,h^{-2}M_\odot\). This difference reflects the distinct mass-to-light ratios and selection limits of the different galaxy populations, and highlights the importance of assessing their completeness separately.

For comparison, we also construct a stellar-mass-complete sample by imposing a redshift-dependent stellar-mass limit. We estimate this limit following the method of Pozzetti et al. \cite{Pozzetti2010}. Specifically, at each redshift, we select the 20 per cent faintest galaxies and estimate the stellar masses they would have if their apparent magnitudes were equal to the survey magnitude limit. The stellar-mass completeness limit is then determined from the 95th percentile of these limiting stellar masses. The black circles in the upper-left panel of Extended Data Fig. \ref{fig:r19.5} show the resulting limits as a function of redshift, and the red dashed line represents their best-fitting relation. We retain only galaxies lying below this redshift-dependent stellar-mass limit to construct the stellar-mass-complete sample.

We then repeat the analysis of the colour--stellar mass and sSFR--stellar mass distributions using this restricted sample, following the same procedure as for the flux-limited sample. The results are shown in the lower-middle and lower-right panels of Extended Data Fig. \ref{fig:r19.5}. Compared with the flux-limited sample, the stellar-mass-complete selection substantially reduces the number of low-mass galaxies in both populations. The resulting distributions become incomplete below approximately \(10^{7.5}\,h^{-2}M_\odot\), with the incompleteness becoming particularly severe below \(10^7\,h^{-2}M_\odot\). Thus, adopting a conventional stellar-mass-complete sample would significantly restrict our ability to investigate the abundance and quenching properties of dwarf galaxies.

Based on these tests, we adopt the flux-limited sample, together with the appropriate \(V_{\rm max}\) and \(N_{\rm max}(M_{\rm h})\) corrections, for our subsequent measurements. This approach allows us to extend the SMFs and CSMFs of blue and star-forming galaxies down to stellar masses of approximately \(10^6\,h^{-2}M_\odot\), while providing reliable measurements of the SMFs, CSMFs, and red and quiescent fractions down to approximately \(10^7\,h^{-2}M_\odot\). We therefore restrict our quantitative analysis of the red and quiescent populations, particularly their fractions, to \(M_\ast\gtrsim10^7\,h^{-2}M_\odot\). Below this limit, the incompleteness of red and quiescent galaxies may lead to an underestimate of their intrinsic abundances and fractions.

\subsection{Measuring the LFs and SMFs of total and central galaxies with different properties}
\label{sec_LF&SMF}

\begin{figure}
    \centering
    \includegraphics[width=0.98\textwidth]
    {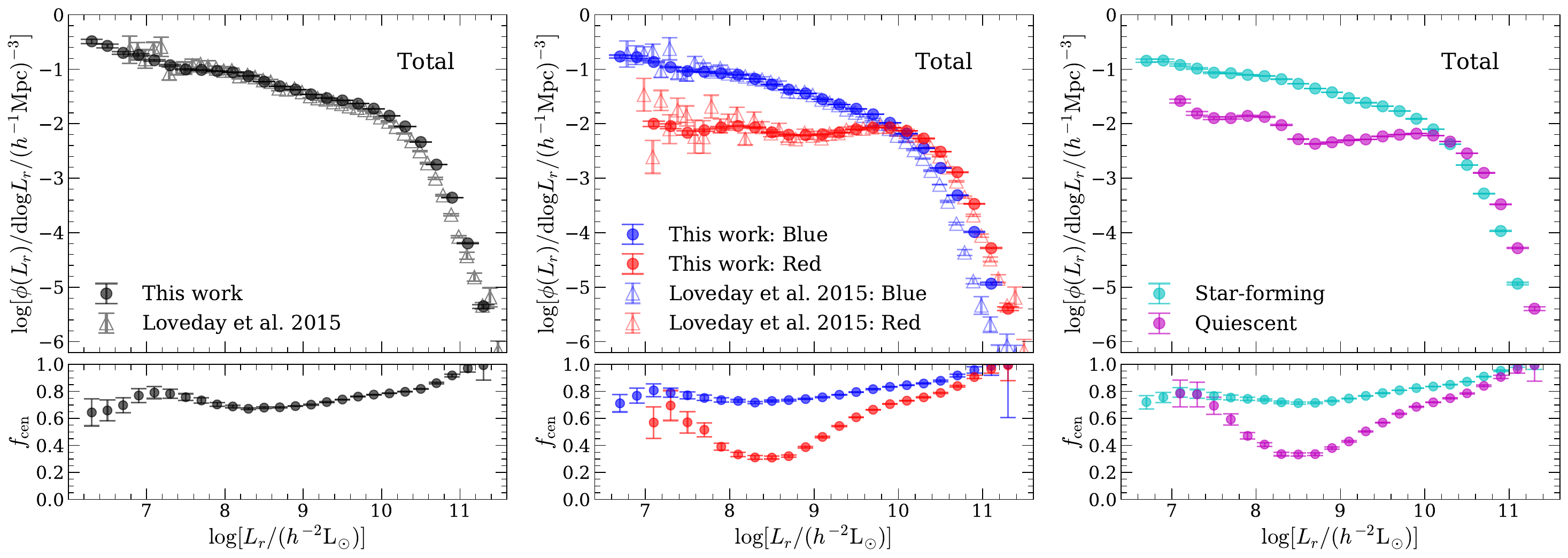}
    \includegraphics[width=0.98\textwidth]
    {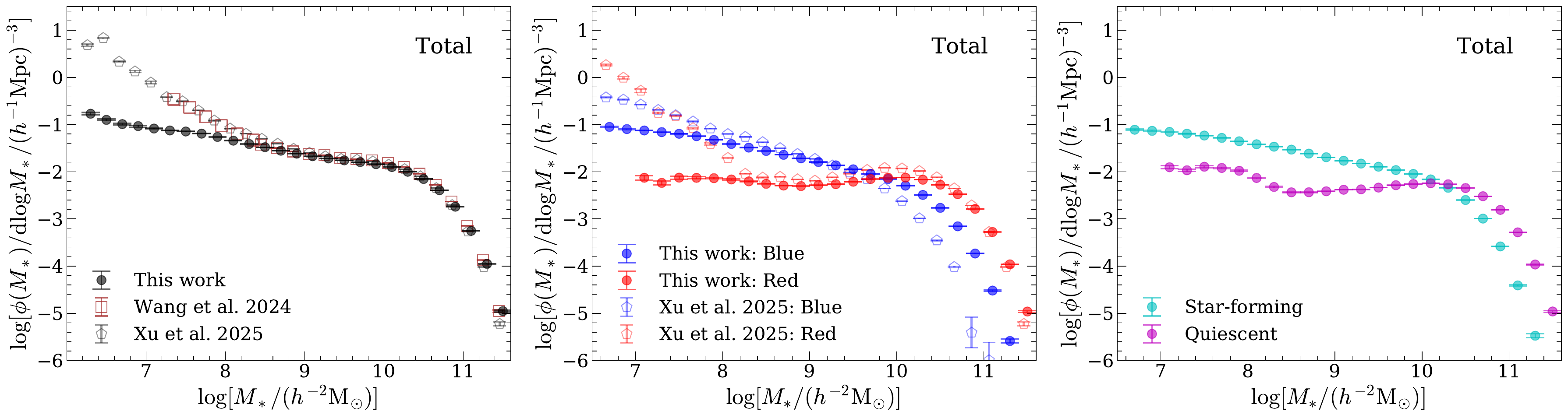}
    \caption{Upper panels: $r$-band luminosity functions (LFs) for all, red/blue, star-forming/quiescent galaxies in the left, middle and right panels, respectively. In each panel, we show the direct measurements with filled circles with  bars that are obtained from the standard deviation of 100 bootstrap resamplings of galaxies. \textbf{Lower panels: similar as the upper panels, but for stellar mass functions (SMFs) of galaxies.} For comparison, the upper panels also include the LFs derived by Loveday et al.\cite{Loveday2015} from the GAMA survey, shown as triangles.  While in the lower panels, the SMFs measured by Wang et al.\cite{Wang2024} are shown as brown squares and Xu et al.\cite{Xu2025} using the Photometric objects Around Cosmic webs (PAC) approach based on DECaLS photometric samples are shown as grey pentagons.}  
    \label{fig:LF_SMF}
\end{figure}

We first measure the LFs and  SMFs of the total galaxy population, as well as those of red, blue, quiescent, and star-forming galaxies, using the standard \(1/V_{\rm max}\) estimator. The resulting \(r\)-band LFs are shown in the upper panels of Extended Data Fig. \ref{fig:LF_SMF}, with the total, color-selected, and sSFR-selected populations presented in the left, middle, and right panels, respectively. At the bright end (\(L_r>10^{10}\,h^{-2}L_\odot\)), the galaxy population is dominated by red and quiescent systems. At intermediate luminosities (\(10^{7.5}<L_r/(h^{-2}L_\odot)<10^{10}\)), blue and star-forming galaxies become increasingly dominant, while the relative contributions of red and quiescent galaxies decrease towards lower luminosities. At the faint end, however, the relative abundance of red and quiescent galaxies increases again, becoming comparable to that of blue and star-forming galaxies in the lowest luminosity bins where their abundances can be reliably constrained.

The corresponding SMFs are shown in the lower panels of Extended Data Fig. \ref{fig:LF_SMF}. They exhibit broadly similar trends to the LFs, with the relative contributions of the different populations changing systematically with stellar mass. Blue and star-forming galaxies dominate over the intermediate stellar-mass range, \(10^{7.5}<M_\ast/(h^{-2}M_\odot)<10^{10}\), whereas red and quiescent galaxies dominate at \(M_\ast>10^{10}\,h^{-2}M_\odot\). Their relative contribution also increases towards the lowest stellar masses probed by our analysis. As discussed in Section \ref{sec:assess_comp}, the abundance of red and quiescent galaxies becomes increasingly affected by incompleteness below approximately \(10^7\,h^{-2}M_\odot\). We therefore interpret their low-mass-end abundance only within the mass range over which these populations can be reliably constrained.

The halo-based group finder further enables us to distinguish central and satellite galaxies, allowing us to investigate the abundance of central galaxies with different properties. We measure their LFs and SMFs using the same \(1/V_{\rm max}\) methodology adopted for the total galaxy population. To quantify the relative contribution of central galaxies, we define the central-to-total ratio, \(f_{\rm cen}\), as the ratio of the central-galaxy LF (or SMF) to the corresponding LF (or SMF) of the total population in each luminosity (or stellar-mass) bin. The resulting ratios are shown in the lower subpanels of the upper panels of Extended Data Fig. \ref{fig:LF_SMF}.

For blue and star-forming galaxies, \(f_{\rm cen}\) is close to unity at the bright end and gradually decreases towards lower luminosities and stellar masses, reaching approximately 80 per cent in the low-mass regime. In contrast, red and quiescent galaxies exhibit a more pronounced variation: \(f_{\rm cen}\) reaches a minimum around \(L_r\sim10^{8.5}\,h^{-2}L_\odot\), before increasing again towards lower luminosities. As demonstrated by our tests using the MGRS mocks, this upturn is primarily associated with the local void (LV) effect, rather than necessarily indicating an intrinsic increase in the central fraction of low-mass red galaxies. We therefore take this selection-related effect into account when interpreting the central-galaxy statistics. The SMFs of central galaxies are also presented explicitly in the upper row of Fig. \ref{fig:CSMF_measure}. These measurements provide the basis for investigating the stellar-mass dependence of quenching among central galaxies, which we discuss further in the following sections.

\subsubsection{Comparison with previous measurements}
\label{subsec:LFSMF_Comp}

Following the LF and SMF measurements presented above, we compare our results with previous studies. We first show the LFs reported by Loveday et al.\cite{Loveday2015} based on the GAMA survey in the upper panels of Extended Data Fig. \ref{fig:LF_SMF}. Their measurements for the total, blue, and red galaxy populations are indicated by the colored triangles. Overall, the GAMA measurements are in good agreement with ours, particularly for the total and red galaxy populations. The most noticeable difference occurs at the bright end of the blue-galaxy LF, where Loveday et al.\cite{Loveday2015} report lower number densities than we do. This difference may partly reflect the different color-selection criteria adopted in the two studies. More recently, Moore et al.\cite{Moore2025} measured LFs in several photometric bands using DESI Year 3 data. Their results are broadly consistent with those of Loveday et al.\cite{Loveday2015} and are also in general agreement with our measurements.

We next compare the galaxy SMFs, as shown in the lower panels of Extended Data Fig. \ref{fig:LF_SMF}. In the lower-left panel, we compare our total SMF with those reported by Wang et al.\cite{Wang2024} and Xu et al.\cite{Xu2025}, shown by solid squares and pentagons, respectively. The measurements are broadly consistent at the massive end, but both previous studies show a steeper low-mass-end slope than ours. Xu et al. derived their SMFs using the Photometric Objects Around Cosmic Webs (PAC) method, which combines galaxy clustering information with photometric data to constrain galaxy abundances beyond the limits of conventional spectroscopic samples. As discussed by Wang et al.\cite{Wang2024}, the difference at the low-mass end may be related to the local-void (LV) effect, which can affect the inferred abundance of low-mass galaxies in the local volume.

The lower-middle panel further compares the SMFs of red and blue galaxies with those from Xu et al.\cite{Xu2025}, shown by the colored dotted lines. The two studies show modest differences at the massive end, which may partly arise from the different color-selection criteria. More notably, Xu et al. infer a substantially higher abundance of red galaxies at the low-mass end than we do. One possible explanation is that their measurements include a larger contribution from red satellite galaxies in massive host haloes. Such systems may be underrepresented in our local sample because of the LV effect, which can affect the abundance and environmental distribution of galaxies in the volume probed by our survey. This possibility is also consistent with the sensitivity of the low-mass red-galaxy statistics to the LV effect identified in our mock-catalogue tests. Nevertheless, differences in sample selection, color classification, and measurement methodology may also contribute to the discrepancy.

\subsection{Measuring the CSMFs of galaxies with different properties}
\label{sec_CSMF}

The CSMF, \(\Phi(M_\ast|M_{\rm h})\), describes the average number of galaxies per dark matter halo as a function of stellar mass, \(M_\ast\), for haloes of a given mass, \(M_{\rm h}\). Unlike the global SMF, which is normalized by the effective survey volume, the CSMF is normalized by the number of galaxy groups in a given halo-mass bin. To account for the different maximum observable redshifts of galaxies with different luminosities, we adopt the \(N_{\max}\) estimator.

Specifically, we select galaxies and groups within the redshift range \(0\leq z\leq0.2\). For each galaxy, we determine its maximum observable redshift, \(z_{\max}\), based on the survey magnitude limit. The maximum number of observable groups, \(N_{\max}\), is then calculated by counting all groups in the same halo-mass bin within the redshift interval
\[
0\leq z\leq \min(0.2,z_{\max}).
\]
The CSMF in a given halo-mass bin is subsequently estimated by weighting each galaxy by the inverse of its corresponding \(N_{\max}\):
\begin{equation}
\Phi(M_\ast|M_{\rm h}) =
\frac{1}{\Delta\log M_\ast}
\sum_i \frac{1}{N_{\max,i}},
\end{equation}
where the sum runs over galaxies in the corresponding stellar-mass and halo-mass bins, and \(\Delta\log M_\ast\) is the width of the stellar-mass bin. This estimator provides the average number of galaxies per group per logarithmic stellar-mass interval.

We apply this method separately to central and satellite galaxies, as well as to different galaxy populations classified by optical colors and specific star formation rates. Specifically, we measure the CSMFs of the total, red, and blue populations, and those of the full, quiescent, and star-forming populations. The resulting CSMFs are presented in Fig. \ref{fig:CSMF_measure}, with the four halo-mass bins spanning \([10^{14.4},10^{14.7}]\), \([10^{13.5},10^{13.8}]\), \([10^{12.9},10^{13.2}]\), and \([10^{12.0},10^{12.3}]\), respectively, in units of \(h^{-1}M_\odot\). We estimate the statistical uncertainties using 100 bootstrap resamplings of galaxy groups and adopt the standard deviation among the resampled measurements as the error on the CSMF.

\begin{figure*}
    \centering
    \includegraphics[width=0.93\textwidth]{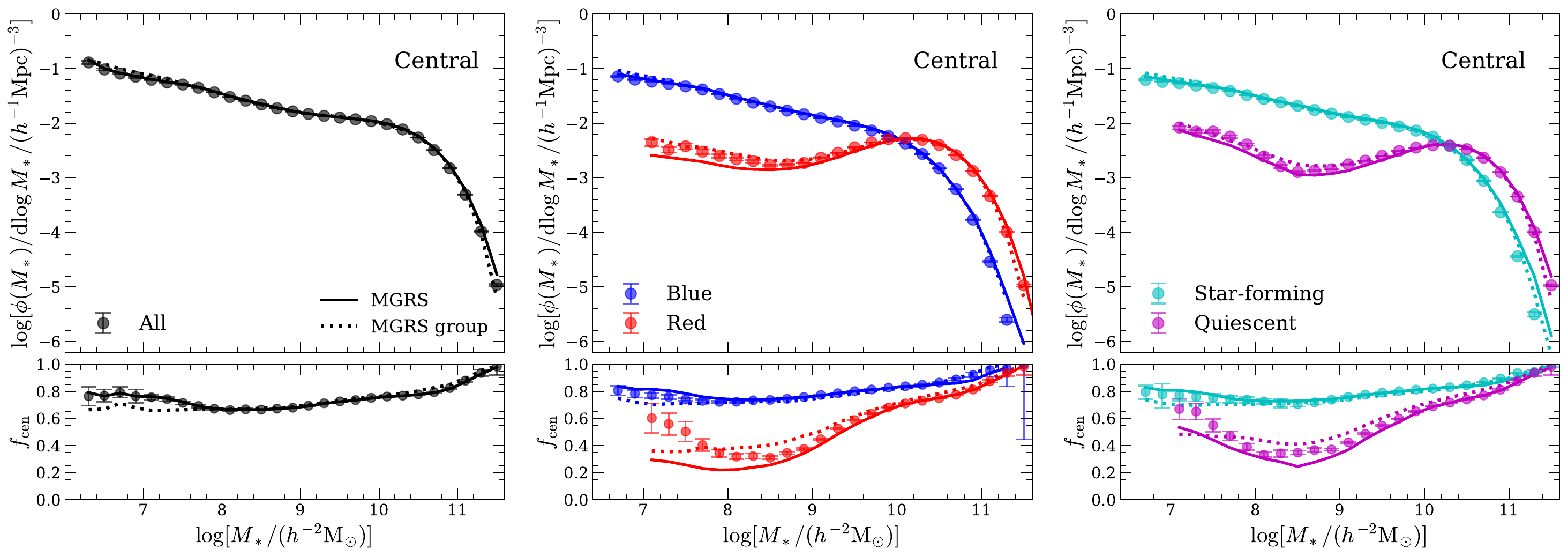}
\includegraphics[width=0.95\textwidth,height=0.4\textwidth]{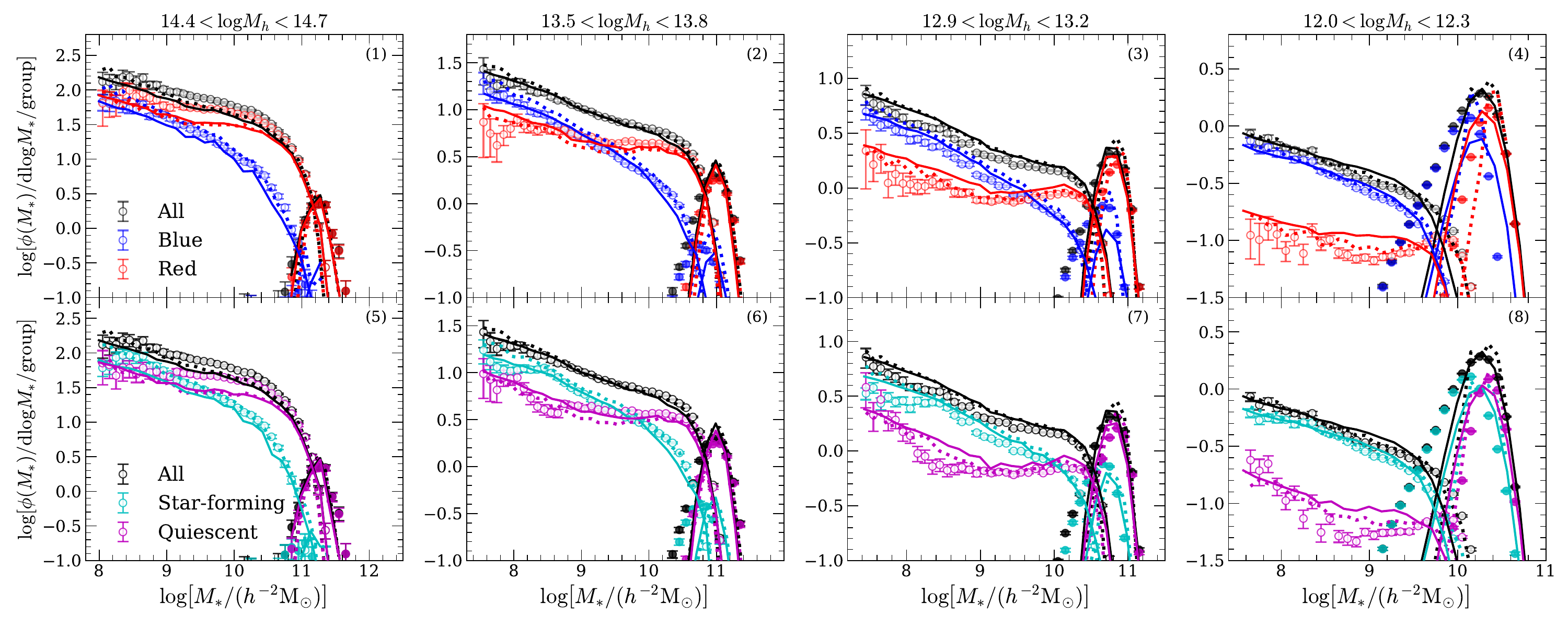}
    \caption{Same as Fig. \ref{fig:CSMF_measure}, except that the comparison is made with the MGRS predictions. In each panel, the points with error bars represent measurements from DESI DR2 observations. The solid curve shows the results directly derived from the MGRS, without applying any observational selection effects. 
     The dotted curves show results obtained from group catalogs that incorporate observational selection effects including the DESI DR2 fiber-assignment effects for both bright and faint samples, with galaxies lacking fiber assignments given photometric redshifts.  
     }
    \label{fig:SMF-check}
\end{figure*}

\subsection{Using a realistic mock galaxy redshift survey to perform end-to-end assessment of our results}

Since our measurements of galaxy quenching rely on the SMFs of central galaxies and the CSMFs of satellite galaxies, their reliability depends critically on the treatment of observational selection effects and the performance of our group-finding algorithm. Two main sources of systematic uncertainty are considered here. The first is observational incompleteness, particularly that arising from DESI fiber assignment, which can reduce the spectroscopic completeness of galaxies in close angular pairs. The second is uncertainty in group identification, including the loss of genuine group members and the contamination of groups by interlopers. These effects can bias the central/satellite classification and, consequently, the inferred SMFs, CSMFs, and galaxy fractions. To quantify these effects, we construct a realistic mock galaxy redshift survey (MGRS) that incorporates the relevant DESI DR2 selection effects and apply the same group-finding pipeline used for the observational data. This enables us to compare the recovered galaxy statistics with their known intrinsic values and to assess the systematic uncertainties in our measurements. Details of the MGRS construction are provided in Methods (Sections \ref{sec_mock} and \ref{sec_mgrs}). Briefly, our MGRS construction proceeds as follows. (1) We start from a high-resolution Jiutian simulation subhalo catalog and assign stellar masses to main and sub-haloes via a central–satellite subhalo abundance matching (CS-SHAM) scheme calibrated against the observed SMFs of central and satellite galaxies. (2) We adopt the $f_{\rm red}$ and $f_{\rm quiescent}$ models derived in this work to assign each galaxy a color and star-formation state. (3) We further assign luminosities to individual galaxies by abundance matching to the appropriate SMFs, using the luminosity functions (LFs) of red and blue galaxies. (4) We place the observer at $(0,0,0)$ and generate the MGRS while accounting for sky coverage, redshift-space distortions (RSD), $K$-corrections, the apparent-magnitude threshold. We apply the DESI DR2 BGS bright and faint fiber-assignment pipeline and yield an MGRS catalog broadly consistent with observations. (5) Finally, we apply our halo-based group finder to the MGRS to construct the corresponding mock group catalogue.

\subsubsection{Generate mock galaxy catalog from Jiutian simulation}
\label{sec_mock}

We utilize host haloes and subhaloes extracted from the highest resolution run in the Jiutian simulation suite\cite{Han2025}, Jiutian-300 to generate our mock galaxy catalog. This is a dark-matter-only simulation executed under the Planck-2018 cosmological framework\cite{Planck2020}, with parameters $\Omega_{M}=0.3111$, $\Omega_\Lambda=0.6889$, $\Omega_b=0.0490$, $\sigma_8=0.8102$, and $n_s=0.9665$. The simulation was carried out using $6144^3$ particles of mass $m_p= 1.005\times10^7 \Msunh$, within a periodic box of $300\mpch$ per side. Dark matter haloes are identified using the Friends-of-Friends (FOF) algorithm\cite{Davis1985}, employing a linking length of $0.2$ times the mean inter-particle separation. Subsequently, these haloes are processed with the Hierarchical Bound-Tracing (\textsc{hbt+}) code\cite{Han2012, Han2018} to determine subhaloes and their evolutionary trajectories. \textsc{hbt+} is a time-domain subhalo finder and tree constructor that operates by tracking the evolution of each halo throughout the simulation. It determines the descendant of a halo at every snapshot as either a distinct halo or a subhalo. The minimum number of particles in a (sub)halo is set to 20, and the corresponding minimum (sub)halo mass is approximately $2.0 \times 10^8 \Msunh$.

We adopted a central–satellite SHAM (CS-SHAM) to assign central galaxies to host haloes and satellite galaxies to subhaloes \cite{Xu2026} using the SMFs for central and satellite galaxies, respectively. Here we use the maximum subhalo mass, $M_{\rm peak}$, over its full evolutionary history for abundance matching. When implementing CS-SHAM, we explicitly include a 0.15 dex scatter in stellar mass at fixed halo/subhalo mass for both central and satellite galaxies, as reported by Yang et al.\cite{yang2008galaxy}. After assigning a stellar mass to each halo and subhalo in the Jiutian-300 simulation, we next assign a color type (red/blue) or star-formation type (quiescent/star-forming) to every object using the quenching model for $f_{\rm red}$ and $f_{\rm quiescent}$ derived in this work, based on each galaxy’s stellar mass and halo mass. The assigned color and star-formation states are constructed to reproduce the observed red and quiescent fractions over the stellar-mass range considered. We then use the mock survey to assess how observational selection effects and group identification affect the recovery of these input galaxy statistics. Our MGRS does not predict galaxy colours or star‑formation properties ab initio from the dark‑matter‑only Jiutian simulation. Instead, we assign galaxy colour and star‑formation states using the best‑fit $f_{\rm red}$ and $f_{\rm quiescent}$ models derived from our DESI observations. The mock is designed purely for end‑to‑end validation of survey selection biases and group‑finder performance, and is not intended as a test of physical feedback models.

Besides galaxy stellar mass, constructing realistic mock galaxy redshift surveys also requires galaxy luminosities. To obtain these, we employ abundance matching between the SMFs and LFs for red and blue galaxies separately. In this way, each host halo and subhalo is populated with a galaxy assigned a stellar mass, $r$-band luminosity, color type, and related properties. 

\subsubsection{Generating mock galaxy redshift survey and group catalog}
\label{sec_mgrs}

Using the mock galaxy catalog produced above, we build a DESI DR2 mock galaxy redshift survey (MGRS) by incorporating a range of survey selection effects. Our overall workflow largely follows \cite{Gu2024, Yang2004}, but with some adjustments. The key changes are the inclusion of a $K$-correction assignment according to observational data and an explicit handling of fiber collisions. Below we briefly summarize the steps used to construct the DESI DR2 MGRS.

\begin{enumerate}
\item We begin by tiling the simulation boxes to fill the survey volume out to redshift $z<0.6$, placing the observer at the center of this volume. We then convert galaxy positions from $x,y,z$ to ra, dec, and $z_{\rm cos}$ coordinates, and discard galaxies that are outside DESI DR2 skycoverage. 

\item  Next, we account for each galaxy’s line-of-sight peculiar velocity and apply a spectroscopic redshift error of $35\,{\rm km/s}$ to derive the observed spectroscopic redshift $z_{\rm spec}$.

\item For each galaxy, we use its color type and $z_{\rm spec}$ to select a DESI galaxy with the same color type and redshift, from which we derive an $r$-band $K$-correction. We then apply this $K$-correction to compute the $r$-band apparent magnitude for these galaxies.

\item We keep only galaxies with apparent magnitude brighter than $r=20.175$ for the subsequent fiber-assignment step, which determines which spectroscopic redshifts are retained. We apply the DESI fiber-assignment code to select galaxies following the DESI DR2 survey strategy for the BGS bright and faint samples, respectively. 

\item In the region that overlaps SDSS DR7, we select all galaxies with $r<17.7$ and use their spectroscopic redshifts. For the remaining galaxies without an assigned fiber, we instead use photometric redshifts. This yields our final MGRS.

\end{enumerate}

Finally, we apply the extended halo-based group finder to the thus constructed MGRS to build the associated group catalog, enabling a comprehensive end-to-end validation of our results.

\subsubsection{End-to-end validation of the measurements}
\label{sec_validate}

We first examine the recovery of the SMFs of central galaxies by comparing the intrinsic input SMFs in the MGRS with those recovered from the mock group catalog. As shown in the upper-left panel of Fig. \ref{fig:SMF-check}, the input and recovered SMFs of the full central-galaxy population agree well, indicating that our group-finding procedure generally recovers the central-galaxy abundance. The SMFs of red and blue centrals, as well as those of quiescent and star-forming centrals, are shown in the upper-middle and upper-right panels, respectively. The input and recovered SMFs are represented by solid and dotted curves, respectively. While the recovered SMFs agree well with the input ones for blue and star-forming centrals, we find a mild overestimation of the abundance of red and quiescent centrals at the low-mass end. This bias arises primarily because some satellite galaxies are incorrectly identified as centrals owing to group-member incompleteness and interloper contamination. Since red and quiescent centrals constitute relatively small fractions of the central population in this mass regime, even a modest level of contamination can produce a noticeable relative bias. We therefore use the discrepancies between the input and recovered SMFs to correct the corresponding measurements of the red and quiescent fractions of central galaxies in the DESI data, as shown in Fig. \ref{fig:fred_Mh}.

We further examine the central-galaxy fraction, \(f_{\rm cen}\), for the full sample and for the red/blue and quiescent/star-forming subsamples. The corresponding measurements are shown in the upper subpanels of Fig. \ref{fig:SMF-check}. For blue and star-forming galaxies, the recovered \(f_{\rm cen}\) values closely reproduce the input values. For red and quiescent galaxies, however, we find a small systematic offset, consistent with the mild overestimation of their central-galaxy SMFs discussed above. In the DESI data, the \(f_{\rm cen}\) of red galaxies exhibits an additional upturn towards low stellar masses that is not reproduced by the MGRS. As discussed in Section \ref{sec_cen_smf}, this feature is likely associated with the LV effect, which can lead to an underrepresentation of low-mass red satellite galaxies in the local volume. This comparison therefore helps distinguish the bias associated with group identification from the additional environmental selection effect affecting the observed sample.

We finally assess the recovery of satellite-galaxy CSMFs by comparing the intrinsic CSMFs in the MGRS with those measured from the mock group catalog. The intrinsic CSMFs, obtained directly by counting satellite galaxies within their true host haloes, are shown by the solid curves in the lower panels of Fig. \ref{fig:SMF-check}. These provide the reference measurements against which the performance of the group-finding procedure can be evaluated. We note that the intrinsic mock CSMFs do not reproduce the DESI measurements exactly: they are slightly lower in the most massive haloes and somewhat higher in lower-mass haloes. This difference reflects the fact that our CS-SHAM model is calibrated primarily to the central and satellite SMFs, rather than being explicitly tuned to reproduce the observed CSMFs in every halo-mass bin. We therefore do not use the agreement between the mock and observed CSMFs as a criterion for validating the intrinsic mock population. Instead, we focus on whether the group-finding pipeline can recover the known input CSMFs after incorporating observational selection effects. The CSMFs recovered from the mock group catalog are shown by the dotted curves. They agree closely with the intrinsic input CSMFs across the stellar-mass and halo-mass ranges considered, generally within the \(1\sigma\) uncertainties. This demonstrates that our extended halo-based group finder, together with the treatment of DESI selection effects, can recover the satellite-galaxy CSMFs with good accuracy in the mock survey.

Taken together, these end-to-end tests quantify the effects of survey selection and group identification on the central-galaxy SMFs, central fractions, and satellite-galaxy CSMFs. They provide a basis for correcting the small systematic biases identified in the central-galaxy statistics and demonstrate the robustness of our group-based measurements against the selection effects and group-identification uncertainties explicitly included in the MGRS.

\backmatter





\bmhead{Acknowledgements}

We sincerely thank Amelie Saintonge, Benjamin Weaver, Rita Tojeiro and Dirk Scholte for useful comments and suggestions that significantly improved the presentation of this paper. 

\section*{Declarations}

\subsection*{Funding}
This work is supported by the National Key R\&D Program of China (2023YFA1607800, 2023YFA1607804, 2023YFA1605600), the National Science Foundation of China (Nos. 12633001, 12373003), ``the Fundamental Research Funds for the Central Universities'', 111 project No. B20019, and Shanghai Natural Science Foundation, grant No.19ZR1466800. We acknowledge the science research grants from the China Manned Space Project with Nos. CMS-CSST-2021-A02 \& CMS-CSST-2025-A04, and Yangyang Development Fund. This project is also supported in part by Office of Science and Technology, Shanghai Municipal Government (grant Nos. 24DX1400100, ZJ2023-ZD-001).
The computations in this paper were run on the Gravity Supercomputer at Shanghai Jiao Tong University. WW is supported by the National Key R\&D Program of China (2023YFA1605600,  2023YFA1605601) and NSFC (12273021).

This research used data obtained with the Dark Energy Spectroscopic Instrument (DESI). DESI construction and operations is managed by the Lawrence Berkeley National Laboratory. This material is based upon work supported by the U.S. Department of Energy, Office of Science, Office of High-Energy Physics, under Contract No. DEAC0205CH11231, and by the National Energy Research Scientific Computing Center, a DOE Office of Science User Facility under the same contract. Additional support for DESI was provided by the U.S. National Science Foundation (NSF), Division of Astronomical Sciences under Contract No. AST-0950945 to the NSF’s National Optical-Infrared Astronomy Research Laboratory; the Science and Technology Facilities Council of the United Kingdom; the Gordon and Betty Moore Foundation; the HeisingSimons Foundation; the French Alternative Energies and Atomic Energy Commission (CEA); the National Council of Science and Technology of Mexico (CONACYT); the Ministry of Science and Innovation of Spain (MICINN), and by the DESI Member Institutions: www.desi.lbl.gov/collaborating-institutions.

The DESI Legacy Imaging Surveys consist of three individual and complementary projects: the Dark Energy Camera Legacy Survey (DECaLS), the Beijing-Arizona Sky Survey (BASS), and the Mayall z-band Legacy Survey (MzLS). DECaLS, BASS and MzLS together include data obtained, respectively, at the Blanco telescope, Cerro Tololo Inter-American Observatory, NSF’s NOIRLab; the Bok telescope, Steward Observatory, University of Arizona; and the Mayall telescope, Kitt Peak National Observatory, NOIRLab. NOIRLab is operated by the Association of Universities for Research in Astronomy (AURA) under a cooperative agreement with the National Science Foundation. Pipeline processing and analyses of the data were supported by NOIRLab and the Lawrence Berkeley National Laboratory (LBNL). Legacy Surveys also uses data products from the Near-Earth Object Wide-field Infrared Survey Explorer (NEOWISE), a project of the Jet Propulsion Laboratory/California Institute of Technology, funded by the National Aeronautics and Space Administration. Legacy Surveys was supported by: the Director, Office of Science, Office of High Energy Physics of the U.S. Department of Energy; the National Energy Research Scientific Computing Center, a DOE Office of Science User Facility; the U.S. National Science Foundation, Division of Astronomical Sciences; the National Astronomical Observatories of China, the Chinese Academy of Sciences and the Chinese National Natural Science Foundation. LBNL is managed by the Regents of the University of California under contract to the U.S. Department of Energy. The complete acknowledgments can be found at https://www.legacysurvey.org/acknowledgment/.
Any opinions, findings, and conclusions or recommendations expressed in this material are those of the author(s) and do not necessarily reflect the views of the U. S. National Science Foundation, the U. S. Department of Energy, or any of the listed funding agencies.
The authors are honored to be permitted to conduct scientific research on I'oligam Du'ag (Kitt Peak), a mountain with particular significance to the Tohono O’odham Nation.

We additionally made use of Astropy, a community-developed core Python package for Astronomy (Astropy Collaboration et al. 2018), IPython (Pérez \& Granger 2007), Matplotlib (Hunter 2007), and Deepseek/ChatGPT for English language polishing. 

\subsection*{Competing Interests} The authors declare that they have no competing financial interests.

\subsection*{Author Contributions} All co-authors read
and commented on the document. QL, YW and YG contributed equally to this work, YW and YG are co-first 
authors of this paper. XY conceived the original idea, initiated the project and supervised the analysis. YW, QL and YG contributed to the analysis of observational data as well as preparing the original draft. JH contributed to the analysis of simulation data and interpretation of the results.

\subsection*{Data and materials availability}
The galaxy group catalog is publicly available at \url{https://gax.sjtu.edu.cn/data/Group.html}. 
The Jiutian Simulation data are available at \url{https://jiutian.sjtu.edu.cn/}.
The simulation data are available through the IllustrisTNG public data release\cite{Nelson2019} at \url{https://www.tng-project.org/}.
The code used in this paper is available at \url{https://github.com/Kambrian/HBTplus}
The LFs and SMFs for all and central galaxies are available at \url{https://gax.sjtu.edu.cn/data/tables/LFs_SMFs_DESI_DR2.tar.gz}


\bibliography{sn-bibliography}


\begin{thebibliography}{118}
\ifx \bisbn   \undefined \def \bisbn  #1{ISBN #1}\fi
\ifx \binits  \undefined \def \binits#1{#1}\fi
\ifx \bauthor  \undefined \def \bauthor#1{#1}\fi
\ifx \batitle  \undefined \def \batitle#1{#1}\fi
\ifx \bjtitle  \undefined \def \bjtitle#1{#1}\fi
\ifx \bvolume  \undefined \def \bvolume#1{\textbf{#1}}\fi
\ifx \byear  \undefined \def \byear#1{#1}\fi
\ifx \bissue  \undefined \def \bissue#1{#1}\fi
\ifx \bfpage  \undefined \def \bfpage#1{#1}\fi
\ifx \blpage  \undefined \def \blpage #1{#1}\fi
\ifx \burl  \undefined \def \burl#1{\textsf{#1}}\fi
\ifx \doiurl  \undefined \def \doiurl#1{\url{https://doi.org/#1}}\fi
\ifx \betal  \undefined \def \betal{\textit{et al.}}\fi
\ifx \binstitute  \undefined \def \binstitute#1{#1}\fi
\ifx \binstitutionaled  \undefined \def \binstitutionaled#1{#1}\fi
\ifx \bctitle  \undefined \def \bctitle#1{#1}\fi
\ifx \beditor  \undefined \def \beditor#1{#1}\fi
\ifx \bpublisher  \undefined \def \bpublisher#1{#1}\fi
\ifx \bbtitle  \undefined \def \bbtitle#1{#1}\fi
\ifx \bedition  \undefined \def \bedition#1{#1}\fi
\ifx \bseriesno  \undefined \def \bseriesno#1{#1}\fi
\ifx \blocation  \undefined \def \blocation#1{#1}\fi
\ifx \bsertitle  \undefined \def \bsertitle#1{#1}\fi
\ifx \bsnm \undefined \def \bsnm#1{#1}\fi
\ifx \bsuffix \undefined \def \bsuffix#1{#1}\fi
\ifx \bparticle \undefined \def \bparticle#1{#1}\fi
\ifx \barticle \undefined \def \barticle#1{#1}\fi
\bibcommenthead
\ifx \bconfdate \undefined \def \bconfdate #1{#1}\fi
\ifx \botherref \undefined \def \botherref #1{#1}\fi
\ifx \url \undefined \def \url#1{\textsf{#1}}\fi
\ifx \bchapter \undefined \def \bchapter#1{#1}\fi
\ifx \bbook \undefined \def \bbook#1{#1}\fi
\ifx \bcomment \undefined \def \bcomment#1{#1}\fi
\ifx \oauthor \undefined \def \oauthor#1{#1}\fi
\ifx \citeauthoryear \undefined \def \citeauthoryear#1{#1}\fi
\ifx \endbibitem  \undefined \def \endbibitem {}\fi
\ifx \bconflocation  \undefined \def \bconflocation#1{#1}\fi
\ifx \arxivurl  \undefined \def \arxivurl#1{\textsf{#1}}\fi
\csname PreBibitemsHook\endcsname

\bibitem[\protect\citeauthoryear{{Mo} et~al.}{2010}]{Mo2010}
\begin{bbook}
\bauthor{\bsnm{{Mo}}, \binits{H.}},
\bauthor{\bsnm{{van den Bosch}}, \binits{F.C.}},
\bauthor{\bsnm{{White}}, \binits{S.}}:
\bbtitle{{Galaxy Formation and Evolution}},
(\byear{2010}).
\doiurl{10.1017/CBO9780511807244}
\end{bbook}
\endbibitem

\bibitem[\protect\citeauthoryear{{Vogelsberger} et~al.}{2014}]{Vogelsberger2014}
\begin{barticle}
\bauthor{\bsnm{{Vogelsberger}}, \binits{M.}},
\bauthor{\bsnm{{Genel}}, \binits{S.}},
\bauthor{\bsnm{{Springel}}, \binits{V.}},
\bauthor{\bsnm{{Torrey}}, \binits{P.}},
\bauthor{\bsnm{{Sijacki}}, \binits{D.}},
\bauthor{\bsnm{{Xu}}, \binits{D.}},
\bauthor{\bsnm{{Snyder}}, \binits{G.}},
\bauthor{\bsnm{{Nelson}}, \binits{D.}},
\bauthor{\bsnm{{Hernquist}}, \binits{L.}}:
\batitle{{Introducing the Illustris Project: simulating the coevolution of dark and visible matter in the Universe}}.
\bjtitle{MNRAS}
\bvolume{444}(\bissue{2}),
\bfpage{1518}--\blpage{1547}
(\byear{2014})
\doiurl{10.1093/mnras/stu1536}
{\href{https://arxiv.org/abs/1405.2921}{{arXiv:1405.2921}}}
{[astro-ph.CO]}
\end{barticle}
\endbibitem

\bibitem[\protect\citeauthoryear{{Nelson} et~al.}{2019}]{Nelson2019}
\begin{barticle}
\bauthor{\bsnm{{Nelson}}, \binits{D.}},
\bauthor{\bsnm{{Springel}}, \binits{V.}},
\bauthor{\bsnm{{Pillepich}}, \binits{A.}},
\bauthor{\bsnm{{Rodriguez-Gomez}}, \binits{V.}},
\bauthor{\bsnm{{Torrey}}, \binits{P.}},
\bauthor{\bsnm{{Genel}}, \binits{S.}},
\bauthor{\bsnm{{Vogelsberger}}, \binits{M.}},
\bauthor{\bsnm{{Pakmor}}, \binits{R.}},
\bauthor{\bsnm{{Marinacci}}, \binits{F.}},
\bauthor{\bsnm{{Weinberger}}, \binits{R.}},
\bauthor{\bsnm{{Kelley}}, \binits{L.}},
\bauthor{\bsnm{{Lovell}}, \binits{M.}},
\bauthor{\bsnm{{Diemer}}, \binits{B.}},
\bauthor{\bsnm{{Hernquist}}, \binits{L.}}:
\batitle{{The IllustrisTNG simulations: public data release}}.
\bjtitle{Computational Astrophysics and Cosmology}
\bvolume{6}(\bissue{1}),
\bfpage{2}
(\byear{2019})
\doiurl{10.1186/s40668-019-0028-x}
{\href{https://arxiv.org/abs/1812.05609}{{arXiv:1812.05609}}}
{[astro-ph.GA]}
\end{barticle}
\endbibitem

\bibitem[\protect\citeauthoryear{{Schaye} et~al.}{2015}]{Schaye2015}
\begin{barticle}
\bauthor{\bsnm{{Schaye}}, \binits{J.}},
\bauthor{\bsnm{{Crain}}, \binits{R.A.}},
\bauthor{\bsnm{{Bower}}, \binits{R.G.}},
\bauthor{\bsnm{{Furlong}}, \binits{M.}},
\bauthor{\bsnm{{Schaller}}, \binits{M.}},
\bauthor{\bsnm{{Theuns}}, \binits{T.}},
\bauthor{\bsnm{{Dalla Vecchia}}, \binits{C.}},
\bauthor{\bsnm{{Frenk}}, \binits{C.S.}},
\bauthor{\bsnm{{McCarthy}}, \binits{I.G.}},
\bauthor{\bsnm{{Helly}}, \binits{J.C.}},
\bauthor{\bsnm{{Jenkins}}, \binits{A.}},
\bauthor{\bsnm{{Rosas-Guevara}}, \binits{Y.M.}},
\bauthor{\bsnm{{White}}, \binits{S.D.M.}},
\bauthor{\bsnm{{Baes}}, \binits{M.}},
\bauthor{\bsnm{{Booth}}, \binits{C.M.}},
\bauthor{\bsnm{{Camps}}, \binits{P.}},
\bauthor{\bsnm{{Navarro}}, \binits{J.F.}},
\bauthor{\bsnm{{Qu}}, \binits{Y.}},
\bauthor{\bsnm{{Rahmati}}, \binits{A.}},
\bauthor{\bsnm{{Sawala}}, \binits{T.}},
\bauthor{\bsnm{{Thomas}}, \binits{P.A.}},
\bauthor{\bsnm{{Trayford}}, \binits{J.}}:
\batitle{{The EAGLE project: simulating the evolution and assembly of galaxies and their environments}}.
\bjtitle{\mnras}
\bvolume{446}(\bissue{1}),
\bfpage{521}--\blpage{554}
(\byear{2015})
\doiurl{10.1093/mnras/stu2058}
{\href{https://arxiv.org/abs/1407.7040}{{arXiv:1407.7040}}}
{[astro-ph.GA]}
\end{barticle}
\endbibitem

\bibitem[\protect\citeauthoryear{{Dav{\'e}} et~al.}{2019}]{Dave2019}
\begin{barticle}
\bauthor{\bsnm{{Dav{\'e}}}, \binits{R.}},
\bauthor{\bsnm{{Angl{\'e}s-Alc{\'a}zar}}, \binits{D.}},
\bauthor{\bsnm{{Narayanan}}, \binits{D.}},
\bauthor{\bsnm{{Li}}, \binits{Q.}},
\bauthor{\bsnm{{Rafieferantsoa}}, \binits{M.H.}},
\bauthor{\bsnm{{Appleby}}, \binits{S.}}:
\batitle{{SIMBA: Cosmological simulations with black hole growth and feedback}}.
\bjtitle{\mnras}
\bvolume{486}(\bissue{2}),
\bfpage{2827}--\blpage{2849}
(\byear{2019})
\doiurl{10.1093/mnras/stz937}
{\href{https://arxiv.org/abs/1901.10203}{{arXiv:1901.10203}}}
{[astro-ph.GA]}
\end{barticle}
\endbibitem

\bibitem[\protect\citeauthoryear{{Dickey} et~al.}{2021}]{Dickey2021}
\begin{barticle}
\bauthor{\bsnm{{Dickey}}, \binits{C.M.}},
\bauthor{\bsnm{{Starkenburg}}, \binits{T.K.}},
\bauthor{\bsnm{{Geha}}, \binits{M.}},
\bauthor{\bsnm{{Hahn}}, \binits{C.}},
\bauthor{\bsnm{{Angl{\'e}s-Alc{\'a}zar}}, \binits{D.}},
\bauthor{\bsnm{{Choi}}, \binits{E.}},
\bauthor{\bsnm{{Dav{\'e}}}, \binits{R.}},
\bauthor{\bsnm{{Genel}}, \binits{S.}},
\bauthor{\bsnm{{Iyer}}, \binits{K.G.}},
\bauthor{\bsnm{{Maller}}, \binits{A.H.}},
\bauthor{\bsnm{{Mandelker}}, \binits{N.}},
\bauthor{\bsnm{{Somerville}}, \binits{R.S.}},
\bauthor{\bsnm{{Yung}}, \binits{L.Y.A.}}:
\batitle{{IQ Collaboratory. II. The Quiescent Fraction of Isolated, Low-mass Galaxies across Simulations and Observations}}.
\bjtitle{\apj}
\bvolume{915}(\bissue{1}),
\bfpage{53}
(\byear{2021})
\doiurl{10.3847/1538-4357/abc014}
{\href{https://arxiv.org/abs/2010.01132}{{arXiv:2010.01132}}}
{[astro-ph.GA]}
\end{barticle}
\endbibitem

\bibitem[\protect\citeauthoryear{{Yang} et~al.}{2009}]{yang2009galaxy}
\begin{barticle}
\bauthor{\bsnm{{Yang}}, \binits{X.}},
\bauthor{\bsnm{{Mo}}, \binits{H.J.}},
\bauthor{\bsnm{{van den Bosch}}, \binits{F.C.}}:
\batitle{{Galaxy Groups in the SDSS DR4. III. The Luminosity and Stellar Mass Functions}}.
\bjtitle{ApJ}
\bvolume{695}(\bissue{2}),
\bfpage{900}--\blpage{916}
(\byear{2009})
\doiurl{10.1088/0004-637X/695/2/900}
{\href{https://arxiv.org/abs/0808.0539}{{arXiv:0808.0539}}}
{[astro-ph]}
\end{barticle}
\endbibitem

\bibitem[\protect\citeauthoryear{{Lan} et~al.}{2016}]{Lan2016}
\begin{barticle}
\bauthor{\bsnm{{Lan}}, \binits{T.-W.}},
\bauthor{\bsnm{{M{\'e}nard}}, \binits{B.}},
\bauthor{\bsnm{{Mo}}, \binits{H.}}:
\batitle{{The galaxy luminosity function in groups and clusters: the faint-end upturn and the connection to the field luminosity function}}.
\bjtitle{MNRAS}
\bvolume{459}(\bissue{4}),
\bfpage{3998}--\blpage{4019}
(\byear{2016})
\doiurl{10.1093/mnras/stw898}
{\href{https://arxiv.org/abs/1509.07870}{{arXiv:1509.07870}}}
{[astro-ph.GA]}
\end{barticle}
\endbibitem

\bibitem[\protect\citeauthoryear{{Blanton} et~al.}{2005}]{Blanton2005}
\begin{barticle}
\bauthor{\bsnm{{Blanton}}, \binits{M.R.}},
\bauthor{\bsnm{{Eisenstein}}, \binits{D.}},
\bauthor{\bsnm{{Hogg}}, \binits{D.W.}},
\bauthor{\bsnm{{Schlegel}}, \binits{D.J.}},
\bauthor{\bsnm{{Brinkmann}}, \binits{J.}}:
\batitle{{Relationship between Environment and the Broadband Optical Properties of Galaxies in the Sloan Digital Sky Survey}}.
\bjtitle{ApJ}
\bvolume{629}(\bissue{1}),
\bfpage{143}--\blpage{157}
(\byear{2005})
\doiurl{10.1086/422897}
{\href{https://arxiv.org/abs/astro-ph/0310453}{{arXiv:astro-ph/0310453}}}
{[astro-ph]}
\end{barticle}
\endbibitem

\bibitem[\protect\citeauthoryear{{Baldry} et~al.}{2006}]{Baldry2006}
\begin{barticle}
\bauthor{\bsnm{{Baldry}}, \binits{I.K.}},
\bauthor{\bsnm{{Balogh}}, \binits{M.L.}},
\bauthor{\bsnm{{Bower}}, \binits{R.G.}},
\bauthor{\bsnm{{Glazebrook}}, \binits{K.}},
\bauthor{\bsnm{{Nichol}}, \binits{R.C.}},
\bauthor{\bsnm{{Bamford}}, \binits{S.P.}},
\bauthor{\bsnm{{Budavari}}, \binits{T.}}:
\batitle{{Galaxy bimodality versus stellar mass and environment}}.
\bjtitle{\mnras}
\bvolume{373}(\bissue{2}),
\bfpage{469}--\blpage{483}
(\byear{2006})
\doiurl{10.1111/j.1365-2966.2006.11081.x}
{\href{https://arxiv.org/abs/astro-ph/0607648}{{arXiv:astro-ph/0607648}}}
{[astro-ph]}
\end{barticle}
\endbibitem

\bibitem[\protect\citeauthoryear{{Conselice}}{2014}]{Conselice2014}
\begin{barticle}
\bauthor{\bsnm{{Conselice}}, \binits{C.J.}}:
\batitle{{The Evolution of Galaxy Structure Over Cosmic Time}}.
\bjtitle{ARA\&A}
\bvolume{52},
\bfpage{291}--\blpage{337}
(\byear{2014})
\doiurl{10.1146/annurev-astro-081913-040037}
{\href{https://arxiv.org/abs/1403.2783}{{arXiv:1403.2783}}}
{[astro-ph.GA]}
\end{barticle}
\endbibitem

\bibitem[\protect\citeauthoryear{{Masters} et~al.}{2010}]{Masters2010}
\begin{barticle}
\bauthor{\bsnm{{Masters}}, \binits{K.L.}},
\bauthor{\bsnm{{Mosleh}}, \binits{M.}},
\bauthor{\bsnm{{Romer}}, \binits{A.K.}},
\bauthor{\bsnm{{Nichol}}, \binits{R.C.}},
\bauthor{\bsnm{{Bamford}}, \binits{S.P.}},
\bauthor{\bsnm{{Schawinski}}, \binits{K.}},
\bauthor{\bsnm{{Lintott}}, \binits{C.J.}},
\bauthor{\bsnm{{Andreescu}}, \binits{D.}},
\bauthor{\bsnm{{Campbell}}, \binits{H.C.}},
\bauthor{\bsnm{{Crowcroft}}, \binits{B.}},
\bauthor{\bsnm{{Doyle}}, \binits{I.}},
\bauthor{\bsnm{{Edmondson}}, \binits{E.M.}},
\bauthor{\bsnm{{Murray}}, \binits{P.}},
\bauthor{\bsnm{{Raddick}}, \binits{M.J.}},
\bauthor{\bsnm{{Slosar}}, \binits{A.}},
\bauthor{\bsnm{{Szalay}}, \binits{A.S.}},
\bauthor{\bsnm{{Vandenberg}}, \binits{J.}}:
\batitle{{Galaxy Zoo: passive red spirals}}.
\bjtitle{MNRAS}
\bvolume{405}(\bissue{2}),
\bfpage{783}--\blpage{799}
(\byear{2010})
\doiurl{10.1111/j.1365-2966.2010.16503.x}
{\href{https://arxiv.org/abs/0910.4113}{{arXiv:0910.4113}}}
{[astro-ph.CO]}
\end{barticle}
\endbibitem

\bibitem[\protect\citeauthoryear{{Strateva} et~al.}{2001}]{Strateva2001}
\begin{barticle}
\bauthor{\bsnm{{Strateva}}, \binits{I.}},
\bauthor{\bsnm{{Ivezi{\'c}}}, \binits{{\v{Z}}.}},
\bauthor{\bsnm{{Knapp}}, \binits{G.R.}},
\bauthor{\bsnm{{Narayanan}}, \binits{V.K.}},
\bauthor{\bsnm{{Strauss}}, \binits{M.A.}},
\bauthor{\bsnm{{Gunn}}, \binits{J.E.}},
\bauthor{\bsnm{{Lupton}}, \binits{R.H.}},
\bauthor{\bsnm{{Schlegel}}, \binits{D.}},
\bauthor{\bsnm{{Bahcall}}, \binits{N.A.}},
\bauthor{\bsnm{{Brinkmann}}, \binits{J.}},
\bauthor{\bsnm{{Brunner}}, \binits{R.J.}},
\bauthor{\bsnm{{Budav{\'a}ri}}, \binits{T.}},
\bauthor{\bsnm{{Csabai}}, \binits{I.}},
\bauthor{\bsnm{{Castander}}, \binits{F.J.}},
\bauthor{\bsnm{{Doi}}, \binits{M.}},
\bauthor{\bsnm{{Fukugita}}, \binits{M.}},
\bauthor{\bsnm{{Gy{\H{o}}ry}}, \binits{Z.}},
\bauthor{\bsnm{{Hamabe}}, \binits{M.}},
\bauthor{\bsnm{{Hennessy}}, \binits{G.}},
\bauthor{\bsnm{{Ichikawa}}, \binits{T.}},
\bauthor{\bsnm{{Kunszt}}, \binits{P.Z.}},
\bauthor{\bsnm{{Lamb}}, \binits{D.Q.}},
\bauthor{\bsnm{{McKay}}, \binits{T.A.}},
\bauthor{\bsnm{{Okamura}}, \binits{S.}},
\bauthor{\bsnm{{Racusin}}, \binits{J.}},
\bauthor{\bsnm{{Sekiguchi}}, \binits{M.}},
\bauthor{\bsnm{{Schneider}}, \binits{D.P.}},
\bauthor{\bsnm{{Shimasaku}}, \binits{K.}},
\bauthor{\bsnm{{York}}, \binits{D.}}:
\batitle{{Color Separation of Galaxy Types in the Sloan Digital Sky Survey Imaging Data}}.
\bjtitle{AJ}
\bvolume{122}(\bissue{4}),
\bfpage{1861}--\blpage{1874}
(\byear{2001})
\doiurl{10.1086/323301}
{\href{https://arxiv.org/abs/astro-ph/0107201}{{arXiv:astro-ph/0107201}}}
{[astro-ph]}
\end{barticle}
\endbibitem

\bibitem[\protect\citeauthoryear{{Baldry} et~al.}{2004}]{Baldry2004}
\begin{barticle}
\bauthor{\bsnm{{Baldry}}, \binits{I.K.}},
\bauthor{\bsnm{{Glazebrook}}, \binits{K.}},
\bauthor{\bsnm{{Brinkmann}}, \binits{J.}},
\bauthor{\bsnm{{Ivezi{\'c}}}, \binits{{\v{Z}}.}},
\bauthor{\bsnm{{Lupton}}, \binits{R.H.}},
\bauthor{\bsnm{{Nichol}}, \binits{R.C.}},
\bauthor{\bsnm{{Szalay}}, \binits{A.S.}}:
\batitle{{Quantifying the Bimodal Color-Magnitude Distribution of Galaxies}}.
\bjtitle{ApJ}
\bvolume{600}(\bissue{2}),
\bfpage{681}--\blpage{694}
(\byear{2004})
\doiurl{10.1086/380092}
{\href{https://arxiv.org/abs/astro-ph/0309710}{{arXiv:astro-ph/0309710}}}
{[astro-ph]}
\end{barticle}
\endbibitem

\bibitem[\protect\citeauthoryear{{Peng} et~al.}{2010}]{Peng2010}
\begin{barticle}
\bauthor{\bsnm{{Peng}}, \binits{Y.-j.}},
\bauthor{\bsnm{{Lilly}}, \binits{S.J.}},
\bauthor{\bsnm{{Kova{\v{c}}}}, \binits{K.}},
\bauthor{\bsnm{{Bolzonella}}, \binits{M.}},
\bauthor{\bsnm{{Pozzetti}}, \binits{L.}},
\bauthor{\bsnm{{Renzini}}, \binits{A.}},
\bauthor{\bsnm{{Zamorani}}, \binits{G.}},
\bauthor{\bsnm{{Ilbert}}, \binits{O.}},
\bauthor{\bsnm{{Knobel}}, \binits{C.}},
\bauthor{\bsnm{{Iovino}}, \binits{A.}},
\bauthor{\bsnm{{Maier}}, \binits{C.}},
\bauthor{\bsnm{{Cucciati}}, \binits{O.}},
\bauthor{\bsnm{{Tasca}}, \binits{L.}},
\bauthor{\bsnm{{Carollo}}, \binits{C.M.}},
\bauthor{\bsnm{{Silverman}}, \binits{J.}},
\bauthor{\bsnm{{Kampczyk}}, \binits{P.}},
\bauthor{\bsnm{{de Ravel}}, \binits{L.}},
\bauthor{\bsnm{{Sanders}}, \binits{D.}},
\bauthor{\bsnm{{Scoville}}, \binits{N.}},
\bauthor{\bsnm{{Contini}}, \binits{T.}},
\bauthor{\bsnm{{Mainieri}}, \binits{V.}},
\bauthor{\bsnm{{Scodeggio}}, \binits{M.}},
\bauthor{\bsnm{{Kneib}}, \binits{J.-P.}},
\bauthor{\bsnm{{Le F{\`e}vre}}, \binits{O.}},
\bauthor{\bsnm{{Bardelli}}, \binits{S.}},
\bauthor{\bsnm{{Bongiorno}}, \binits{A.}},
\bauthor{\bsnm{{Caputi}}, \binits{K.}},
\bauthor{\bsnm{{Coppa}}, \binits{G.}},
\bauthor{\bsnm{{de la Torre}}, \binits{S.}},
\bauthor{\bsnm{{Franzetti}}, \binits{P.}},
\bauthor{\bsnm{{Garilli}}, \binits{B.}},
\bauthor{\bsnm{{Lamareille}}, \binits{F.}},
\bauthor{\bsnm{{Le Borgne}}, \binits{J.-F.}},
\bauthor{\bsnm{{Le Brun}}, \binits{V.}},
\bauthor{\bsnm{{Mignoli}}, \binits{M.}},
\bauthor{\bsnm{{Perez Montero}}, \binits{E.}},
\bauthor{\bsnm{{Pello}}, \binits{R.}},
\bauthor{\bsnm{{Ricciardelli}}, \binits{E.}},
\bauthor{\bsnm{{Tanaka}}, \binits{M.}},
\bauthor{\bsnm{{Tresse}}, \binits{L.}},
\bauthor{\bsnm{{Vergani}}, \binits{D.}},
\bauthor{\bsnm{{Welikala}}, \binits{N.}},
\bauthor{\bsnm{{Zucca}}, \binits{E.}},
\bauthor{\bsnm{{Oesch}}, \binits{P.}},
\bauthor{\bsnm{{Abbas}}, \binits{U.}},
\bauthor{\bsnm{{Barnes}}, \binits{L.}},
\bauthor{\bsnm{{Bordoloi}}, \binits{R.}},
\bauthor{\bsnm{{Bottini}}, \binits{D.}},
\bauthor{\bsnm{{Cappi}}, \binits{A.}},
\bauthor{\bsnm{{Cassata}}, \binits{P.}},
\bauthor{\bsnm{{Cimatti}}, \binits{A.}},
\bauthor{\bsnm{{Fumana}}, \binits{M.}},
\bauthor{\bsnm{{Hasinger}}, \binits{G.}},
\bauthor{\bsnm{{Koekemoer}}, \binits{A.}},
\bauthor{\bsnm{{Leauthaud}}, \binits{A.}},
\bauthor{\bsnm{{Maccagni}}, \binits{D.}},
\bauthor{\bsnm{{Marinoni}}, \binits{C.}},
\bauthor{\bsnm{{McCracken}}, \binits{H.}},
\bauthor{\bsnm{{Memeo}}, \binits{P.}},
\bauthor{\bsnm{{Meneux}}, \binits{B.}},
\bauthor{\bsnm{{Nair}}, \binits{P.}},
\bauthor{\bsnm{{Porciani}}, \binits{C.}},
\bauthor{\bsnm{{Presotto}}, \binits{V.}},
\bauthor{\bsnm{{Scaramella}}, \binits{R.}}:
\batitle{{Mass and Environment as Drivers of Galaxy Evolution in SDSS and zCOSMOS and the Origin of the Schechter Function}}.
\bjtitle{ApJ}
\bvolume{721}(\bissue{1}),
\bfpage{193}--\blpage{221}
(\byear{2010})
\doiurl{10.1088/0004-637X/721/1/193}
{\href{https://arxiv.org/abs/1003.4747}{{arXiv:1003.4747}}}
{[astro-ph.CO]}
\end{barticle}
\endbibitem

\bibitem[\protect\citeauthoryear{{Wang} et~al.}{2018}]{Wang2018}
\begin{barticle}
\bauthor{\bsnm{{Wang}}, \binits{H.}},
\bauthor{\bsnm{{Mo}}, \binits{H.J.}},
\bauthor{\bsnm{{Chen}}, \binits{S.}},
\bauthor{\bsnm{{Yang}}, \binits{Y.}},
\bauthor{\bsnm{{Yang}}, \binits{X.}},
\bauthor{\bsnm{{Wang}}, \binits{E.}},
\bauthor{\bsnm{{van den Bosch}}, \binits{F.C.}},
\bauthor{\bsnm{{Jing}}, \binits{Y.}},
\bauthor{\bsnm{{Kang}}, \binits{X.}},
\bauthor{\bsnm{{Lin}}, \binits{W.}},
\bauthor{\bsnm{{Lim}}, \binits{S.H.}},
\bauthor{\bsnm{{Huang}}, \binits{S.}},
\bauthor{\bsnm{{Lu}}, \binits{Y.}},
\bauthor{\bsnm{{Li}}, \binits{S.}},
\bauthor{\bsnm{{Cui}}, \binits{W.}},
\bauthor{\bsnm{{Zhang}}, \binits{Y.}},
\bauthor{\bsnm{{Tweed}}, \binits{D.}},
\bauthor{\bsnm{{Wei}}, \binits{C.}},
\bauthor{\bsnm{{Li}}, \binits{G.}},
\bauthor{\bsnm{{Shi}}, \binits{F.}}:
\batitle{{ELUCID. IV. Galaxy Quenching and its Relation to Halo Mass, Environment, and Assembly Bias}}.
\bjtitle{ApJ}
\bvolume{852}(\bissue{1}),
\bfpage{31}
(\byear{2018})
\doiurl{10.3847/1538-4357/aa9e01}
{\href{https://arxiv.org/abs/1707.09002}{{arXiv:1707.09002}}}
{[astro-ph.GA]}
\end{barticle}
\endbibitem

\bibitem[\protect\citeauthoryear{{Yang} et~al.}{2007}]{yang2007galaxy}
\begin{barticle}
\bauthor{\bsnm{{Yang}}, \binits{X.}},
\bauthor{\bsnm{{Mo}}, \binits{H.J.}},
\bauthor{\bsnm{{van den Bosch}}, \binits{F.C.}},
\bauthor{\bsnm{{Pasquali}}, \binits{A.}},
\bauthor{\bsnm{{Li}}, \binits{C.}},
\bauthor{\bsnm{{Barden}}, \binits{M.}}:
\batitle{{Galaxy Groups in the SDSS DR4. I. The Catalog and Basic Properties}}.
\bjtitle{ApJ}
\bvolume{671}(\bissue{1}),
\bfpage{153}--\blpage{170}
(\byear{2007})
\doiurl{10.1086/522027}
{\href{https://arxiv.org/abs/0707.4640}{{arXiv:0707.4640}}}
{[astro-ph]}
\end{barticle}
\endbibitem

\bibitem[\protect\citeauthoryear{{Donnari} et~al.}{2021}]{Donnari2021}
\begin{barticle}
\bauthor{\bsnm{{Donnari}}, \binits{M.}},
\bauthor{\bsnm{{Pillepich}}, \binits{A.}},
\bauthor{\bsnm{{Nelson}}, \binits{D.}},
\bauthor{\bsnm{{Marinacci}}, \binits{F.}},
\bauthor{\bsnm{{Vogelsberger}}, \binits{M.}},
\bauthor{\bsnm{{Hernquist}}, \binits{L.}}:
\batitle{{Quenched fractions in the IllustrisTNG simulations: comparison with observations and other theoretical models}}.
\bjtitle{\mnras}
\bvolume{506}(\bissue{4}),
\bfpage{4760}--\blpage{4780}
(\byear{2021})
\doiurl{10.1093/mnras/stab1950}
{\href{https://arxiv.org/abs/2008.00004}{{arXiv:2008.00004}}}
{[astro-ph.GA]}
\end{barticle}
\endbibitem

\bibitem[\protect\citeauthoryear{{Geha} et~al.}{2012}]{Geha2012}
\begin{barticle}
\bauthor{\bsnm{{Geha}}, \binits{M.}},
\bauthor{\bsnm{{Blanton}}, \binits{M.R.}},
\bauthor{\bsnm{{Yan}}, \binits{R.}},
\bauthor{\bsnm{{Tinker}}, \binits{J.L.}}:
\batitle{{A Stellar Mass Threshold for Quenching of Field Galaxies}}.
\bjtitle{\apj}
\bvolume{757}(\bissue{1}),
\bfpage{85}
(\byear{2012})
\doiurl{10.1088/0004-637X/757/1/85}
{\href{https://arxiv.org/abs/1206.3573}{{arXiv:1206.3573}}}
{[astro-ph.CO]}
\end{barticle}
\endbibitem

\bibitem[\protect\citeauthoryear{{Schaefer} et~al.}{2019}]{Schaefer2019}
\begin{barticle}
\bauthor{\bsnm{{Schaefer}}, \binits{A.L.}},
\bauthor{\bsnm{{Croom}}, \binits{S.M.}},
\bauthor{\bsnm{{Scott}}, \binits{N.}},
\bauthor{\bsnm{{Brough}}, \binits{S.}},
\bauthor{\bsnm{{Allen}}, \binits{J.T.}},
\bauthor{\bsnm{{Bekki}}, \binits{K.}},
\bauthor{\bsnm{{Bland-Hawthorn}}, \binits{J.}},
\bauthor{\bsnm{{Bloom}}, \binits{J.V.}},
\bauthor{\bsnm{{Bryant}}, \binits{J.J.}},
\bauthor{\bsnm{{Cortese}}, \binits{L.}},
\bauthor{\bsnm{{Davies}}, \binits{L.J.M.}},
\bauthor{\bsnm{{Federrath}}, \binits{C.}},
\bauthor{\bsnm{{Fogarty}}, \binits{L.M.R.}},
\bauthor{\bsnm{{Green}}, \binits{A.W.}},
\bauthor{\bsnm{{Groves}}, \binits{B.}},
\bauthor{\bsnm{{Hopkins}}, \binits{A.M.}},
\bauthor{\bsnm{{Konstantopoulos}}, \binits{I.S.}},
\bauthor{\bsnm{{L{\'o}pez-S{\'a}nchez}}, \binits{A.R.}},
\bauthor{\bsnm{{Lawrence}}, \binits{J.S.}},
\bauthor{\bsnm{{McElroy}}, \binits{R.E.}},
\bauthor{\bsnm{{Medling}}, \binits{A.M.}},
\bauthor{\bsnm{{Owers}}, \binits{M.S.}},
\bauthor{\bsnm{{Pracy}}, \binits{M.B.}},
\bauthor{\bsnm{{Richards}}, \binits{S.N.}},
\bauthor{\bsnm{{Robotham}}, \binits{A.S.G.}},
\bauthor{\bsnm{{van de Sande}}, \binits{J.}},
\bauthor{\bsnm{{Tonini}}, \binits{C.}},
\bauthor{\bsnm{{Yi}}, \binits{S.K.}}:
\batitle{{The SAMI Galaxy Survey: observing the environmental quenching of star formation in GAMA groups}}.
\bjtitle{\mnras}
\bvolume{483}(\bissue{3}),
\bfpage{2851}--\blpage{2870}
(\byear{2019})
\doiurl{10.1093/mnras/sty3258}
{\href{https://arxiv.org/abs/1811.11676}{{arXiv:1811.11676}}}
{[astro-ph.GA]}
\end{barticle}
\endbibitem

\bibitem[\protect\citeauthoryear{{Yang} et~al.}{2021}]{yang2021extended}
\begin{barticle}
\bauthor{\bsnm{{Yang}}, \binits{X.}},
\bauthor{\bsnm{{Xu}}, \binits{H.}},
\bauthor{\bsnm{{He}}, \binits{M.}},
\bauthor{\bsnm{{Gu}}, \binits{Y.}},
\bauthor{\bsnm{{Katsianis}}, \binits{A.}},
\bauthor{\bsnm{{Meng}}, \binits{J.}},
\bauthor{\bsnm{{Shi}}, \binits{F.}},
\bauthor{\bsnm{{Zou}}, \binits{H.}},
\bauthor{\bsnm{{Zhang}}, \binits{Y.}},
\bauthor{\bsnm{{Liu}}, \binits{C.}},
\bauthor{\bsnm{{Wang}}, \binits{Z.}},
\bauthor{\bsnm{{Dong}}, \binits{F.}},
\bauthor{\bsnm{{Lu}}, \binits{Y.}},
\bauthor{\bsnm{{Li}}, \binits{Q.}},
\bauthor{\bsnm{{Chen}}, \binits{Y.}},
\bauthor{\bsnm{{Wang}}, \binits{H.}},
\bauthor{\bsnm{{Mo}}, \binits{H.}},
\bauthor{\bsnm{{Fu}}, \binits{J.}},
\bauthor{\bsnm{{Guo}}, \binits{H.}},
\bauthor{\bsnm{{Leauthaud}}, \binits{A.}},
\bauthor{\bsnm{{Luo}}, \binits{Y.}},
\bauthor{\bsnm{{Zhang}}, \binits{J.}},
\bauthor{\bsnm{{Zu}}, \binits{Y.}}:
\batitle{{An Extended Halo-based Group/Cluster Finder: Application to the DESI Legacy Imaging Surveys DR8}}.
\bjtitle{ApJ}
\bvolume{909}(\bissue{2}),
\bfpage{143}
(\byear{2021})
\doiurl{10.3847/1538-4357/abddb2}
{\href{https://arxiv.org/abs/2012.14998}{{arXiv:2012.14998}}}
{[astro-ph.GA]}
\end{barticle}
\endbibitem

\bibitem[\protect\citeauthoryear{{van den Bosch} et~al.}{2008}]{van2008}
\begin{barticle}
\bauthor{\bsnm{{van den Bosch}}, \binits{F.C.}},
\bauthor{\bsnm{{Aquino}}, \binits{D.}},
\bauthor{\bsnm{{Yang}}, \binits{X.}},
\bauthor{\bsnm{{Mo}}, \binits{H.J.}},
\bauthor{\bsnm{{Pasquali}}, \binits{A.}},
\bauthor{\bsnm{{McIntosh}}, \binits{D.H.}},
\bauthor{\bsnm{{Weinmann}}, \binits{S.M.}},
\bauthor{\bsnm{{Kang}}, \binits{X.}}:
\batitle{{The importance of satellite quenching for the build-up of the red sequence of present-day galaxies}}.
\bjtitle{\mnras}
\bvolume{387}(\bissue{1}),
\bfpage{79}--\blpage{91}
(\byear{2008})
\doiurl{10.1111/j.1365-2966.2008.13230.x}
{\href{https://arxiv.org/abs/0710.3164}{{arXiv:0710.3164}}}
{[astro-ph]}
\end{barticle}
\endbibitem

\bibitem[\protect\citeauthoryear{Tully et~al.}{2019}]{Tully2019}
\begin{barticle}
\bauthor{\bsnm{Tully}, \binits{R.B.}},
\bauthor{\bsnm{Pomarède}, \binits{D.}},
\bauthor{\bsnm{Graziani}, \binits{R.}},
\bauthor{\bsnm{Courtois}, \binits{H.M.}},
\bauthor{\bsnm{Hoffman}, \binits{Y.}},
\bauthor{\bsnm{Shaya}, \binits{E.J.}}:
\batitle{Cosmicflows-3: Cosmography of the local void}.
\bjtitle{The Astrophysical Journal}
\bvolume{880}(\bissue{1}),
\bfpage{24}
(\byear{2019})
\doiurl{10.3847/1538-4357/ab2597}
\end{barticle}
\endbibitem

\bibitem[\protect\citeauthoryear{{Wang} et~al.}{2016}]{Wang2016}
\begin{barticle}
\bauthor{\bsnm{{Wang}}, \binits{H.}},
\bauthor{\bsnm{{Mo}}, \binits{H.J.}},
\bauthor{\bsnm{{Yang}}, \binits{X.}},
\bauthor{\bsnm{{Zhang}}, \binits{Y.}},
\bauthor{\bsnm{{Shi}}, \binits{J.}},
\bauthor{\bsnm{{Jing}}, \binits{Y.P.}},
\bauthor{\bsnm{{Liu}}, \binits{C.}},
\bauthor{\bsnm{{Li}}, \binits{S.}},
\bauthor{\bsnm{{Kang}}, \binits{X.}},
\bauthor{\bsnm{{Gao}}, \binits{Y.}}:
\batitle{{ELUCID - Exploring the Local Universe with ReConstructed Initial Density Field III: Constrained Simulation in the SDSS Volume}}.
\bjtitle{\apj}
\bvolume{831}(\bissue{2}),
\bfpage{164}
(\byear{2016})
\doiurl{10.3847/0004-637X/831/2/164}
{\href{https://arxiv.org/abs/1608.01763}{{arXiv:1608.01763}}}
{[astro-ph.CO]}
\end{barticle}
\endbibitem

\bibitem[\protect\citeauthoryear{{Yang} et~al.}{2005}]{yang2005galoccup}
\begin{barticle}
\bauthor{\bsnm{{Yang}}, \binits{X.}},
\bauthor{\bsnm{{Mo}}, \binits{H.J.}},
\bauthor{\bsnm{{Jing}}, \binits{Y.P.}},
\bauthor{\bsnm{{van den Bosch}}, \binits{F.C.}}:
\batitle{{Galaxy occupation statistics of dark matter haloes: observational results}}.
\bjtitle{MNRAS}
\bvolume{358}(\bissue{1}),
\bfpage{217}--\blpage{232}
(\byear{2005})
\doiurl{10.1111/j.1365-2966.2005.08801.x}
{\href{https://arxiv.org/abs/astro-ph/0410114}{{arXiv:astro-ph/0410114}}}
{[astro-ph]}
\end{barticle}
\endbibitem

\bibitem[\protect\citeauthoryear{{Wang} et~al.}{2024}]{Wang2024}
\begin{barticle}
\bauthor{\bsnm{{Wang}}, \binits{Y.}},
\bauthor{\bsnm{{Yang}}, \binits{X.}},
\bauthor{\bsnm{{Gu}}, \binits{Y.}},
\bauthor{\bsnm{{Xu}}, \binits{X.}},
\bauthor{\bsnm{{Xu}}, \binits{H.}},
\bauthor{\bsnm{{Wang}}, \binits{Y.}},
\bauthor{\bsnm{{Katsianis}}, \binits{A.}},
\bauthor{\bsnm{{Han}}, \binits{J.}},
\bauthor{\bsnm{{He}}, \binits{M.}},
\bauthor{\bsnm{{Zheng}}, \binits{Y.}},
\bauthor{\bsnm{{Li}}, \binits{Q.}},
\bauthor{\bsnm{{Wang}}, \binits{Y.}},
\bauthor{\bsnm{{Hong}}, \binits{W.}},
\bauthor{\bsnm{{Wang}}, \binits{J.}},
\bauthor{\bsnm{{Tan}}, \binits{Z.}},
\bauthor{\bsnm{{Zou}}, \binits{H.}},
\bauthor{\bsnm{{Lange}}, \binits{J.U.}},
\bauthor{\bsnm{{Hahn}}, \binits{C.}},
\bauthor{\bsnm{{Behroozi}}, \binits{P.}},
\bauthor{\bsnm{{Aguilar}}, \binits{J.N.}},
\bauthor{\bsnm{{Ahlen}}, \binits{S.}},
\bauthor{\bsnm{{Brooks}}, \binits{D.}},
\bauthor{\bsnm{{Claybaugh}}, \binits{T.}},
\bauthor{\bsnm{{Cole}}, \binits{S.}},
\bauthor{\bsnm{{de la Macorra}}, \binits{A.}},
\bauthor{\bsnm{{Dey}}, \binits{B.}},
\bauthor{\bsnm{{Doel}}, \binits{P.}},
\bauthor{\bsnm{{Forero-Romero}}, \binits{J.E.}},
\bauthor{\bsnm{{Honscheid}}, \binits{K.}},
\bauthor{\bsnm{{Kehoe}}, \binits{R.}},
\bauthor{\bsnm{{Kisner}}, \binits{T.}},
\bauthor{\bsnm{{Lambert}}, \binits{A.}},
\bauthor{\bsnm{{Manera}}, \binits{M.}},
\bauthor{\bsnm{{Meisner}}, \binits{A.}},
\bauthor{\bsnm{{Miquel}}, \binits{R.}},
\bauthor{\bsnm{{Moustakas}}, \binits{J.}},
\bauthor{\bsnm{{Nie}}, \binits{J.}},
\bauthor{\bsnm{{Poppett}}, \binits{C.}},
\bauthor{\bsnm{{Rezaie}}, \binits{M.}},
\bauthor{\bsnm{{Rossi}}, \binits{G.}},
\bauthor{\bsnm{{Sanchez}}, \binits{E.}},
\bauthor{\bsnm{{Schubnell}}, \binits{M.}},
\bauthor{\bsnm{{Tarl{\'e}}}, \binits{G.}},
\bauthor{\bsnm{{Weaver}}, \binits{B.A.}},
\bauthor{\bsnm{{Zhou}}, \binits{Z.}}:
\batitle{{Measuring the Conditional Luminosity and Stellar Mass Functions of Galaxies by Combining the Dark Energy Spectroscopic Instrument Legacy Imaging Surveys Data Release 9, Survey Validation 3, and Year 1 Data}}.
\bjtitle{ApJ}
\bvolume{971}(\bissue{1}),
\bfpage{119}
(\byear{2024})
\doiurl{10.3847/1538-4357/ad5294}
{\href{https://arxiv.org/abs/2312.17459}{{arXiv:2312.17459}}}
{[astro-ph.GA]}
\end{barticle}
\endbibitem

\bibitem[\protect\citeauthoryear{{Wang} et~al.}{2025}]{Wang2025}
\begin{barticle}
\bauthor{\bsnm{{Wang}}, \binits{W.}},
\bauthor{\bsnm{{Yang}}, \binits{X.}},
\bauthor{\bsnm{{Jing}}, \binits{Y.}},
\bauthor{\bsnm{{Ross}}, \binits{A.J.}},
\bauthor{\bsnm{{Siudek}}, \binits{M.}},
\bauthor{\bsnm{{Moustakas}}, \binits{J.}},
\bauthor{\bsnm{{Moore}}, \binits{S.G.}},
\bauthor{\bsnm{{Cole}}, \binits{S.}},
\bauthor{\bsnm{{Frenk}}, \binits{C.}},
\bauthor{\bsnm{{Yu}}, \binits{J.}},
\bauthor{\bsnm{{Koposov}}, \binits{S.E.}},
\bauthor{\bsnm{{Han}}, \binits{J.}},
\bauthor{\bsnm{{Tan}}, \binits{Z.}},
\bauthor{\bsnm{{Xu}}, \binits{K.}},
\bauthor{\bsnm{{Gu}}, \binits{Y.}},
\bauthor{\bsnm{{Wang}}, \binits{Y.}},
\bauthor{\bsnm{{Gnedin}}, \binits{O.Y.}},
\bauthor{\bsnm{{Aguilar}}, \binits{J.N.}},
\bauthor{\bsnm{{Ahlen}}, \binits{S.}},
\bauthor{\bsnm{{Bianchi}}, \binits{D.}},
\bauthor{\bsnm{{Brooks}}, \binits{D.}},
\bauthor{\bsnm{{Claybaugh}}, \binits{T.}},
\bauthor{\bsnm{{de la Macorra}}, \binits{A.}},
\bauthor{\bsnm{{Dey}}, \binits{A.}},
\bauthor{\bsnm{{Doel}}, \binits{P.}},
\bauthor{\bsnm{{Forero-Romero}}, \binits{J.E.}},
\bauthor{\bsnm{{Gazta{\~n}aga}}, \binits{E.}},
\bauthor{\bsnm{{Gontcho A Gontcho}}, \binits{S.}},
\bauthor{\bsnm{{Gutierrez}}, \binits{G.}},
\bauthor{\bsnm{{Honscheid}}, \binits{K.}},
\bauthor{\bsnm{{Ishak}}, \binits{M.}},
\bauthor{\bsnm{{Kisner}}, \binits{T.}},
\bauthor{\bsnm{{Landriau}}, \binits{M.}},
\bauthor{\bsnm{{Le Guillou}}, \binits{L.}},
\bauthor{\bsnm{{Manera}}, \binits{M.}},
\bauthor{\bsnm{{Meisner}}, \binits{A.}},
\bauthor{\bsnm{{Miquel}}, \binits{R.}},
\bauthor{\bsnm{{Nadathur}}, \binits{S.}},
\bauthor{\bsnm{{Poppett}}, \binits{C.}},
\bauthor{\bsnm{{Prada}}, \binits{F.}},
\bauthor{\bsnm{{P{\'e}rez-R{\`a}fols}}, \binits{I.}},
\bauthor{\bsnm{{Rossi}}, \binits{G.}},
\bauthor{\bsnm{{Sanchez}}, \binits{E.}},
\bauthor{\bsnm{{Schlegel}}, \binits{D.}},
\bauthor{\bsnm{{Seo}}, \binits{H.-J.}},
\bauthor{\bsnm{{Silber}}, \binits{J.H.}},
\bauthor{\bsnm{{Sprayberry}}, \binits{D.}},
\bauthor{\bsnm{{Tarl{\'e}}}, \binits{G.}},
\bauthor{\bsnm{{Weaver}}, \binits{B.A.}},
\bauthor{\bsnm{{Zou}}, \binits{H.}}:
\batitle{{Luminosity and Stellar Mass Functions of Faint Photometric Satellites around Spectroscopic Central Galaxies from DESI Year-1 Bright Galaxy Survey}}.
\bjtitle{\apj}
\bvolume{986}(\bissue{2}),
\bfpage{218}
(\byear{2025})
\doiurl{10.3847/1538-4357/add5df}
{\href{https://arxiv.org/abs/2503.03317}{{arXiv:2503.03317}}}
{[astro-ph.GA]}
\end{barticle}
\endbibitem

\bibitem[\protect\citeauthoryear{{Vale} and {Ostriker}}{2004}]{Vale2004}
\begin{barticle}
\bauthor{\bsnm{{Vale}}, \binits{A.}},
\bauthor{\bsnm{{Ostriker}}, \binits{J.P.}}:
\batitle{{Linking halo mass to galaxy luminosity}}.
\bjtitle{MNRAS}
\bvolume{353}(\bissue{1}),
\bfpage{189}--\blpage{200}
(\byear{2004})
\doiurl{10.1111/j.1365-2966.2004.08059.x}
{\href{https://arxiv.org/abs/astro-ph/0402500}{{arXiv:astro-ph/0402500}}}
{[astro-ph]}
\end{barticle}
\endbibitem

\bibitem[\protect\citeauthoryear{{Wechsler} and {Tinker}}{2018}]{Wechsler2018}
\begin{barticle}
\bauthor{\bsnm{{Wechsler}}, \binits{R.H.}},
\bauthor{\bsnm{{Tinker}}, \binits{J.L.}}:
\batitle{{The Connection Between Galaxies and Their Dark Matter Halos}}.
\bjtitle{ARA\&A}
\bvolume{56},
\bfpage{435}--\blpage{487}
(\byear{2018})
\doiurl{10.1146/annurev-astro-081817-051756}
{\href{https://arxiv.org/abs/1804.03097}{{arXiv:1804.03097}}}
{[astro-ph.GA]}
\end{barticle}
\endbibitem

\bibitem[\protect\citeauthoryear{{Chen} et~al.}{2019}]{Chen2019}
\begin{barticle}
\bauthor{\bsnm{{Chen}}, \binits{Y.}},
\bauthor{\bsnm{{Mo}}, \binits{H.J.}},
\bauthor{\bsnm{{Li}}, \binits{C.}},
\bauthor{\bsnm{{Wang}}, \binits{H.}},
\bauthor{\bsnm{{Yang}}, \binits{X.}},
\bauthor{\bsnm{{Zhou}}, \binits{S.}},
\bauthor{\bsnm{{Zhang}}, \binits{Y.}}:
\batitle{{ELUCID. VI. Cosmic Variance of the Galaxy Distribution in the Local Universe}}.
\bjtitle{ApJ}
\bvolume{872}(\bissue{2}),
\bfpage{180}
(\byear{2019})
\doiurl{10.3847/1538-4357/ab0208}
{\href{https://arxiv.org/abs/1809.00523}{{arXiv:1809.00523}}}
{[astro-ph.GA]}
\end{barticle}
\endbibitem

\bibitem[\protect\citeauthoryear{{Colless}}{1999}]{Colless1999}
\begin{barticle}
\bauthor{\bsnm{{Colless}}, \binits{M.}}:
\batitle{{First results from the 2dF Galaxy Redshift Survey}}.
\bjtitle{Philosophical Transactions of the Royal Society of London Series A}
\bvolume{357}(\bissue{1750}),
\bfpage{105}
(\byear{1999})
\doiurl{10.1098/rsta.1999.0317}
{\href{https://arxiv.org/abs/astro-ph/9804079}{{arXiv:astro-ph/9804079}}}
{[astro-ph]}
\end{barticle}
\endbibitem

\bibitem[\protect\citeauthoryear{{York} et~al.}{2000}]{York2000}
\begin{barticle}
\bauthor{\bsnm{{York}}, \binits{D.G.}},
\bauthor{\bsnm{{Adelman}}, \binits{J.}},
\bauthor{\bsnm{{Anderson}}, \binits{J.} \bsuffix{John~E.}},
\bauthor{\bsnm{{Anderson}}, \binits{S.F.}},
\bauthor{\bsnm{{Annis}}, \binits{J.}},
\bauthor{\bsnm{{Bahcall}}, \binits{N.A.}},
\bauthor{\bsnm{{Bakken}}, \binits{J.A.}},
\bauthor{\bsnm{{Barkhouser}}, \binits{R.}},
\bauthor{\bsnm{{Bastian}}, \binits{S.}},
\bauthor{\bsnm{{Berman}}, \binits{E.}},
\bauthor{\bsnm{{Boroski}}, \binits{W.N.}},
\bauthor{\bsnm{{Bracker}}, \binits{S.}},
\bauthor{\bsnm{{Briegel}}, \binits{C.}},
\bauthor{\bsnm{{Briggs}}, \binits{J.W.}},
\bauthor{\bsnm{{Brinkmann}}, \binits{J.}},
\bauthor{\bsnm{{Brunner}}, \binits{R.}},
\bauthor{\bsnm{{Burles}}, \binits{S.}},
\bauthor{\bsnm{{Carey}}, \binits{L.}},
\bauthor{\bsnm{{Carr}}, \binits{M.A.}},
\bauthor{\bsnm{{Castander}}, \binits{F.J.}},
\bauthor{\bsnm{{Chen}}, \binits{B.}},
\bauthor{\bsnm{{Colestock}}, \binits{P.L.}},
\bauthor{\bsnm{{Connolly}}, \binits{A.J.}},
\bauthor{\bsnm{{Crocker}}, \binits{J.H.}},
\bauthor{\bsnm{{Csabai}}, \binits{I.}},
\bauthor{\bsnm{{Czarapata}}, \binits{P.C.}},
\bauthor{\bsnm{{Davis}}, \binits{J.E.}},
\bauthor{\bsnm{{Doi}}, \binits{M.}},
\bauthor{\bsnm{{Dombeck}}, \binits{T.}},
\bauthor{\bsnm{{Eisenstein}}, \binits{D.}},
\bauthor{\bsnm{{Ellman}}, \binits{N.}},
\bauthor{\bsnm{{Elms}}, \binits{B.R.}},
\bauthor{\bsnm{{Evans}}, \binits{M.L.}},
\bauthor{\bsnm{{Fan}}, \binits{X.}},
\bauthor{\bsnm{{Federwitz}}, \binits{G.R.}},
\bauthor{\bsnm{{Fiscelli}}, \binits{L.}},
\bauthor{\bsnm{{Friedman}}, \binits{S.}},
\bauthor{\bsnm{{Frieman}}, \binits{J.A.}},
\bauthor{\bsnm{{Fukugita}}, \binits{M.}},
\bauthor{\bsnm{{Gillespie}}, \binits{B.}},
\bauthor{\bsnm{{Gunn}}, \binits{J.E.}},
\bauthor{\bsnm{{Gurbani}}, \binits{V.K.}},
\bauthor{\bsnm{{de Haas}}, \binits{E.}},
\bauthor{\bsnm{{Haldeman}}, \binits{M.}},
\bauthor{\bsnm{{Harris}}, \binits{F.H.}},
\bauthor{\bsnm{{Hayes}}, \binits{J.}},
\bauthor{\bsnm{{Heckman}}, \binits{T.M.}},
\bauthor{\bsnm{{Hennessy}}, \binits{G.S.}},
\bauthor{\bsnm{{Hindsley}}, \binits{R.B.}},
\bauthor{\bsnm{{Holm}}, \binits{S.}},
\bauthor{\bsnm{{Holmgren}}, \binits{D.J.}},
\bauthor{\bsnm{{Huang}}, \binits{C.-h.}},
\bauthor{\bsnm{{Hull}}, \binits{C.}},
\bauthor{\bsnm{{Husby}}, \binits{D.}},
\bauthor{\bsnm{{Ichikawa}}, \binits{S.-I.}},
\bauthor{\bsnm{{Ichikawa}}, \binits{T.}},
\bauthor{\bsnm{{Ivezi{\'c}}}, \binits{{\v{Z}}.}},
\bauthor{\bsnm{{Kent}}, \binits{S.}},
\bauthor{\bsnm{{Kim}}, \binits{R.S.J.}},
\bauthor{\bsnm{{Kinney}}, \binits{E.}},
\bauthor{\bsnm{{Klaene}}, \binits{M.}},
\bauthor{\bsnm{{Kleinman}}, \binits{A.N.}},
\bauthor{\bsnm{{Kleinman}}, \binits{S.}},
\bauthor{\bsnm{{Knapp}}, \binits{G.R.}},
\bauthor{\bsnm{{Korienek}}, \binits{J.}},
\bauthor{\bsnm{{Kron}}, \binits{R.G.}},
\bauthor{\bsnm{{Kunszt}}, \binits{P.Z.}},
\bauthor{\bsnm{{Lamb}}, \binits{D.Q.}},
\bauthor{\bsnm{{Lee}}, \binits{B.}},
\bauthor{\bsnm{{Leger}}, \binits{R.F.}},
\bauthor{\bsnm{{Limmongkol}}, \binits{S.}},
\bauthor{\bsnm{{Lindenmeyer}}, \binits{C.}},
\bauthor{\bsnm{{Long}}, \binits{D.C.}},
\bauthor{\bsnm{{Loomis}}, \binits{C.}},
\bauthor{\bsnm{{Loveday}}, \binits{J.}},
\bauthor{\bsnm{{Lucinio}}, \binits{R.}},
\bauthor{\bsnm{{Lupton}}, \binits{R.H.}},
\bauthor{\bsnm{{MacKinnon}}, \binits{B.}},
\bauthor{\bsnm{{Mannery}}, \binits{E.J.}},
\bauthor{\bsnm{{Mantsch}}, \binits{P.M.}},
\bauthor{\bsnm{{Margon}}, \binits{B.}},
\bauthor{\bsnm{{McGehee}}, \binits{P.}},
\bauthor{\bsnm{{McKay}}, \binits{T.A.}},
\bauthor{\bsnm{{Meiksin}}, \binits{A.}},
\bauthor{\bsnm{{Merelli}}, \binits{A.}},
\bauthor{\bsnm{{Monet}}, \binits{D.G.}},
\bauthor{\bsnm{{Munn}}, \binits{J.A.}},
\bauthor{\bsnm{{Narayanan}}, \binits{V.K.}},
\bauthor{\bsnm{{Nash}}, \binits{T.}},
\bauthor{\bsnm{{Neilsen}}, \binits{E.}},
\bauthor{\bsnm{{Neswold}}, \binits{R.}},
\bauthor{\bsnm{{Newberg}}, \binits{H.J.}},
\bauthor{\bsnm{{Nichol}}, \binits{R.C.}},
\bauthor{\bsnm{{Nicinski}}, \binits{T.}},
\bauthor{\bsnm{{Nonino}}, \binits{M.}},
\bauthor{\bsnm{{Okada}}, \binits{N.}},
\bauthor{\bsnm{{Okamura}}, \binits{S.}},
\bauthor{\bsnm{{Ostriker}}, \binits{J.P.}},
\bauthor{\bsnm{{Owen}}, \binits{R.}},
\bauthor{\bsnm{{Pauls}}, \binits{A.G.}},
\bauthor{\bsnm{{Peoples}}, \binits{J.}},
\bauthor{\bsnm{{Peterson}}, \binits{R.L.}},
\bauthor{\bsnm{{Petravick}}, \binits{D.}},
\bauthor{\bsnm{{Pier}}, \binits{J.R.}},
\bauthor{\bsnm{{Pope}}, \binits{A.}},
\bauthor{\bsnm{{Pordes}}, \binits{R.}},
\bauthor{\bsnm{{Prosapio}}, \binits{A.}},
\bauthor{\bsnm{{Rechenmacher}}, \binits{R.}},
\bauthor{\bsnm{{Quinn}}, \binits{T.R.}},
\bauthor{\bsnm{{Richards}}, \binits{G.T.}},
\bauthor{\bsnm{{Richmond}}, \binits{M.W.}},
\bauthor{\bsnm{{Rivetta}}, \binits{C.H.}},
\bauthor{\bsnm{{Rockosi}}, \binits{C.M.}},
\bauthor{\bsnm{{Ruthmansdorfer}}, \binits{K.}},
\bauthor{\bsnm{{Sandford}}, \binits{D.}},
\bauthor{\bsnm{{Schlegel}}, \binits{D.J.}},
\bauthor{\bsnm{{Schneider}}, \binits{D.P.}},
\bauthor{\bsnm{{Sekiguchi}}, \binits{M.}},
\bauthor{\bsnm{{Sergey}}, \binits{G.}},
\bauthor{\bsnm{{Shimasaku}}, \binits{K.}},
\bauthor{\bsnm{{Siegmund}}, \binits{W.A.}},
\bauthor{\bsnm{{Smee}}, \binits{S.}},
\bauthor{\bsnm{{Smith}}, \binits{J.A.}},
\bauthor{\bsnm{{Snedden}}, \binits{S.}},
\bauthor{\bsnm{{Stone}}, \binits{R.}},
\bauthor{\bsnm{{Stoughton}}, \binits{C.}},
\bauthor{\bsnm{{Strauss}}, \binits{M.A.}},
\bauthor{\bsnm{{Stubbs}}, \binits{C.}},
\bauthor{\bsnm{{SubbaRao}}, \binits{M.}},
\bauthor{\bsnm{{Szalay}}, \binits{A.S.}},
\bauthor{\bsnm{{Szapudi}}, \binits{I.}},
\bauthor{\bsnm{{Szokoly}}, \binits{G.P.}},
\bauthor{\bsnm{{Thakar}}, \binits{A.R.}},
\bauthor{\bsnm{{Tremonti}}, \binits{C.}},
\bauthor{\bsnm{{Tucker}}, \binits{D.L.}},
\bauthor{\bsnm{{Uomoto}}, \binits{A.}},
\bauthor{\bsnm{{Vanden Berk}}, \binits{D.}},
\bauthor{\bsnm{{Vogeley}}, \binits{M.S.}},
\bauthor{\bsnm{{Waddell}}, \binits{P.}},
\bauthor{\bsnm{{Wang}}, \binits{S.-i.}},
\bauthor{\bsnm{{Watanabe}}, \binits{M.}},
\bauthor{\bsnm{{Weinberg}}, \binits{D.H.}},
\bauthor{\bsnm{{Yanny}}, \binits{B.}},
\bauthor{\bsnm{{Yasuda}}, \binits{N.}},
\bauthor{\bsnm{{SDSS Collaboration}}}:
\batitle{{The Sloan Digital Sky Survey: Technical Summary}}.
\bjtitle{AJ}
\bvolume{120}(\bissue{3}),
\bfpage{1579}--\blpage{1587}
(\byear{2000})
\doiurl{10.1086/301513}
{\href{https://arxiv.org/abs/astro-ph/0006396}{{arXiv:astro-ph/0006396}}}
{[astro-ph]}
\end{barticle}
\endbibitem

\bibitem[\protect\citeauthoryear{{Norberg} et~al.}{2002}]{Norberg2002}
\begin{barticle}
\bauthor{\bsnm{{Norberg}}, \binits{P.}},
\bauthor{\bsnm{{Baugh}}, \binits{C.M.}},
\bauthor{\bsnm{{Hawkins}}, \binits{E.}},
\bauthor{\bsnm{{Maddox}}, \binits{S.}},
\bauthor{\bsnm{{Madgwick}}, \binits{D.}},
\bauthor{\bsnm{{Lahav}}, \binits{O.}},
\bauthor{\bsnm{{Cole}}, \binits{S.}},
\bauthor{\bsnm{{Frenk}}, \binits{C.S.}},
\bauthor{\bsnm{{Baldry}}, \binits{I.}},
\bauthor{\bsnm{{Bland-Hawthorn}}, \binits{J.}},
\bauthor{\bsnm{{Bridges}}, \binits{T.}},
\bauthor{\bsnm{{Cannon}}, \binits{R.}},
\bauthor{\bsnm{{Colless}}, \binits{M.}},
\bauthor{\bsnm{{Collins}}, \binits{C.}},
\bauthor{\bsnm{{Couch}}, \binits{W.}},
\bauthor{\bsnm{{Dalton}}, \binits{G.}},
\bauthor{\bsnm{{De Propris}}, \binits{R.}},
\bauthor{\bsnm{{Driver}}, \binits{S.P.}},
\bauthor{\bsnm{{Efstathiou}}, \binits{G.}},
\bauthor{\bsnm{{Ellis}}, \binits{R.S.}},
\bauthor{\bsnm{{Glazebrook}}, \binits{K.}},
\bauthor{\bsnm{{Jackson}}, \binits{C.}},
\bauthor{\bsnm{{Lewis}}, \binits{I.}},
\bauthor{\bsnm{{Lumsden}}, \binits{S.}},
\bauthor{\bsnm{{Peacock}}, \binits{J.A.}},
\bauthor{\bsnm{{Peterson}}, \binits{B.A.}},
\bauthor{\bsnm{{Sutherland}}, \binits{W.}},
\bauthor{\bsnm{{Taylor}}, \binits{K.}}:
\batitle{{The 2dF Galaxy Redshift Survey: the dependence of galaxy clustering on luminosity and spectral type}}.
\bjtitle{MNRAS}
\bvolume{332}(\bissue{4}),
\bfpage{827}--\blpage{838}
(\byear{2002})
\doiurl{10.1046/j.1365-8711.2002.05348.x}
{\href{https://arxiv.org/abs/astro-ph/0112043}{{arXiv:astro-ph/0112043}}}
{[astro-ph]}
\end{barticle}
\endbibitem

\bibitem[\protect\citeauthoryear{{Blanton} et~al.}{2003}]{blanton2003galaxy}
\begin{barticle}
\bauthor{\bsnm{{Blanton}}, \binits{M.R.}},
\bauthor{\bsnm{{Hogg}}, \binits{D.W.}},
\bauthor{\bsnm{{Bahcall}}, \binits{N.A.}},
\bauthor{\bsnm{{Brinkmann}}, \binits{J.}},
\bauthor{\bsnm{{Britton}}, \binits{M.}},
\bauthor{\bsnm{{Connolly}}, \binits{A.J.}},
\bauthor{\bsnm{{Csabai}}, \binits{I.}},
\bauthor{\bsnm{{Fukugita}}, \binits{M.}},
\bauthor{\bsnm{{Loveday}}, \binits{J.}},
\bauthor{\bsnm{{Meiksin}}, \binits{A.}},
\bauthor{\bsnm{{Munn}}, \binits{J.A.}},
\bauthor{\bsnm{{Nichol}}, \binits{R.C.}},
\bauthor{\bsnm{{Okamura}}, \binits{S.}},
\bauthor{\bsnm{{Quinn}}, \binits{T.}},
\bauthor{\bsnm{{Schneider}}, \binits{D.P.}},
\bauthor{\bsnm{{Shimasaku}}, \binits{K.}},
\bauthor{\bsnm{{Strauss}}, \binits{M.A.}},
\bauthor{\bsnm{{Tegmark}}, \binits{M.}},
\bauthor{\bsnm{{Vogeley}}, \binits{M.S.}},
\bauthor{\bsnm{{Weinberg}}, \binits{D.H.}}:
\batitle{{The Galaxy Luminosity Function and Luminosity Density at Redshift z = 0.1}}.
\bjtitle{ApJ}
\bvolume{592}(\bissue{2}),
\bfpage{819}--\blpage{838}
(\byear{2003})
\doiurl{10.1086/375776}
{\href{https://arxiv.org/abs/astro-ph/0210215}{{arXiv:astro-ph/0210215}}}
{[astro-ph]}
\end{barticle}
\endbibitem

\bibitem[\protect\citeauthoryear{{Zehavi} et~al.}{2005}]{zehavi2005luminosity}
\begin{barticle}
\bauthor{\bsnm{{Zehavi}}, \binits{I.}},
\bauthor{\bsnm{{Zheng}}, \binits{Z.}},
\bauthor{\bsnm{{Weinberg}}, \binits{D.H.}},
\bauthor{\bsnm{{Frieman}}, \binits{J.A.}},
\bauthor{\bsnm{{Berlind}}, \binits{A.A.}},
\bauthor{\bsnm{{Blanton}}, \binits{M.R.}},
\bauthor{\bsnm{{Scoccimarro}}, \binits{R.}},
\bauthor{\bsnm{{Sheth}}, \binits{R.K.}},
\bauthor{\bsnm{{Strauss}}, \binits{M.A.}},
\bauthor{\bsnm{{Kayo}}, \binits{I.}},
\bauthor{\bsnm{{Suto}}, \binits{Y.}},
\bauthor{\bsnm{{Fukugita}}, \binits{M.}},
\bauthor{\bsnm{{Nakamura}}, \binits{O.}},
\bauthor{\bsnm{{Bahcall}}, \binits{N.A.}},
\bauthor{\bsnm{{Brinkmann}}, \binits{J.}},
\bauthor{\bsnm{{Gunn}}, \binits{J.E.}},
\bauthor{\bsnm{{Hennessy}}, \binits{G.S.}},
\bauthor{\bsnm{{Ivezi{\'c}}}, \binits{{\v{Z}}.}},
\bauthor{\bsnm{{Knapp}}, \binits{G.R.}},
\bauthor{\bsnm{{Loveday}}, \binits{J.}},
\bauthor{\bsnm{{Meiksin}}, \binits{A.}},
\bauthor{\bsnm{{Schlegel}}, \binits{D.J.}},
\bauthor{\bsnm{{Schneider}}, \binits{D.P.}},
\bauthor{\bsnm{{Szapudi}}, \binits{I.}},
\bauthor{\bsnm{{Tegmark}}, \binits{M.}},
\bauthor{\bsnm{{Vogeley}}, \binits{M.S.}},
\bauthor{\bsnm{{York}}, \binits{D.G.}},
\bauthor{\bsnm{{SDSS Collaboration}}}:
\batitle{{The Luminosity and Color Dependence of the Galaxy Correlation Function}}.
\bjtitle{ApJ}
\bvolume{630}(\bissue{1}),
\bfpage{1}--\blpage{27}
(\byear{2005})
\doiurl{10.1086/431891}
{\href{https://arxiv.org/abs/astro-ph/0408569}{{arXiv:astro-ph/0408569}}}
{[astro-ph]}
\end{barticle}
\endbibitem

\bibitem[\protect\citeauthoryear{{Wang} et~al.}{2007}]{wang2007cross}
\begin{barticle}
\bauthor{\bsnm{{Wang}}, \binits{Y.}},
\bauthor{\bsnm{{Yang}}, \binits{X.}},
\bauthor{\bsnm{{Mo}}, \binits{H.J.}},
\bauthor{\bsnm{{van den Bosch}}, \binits{F.C.}}:
\batitle{{The Cross-Correlation between Galaxies of Different Luminosities and Colors}}.
\bjtitle{ApJ}
\bvolume{664}(\bissue{2}),
\bfpage{608}--\blpage{632}
(\byear{2007})
\doiurl{10.1086/519245}
{\href{https://arxiv.org/abs/astro-ph/0703253}{{arXiv:astro-ph/0703253}}}
{[astro-ph]}
\end{barticle}
\endbibitem

\bibitem[\protect\citeauthoryear{{Li} and {White}}{2009}]{li2009distribution}
\begin{barticle}
\bauthor{\bsnm{{Li}}, \binits{C.}},
\bauthor{\bsnm{{White}}, \binits{S.D.M.}}:
\batitle{{The distribution of stellar mass in the low-redshift Universe}}.
\bjtitle{MNRAS}
\bvolume{398}(\bissue{4}),
\bfpage{2177}--\blpage{2187}
(\byear{2009})
\doiurl{10.1111/j.1365-2966.2009.15268.x}
{\href{https://arxiv.org/abs/0901.0706}{{arXiv:0901.0706}}}
{[astro-ph.CO]}
\end{barticle}
\endbibitem

\bibitem[\protect\citeauthoryear{{Zehavi} et~al.}{2011}]{zehavi2011galaxy}
\begin{barticle}
\bauthor{\bsnm{{Zehavi}}, \binits{I.}},
\bauthor{\bsnm{{Zheng}}, \binits{Z.}},
\bauthor{\bsnm{{Weinberg}}, \binits{D.H.}},
\bauthor{\bsnm{{Blanton}}, \binits{M.R.}},
\bauthor{\bsnm{{Bahcall}}, \binits{N.A.}},
\bauthor{\bsnm{{Berlind}}, \binits{A.A.}},
\bauthor{\bsnm{{Brinkmann}}, \binits{J.}},
\bauthor{\bsnm{{Frieman}}, \binits{J.A.}},
\bauthor{\bsnm{{Gunn}}, \binits{J.E.}},
\bauthor{\bsnm{{Lupton}}, \binits{R.H.}},
\bauthor{\bsnm{{Nichol}}, \binits{R.C.}},
\bauthor{\bsnm{{Percival}}, \binits{W.J.}},
\bauthor{\bsnm{{Schneider}}, \binits{D.P.}},
\bauthor{\bsnm{{Skibba}}, \binits{R.A.}},
\bauthor{\bsnm{{Strauss}}, \binits{M.A.}},
\bauthor{\bsnm{{Tegmark}}, \binits{M.}},
\bauthor{\bsnm{{York}}, \binits{D.G.}}:
\batitle{{Galaxy Clustering in the Completed SDSS Redshift Survey: The Dependence on Color and Luminosity}}.
\bjtitle{ApJ}
\bvolume{736}(\bissue{1}),
\bfpage{59}
(\byear{2011})
\doiurl{10.1088/0004-637X/736/1/59}
{\href{https://arxiv.org/abs/1005.2413}{{arXiv:1005.2413}}}
{[astro-ph.CO]}
\end{barticle}
\endbibitem

\bibitem[\protect\citeauthoryear{{Moustakas} et~al.}{2013}]{2013ApJ...767...50M}
\begin{barticle}
\bauthor{\bsnm{{Moustakas}}, \binits{J.}},
\bauthor{\bsnm{{Coil}}, \binits{A.L.}},
\bauthor{\bsnm{{Aird}}, \binits{J.}},
\bauthor{\bsnm{{Blanton}}, \binits{M.R.}},
\bauthor{\bsnm{{Cool}}, \binits{R.J.}},
\bauthor{\bsnm{{Eisenstein}}, \binits{D.J.}},
\bauthor{\bsnm{{Mendez}}, \binits{A.J.}},
\bauthor{\bsnm{{Wong}}, \binits{K.C.}},
\bauthor{\bsnm{{Zhu}}, \binits{G.}},
\bauthor{\bsnm{{Arnouts}}, \binits{S.}}:
\batitle{{PRIMUS: Constraints on Star Formation Quenching and Galaxy Merging, and the Evolution of the Stellar Mass Function from z = 0-1}}.
\bjtitle{ApJ}
\bvolume{767}(\bissue{1}),
\bfpage{50}
(\byear{2013})
\doiurl{10.1088/0004-637X/767/1/50}
{\href{https://arxiv.org/abs/1301.1688}{{arXiv:1301.1688}}}
{[astro-ph.CO]}
\end{barticle}
\endbibitem

\bibitem[\protect\citeauthoryear{{Zhao} et~al.}{2020}]{zhao2000}
\begin{barticle}
\bauthor{\bsnm{{Zhao}}, \binits{P.}},
\bauthor{\bsnm{{Xu}}, \binits{H.}},
\bauthor{\bsnm{{Katsianis}}, \binits{A.}},
\bauthor{\bsnm{{Yang}}, \binits{X.-H.}}:
\batitle{{The intrinsic SFRF and sSFRF of galaxies: comparing SDSS observation with IllustrisTNG simulation}}.
\bjtitle{Research in Astronomy and Astrophysics}
\bvolume{20}(\bissue{12}),
\bfpage{195}
(\byear{2020})
\doiurl{10.1088/1674-4527/20/12/195}
{\href{https://arxiv.org/abs/2006.08634}{{arXiv:2006.08634}}}
{[astro-ph.GA]}
\end{barticle}
\endbibitem

\bibitem[\protect\citeauthoryear{{Wang} et~al.}{2021a}]{Wang2021}
\begin{barticle}
\bauthor{\bsnm{{Wang}}, \binits{Z.}},
\bauthor{\bsnm{{Xu}}, \binits{H.}},
\bauthor{\bsnm{{Yang}}, \binits{X.}},
\bauthor{\bsnm{{Jing}}, \binits{Y.}},
\bauthor{\bsnm{{Wang}}, \binits{K.}},
\bauthor{\bsnm{{Guo}}, \binits{H.}},
\bauthor{\bsnm{{Dong}}, \binits{F.}},
\bauthor{\bsnm{{He}}, \binits{M.}}:
\batitle{{The clustering of galaxies in the DESI imaging legacy surveys DR8: I. The luminosity and color dependent intrinsic clustering}}.
\bjtitle{Science China Physics, Mechanics, and Astronomy}
\bvolume{64}(\bissue{8}),
\bfpage{289811}
(\byear{2021})
\doiurl{10.1007/s11433-021-1707-6}
{\href{https://arxiv.org/abs/2106.14159}{{arXiv:2106.14159}}}
{[astro-ph.GA]}
\end{barticle}
\endbibitem

\bibitem[\protect\citeauthoryear{{Wang} et~al.}{2021b}]{2021MNRAS.500.3776W}
\begin{barticle}
\bauthor{\bsnm{{Wang}}, \binits{W.}},
\bauthor{\bsnm{{Takada}}, \binits{M.}},
\bauthor{\bsnm{{Li}}, \binits{X.}},
\bauthor{\bsnm{{Carlsten}}, \binits{S.G.}},
\bauthor{\bsnm{{Lan}}, \binits{T.-W.}},
\bauthor{\bsnm{{Shi}}, \binits{J.}},
\bauthor{\bsnm{{Miyatake}}, \binits{H.}},
\bauthor{\bsnm{{More}}, \binits{S.}},
\bauthor{\bsnm{{Beaton}}, \binits{R.L.}},
\bauthor{\bsnm{{Lupton}}, \binits{R.}},
\bauthor{\bsnm{{Lin}}, \binits{Y.-T.}},
\bauthor{\bsnm{{Qiu}}, \binits{T.}},
\bauthor{\bsnm{{Luo}}, \binits{W.}}:
\batitle{{A comparative study of satellite galaxies in Milky Way-like galaxies from HSC, DECaLS, and SDSS}}.
\bjtitle{MNRAS}
\bvolume{500}(\bissue{3}),
\bfpage{3776}--\blpage{3801}
(\byear{2021})
\doiurl{10.1093/mnras/staa3495}
{\href{https://arxiv.org/abs/2009.06882}{{arXiv:2009.06882}}}
{[astro-ph.GA]}
\end{barticle}
\endbibitem

\bibitem[\protect\citeauthoryear{{Wang} et~al.}{2025}]{2025ApJ...986..218W}
\begin{barticle}
\bauthor{\bsnm{{Wang}}, \binits{W.}},
\bauthor{\bsnm{{Yang}}, \binits{X.}},
\bauthor{\bsnm{{Jing}}, \binits{Y.}},
\bauthor{\bsnm{{Ross}}, \binits{A.J.}},
\bauthor{\bsnm{{Siudek}}, \binits{M.}},
\bauthor{\bsnm{{Moustakas}}, \binits{J.}},
\bauthor{\bsnm{{Moore}}, \binits{S.G.}},
\bauthor{\bsnm{{Cole}}, \binits{S.}},
\bauthor{\bsnm{{Frenk}}, \binits{C.}},
\bauthor{\bsnm{{Yu}}, \binits{J.}},
\bauthor{\bsnm{{Koposov}}, \binits{S.E.}},
\bauthor{\bsnm{{Han}}, \binits{J.}},
\bauthor{\bsnm{{Tan}}, \binits{Z.}},
\bauthor{\bsnm{{Xu}}, \binits{K.}},
\bauthor{\bsnm{{Gu}}, \binits{Y.}},
\bauthor{\bsnm{{Wang}}, \binits{Y.}},
\bauthor{\bsnm{{Gnedin}}, \binits{O.Y.}},
\bauthor{\bsnm{{Aguilar}}, \binits{J.N.}},
\bauthor{\bsnm{{Ahlen}}, \binits{S.}},
\bauthor{\bsnm{{Bianchi}}, \binits{D.}},
\bauthor{\bsnm{{Brooks}}, \binits{D.}},
\bauthor{\bsnm{{Claybaugh}}, \binits{T.}},
\bauthor{\bsnm{{de la Macorra}}, \binits{A.}},
\bauthor{\bsnm{{Dey}}, \binits{A.}},
\bauthor{\bsnm{{Doel}}, \binits{P.}},
\bauthor{\bsnm{{Forero-Romero}}, \binits{J.E.}},
\bauthor{\bsnm{{Gazta{\~n}aga}}, \binits{E.}},
\bauthor{\bsnm{{Gontcho A Gontcho}}, \binits{S.}},
\bauthor{\bsnm{{Gutierrez}}, \binits{G.}},
\bauthor{\bsnm{{Honscheid}}, \binits{K.}},
\bauthor{\bsnm{{Ishak}}, \binits{M.}},
\bauthor{\bsnm{{Kisner}}, \binits{T.}},
\bauthor{\bsnm{{Landriau}}, \binits{M.}},
\bauthor{\bsnm{{Le Guillou}}, \binits{L.}},
\bauthor{\bsnm{{Manera}}, \binits{M.}},
\bauthor{\bsnm{{Meisner}}, \binits{A.}},
\bauthor{\bsnm{{Miquel}}, \binits{R.}},
\bauthor{\bsnm{{Nadathur}}, \binits{S.}},
\bauthor{\bsnm{{Poppett}}, \binits{C.}},
\bauthor{\bsnm{{Prada}}, \binits{F.}},
\bauthor{\bsnm{{P{\'e}rez-R{\`a}fols}}, \binits{I.}},
\bauthor{\bsnm{{Rossi}}, \binits{G.}},
\bauthor{\bsnm{{Sanchez}}, \binits{E.}},
\bauthor{\bsnm{{Schlegel}}, \binits{D.}},
\bauthor{\bsnm{{Seo}}, \binits{H.-J.}},
\bauthor{\bsnm{{Silber}}, \binits{J.H.}},
\bauthor{\bsnm{{Sprayberry}}, \binits{D.}},
\bauthor{\bsnm{{Tarl{\'e}}}, \binits{G.}},
\bauthor{\bsnm{{Weaver}}, \binits{B.A.}},
\bauthor{\bsnm{{Zou}}, \binits{H.}}:
\batitle{{Luminosity and Stellar Mass Functions of Faint Photometric Satellites around Spectroscopic Central Galaxies from DESI Year-1 Bright Galaxy Survey}}.
\bjtitle{ApJ}
\bvolume{986}(\bissue{2}),
\bfpage{218}
(\byear{2025})
\doiurl{10.3847/1538-4357/add5df}
{\href{https://arxiv.org/abs/2503.03317}{{arXiv:2503.03317}}}
{[astro-ph.GA]}
\end{barticle}
\endbibitem

\bibitem[\protect\citeauthoryear{{Naab} and {Ostriker}}{2017}]{Naab2017}
\begin{barticle}
\bauthor{\bsnm{{Naab}}, \binits{T.}},
\bauthor{\bsnm{{Ostriker}}, \binits{J.P.}}:
\batitle{{Theoretical Challenges in Galaxy Formation}}.
\bjtitle{ARA\&A}
\bvolume{55}(\bissue{1}),
\bfpage{59}--\blpage{109}
(\byear{2017})
\doiurl{10.1146/annurev-astro-081913-040019}
{\href{https://arxiv.org/abs/1612.06891}{{arXiv:1612.06891}}}
{[astro-ph.GA]}
\end{barticle}
\endbibitem

\bibitem[\protect\citeauthoryear{{Smercina} et~al.}{2018}]{Smercina2018}
\begin{barticle}
\bauthor{\bsnm{{Smercina}}, \binits{A.}},
\bauthor{\bsnm{{Bell}}, \binits{E.F.}},
\bauthor{\bsnm{{Price}}, \binits{P.A.}},
\bauthor{\bsnm{{D'Souza}}, \binits{R.}},
\bauthor{\bsnm{{Slater}}, \binits{C.T.}},
\bauthor{\bsnm{{Bailin}}, \binits{J.}},
\bauthor{\bsnm{{Monachesi}}, \binits{A.}},
\bauthor{\bsnm{{Nidever}}, \binits{D.}}:
\batitle{{A Lonely Giant: The Sparse Satellite Population of M94 Challenges Galaxy Formation}}.
\bjtitle{ApJ}
\bvolume{863}(\bissue{2}),
\bfpage{152}
(\byear{2018})
\doiurl{10.3847/1538-4357/aad2d6}
{\href{https://arxiv.org/abs/1807.03779}{{arXiv:1807.03779}}}
{[astro-ph.GA]}
\end{barticle}
\endbibitem

\bibitem[\protect\citeauthoryear{{Katsianis} et~al.}{2021}]{Katsianis2021}
\begin{barticle}
\bauthor{\bsnm{{Katsianis}}, \binits{A.}},
\bauthor{\bsnm{{Xu}}, \binits{H.}},
\bauthor{\bsnm{{Yang}}, \binits{X.}},
\bauthor{\bsnm{{Luo}}, \binits{Y.}},
\bauthor{\bsnm{{Cui}}, \binits{W.}},
\bauthor{\bsnm{{Dav{\'e}}}, \binits{R.}},
\bauthor{\bsnm{{Lagos}}, \binits{C.D.P.}},
\bauthor{\bsnm{{Zheng}}, \binits{X.}},
\bauthor{\bsnm{{Zhao}}, \binits{P.}}:
\batitle{{The specific star formation rate function at different mass scales and quenching: a comparison between cosmological models and SDSS}}.
\bjtitle{MNRAS}
\bvolume{500}(\bissue{2}),
\bfpage{2036}--\blpage{2048}
(\byear{2021})
\doiurl{10.1093/mnras/staa3236}
{\href{https://arxiv.org/abs/2010.08173}{{arXiv:2010.08173}}}
{[astro-ph.GA]}
\end{barticle}
\endbibitem

\bibitem[\protect\citeauthoryear{{Sales} et~al.}{2022}]{Sales2022}
\begin{barticle}
\bauthor{\bsnm{{Sales}}, \binits{L.V.}},
\bauthor{\bsnm{{Wetzel}}, \binits{A.}},
\bauthor{\bsnm{{Fattahi}}, \binits{A.}}:
\batitle{{Baryonic solutions and challenges for cosmological models of dwarf galaxies}}.
\bjtitle{Nature Astronomy}
\bvolume{6},
\bfpage{897}--\blpage{910}
(\byear{2022})
\doiurl{10.1038/s41550-022-01689-w}
{\href{https://arxiv.org/abs/2206.05295}{{arXiv:2206.05295}}}
{[astro-ph.GA]}
\end{barticle}
\endbibitem

\bibitem[\protect\citeauthoryear{{Jing} et~al.}{1998}]{jing1998spatial}
\begin{barticle}
\bauthor{\bsnm{{Jing}}, \binits{Y.P.}},
\bauthor{\bsnm{{Mo}}, \binits{H.J.}},
\bauthor{\bsnm{{B{\"o}rner}}, \binits{G.}}:
\batitle{{Spatial Correlation Function and Pairwise Velocity Dispersion of Galaxies: Cold Dark Matter Models versus the Las Campanas Survey}}.
\bjtitle{ApJ}
\bvolume{494}(\bissue{1}),
\bfpage{1}--\blpage{12}
(\byear{1998})
\doiurl{10.1086/305209}
{\href{https://arxiv.org/abs/astro-ph/9707106}{{arXiv:astro-ph/9707106}}}
{[astro-ph]}
\end{barticle}
\endbibitem

\bibitem[\protect\citeauthoryear{{Peacock} and {Smith}}{2000}]{peacock2000halo}
\begin{barticle}
\bauthor{\bsnm{{Peacock}}, \binits{J.A.}},
\bauthor{\bsnm{{Smith}}, \binits{R.E.}}:
\batitle{{Halo occupation numbers and galaxy bias}}.
\bjtitle{MNRAS}
\bvolume{318}(\bissue{4}),
\bfpage{1144}--\blpage{1156}
(\byear{2000})
\doiurl{10.1046/j.1365-8711.2000.03779.x}
{\href{https://arxiv.org/abs/astro-ph/0005010}{{arXiv:astro-ph/0005010}}}
{[astro-ph]}
\end{barticle}
\endbibitem

\bibitem[\protect\citeauthoryear{{Tinker} et~al.}{2005}]{tinker2005mass}
\begin{barticle}
\bauthor{\bsnm{{Tinker}}, \binits{J.L.}},
\bauthor{\bsnm{{Weinberg}}, \binits{D.H.}},
\bauthor{\bsnm{{Zheng}}, \binits{Z.}},
\bauthor{\bsnm{{Zehavi}}, \binits{I.}}:
\batitle{{On the Mass-to-Light Ratio of Large-Scale Structure}}.
\bjtitle{ApJ}
\bvolume{631}(\bissue{1}),
\bfpage{41}--\blpage{58}
(\byear{2005})
\doiurl{10.1086/432084}
{\href{https://arxiv.org/abs/astro-ph/0411777}{{arXiv:astro-ph/0411777}}}
{[astro-ph]}
\end{barticle}
\endbibitem

\bibitem[\protect\citeauthoryear{{Zheng} et~al.}{2005}]{Zheng2005}
\begin{barticle}
\bauthor{\bsnm{{Zheng}}, \binits{Z.}},
\bauthor{\bsnm{{Berlind}}, \binits{A.A.}},
\bauthor{\bsnm{{Weinberg}}, \binits{D.H.}},
\bauthor{\bsnm{{Benson}}, \binits{A.J.}},
\bauthor{\bsnm{{Baugh}}, \binits{C.M.}},
\bauthor{\bsnm{{Cole}}, \binits{S.}},
\bauthor{\bsnm{{Dav{\'e}}}, \binits{R.}},
\bauthor{\bsnm{{Frenk}}, \binits{C.S.}},
\bauthor{\bsnm{{Katz}}, \binits{N.}},
\bauthor{\bsnm{{Lacey}}, \binits{C.G.}}:
\batitle{{Theoretical Models of the Halo Occupation Distribution: Separating Central and Satellite Galaxies}}.
\bjtitle{ApJ}
\bvolume{633}(\bissue{2}),
\bfpage{791}--\blpage{809}
(\byear{2005})
\doiurl{10.1086/466510}
{\href{https://arxiv.org/abs/astro-ph/0408564}{{arXiv:astro-ph/0408564}}}
{[astro-ph]}
\end{barticle}
\endbibitem

\bibitem[\protect\citeauthoryear{{Zheng} et~al.}{2009}]{Zheng2009}
\begin{barticle}
\bauthor{\bsnm{{Zheng}}, \binits{Z.}},
\bauthor{\bsnm{{Zehavi}}, \binits{I.}},
\bauthor{\bsnm{{Eisenstein}}, \binits{D.J.}},
\bauthor{\bsnm{{Weinberg}}, \binits{D.H.}},
\bauthor{\bsnm{{Jing}}, \binits{Y.P.}}:
\batitle{{Halo Occupation Distribution Modeling of Clustering of Luminous Red Galaxies}}.
\bjtitle{ApJ}
\bvolume{707}(\bissue{1}),
\bfpage{554}--\blpage{572}
(\byear{2009})
\doiurl{10.1088/0004-637X/707/1/554}
{\href{https://arxiv.org/abs/0809.1868}{{arXiv:0809.1868}}}
{[astro-ph]}
\end{barticle}
\endbibitem

\bibitem[\protect\citeauthoryear{{Brown} et~al.}{2008}]{brown2008red}
\begin{barticle}
\bauthor{\bsnm{{Brown}}, \binits{M.J.I.}},
\bauthor{\bsnm{{Zheng}}, \binits{Z.}},
\bauthor{\bsnm{{White}}, \binits{M.}},
\bauthor{\bsnm{{Dey}}, \binits{A.}},
\bauthor{\bsnm{{Jannuzi}}, \binits{B.T.}},
\bauthor{\bsnm{{Benson}}, \binits{A.J.}},
\bauthor{\bsnm{{Brand}}, \binits{K.}},
\bauthor{\bsnm{{Brodwin}}, \binits{M.}},
\bauthor{\bsnm{{Croton}}, \binits{D.J.}}:
\batitle{{Red Galaxy Growth and the Halo Occupation Distribution}}.
\bjtitle{ApJ}
\bvolume{682}(\bissue{2}),
\bfpage{937}--\blpage{963}
(\byear{2008})
\doiurl{10.1086/589538}
{\href{https://arxiv.org/abs/0804.2293}{{arXiv:0804.2293}}}
{[astro-ph]}
\end{barticle}
\endbibitem

\bibitem[\protect\citeauthoryear{{Zu} and {Mandelbaum}}{2015}]{2015Zu}
\begin{barticle}
\bauthor{\bsnm{{Zu}}, \binits{Y.}},
\bauthor{\bsnm{{Mandelbaum}}, \binits{R.}}:
\batitle{{Mapping stellar content to dark matter haloes using galaxy clustering and galaxy-galaxy lensing in the SDSS DR7}}.
\bjtitle{MNRAS}
\bvolume{454}(\bissue{2}),
\bfpage{1161}--\blpage{1191}
(\byear{2015})
\doiurl{10.1093/mnras/stv2062}
{\href{https://arxiv.org/abs/1505.02781}{{arXiv:1505.02781}}}
{[astro-ph.CO]}
\end{barticle}
\endbibitem

\bibitem[\protect\citeauthoryear{{Zu} and {Mandelbaum}}{2016}]{2016Zu}
\begin{barticle}
\bauthor{\bsnm{{Zu}}, \binits{Y.}},
\bauthor{\bsnm{{Mandelbaum}}, \binits{R.}}:
\batitle{{Mapping stellar content to dark matter haloes - II. Halo mass is the main driver of galaxy quenching}}.
\bjtitle{MNRAS}
\bvolume{457}(\bissue{4}),
\bfpage{4360}--\blpage{4383}
(\byear{2016})
\doiurl{10.1093/mnras/stw221}
{\href{https://arxiv.org/abs/1509.06758}{{arXiv:1509.06758}}}
{[astro-ph.GA]}
\end{barticle}
\endbibitem

\bibitem[\protect\citeauthoryear{{Zu} and {Mandelbaum}}{2018}]{2018Zu}
\begin{barticle}
\bauthor{\bsnm{{Zu}}, \binits{Y.}},
\bauthor{\bsnm{{Mandelbaum}}, \binits{R.}}:
\batitle{{Mapping stellar content to dark matter haloes - III. Environmental dependence and conformity of galaxy colours}}.
\bjtitle{MNRAS}
\bvolume{476}(\bissue{2}),
\bfpage{1637}--\blpage{1653}
(\byear{2018})
\doiurl{10.1093/mnras/sty279}
{\href{https://arxiv.org/abs/1703.09219}{{arXiv:1703.09219}}}
{[astro-ph.GA]}
\end{barticle}
\endbibitem

\bibitem[\protect\citeauthoryear{{Wang} et~al.}{2019}]{2019Wang}
\begin{barticle}
\bauthor{\bsnm{{Wang}}, \binits{Z.}},
\bauthor{\bsnm{{Xu}}, \binits{H.}},
\bauthor{\bsnm{{Yang}}, \binits{X.}},
\bauthor{\bsnm{{Jing}}, \binits{Y.P.}},
\bauthor{\bsnm{{Guo}}, \binits{H.}},
\bauthor{\bsnm{{Zheng}}, \binits{Z.}},
\bauthor{\bsnm{{Zu}}, \binits{Y.}},
\bauthor{\bsnm{{Li}}, \binits{Z.}},
\bauthor{\bsnm{{Liu}}, \binits{C.}}:
\batitle{{Accurate Modeling of the Projected Galaxy Clustering in Photometric Surveys. I. Tests with Mock Catalogs}}.
\bjtitle{ApJ}
\bvolume{879}(\bissue{2}),
\bfpage{71}
(\byear{2019})
\doiurl{10.3847/1538-4357/ab25e6}
{\href{https://arxiv.org/abs/1906.00516}{{arXiv:1906.00516}}}
{[astro-ph.CO]}
\end{barticle}
\endbibitem

\bibitem[\protect\citeauthoryear{{Alam} et~al.}{2020}]{2020Alam}
\begin{barticle}
\bauthor{\bsnm{{Alam}}, \binits{S.}},
\bauthor{\bsnm{{Peacock}}, \binits{J.A.}},
\bauthor{\bsnm{{Kraljic}}, \binits{K.}},
\bauthor{\bsnm{{Ross}}, \binits{A.J.}},
\bauthor{\bsnm{{Comparat}}, \binits{J.}}:
\batitle{{Multitracer extension of the halo model: probing quenching and conformity in eBOSS}}.
\bjtitle{MNRAS}
\bvolume{497}(\bissue{1}),
\bfpage{581}--\blpage{595}
(\byear{2020})
\doiurl{10.1093/mnras/staa1956}
{\href{https://arxiv.org/abs/1910.05095}{{arXiv:1910.05095}}}
{[astro-ph.CO]}
\end{barticle}
\endbibitem

\bibitem[\protect\citeauthoryear{{Yuan} et~al.}{2018}]{2018Yuan}
\begin{barticle}
\bauthor{\bsnm{{Yuan}}, \binits{S.}},
\bauthor{\bsnm{{Eisenstein}}, \binits{D.J.}},
\bauthor{\bsnm{{Garrison}}, \binits{L.H.}}:
\batitle{{Exploring the squeezed three-point galaxy correlation function with generalized halo occupation distribution models}}.
\bjtitle{MNRAS}
\bvolume{478}(\bissue{2}),
\bfpage{2019}--\blpage{2033}
(\byear{2018})
\doiurl{10.1093/mnras/sty1089}
{\href{https://arxiv.org/abs/1802.10115}{{arXiv:1802.10115}}}
{[astro-ph.CO]}
\end{barticle}
\endbibitem

\bibitem[\protect\citeauthoryear{{Yuan} et~al.}{2022}]{2022Yuan}
\begin{barticle}
\bauthor{\bsnm{{Yuan}}, \binits{S.}},
\bauthor{\bsnm{{Garrison}}, \binits{L.H.}},
\bauthor{\bsnm{{Hadzhiyska}}, \binits{B.}},
\bauthor{\bsnm{{Bose}}, \binits{S.}},
\bauthor{\bsnm{{Eisenstein}}, \binits{D.J.}}:
\batitle{{ABACUSHOD: a highly efficient extended multitracer HOD framework and its application to BOSS and eBOSS data}}.
\bjtitle{MNRAS}
\bvolume{510}(\bissue{3}),
\bfpage{3301}--\blpage{3320}
(\byear{2022})
\doiurl{10.1093/mnras/stab3355}
{\href{https://arxiv.org/abs/2110.11412}{{arXiv:2110.11412}}}
{[astro-ph.CO]}
\end{barticle}
\endbibitem

\bibitem[\protect\citeauthoryear{{Yang} et~al.}{2003}]{yang2003constraining}
\begin{barticle}
\bauthor{\bsnm{{Yang}}, \binits{X.}},
\bauthor{\bsnm{{Mo}}, \binits{H.J.}},
\bauthor{\bsnm{{van den Bosch}}, \binits{F.C.}}:
\batitle{{Constraining galaxy formation and cosmology with the conditional luminosity function of galaxies}}.
\bjtitle{MNRAS}
\bvolume{339}(\bissue{4}),
\bfpage{1057}--\blpage{1080}
(\byear{2003})
\doiurl{10.1046/j.1365-8711.2003.06254.x}
{\href{https://arxiv.org/abs/astro-ph/0207019}{{arXiv:astro-ph/0207019}}}
{[astro-ph]}
\end{barticle}
\endbibitem

\bibitem[\protect\citeauthoryear{{van den Bosch} et~al.}{2003}]{van2003linking}
\begin{barticle}
\bauthor{\bsnm{{van den Bosch}}, \binits{F.C.}},
\bauthor{\bsnm{{Yang}}, \binits{X.}},
\bauthor{\bsnm{{Mo}}, \binits{H.J.}}:
\batitle{{Linking early- and late-type galaxies to their dark matter haloes}}.
\bjtitle{MNRAS}
\bvolume{340}(\bissue{3}),
\bfpage{771}--\blpage{792}
(\byear{2003})
\doiurl{10.1046/j.1365-8711.2003.06335.x}
{\href{https://arxiv.org/abs/astro-ph/0210495}{{arXiv:astro-ph/0210495}}}
{[astro-ph]}
\end{barticle}
\endbibitem

\bibitem[\protect\citeauthoryear{{van den Bosch} et~al.}{2007}]{van2007towards}
\begin{barticle}
\bauthor{\bsnm{{van den Bosch}}, \binits{F.C.}},
\bauthor{\bsnm{{Yang}}, \binits{X.}},
\bauthor{\bsnm{{Mo}}, \binits{H.J.}},
\bauthor{\bsnm{{Weinmann}}, \binits{S.M.}},
\bauthor{\bsnm{{Macci{\`o}}}, \binits{A.V.}},
\bauthor{\bsnm{{More}}, \binits{S.}},
\bauthor{\bsnm{{Cacciato}}, \binits{M.}},
\bauthor{\bsnm{{Skibba}}, \binits{R.}},
\bauthor{\bsnm{{Kang}}, \binits{X.}}:
\batitle{{Towards a concordant model of halo occupation statistics}}.
\bjtitle{MNRAS}
\bvolume{376}(\bissue{2}),
\bfpage{841}--\blpage{860}
(\byear{2007})
\doiurl{10.1111/j.1365-2966.2007.11493.x}
{\href{https://arxiv.org/abs/astro-ph/0610686}{{arXiv:astro-ph/0610686}}}
{[astro-ph]}
\end{barticle}
\endbibitem

\bibitem[\protect\citeauthoryear{{Yang} et~al.}{2008}]{yang2008galaxy}
\begin{barticle}
\bauthor{\bsnm{{Yang}}, \binits{X.}},
\bauthor{\bsnm{{Mo}}, \binits{H.J.}},
\bauthor{\bsnm{{van den Bosch}}, \binits{F.C.}}:
\batitle{{Galaxy Groups in the SDSS DR4. II. Halo Occupation Statistics}}.
\bjtitle{ApJ}
\bvolume{676}(\bissue{1}),
\bfpage{248}--\blpage{261}
(\byear{2008})
\doiurl{10.1086/528954}
{\href{https://arxiv.org/abs/0710.5096}{{arXiv:0710.5096}}}
{[astro-ph]}
\end{barticle}
\endbibitem

\bibitem[\protect\citeauthoryear{{Cacciato} et~al.}{2009}]{cacciato2009galaxy}
\begin{barticle}
\bauthor{\bsnm{{Cacciato}}, \binits{M.}},
\bauthor{\bsnm{{van den Bosch}}, \binits{F.C.}},
\bauthor{\bsnm{{More}}, \binits{S.}},
\bauthor{\bsnm{{Li}}, \binits{R.}},
\bauthor{\bsnm{{Mo}}, \binits{H.J.}},
\bauthor{\bsnm{{Yang}}, \binits{X.}}:
\batitle{{Galaxy clustering and galaxy-galaxy lensing: a promising union to constrain cosmological parameters}}.
\bjtitle{MNRAS}
\bvolume{394}(\bissue{2}),
\bfpage{929}--\blpage{946}
(\byear{2009})
\doiurl{10.1111/j.1365-2966.2008.14362.x}
{\href{https://arxiv.org/abs/0807.4932}{{arXiv:0807.4932}}}
{[astro-ph]}
\end{barticle}
\endbibitem

\bibitem[\protect\citeauthoryear{{Yang} et~al.}{2012}]{Yang2012}
\begin{barticle}
\bauthor{\bsnm{{Yang}}, \binits{X.}},
\bauthor{\bsnm{{Mo}}, \binits{H.J.}},
\bauthor{\bsnm{{van den Bosch}}, \binits{F.C.}},
\bauthor{\bsnm{{Zhang}}, \binits{Y.}},
\bauthor{\bsnm{{Han}}, \binits{J.}}:
\batitle{{Evolution of the Galaxy-Dark Matter Connection and the Assembly of Galaxies in Dark Matter Halos}}.
\bjtitle{ApJ}
\bvolume{752}(\bissue{1}),
\bfpage{41}
(\byear{2012})
\doiurl{10.1088/0004-637X/752/1/41}
{\href{https://arxiv.org/abs/1110.1420}{{arXiv:1110.1420}}}
{[astro-ph.CO]}
\end{barticle}
\endbibitem

\bibitem[\protect\citeauthoryear{{Conroy} et~al.}{2006}]{conroy2006modeling}
\begin{barticle}
\bauthor{\bsnm{{Conroy}}, \binits{C.}},
\bauthor{\bsnm{{Wechsler}}, \binits{R.H.}},
\bauthor{\bsnm{{Kravtsov}}, \binits{A.V.}}:
\batitle{{Modeling Luminosity-dependent Galaxy Clustering through Cosmic Time}}.
\bjtitle{ApJ}
\bvolume{647}(\bissue{1}),
\bfpage{201}--\blpage{214}
(\byear{2006})
\doiurl{10.1086/503602}
{\href{https://arxiv.org/abs/astro-ph/0512234}{{arXiv:astro-ph/0512234}}}
{[astro-ph]}
\end{barticle}
\endbibitem

\bibitem[\protect\citeauthoryear{{Behroozi} et~al.}{2010}]{Behroozi2010}
\begin{barticle}
\bauthor{\bsnm{{Behroozi}}, \binits{P.S.}},
\bauthor{\bsnm{{Conroy}}, \binits{C.}},
\bauthor{\bsnm{{Wechsler}}, \binits{R.H.}}:
\batitle{{A Comprehensive Analysis of Uncertainties Affecting the Stellar Mass-Halo Mass Relation for 0 < z < 4}}.
\bjtitle{ApJ}
\bvolume{717}(\bissue{1}),
\bfpage{379}--\blpage{403}
(\byear{2010})
\doiurl{10.1088/0004-637X/717/1/379}
{\href{https://arxiv.org/abs/1001.0015}{{arXiv:1001.0015}}}
{[astro-ph.CO]}
\end{barticle}
\endbibitem

\bibitem[\protect\citeauthoryear{{Neistein} et~al.}{2011}]{neistein2011tale}
\begin{barticle}
\bauthor{\bsnm{{Neistein}}, \binits{E.}},
\bauthor{\bsnm{{Li}}, \binits{C.}},
\bauthor{\bsnm{{Khochfar}}, \binits{S.}},
\bauthor{\bsnm{{Weinmann}}, \binits{S.M.}},
\bauthor{\bsnm{{Shankar}}, \binits{F.}},
\bauthor{\bsnm{{Boylan-Kolchin}}, \binits{M.}}:
\batitle{{A tale of two populations: the stellar mass of central and satellite galaxies}}.
\bjtitle{MNRAS}
\bvolume{416}(\bissue{2}),
\bfpage{1486}--\blpage{1499}
(\byear{2011})
\doiurl{10.1111/j.1365-2966.2011.19145.x}
{\href{https://arxiv.org/abs/1103.3272}{{arXiv:1103.3272}}}
{[astro-ph.CO]}
\end{barticle}
\endbibitem

\bibitem[\protect\citeauthoryear{{Guo} et~al.}{2016}]{Guo2016}
\begin{barticle}
\bauthor{\bsnm{{Guo}}, \binits{H.}},
\bauthor{\bsnm{{Zheng}}, \binits{Z.}},
\bauthor{\bsnm{{Behroozi}}, \binits{P.S.}},
\bauthor{\bsnm{{Zehavi}}, \binits{I.}},
\bauthor{\bsnm{{Chuang}}, \binits{C.-H.}},
\bauthor{\bsnm{{Comparat}}, \binits{J.}},
\bauthor{\bsnm{{Favole}}, \binits{G.}},
\bauthor{\bsnm{{Gottloeber}}, \binits{S.}},
\bauthor{\bsnm{{Klypin}}, \binits{A.}},
\bauthor{\bsnm{{Prada}}, \binits{F.}},
\bauthor{\bsnm{{Rodr{\'\i}guez-Torres}}, \binits{S.A.}},
\bauthor{\bsnm{{Weinberg}}, \binits{D.H.}},
\bauthor{\bsnm{{Yepes}}, \binits{G.}}:
\batitle{{Modelling galaxy clustering: halo occupation distribution versus subhalo matching}}.
\bjtitle{MNRAS}
\bvolume{459}(\bissue{3}),
\bfpage{3040}--\blpage{3058}
(\byear{2016})
\doiurl{10.1093/mnras/stw845}
{\href{https://arxiv.org/abs/1508.07012}{{arXiv:1508.07012}}}
{[astro-ph.CO]}
\end{barticle}
\endbibitem

\bibitem[\protect\citeauthoryear{{Katsianis} et~al.}{2023}]{Katsianis2023}
\begin{barticle}
\bauthor{\bsnm{{Katsianis}}, \binits{A.}},
\bauthor{\bsnm{{Yang}}, \binits{X.}},
\bauthor{\bsnm{{Fong}}, \binits{M.}},
\bauthor{\bsnm{{Wang}}, \binits{J.}}:
\batitle{{Modelling the mass accretion histories of dark matter haloes using a gamma formalism}}.
\bjtitle{MNRAS}
\bvolume{523}(\bissue{1}),
\bfpage{1538}--\blpage{1555}
(\byear{2023})
\doiurl{10.1093/mnras/stad1498}
{\href{https://arxiv.org/abs/2305.17772}{{arXiv:2305.17772}}}
{[astro-ph.GA]}
\end{barticle}
\endbibitem

\bibitem[\protect\citeauthoryear{{Yang} et~al.}{2005}]{yang2005halo}
\begin{barticle}
\bauthor{\bsnm{{Yang}}, \binits{X.}},
\bauthor{\bsnm{{Mo}}, \binits{H.J.}},
\bauthor{\bsnm{{van den Bosch}}, \binits{F.C.}},
\bauthor{\bsnm{{Jing}}, \binits{Y.P.}}:
\batitle{{A halo-based galaxy group finder: calibration and application to the 2dFGRS}}.
\bjtitle{MNRAS}
\bvolume{356}(\bissue{4}),
\bfpage{1293}--\blpage{1307}
(\byear{2005})
\doiurl{10.1111/j.1365-2966.2005.08560.x}
{\href{https://arxiv.org/abs/astro-ph/0405234}{{arXiv:astro-ph/0405234}}}
{[astro-ph]}
\end{barticle}
\endbibitem

\bibitem[\protect\citeauthoryear{{To} et~al.}{2020}]{To2020}
\begin{barticle}
\bauthor{\bsnm{{To}}, \binits{C.-H.}},
\bauthor{\bsnm{{Reddick}}, \binits{R.M.}},
\bauthor{\bsnm{{Rozo}}, \binits{E.}},
\bauthor{\bsnm{{Rykoff}}, \binits{E.}},
\bauthor{\bsnm{{Wechsler}}, \binits{R.H.}}:
\batitle{{RedMaPPer: Evolution and Mass Dependence of the Conditional Luminosity Functions of Red Galaxies in Galaxy Clusters}}.
\bjtitle{ApJ}
\bvolume{897}(\bissue{1}),
\bfpage{15}
(\byear{2020})
\doiurl{10.3847/1538-4357/ab9636}
{\href{https://arxiv.org/abs/1910.01656}{{arXiv:1910.01656}}}
{[astro-ph.GA]}
\end{barticle}
\endbibitem

\bibitem[\protect\citeauthoryear{{Meng} et~al.}{2023}]{Meng2023}
\begin{barticle}
\bauthor{\bsnm{{Meng}}, \binits{J.}},
\bauthor{\bsnm{{Li}}, \binits{C.}},
\bauthor{\bsnm{{Mo}}, \binits{H.J.}},
\bauthor{\bsnm{{Chen}}, \binits{Y.}},
\bauthor{\bsnm{{Jiang}}, \binits{Z.}},
\bauthor{\bsnm{{Xie}}, \binits{L.}}:
\batitle{{Galaxy Populations in Groups and Clusters: Evidence for a Characteristic Stellar Mass Scale at M $_{{\ensuremath{*}}}$ 10$^{9.5}$ M $_{{\ensuremath{\odot}}}$}}.
\bjtitle{ApJ}
\bvolume{944}(\bissue{1}),
\bfpage{75}
(\byear{2023})
\doiurl{10.3847/1538-4357/acae86}
{\href{https://arxiv.org/abs/2210.17186}{{arXiv:2210.17186}}}
{[astro-ph.GA]}
\end{barticle}
\endbibitem

\bibitem[\protect\citeauthoryear{{Golden-Marx} et~al.}{2022}]{Golden-Marx2022}
\begin{botherref}
\oauthor{\bsnm{{Golden-Marx}}, \binits{J.B.}},
\oauthor{\bsnm{{Zu}}, \binits{Y.}},
\oauthor{\bsnm{{Wang}}, \binits{J.}},
\oauthor{\bsnm{{Li}}, \binits{H.}},
\oauthor{\bsnm{{Zhang}}, \binits{J.}},
\oauthor{\bsnm{{Yang}}, \binits{X.}}:
{Satellite Content and Halo Mass of Galaxy Clusters: Comparison between Red-Sequence and Halo-based Optical Cluster Finders}.
arXiv e-prints,
2212--13270
(2022)
\doiurl{10.48550/arXiv.2212.13270}
{\href{https://arxiv.org/abs/2212.13270}{{arXiv:2212.13270}}}
{[astro-ph.GA]}
\end{botherref}
\endbibitem

\bibitem[\protect\citeauthoryear{{Tinker} et~al.}{2021}]{Tinker2021}
\begin{barticle}
\bauthor{\bsnm{{Tinker}}, \binits{J.L.}},
\bauthor{\bsnm{{Cao}}, \binits{J.}},
\bauthor{\bsnm{{Alpaslan}}, \binits{M.}},
\bauthor{\bsnm{{DeRose}}, \binits{J.}},
\bauthor{\bsnm{{Mao}}, \binits{Y.-Y.}},
\bauthor{\bsnm{{Wechsler}}, \binits{R.H.}}:
\batitle{{Probing the galaxy-halo connection with total satellite luminosity}}.
\bjtitle{MNRAS}
\bvolume{505}(\bissue{4}),
\bfpage{5370}--\blpage{5388}
(\byear{2021})
\doiurl{10.1093/mnras/stab1576}
{\href{https://arxiv.org/abs/1911.04507}{{arXiv:1911.04507}}}
{[astro-ph.GA]}
\end{barticle}
\endbibitem

\bibitem[\protect\citeauthoryear{{Boquien} et~al.}{2019}]{Boquien2019}
\begin{barticle}
\bauthor{\bsnm{{Boquien}}, \binits{M.}},
\bauthor{\bsnm{{Burgarella}}, \binits{D.}},
\bauthor{\bsnm{{Roehlly}}, \binits{Y.}},
\bauthor{\bsnm{{Buat}}, \binits{V.}},
\bauthor{\bsnm{{Ciesla}}, \binits{L.}},
\bauthor{\bsnm{{Corre}}, \binits{D.}},
\bauthor{\bsnm{{Inoue}}, \binits{A.K.}},
\bauthor{\bsnm{{Salas}}, \binits{H.}}:
\batitle{{CIGALE: a python Code Investigating GALaxy Emission}}.
\bjtitle{A\&A}
\bvolume{622},
\bfpage{103}
(\byear{2019})
\doiurl{10.1051/0004-6361/201834156}
{\href{https://arxiv.org/abs/1811.03094}{{arXiv:1811.03094}}}
{[astro-ph.GA]}
\end{barticle}
\endbibitem

\bibitem[\protect\citeauthoryear{{Yang} et~al.}{2020}]{Yang2020}
\begin{barticle}
\bauthor{\bsnm{{Yang}}, \binits{G.}},
\bauthor{\bsnm{{Boquien}}, \binits{M.}},
\bauthor{\bsnm{{Buat}}, \binits{V.}},
\bauthor{\bsnm{{Burgarella}}, \binits{D.}},
\bauthor{\bsnm{{Ciesla}}, \binits{L.}},
\bauthor{\bsnm{{Duras}}, \binits{F.}},
\bauthor{\bsnm{{Stalevski}}, \binits{M.}},
\bauthor{\bsnm{{Brandt}}, \binits{W.N.}},
\bauthor{\bsnm{{Papovich}}, \binits{C.}}:
\batitle{{X-CIGALE: Fitting AGN/galaxy SEDs from X-ray to infrared}}.
\bjtitle{MNRAS}
\bvolume{491}(\bissue{1}),
\bfpage{740}--\blpage{757}
(\byear{2020})
\doiurl{10.1093/mnras/stz3001}
{\href{https://arxiv.org/abs/2001.08263}{{arXiv:2001.08263}}}
{[astro-ph.GA]}
\end{barticle}
\endbibitem

\bibitem[\protect\citeauthoryear{{Yang} et~al.}{2022}]{Yang2022}
\begin{barticle}
\bauthor{\bsnm{{Yang}}, \binits{G.}},
\bauthor{\bsnm{{Boquien}}, \binits{M.}},
\bauthor{\bsnm{{Brandt}}, \binits{W.N.}},
\bauthor{\bsnm{{Buat}}, \binits{V.}},
\bauthor{\bsnm{{Burgarella}}, \binits{D.}},
\bauthor{\bsnm{{Ciesla}}, \binits{L.}},
\bauthor{\bsnm{{Lehmer}}, \binits{B.D.}},
\bauthor{\bsnm{{Ma{\l}ek}}, \binits{K.}},
\bauthor{\bsnm{{Mountrichas}}, \binits{G.}},
\bauthor{\bsnm{{Papovich}}, \binits{C.}},
\bauthor{\bsnm{{Pons}}, \binits{E.}},
\bauthor{\bsnm{{Stalevski}}, \binits{M.}},
\bauthor{\bsnm{{Theul{\'e}}}, \binits{P.}},
\bauthor{\bsnm{{Zhu}}, \binits{S.}}:
\batitle{{Fitting AGN/Galaxy X-Ray-to-radio SEDs with CIGALE and Improvement of the Code}}.
\bjtitle{ApJ}
\bvolume{927}(\bissue{2}),
\bfpage{192}
(\byear{2022})
\doiurl{10.3847/1538-4357/ac4971}
{\href{https://arxiv.org/abs/2201.03718}{{arXiv:2201.03718}}}
{[astro-ph.GA]}
\end{barticle}
\endbibitem

\bibitem[\protect\citeauthoryear{{Noll} et~al.}{2009}]{Noll2009}
\begin{barticle}
\bauthor{\bsnm{{Noll}}, \binits{S.}},
\bauthor{\bsnm{{Burgarella}}, \binits{D.}},
\bauthor{\bsnm{{Giovannoli}}, \binits{E.}},
\bauthor{\bsnm{{Buat}}, \binits{V.}},
\bauthor{\bsnm{{Marcillac}}, \binits{D.}},
\bauthor{\bsnm{{Mu{\~n}oz-Mateos}}, \binits{J.C.}}:
\batitle{{Analysis of galaxy spectral energy distributions from far-UV to far-IR with CIGALE: studying a SINGS test sample}}.
\bjtitle{A\&A}
\bvolume{507}(\bissue{3}),
\bfpage{1793}--\blpage{1813}
(\byear{2009})
\doiurl{10.1051/0004-6361/200912497}
{\href{https://arxiv.org/abs/0909.5439}{{arXiv:0909.5439}}}
{[astro-ph.CO]}
\end{barticle}
\endbibitem

\bibitem[\protect\citeauthoryear{{Guy} et~al.}{2023}]{Guy2023}
\begin{barticle}
\bauthor{\bsnm{{Guy}}, \binits{J.}},
\bauthor{\bsnm{{Bailey}}, \binits{S.}},
\bauthor{\bsnm{{Kremin}}, \binits{A.}},
\bauthor{\bsnm{{Alam}}, \binits{S.}},
\bauthor{\bsnm{{Alexander}}, \binits{D.M.}},
\bauthor{\bsnm{{Allende Prieto}}, \binits{C.}},
\bauthor{\bsnm{{BenZvi}}, \binits{S.}},
\bauthor{\bsnm{{Bolton}}, \binits{A.S.}},
\bauthor{\bsnm{{Brooks}}, \binits{D.}},
\bauthor{\bsnm{{Chaussidon}}, \binits{E.}},
\bauthor{\bsnm{{Cooper}}, \binits{A.P.}},
\bauthor{\bsnm{{Dawson}}, \binits{K.}},
\bauthor{\bsnm{{de la Macorra}}, \binits{A.}},
\bauthor{\bsnm{{Dey}}, \binits{A.}},
\bauthor{\bsnm{{Dey}}, \binits{B.}},
\bauthor{\bsnm{{Dhungana}}, \binits{G.}},
\bauthor{\bsnm{{Eisenstein}}, \binits{D.J.}},
\bauthor{\bsnm{{Font-Ribera}}, \binits{A.}},
\bauthor{\bsnm{{Forero-Romero}}, \binits{J.E.}},
\bauthor{\bsnm{{Gazta{\~n}aga}}, \binits{E.}},
\bauthor{\bsnm{{Gontcho A Gontcho}}, \binits{S.}},
\bauthor{\bsnm{{Green}}, \binits{D.}},
\bauthor{\bsnm{{Honscheid}}, \binits{K.}},
\bauthor{\bsnm{{Ishak}}, \binits{M.}},
\bauthor{\bsnm{{Kehoe}}, \binits{R.}},
\bauthor{\bsnm{{Kirkby}}, \binits{D.}},
\bauthor{\bsnm{{Kisner}}, \binits{T.}},
\bauthor{\bsnm{{Koposov}}, \binits{S.E.}},
\bauthor{\bsnm{{Lan}}, \binits{T.-W.}},
\bauthor{\bsnm{{Landriau}}, \binits{M.}},
\bauthor{\bsnm{{Le Guillou}}, \binits{L.}},
\bauthor{\bsnm{{Levi}}, \binits{M.E.}},
\bauthor{\bsnm{{Magneville}}, \binits{C.}},
\bauthor{\bsnm{{Manser}}, \binits{C.J.}},
\bauthor{\bsnm{{Martini}}, \binits{P.}},
\bauthor{\bsnm{{Meisner}}, \binits{A.M.}},
\bauthor{\bsnm{{Miquel}}, \binits{R.}},
\bauthor{\bsnm{{Moustakas}}, \binits{J.}},
\bauthor{\bsnm{{Myers}}, \binits{A.D.}},
\bauthor{\bsnm{{Newman}}, \binits{J.A.}},
\bauthor{\bsnm{{Nie}}, \binits{J.}},
\bauthor{\bsnm{{Palanque-Delabrouille}}, \binits{N.}},
\bauthor{\bsnm{{Percival}}, \binits{W.J.}},
\bauthor{\bsnm{{Poppett}}, \binits{C.}},
\bauthor{\bsnm{{Prada}}, \binits{F.}},
\bauthor{\bsnm{{Raichoor}}, \binits{A.}},
\bauthor{\bsnm{{Ravoux}}, \binits{C.}},
\bauthor{\bsnm{{Ross}}, \binits{A.J.}},
\bauthor{\bsnm{{Schlafly}}, \binits{E.F.}},
\bauthor{\bsnm{{Schlegel}}, \binits{D.}},
\bauthor{\bsnm{{Schubnell}}, \binits{M.}},
\bauthor{\bsnm{{Sharples}}, \binits{R.M.}},
\bauthor{\bsnm{{Tarl{\'e}}}, \binits{G.}},
\bauthor{\bsnm{{Weaver}}, \binits{B.A.}},
\bauthor{\bsnm{{Y{\'e}che}}, \binits{C.}},
\bauthor{\bsnm{{Zhou}}, \binits{R.}},
\bauthor{\bsnm{{Zhou}}, \binits{Z.}},
\bauthor{\bsnm{{Zou}}, \binits{H.}}:
\batitle{{The Spectroscopic Data Processing Pipeline for the Dark Energy Spectroscopic Instrument}}.
\bjtitle{AJ}
\bvolume{165}(\bissue{4}),
\bfpage{144}
(\byear{2023})
\doiurl{10.3847/1538-3881/acb212}
{\href{https://arxiv.org/abs/2209.14482}{{arXiv:2209.14482}}}
{[astro-ph.IM]}
\end{barticle}
\endbibitem

\bibitem[\protect\citeauthoryear{{Brodzeller} et~al.}{2023}]{Brodzeller2023}
\begin{barticle}
\bauthor{\bsnm{{Brodzeller}}, \binits{A.}},
\bauthor{\bsnm{{Dawson}}, \binits{K.}},
\bauthor{\bsnm{{Bailey}}, \binits{S.}},
\bauthor{\bsnm{{Yu}}, \binits{J.}},
\bauthor{\bsnm{{Ross}}, \binits{A.J.}},
\bauthor{\bsnm{{Bault}}, \binits{A.}},
\bauthor{\bsnm{{Filbert}}, \binits{S.}},
\bauthor{\bsnm{{Aguilar}}, \binits{J.}},
\bauthor{\bsnm{{Ahlen}}, \binits{S.}},
\bauthor{\bsnm{{Alexander}}, \binits{D.M.}},
\bauthor{\bsnm{{Armengaud}}, \binits{E.}},
\bauthor{\bsnm{{Berti}}, \binits{A.}},
\bauthor{\bsnm{{Brooks}}, \binits{D.}},
\bauthor{\bsnm{{Chaussidon}}, \binits{E.}},
\bauthor{\bsnm{{de la Macorra}}, \binits{A.}},
\bauthor{\bsnm{{Doel}}, \binits{P.}},
\bauthor{\bsnm{{Fanning}}, \binits{K.}},
\bauthor{\bsnm{{Fawcett}}, \binits{V.A.}},
\bauthor{\bsnm{{Font-Ribera}}, \binits{A.}},
\bauthor{\bsnm{{A Gontcho}}, \binits{S.G.}},
\bauthor{\bsnm{{Guy}}, \binits{J.}},
\bauthor{\bsnm{{Honscheid}}, \binits{K.}},
\bauthor{\bsnm{{Juneau}}, \binits{S.}},
\bauthor{\bsnm{{Kehoe}}, \binits{R.}},
\bauthor{\bsnm{{Kisner}}, \binits{T.}},
\bauthor{\bsnm{{Kremin}}, \binits{A.}},
\bauthor{\bsnm{{Lan}}, \binits{T.-W.}},
\bauthor{\bsnm{{Landriau}}, \binits{M.}},
\bauthor{\bsnm{{Levi}}, \binits{M.E.}},
\bauthor{\bsnm{{Magneville}}, \binits{C.}},
\bauthor{\bsnm{{Martini}}, \binits{P.}},
\bauthor{\bsnm{{Meisner}}, \binits{A.M.}},
\bauthor{\bsnm{{Miquel}}, \binits{R.}},
\bauthor{\bsnm{{Moustakas}}, \binits{J.}},
\bauthor{\bsnm{{Palanque-Delabrouille}}, \binits{N.}},
\bauthor{\bsnm{{Percival}}, \binits{W.J.}},
\bauthor{\bsnm{{Prada}}, \binits{F.}},
\bauthor{\bsnm{{Ravoux}}, \binits{C.}},
\bauthor{\bsnm{{Rossi}}, \binits{G.}},
\bauthor{\bsnm{{Saulder}}, \binits{C.}},
\bauthor{\bsnm{{Siudek}}, \binits{M.}},
\bauthor{\bsnm{{Tarl{\'e}}}, \binits{G.}},
\bauthor{\bsnm{{Weaver}}, \binits{B.A.}},
\bauthor{\bsnm{{Youles}}, \binits{S.}},
\bauthor{\bsnm{{Zheng}}, \binits{Z.}},
\bauthor{\bsnm{{Zhou}}, \binits{R.}},
\bauthor{\bsnm{{Zhou}}, \binits{Z.}}:
\batitle{{Performance of the Quasar Spectral Templates for the Dark Energy Spectroscopic Instrument}}.
\bjtitle{AJ}
\bvolume{166}(\bissue{2}),
\bfpage{66}
(\byear{2023})
\doiurl{10.3847/1538-3881/ace35d}
{\href{https://arxiv.org/abs/2305.10426}{{arXiv:2305.10426}}}
{[astro-ph.IM]}
\end{barticle}
\endbibitem

\bibitem[\protect\citeauthoryear{{Moustakas}}{2023}]{Moustakas2023}
\begin{botherref}
\oauthor{\bsnm{{Moustakas}}, \binits{J.}}:
{FastSpecFit: Fast spectral synthesis and emission-line fitting of DESI spectra}.
Astrophysics Source Code Library, record ascl:2308.005
(2023)
\end{botherref}
\endbibitem

\bibitem[\protect\citeauthoryear{{Schlafly} et~al.}{2023}]{Schlafly2023}
\begin{botherref}
\oauthor{\bsnm{{Schlafly}}, \binits{E.F.}},
\oauthor{\bsnm{{Kirkby}}, \binits{D.}},
\oauthor{\bsnm{{Schlegel}}, \binits{D.J.}},
\oauthor{\bsnm{{Myers}}, \binits{A.D.}},
\oauthor{\bsnm{{Raichoor}}, \binits{A.}},
\oauthor{\bsnm{{Dawson}}, \binits{K.}},
\oauthor{\bsnm{{Aguilar}}, \binits{J.}},
\oauthor{\bsnm{{Allende Prieto}}, \binits{C.}},
\oauthor{\bsnm{{Bailey}}, \binits{S.}},
\oauthor{\bsnm{{BenZvi}}, \binits{S.}},
\oauthor{\bsnm{{Bermejo-Climent}}, \binits{J.}},
\oauthor{\bsnm{{Brooks}}, \binits{D.}},
\oauthor{\bsnm{{de la Macorra}}, \binits{A.}},
\oauthor{\bsnm{{Dey}}, \binits{A.}},
\oauthor{\bsnm{{Doel}}, \binits{P.}},
\oauthor{\bsnm{{Fanning}}, \binits{K.}},
\oauthor{\bsnm{{Font-Ribera}}, \binits{A.}},
\oauthor{\bsnm{{Forero-Romero}}, \binits{J.E.}},
\oauthor{\bsnm{{Garc{\'\i}a-Bellido}}, \binits{J.}},
\oauthor{\bsnm{{Gontcho}}, \binits{S.G.A.}},
\oauthor{\bsnm{{Guy}}, \binits{J.}},
\oauthor{\bsnm{{Hahn}}, \binits{C.}},
\oauthor{\bsnm{{Honscheid}}, \binits{K.}},
\oauthor{\bsnm{{Ishak}}, \binits{M.}},
\oauthor{\bsnm{{Juneau}}, \binits{S.}},
\oauthor{\bsnm{{Kehoe}}, \binits{R.}},
\oauthor{\bsnm{{Kisner}}, \binits{T.}},
\oauthor{\bsnm{{Kremin}}, \binits{A.}},
\oauthor{\bsnm{{Landriau}}, \binits{M.}},
\oauthor{\bsnm{{Lang}}, \binits{D.A.}},
\oauthor{\bsnm{{Lasker}}, \binits{J.}},
\oauthor{\bsnm{{Levi}}, \binits{M.E.}},
\oauthor{\bsnm{{Magneville}}, \binits{C.}},
\oauthor{\bsnm{{Manser}}, \binits{C.J.}},
\oauthor{\bsnm{{Martini}}, \binits{P.}},
\oauthor{\bsnm{{Meisner}}, \binits{A.M.}},
\oauthor{\bsnm{{Miquel}}, \binits{R.}},
\oauthor{\bsnm{{Moustakas}}, \binits{J.}},
\oauthor{\bsnm{{Newman}}, \binits{J.A.}},
\oauthor{\bsnm{{Nie}}, \binits{J.}},
\oauthor{\bsnm{{Palanque-Delabrouille}}, \binits{N.}},
\oauthor{\bsnm{{Percival}}, \binits{W.J.}},
\oauthor{\bsnm{{Poppett}}, \binits{C.}},
\oauthor{\bsnm{{Rockosi}}, \binits{C.}},
\oauthor{\bsnm{{Ross}}, \binits{A.J.}},
\oauthor{\bsnm{{Rossi}}, \binits{G.}},
\oauthor{\bsnm{{Tarl{\'e}}}, \binits{G.}},
\oauthor{\bsnm{{Weaver}}, \binits{B.A.}},
\oauthor{\bsnm{{Y{\`e}che}}, \binits{C.}},
\oauthor{\bsnm{{Zhou}}, \binits{R.}}:
{Survey Operations for the Dark Energy Spectroscopic Instrument}.
arXiv e-prints,
2306--06309
(2023)
\doiurl{10.48550/arXiv.2306.06309}
{\href{https://arxiv.org/abs/2306.06309}{{arXiv:2306.06309}}}
{[astro-ph.CO]}
\end{botherref}
\endbibitem

\bibitem[\protect\citeauthoryear{{DESI Collaboration} et~al.}{2016a}]{DESI_Collaboration2016a}
\begin{botherref}
\oauthor{\bsnm{{DESI Collaboration}}},
\oauthor{\bsnm{{Aghamousa}}, \binits{A.}},
\oauthor{\bsnm{{Aguilar}}, \binits{J.}},
\oauthor{\bsnm{{Ahlen}}, \binits{S.}},
\oauthor{\bsnm{{Alam}}, \binits{S.}},
\oauthor{\bsnm{{Allen}}, \binits{L.E.}},
\oauthor{\bsnm{{Allende Prieto}}, \binits{C.}},
\oauthor{\bsnm{{Annis}}, \binits{J.}},
\oauthor{\bsnm{{Bailey}}, \binits{S.}},
\oauthor{\bsnm{{Balland}}, \binits{C.}},
\oauthor{\bsnm{{Ballester}}, \binits{O.}},
\oauthor{\bsnm{{Baltay}}, \binits{C.}},
\oauthor{\bsnm{{Beaufore}}, \binits{L.}},
\oauthor{\bsnm{{Bebek}}, \binits{C.}},
\oauthor{\bsnm{{Beers}}, \binits{T.C.}},
\oauthor{\bsnm{{Bell}}, \binits{E.F.}},
\oauthor{\bsnm{{Bernal}}, \binits{J.L.}},
\oauthor{\bsnm{{Besuner}}, \binits{R.}},
\oauthor{\bsnm{{Beutler}}, \binits{F.}},
\oauthor{\bsnm{{Blake}}, \binits{C.}},
\oauthor{\bsnm{{Bleuler}}, \binits{H.}},
\oauthor{\bsnm{{Blomqvist}}, \binits{M.}},
\oauthor{\bsnm{{Blum}}, \binits{R.}},
\oauthor{\bsnm{{Bolton}}, \binits{A.S.}},
\oauthor{\bsnm{{Briceno}}, \binits{C.}},
\oauthor{\bsnm{{Brooks}}, \binits{D.}},
\oauthor{\bsnm{{Brownstein}}, \binits{J.R.}},
\oauthor{\bsnm{{Buckley-Geer}}, \binits{E.}},
\oauthor{\bsnm{{Burden}}, \binits{A.}},
\oauthor{\bsnm{{Burtin}}, \binits{E.}},
\oauthor{\bsnm{{Busca}}, \binits{N.G.}},
\oauthor{\bsnm{{Cahn}}, \binits{R.N.}},
\oauthor{\bsnm{{Cai}}, \binits{Y.-C.}},
\oauthor{\bsnm{{Cardiel-Sas}}, \binits{L.}},
\oauthor{\bsnm{{Carlberg}}, \binits{R.G.}},
\oauthor{\bsnm{{Carton}}, \binits{P.-H.}},
\oauthor{\bsnm{{Casas}}, \binits{R.}},
\oauthor{\bsnm{{Castander}}, \binits{F.J.}},
\oauthor{\bsnm{{Cervantes-Cota}}, \binits{J.L.}},
\oauthor{\bsnm{{Claybaugh}}, \binits{T.M.}},
\oauthor{\bsnm{{Close}}, \binits{M.}},
\oauthor{\bsnm{{Coker}}, \binits{C.T.}},
\oauthor{\bsnm{{Cole}}, \binits{S.}},
\oauthor{\bsnm{{Comparat}}, \binits{J.}},
\oauthor{\bsnm{{Cooper}}, \binits{A.P.}},
\oauthor{\bsnm{{Cousinou}}, \binits{M.-C.}},
\oauthor{\bsnm{{Crocce}}, \binits{M.}},
\oauthor{\bsnm{{Cuby}}, \binits{J.-G.}},
\oauthor{\bsnm{{Cunningham}}, \binits{D.P.}},
\oauthor{\bsnm{{Davis}}, \binits{T.M.}},
\oauthor{\bsnm{{Dawson}}, \binits{K.S.}},
\oauthor{\bsnm{{de la Macorra}}, \binits{A.}},
\oauthor{\bsnm{{De Vicente}}, \binits{J.}},
\oauthor{\bsnm{{Delubac}}, \binits{T.}},
\oauthor{\bsnm{{Derwent}}, \binits{M.}},
\oauthor{\bsnm{{Dey}}, \binits{A.}},
\oauthor{\bsnm{{Dhungana}}, \binits{G.}},
\oauthor{\bsnm{{Ding}}, \binits{Z.}},
\oauthor{\bsnm{{Doel}}, \binits{P.}},
\oauthor{\bsnm{{Duan}}, \binits{Y.T.}},
\oauthor{\bsnm{{Ealet}}, \binits{A.}},
\oauthor{\bsnm{{Edelstein}}, \binits{J.}},
\oauthor{\bsnm{{Eftekharzadeh}}, \binits{S.}},
\oauthor{\bsnm{{Eisenstein}}, \binits{D.J.}},
\oauthor{\bsnm{{Elliott}}, \binits{A.}},
\oauthor{\bsnm{{Escoffier}}, \binits{S.}},
\oauthor{\bsnm{{Evatt}}, \binits{M.}},
\oauthor{\bsnm{{Fagrelius}}, \binits{P.}},
\oauthor{\bsnm{{Fan}}, \binits{X.}},
\oauthor{\bsnm{{Fanning}}, \binits{K.}},
\oauthor{\bsnm{{Farahi}}, \binits{A.}},
\oauthor{\bsnm{{Farihi}}, \binits{J.}},
\oauthor{\bsnm{{Favole}}, \binits{G.}},
\oauthor{\bsnm{{Feng}}, \binits{Y.}},
\oauthor{\bsnm{{Fernandez}}, \binits{E.}},
\oauthor{\bsnm{{Findlay}}, \binits{J.R.}},
\oauthor{\bsnm{{Finkbeiner}}, \binits{D.P.}},
\oauthor{\bsnm{{Fitzpatrick}}, \binits{M.J.}},
\oauthor{\bsnm{{Flaugher}}, \binits{B.}},
\oauthor{\bsnm{{Flender}}, \binits{S.}},
\oauthor{\bsnm{{Font-Ribera}}, \binits{A.}},
\oauthor{\bsnm{{Forero-Romero}}, \binits{J.E.}},
\oauthor{\bsnm{{Fosalba}}, \binits{P.}},
\oauthor{\bsnm{{Frenk}}, \binits{C.S.}},
\oauthor{\bsnm{{Fumagalli}}, \binits{M.}},
\oauthor{\bsnm{{Gaensicke}}, \binits{B.T.}},
\oauthor{\bsnm{{Gallo}}, \binits{G.}},
\oauthor{\bsnm{{Garcia-Bellido}}, \binits{J.}},
\oauthor{\bsnm{{Gaztanaga}}, \binits{E.}},
\oauthor{\bsnm{{Pietro Gentile Fusillo}}, \binits{N.}},
\oauthor{\bsnm{{Gerard}}, \binits{T.}},
\oauthor{\bsnm{{Gershkovich}}, \binits{I.}},
\oauthor{\bsnm{{Giannantonio}}, \binits{T.}},
\oauthor{\bsnm{{Gillet}}, \binits{D.}},
\oauthor{\bsnm{{Gonzalez-de-Rivera}}, \binits{G.}},
\oauthor{\bsnm{{Gonzalez-Perez}}, \binits{V.}},
\oauthor{\bsnm{{Gott}}, \binits{S.}},
\oauthor{\bsnm{{Graur}}, \binits{O.}},
\oauthor{\bsnm{{Gutierrez}}, \binits{G.}},
\oauthor{\bsnm{{Guy}}, \binits{J.}},
\oauthor{\bsnm{{Habib}}, \binits{S.}},
\oauthor{\bsnm{{Heetderks}}, \binits{H.}},
\oauthor{\bsnm{{Heetderks}}, \binits{I.}},
\oauthor{\bsnm{{Heitmann}}, \binits{K.}},
\oauthor{\bsnm{{Hellwing}}, \binits{W.A.}},
\oauthor{\bsnm{{Herrera}}, \binits{D.A.}},
\oauthor{\bsnm{{Ho}}, \binits{S.}},
\oauthor{\bsnm{{Holland}}, \binits{S.}},
\oauthor{\bsnm{{Honscheid}}, \binits{K.}},
\oauthor{\bsnm{{Huff}}, \binits{E.}},
\oauthor{\bsnm{{Hutchinson}}, \binits{T.A.}},
\oauthor{\bsnm{{Huterer}}, \binits{D.}},
\oauthor{\bsnm{{Hwang}}, \binits{H.S.}},
\oauthor{\bsnm{{Illa Laguna}}, \binits{J.M.}},
\oauthor{\bsnm{{Ishikawa}}, \binits{Y.}},
\oauthor{\bsnm{{Jacobs}}, \binits{D.}},
\oauthor{\bsnm{{Jeffrey}}, \binits{N.}},
\oauthor{\bsnm{{Jelinsky}}, \binits{P.}},
\oauthor{\bsnm{{Jennings}}, \binits{E.}},
\oauthor{\bsnm{{Jiang}}, \binits{L.}},
\oauthor{\bsnm{{Jimenez}}, \binits{J.}},
\oauthor{\bsnm{{Johnson}}, \binits{J.}},
\oauthor{\bsnm{{Joyce}}, \binits{R.}},
\oauthor{\bsnm{{Jullo}}, \binits{E.}},
\oauthor{\bsnm{{Juneau}}, \binits{S.}},
\oauthor{\bsnm{{Kama}}, \binits{S.}},
\oauthor{\bsnm{{Karcher}}, \binits{A.}},
\oauthor{\bsnm{{Karkar}}, \binits{S.}},
\oauthor{\bsnm{{Kehoe}}, \binits{R.}},
\oauthor{\bsnm{{Kennamer}}, \binits{N.}},
\oauthor{\bsnm{{Kent}}, \binits{S.}},
\oauthor{\bsnm{{Kilbinger}}, \binits{M.}},
\oauthor{\bsnm{{Kim}}, \binits{A.G.}},
\oauthor{\bsnm{{Kirkby}}, \binits{D.}},
\oauthor{\bsnm{{Kisner}}, \binits{T.}},
\oauthor{\bsnm{{Kitanidis}}, \binits{E.}},
\oauthor{\bsnm{{Kneib}}, \binits{J.-P.}},
\oauthor{\bsnm{{Koposov}}, \binits{S.}},
\oauthor{\bsnm{{Kovacs}}, \binits{E.}},
\oauthor{\bsnm{{Koyama}}, \binits{K.}},
\oauthor{\bsnm{{Kremin}}, \binits{A.}},
\oauthor{\bsnm{{Kron}}, \binits{R.}},
\oauthor{\bsnm{{Kronig}}, \binits{L.}},
\oauthor{\bsnm{{Kueter-Young}}, \binits{A.}},
\oauthor{\bsnm{{Lacey}}, \binits{C.G.}},
\oauthor{\bsnm{{Lafever}}, \binits{R.}},
\oauthor{\bsnm{{Lahav}}, \binits{O.}},
\oauthor{\bsnm{{Lambert}}, \binits{A.}},
\oauthor{\bsnm{{Lampton}}, \binits{M.}},
\oauthor{\bsnm{{Landriau}}, \binits{M.}},
\oauthor{\bsnm{{Lang}}, \binits{D.}},
\oauthor{\bsnm{{Lauer}}, \binits{T.R.}},
\oauthor{\bsnm{{Le Goff}}, \binits{J.-M.}},
\oauthor{\bsnm{{Le Guillou}}, \binits{L.}},
\oauthor{\bsnm{{Le Van Suu}}, \binits{A.}},
\oauthor{\bsnm{{Lee}}, \binits{J.H.}},
\oauthor{\bsnm{{Lee}}, \binits{S.-J.}},
\oauthor{\bsnm{{Leitner}}, \binits{D.}},
\oauthor{\bsnm{{Lesser}}, \binits{M.}},
\oauthor{\bsnm{{Levi}}, \binits{M.E.}},
\oauthor{\bsnm{{L'Huillier}}, \binits{B.}},
\oauthor{\bsnm{{Li}}, \binits{B.}},
\oauthor{\bsnm{{Liang}}, \binits{M.}},
\oauthor{\bsnm{{Lin}}, \binits{H.}},
\oauthor{\bsnm{{Linder}}, \binits{E.}},
\oauthor{\bsnm{{Loebman}}, \binits{S.R.}},
\oauthor{\bsnm{{Luki{\'c}}}, \binits{Z.}},
\oauthor{\bsnm{{Ma}}, \binits{J.}},
\oauthor{\bsnm{{MacCrann}}, \binits{N.}},
\oauthor{\bsnm{{Magneville}}, \binits{C.}},
\oauthor{\bsnm{{Makarem}}, \binits{L.}},
\oauthor{\bsnm{{Manera}}, \binits{M.}},
\oauthor{\bsnm{{Manser}}, \binits{C.J.}},
\oauthor{\bsnm{{Marshall}}, \binits{R.}},
\oauthor{\bsnm{{Martini}}, \binits{P.}},
\oauthor{\bsnm{{Massey}}, \binits{R.}},
\oauthor{\bsnm{{Matheson}}, \binits{T.}},
\oauthor{\bsnm{{McCauley}}, \binits{J.}},
\oauthor{\bsnm{{McDonald}}, \binits{P.}},
\oauthor{\bsnm{{McGreer}}, \binits{I.D.}},
\oauthor{\bsnm{{Meisner}}, \binits{A.}},
\oauthor{\bsnm{{Metcalfe}}, \binits{N.}},
\oauthor{\bsnm{{Miller}}, \binits{T.N.}},
\oauthor{\bsnm{{Miquel}}, \binits{R.}},
\oauthor{\bsnm{{Moustakas}}, \binits{J.}},
\oauthor{\bsnm{{Myers}}, \binits{A.}},
\oauthor{\bsnm{{Naik}}, \binits{M.}},
\oauthor{\bsnm{{Newman}}, \binits{J.A.}},
\oauthor{\bsnm{{Nichol}}, \binits{R.C.}},
\oauthor{\bsnm{{Nicola}}, \binits{A.}},
\oauthor{\bsnm{{Nicolati da Costa}}, \binits{L.}},
\oauthor{\bsnm{{Nie}}, \binits{J.}},
\oauthor{\bsnm{{Niz}}, \binits{G.}},
\oauthor{\bsnm{{Norberg}}, \binits{P.}},
\oauthor{\bsnm{{Nord}}, \binits{B.}},
\oauthor{\bsnm{{Norman}}, \binits{D.}},
\oauthor{\bsnm{{Nugent}}, \binits{P.}},
\oauthor{\bsnm{{O'Brien}}, \binits{T.}},
\oauthor{\bsnm{{Oh}}, \binits{M.}},
\oauthor{\bsnm{{Olsen}}, \binits{K.A.G.}},
\oauthor{\bsnm{{Padilla}}, \binits{C.}},
\oauthor{\bsnm{{Padmanabhan}}, \binits{H.}},
\oauthor{\bsnm{{Padmanabhan}}, \binits{N.}},
\oauthor{\bsnm{{Palanque-Delabrouille}}, \binits{N.}},
\oauthor{\bsnm{{Palmese}}, \binits{A.}},
\oauthor{\bsnm{{Pappalardo}}, \binits{D.}},
\oauthor{\bsnm{{P{\^a}ris}}, \binits{I.}},
\oauthor{\bsnm{{Park}}, \binits{C.}},
\oauthor{\bsnm{{Patej}}, \binits{A.}},
\oauthor{\bsnm{{Peacock}}, \binits{J.A.}},
\oauthor{\bsnm{{Peiris}}, \binits{H.V.}},
\oauthor{\bsnm{{Peng}}, \binits{X.}},
\oauthor{\bsnm{{Percival}}, \binits{W.J.}},
\oauthor{\bsnm{{Perruchot}}, \binits{S.}},
\oauthor{\bsnm{{Pieri}}, \binits{M.M.}},
\oauthor{\bsnm{{Pogge}}, \binits{R.}},
\oauthor{\bsnm{{Pollack}}, \binits{J.E.}},
\oauthor{\bsnm{{Poppett}}, \binits{C.}},
\oauthor{\bsnm{{Prada}}, \binits{F.}},
\oauthor{\bsnm{{Prakash}}, \binits{A.}},
\oauthor{\bsnm{{Probst}}, \binits{R.G.}},
\oauthor{\bsnm{{Rabinowitz}}, \binits{D.}},
\oauthor{\bsnm{{Raichoor}}, \binits{A.}},
\oauthor{\bsnm{{Ree}}, \binits{C.H.}},
\oauthor{\bsnm{{Refregier}}, \binits{A.}},
\oauthor{\bsnm{{Regal}}, \binits{X.}},
\oauthor{\bsnm{{Reid}}, \binits{B.}},
\oauthor{\bsnm{{Reil}}, \binits{K.}},
\oauthor{\bsnm{{Rezaie}}, \binits{M.}},
\oauthor{\bsnm{{Rockosi}}, \binits{C.M.}},
\oauthor{\bsnm{{Roe}}, \binits{N.}},
\oauthor{\bsnm{{Ronayette}}, \binits{S.}},
\oauthor{\bsnm{{Roodman}}, \binits{A.}},
\oauthor{\bsnm{{Ross}}, \binits{A.J.}},
\oauthor{\bsnm{{Ross}}, \binits{N.P.}},
\oauthor{\bsnm{{Rossi}}, \binits{G.}},
\oauthor{\bsnm{{Rozo}}, \binits{E.}},
\oauthor{\bsnm{{Ruhlmann-Kleider}}, \binits{V.}},
\oauthor{\bsnm{{Rykoff}}, \binits{E.S.}},
\oauthor{\bsnm{{Sabiu}}, \binits{C.}},
\oauthor{\bsnm{{Samushia}}, \binits{L.}},
\oauthor{\bsnm{{Sanchez}}, \binits{E.}},
\oauthor{\bsnm{{Sanchez}}, \binits{J.}},
\oauthor{\bsnm{{Schlegel}}, \binits{D.J.}},
\oauthor{\bsnm{{Schneider}}, \binits{M.}},
\oauthor{\bsnm{{Schubnell}}, \binits{M.}},
\oauthor{\bsnm{{Secroun}}, \binits{A.}},
\oauthor{\bsnm{{Seljak}}, \binits{U.}},
\oauthor{\bsnm{{Seo}}, \binits{H.-J.}},
\oauthor{\bsnm{{Serrano}}, \binits{S.}},
\oauthor{\bsnm{{Shafieloo}}, \binits{A.}},
\oauthor{\bsnm{{Shan}}, \binits{H.}},
\oauthor{\bsnm{{Sharples}}, \binits{R.}},
\oauthor{\bsnm{{Sholl}}, \binits{M.J.}},
\oauthor{\bsnm{{Shourt}}, \binits{W.V.}},
\oauthor{\bsnm{{Silber}}, \binits{J.H.}},
\oauthor{\bsnm{{Silva}}, \binits{D.R.}},
\oauthor{\bsnm{{Sirk}}, \binits{M.M.}},
\oauthor{\bsnm{{Slosar}}, \binits{A.}},
\oauthor{\bsnm{{Smith}}, \binits{A.}},
\oauthor{\bsnm{{Smoot}}, \binits{G.F.}},
\oauthor{\bsnm{{Som}}, \binits{D.}},
\oauthor{\bsnm{{Song}}, \binits{Y.-S.}},
\oauthor{\bsnm{{Sprayberry}}, \binits{D.}},
\oauthor{\bsnm{{Staten}}, \binits{R.}},
\oauthor{\bsnm{{Stefanik}}, \binits{A.}},
\oauthor{\bsnm{{Tarle}}, \binits{G.}},
\oauthor{\bsnm{{Sien Tie}}, \binits{S.}},
\oauthor{\bsnm{{Tinker}}, \binits{J.L.}},
\oauthor{\bsnm{{Tojeiro}}, \binits{R.}},
\oauthor{\bsnm{{Valdes}}, \binits{F.}},
\oauthor{\bsnm{{Valenzuela}}, \binits{O.}},
\oauthor{\bsnm{{Valluri}}, \binits{M.}},
\oauthor{\bsnm{{Vargas-Magana}}, \binits{M.}},
\oauthor{\bsnm{{Verde}}, \binits{L.}},
\oauthor{\bsnm{{Walker}}, \binits{A.R.}},
\oauthor{\bsnm{{Wang}}, \binits{J.}},
\oauthor{\bsnm{{Wang}}, \binits{Y.}},
\oauthor{\bsnm{{Weaver}}, \binits{B.A.}},
\oauthor{\bsnm{{Weaverdyck}}, \binits{C.}},
\oauthor{\bsnm{{Wechsler}}, \binits{R.H.}},
\oauthor{\bsnm{{Weinberg}}, \binits{D.H.}},
\oauthor{\bsnm{{White}}, \binits{M.}},
\oauthor{\bsnm{{Yang}}, \binits{Q.}},
\oauthor{\bsnm{{Yeche}}, \binits{C.}},
\oauthor{\bsnm{{Zhang}}, \binits{T.}},
\oauthor{\bsnm{{Zhao}}, \binits{G.-B.}},
\oauthor{\bsnm{{Zheng}}, \binits{Y.}},
\oauthor{\bsnm{{Zhou}}, \binits{X.}},
\oauthor{\bsnm{{Zhou}}, \binits{Z.}},
\oauthor{\bsnm{{Zhu}}, \binits{Y.}},
\oauthor{\bsnm{{Zou}}, \binits{H.}},
\oauthor{\bsnm{{Zu}}, \binits{Y.}}:
{The DESI Experiment Part I: Science,Targeting, and Survey Design}.
arXiv e-prints,
1611--00036
(2016)
{\href{https://arxiv.org/abs/1611.00036}{{arXiv:1611.00036}}}
{[astro-ph.IM]}
\end{botherref}
\endbibitem

\bibitem[\protect\citeauthoryear{{DESI Collaboration} et~al.}{2016b}]{DESI_Collaboration2016b}
\begin{botherref}
\oauthor{\bsnm{{DESI Collaboration}}},
\oauthor{\bsnm{{Aghamousa}}, \binits{A.}},
\oauthor{\bsnm{{Aguilar}}, \binits{J.}},
\oauthor{\bsnm{{Ahlen}}, \binits{S.}},
\oauthor{\bsnm{{Alam}}, \binits{S.}},
\oauthor{\bsnm{{Allen}}, \binits{L.E.}},
\oauthor{\bsnm{{Allende Prieto}}, \binits{C.}},
\oauthor{\bsnm{{Annis}}, \binits{J.}},
\oauthor{\bsnm{{Bailey}}, \binits{S.}},
\oauthor{\bsnm{{Balland}}, \binits{C.}},
\oauthor{\bsnm{{Ballester}}, \binits{O.}},
\oauthor{\bsnm{{Baltay}}, \binits{C.}},
\oauthor{\bsnm{{Beaufore}}, \binits{L.}},
\oauthor{\bsnm{{Bebek}}, \binits{C.}},
\oauthor{\bsnm{{Beers}}, \binits{T.C.}},
\oauthor{\bsnm{{Bell}}, \binits{E.F.}},
\oauthor{\bsnm{{Bernal}}, \binits{J.L.}},
\oauthor{\bsnm{{Besuner}}, \binits{R.}},
\oauthor{\bsnm{{Beutler}}, \binits{F.}},
\oauthor{\bsnm{{Blake}}, \binits{C.}},
\oauthor{\bsnm{{Bleuler}}, \binits{H.}},
\oauthor{\bsnm{{Blomqvist}}, \binits{M.}},
\oauthor{\bsnm{{Blum}}, \binits{R.}},
\oauthor{\bsnm{{Bolton}}, \binits{A.S.}},
\oauthor{\bsnm{{Briceno}}, \binits{C.}},
\oauthor{\bsnm{{Brooks}}, \binits{D.}},
\oauthor{\bsnm{{Brownstein}}, \binits{J.R.}},
\oauthor{\bsnm{{Buckley-Geer}}, \binits{E.}},
\oauthor{\bsnm{{Burden}}, \binits{A.}},
\oauthor{\bsnm{{Burtin}}, \binits{E.}},
\oauthor{\bsnm{{Busca}}, \binits{N.G.}},
\oauthor{\bsnm{{Cahn}}, \binits{R.N.}},
\oauthor{\bsnm{{Cai}}, \binits{Y.-C.}},
\oauthor{\bsnm{{Cardiel-Sas}}, \binits{L.}},
\oauthor{\bsnm{{Carlberg}}, \binits{R.G.}},
\oauthor{\bsnm{{Carton}}, \binits{P.-H.}},
\oauthor{\bsnm{{Casas}}, \binits{R.}},
\oauthor{\bsnm{{Castander}}, \binits{F.J.}},
\oauthor{\bsnm{{Cervantes-Cota}}, \binits{J.L.}},
\oauthor{\bsnm{{Claybaugh}}, \binits{T.M.}},
\oauthor{\bsnm{{Close}}, \binits{M.}},
\oauthor{\bsnm{{Coker}}, \binits{C.T.}},
\oauthor{\bsnm{{Cole}}, \binits{S.}},
\oauthor{\bsnm{{Comparat}}, \binits{J.}},
\oauthor{\bsnm{{Cooper}}, \binits{A.P.}},
\oauthor{\bsnm{{Cousinou}}, \binits{M.-C.}},
\oauthor{\bsnm{{Crocce}}, \binits{M.}},
\oauthor{\bsnm{{Cuby}}, \binits{J.-G.}},
\oauthor{\bsnm{{Cunningham}}, \binits{D.P.}},
\oauthor{\bsnm{{Davis}}, \binits{T.M.}},
\oauthor{\bsnm{{Dawson}}, \binits{K.S.}},
\oauthor{\bsnm{{de la Macorra}}, \binits{A.}},
\oauthor{\bsnm{{De Vicente}}, \binits{J.}},
\oauthor{\bsnm{{Delubac}}, \binits{T.}},
\oauthor{\bsnm{{Derwent}}, \binits{M.}},
\oauthor{\bsnm{{Dey}}, \binits{A.}},
\oauthor{\bsnm{{Dhungana}}, \binits{G.}},
\oauthor{\bsnm{{Ding}}, \binits{Z.}},
\oauthor{\bsnm{{Doel}}, \binits{P.}},
\oauthor{\bsnm{{Duan}}, \binits{Y.T.}},
\oauthor{\bsnm{{Ealet}}, \binits{A.}},
\oauthor{\bsnm{{Edelstein}}, \binits{J.}},
\oauthor{\bsnm{{Eftekharzadeh}}, \binits{S.}},
\oauthor{\bsnm{{Eisenstein}}, \binits{D.J.}},
\oauthor{\bsnm{{Elliott}}, \binits{A.}},
\oauthor{\bsnm{{Escoffier}}, \binits{S.}},
\oauthor{\bsnm{{Evatt}}, \binits{M.}},
\oauthor{\bsnm{{Fagrelius}}, \binits{P.}},
\oauthor{\bsnm{{Fan}}, \binits{X.}},
\oauthor{\bsnm{{Fanning}}, \binits{K.}},
\oauthor{\bsnm{{Farahi}}, \binits{A.}},
\oauthor{\bsnm{{Farihi}}, \binits{J.}},
\oauthor{\bsnm{{Favole}}, \binits{G.}},
\oauthor{\bsnm{{Feng}}, \binits{Y.}},
\oauthor{\bsnm{{Fernandez}}, \binits{E.}},
\oauthor{\bsnm{{Findlay}}, \binits{J.R.}},
\oauthor{\bsnm{{Finkbeiner}}, \binits{D.P.}},
\oauthor{\bsnm{{Fitzpatrick}}, \binits{M.J.}},
\oauthor{\bsnm{{Flaugher}}, \binits{B.}},
\oauthor{\bsnm{{Flender}}, \binits{S.}},
\oauthor{\bsnm{{Font-Ribera}}, \binits{A.}},
\oauthor{\bsnm{{Forero-Romero}}, \binits{J.E.}},
\oauthor{\bsnm{{Fosalba}}, \binits{P.}},
\oauthor{\bsnm{{Frenk}}, \binits{C.S.}},
\oauthor{\bsnm{{Fumagalli}}, \binits{M.}},
\oauthor{\bsnm{{Gaensicke}}, \binits{B.T.}},
\oauthor{\bsnm{{Gallo}}, \binits{G.}},
\oauthor{\bsnm{{Garcia-Bellido}}, \binits{J.}},
\oauthor{\bsnm{{Gaztanaga}}, \binits{E.}},
\oauthor{\bsnm{{Pietro Gentile Fusillo}}, \binits{N.}},
\oauthor{\bsnm{{Gerard}}, \binits{T.}},
\oauthor{\bsnm{{Gershkovich}}, \binits{I.}},
\oauthor{\bsnm{{Giannantonio}}, \binits{T.}},
\oauthor{\bsnm{{Gillet}}, \binits{D.}},
\oauthor{\bsnm{{Gonzalez-de-Rivera}}, \binits{G.}},
\oauthor{\bsnm{{Gonzalez-Perez}}, \binits{V.}},
\oauthor{\bsnm{{Gott}}, \binits{S.}},
\oauthor{\bsnm{{Graur}}, \binits{O.}},
\oauthor{\bsnm{{Gutierrez}}, \binits{G.}},
\oauthor{\bsnm{{Guy}}, \binits{J.}},
\oauthor{\bsnm{{Habib}}, \binits{S.}},
\oauthor{\bsnm{{Heetderks}}, \binits{H.}},
\oauthor{\bsnm{{Heetderks}}, \binits{I.}},
\oauthor{\bsnm{{Heitmann}}, \binits{K.}},
\oauthor{\bsnm{{Hellwing}}, \binits{W.A.}},
\oauthor{\bsnm{{Herrera}}, \binits{D.A.}},
\oauthor{\bsnm{{Ho}}, \binits{S.}},
\oauthor{\bsnm{{Holland}}, \binits{S.}},
\oauthor{\bsnm{{Honscheid}}, \binits{K.}},
\oauthor{\bsnm{{Huff}}, \binits{E.}},
\oauthor{\bsnm{{Hutchinson}}, \binits{T.A.}},
\oauthor{\bsnm{{Huterer}}, \binits{D.}},
\oauthor{\bsnm{{Hwang}}, \binits{H.S.}},
\oauthor{\bsnm{{Illa Laguna}}, \binits{J.M.}},
\oauthor{\bsnm{{Ishikawa}}, \binits{Y.}},
\oauthor{\bsnm{{Jacobs}}, \binits{D.}},
\oauthor{\bsnm{{Jeffrey}}, \binits{N.}},
\oauthor{\bsnm{{Jelinsky}}, \binits{P.}},
\oauthor{\bsnm{{Jennings}}, \binits{E.}},
\oauthor{\bsnm{{Jiang}}, \binits{L.}},
\oauthor{\bsnm{{Jimenez}}, \binits{J.}},
\oauthor{\bsnm{{Johnson}}, \binits{J.}},
\oauthor{\bsnm{{Joyce}}, \binits{R.}},
\oauthor{\bsnm{{Jullo}}, \binits{E.}},
\oauthor{\bsnm{{Juneau}}, \binits{S.}},
\oauthor{\bsnm{{Kama}}, \binits{S.}},
\oauthor{\bsnm{{Karcher}}, \binits{A.}},
\oauthor{\bsnm{{Karkar}}, \binits{S.}},
\oauthor{\bsnm{{Kehoe}}, \binits{R.}},
\oauthor{\bsnm{{Kennamer}}, \binits{N.}},
\oauthor{\bsnm{{Kent}}, \binits{S.}},
\oauthor{\bsnm{{Kilbinger}}, \binits{M.}},
\oauthor{\bsnm{{Kim}}, \binits{A.G.}},
\oauthor{\bsnm{{Kirkby}}, \binits{D.}},
\oauthor{\bsnm{{Kisner}}, \binits{T.}},
\oauthor{\bsnm{{Kitanidis}}, \binits{E.}},
\oauthor{\bsnm{{Kneib}}, \binits{J.-P.}},
\oauthor{\bsnm{{Koposov}}, \binits{S.}},
\oauthor{\bsnm{{Kovacs}}, \binits{E.}},
\oauthor{\bsnm{{Koyama}}, \binits{K.}},
\oauthor{\bsnm{{Kremin}}, \binits{A.}},
\oauthor{\bsnm{{Kron}}, \binits{R.}},
\oauthor{\bsnm{{Kronig}}, \binits{L.}},
\oauthor{\bsnm{{Kueter-Young}}, \binits{A.}},
\oauthor{\bsnm{{Lacey}}, \binits{C.G.}},
\oauthor{\bsnm{{Lafever}}, \binits{R.}},
\oauthor{\bsnm{{Lahav}}, \binits{O.}},
\oauthor{\bsnm{{Lambert}}, \binits{A.}},
\oauthor{\bsnm{{Lampton}}, \binits{M.}},
\oauthor{\bsnm{{Landriau}}, \binits{M.}},
\oauthor{\bsnm{{Lang}}, \binits{D.}},
\oauthor{\bsnm{{Lauer}}, \binits{T.R.}},
\oauthor{\bsnm{{Le Goff}}, \binits{J.-M.}},
\oauthor{\bsnm{{Le Guillou}}, \binits{L.}},
\oauthor{\bsnm{{Le Van Suu}}, \binits{A.}},
\oauthor{\bsnm{{Lee}}, \binits{J.H.}},
\oauthor{\bsnm{{Lee}}, \binits{S.-J.}},
\oauthor{\bsnm{{Leitner}}, \binits{D.}},
\oauthor{\bsnm{{Lesser}}, \binits{M.}},
\oauthor{\bsnm{{Levi}}, \binits{M.E.}},
\oauthor{\bsnm{{L'Huillier}}, \binits{B.}},
\oauthor{\bsnm{{Li}}, \binits{B.}},
\oauthor{\bsnm{{Liang}}, \binits{M.}},
\oauthor{\bsnm{{Lin}}, \binits{H.}},
\oauthor{\bsnm{{Linder}}, \binits{E.}},
\oauthor{\bsnm{{Loebman}}, \binits{S.R.}},
\oauthor{\bsnm{{Luki{\'c}}}, \binits{Z.}},
\oauthor{\bsnm{{Ma}}, \binits{J.}},
\oauthor{\bsnm{{MacCrann}}, \binits{N.}},
\oauthor{\bsnm{{Magneville}}, \binits{C.}},
\oauthor{\bsnm{{Makarem}}, \binits{L.}},
\oauthor{\bsnm{{Manera}}, \binits{M.}},
\oauthor{\bsnm{{Manser}}, \binits{C.J.}},
\oauthor{\bsnm{{Marshall}}, \binits{R.}},
\oauthor{\bsnm{{Martini}}, \binits{P.}},
\oauthor{\bsnm{{Massey}}, \binits{R.}},
\oauthor{\bsnm{{Matheson}}, \binits{T.}},
\oauthor{\bsnm{{McCauley}}, \binits{J.}},
\oauthor{\bsnm{{McDonald}}, \binits{P.}},
\oauthor{\bsnm{{McGreer}}, \binits{I.D.}},
\oauthor{\bsnm{{Meisner}}, \binits{A.}},
\oauthor{\bsnm{{Metcalfe}}, \binits{N.}},
\oauthor{\bsnm{{Miller}}, \binits{T.N.}},
\oauthor{\bsnm{{Miquel}}, \binits{R.}},
\oauthor{\bsnm{{Moustakas}}, \binits{J.}},
\oauthor{\bsnm{{Myers}}, \binits{A.}},
\oauthor{\bsnm{{Naik}}, \binits{M.}},
\oauthor{\bsnm{{Newman}}, \binits{J.A.}},
\oauthor{\bsnm{{Nichol}}, \binits{R.C.}},
\oauthor{\bsnm{{Nicola}}, \binits{A.}},
\oauthor{\bsnm{{Nicolati da Costa}}, \binits{L.}},
\oauthor{\bsnm{{Nie}}, \binits{J.}},
\oauthor{\bsnm{{Niz}}, \binits{G.}},
\oauthor{\bsnm{{Norberg}}, \binits{P.}},
\oauthor{\bsnm{{Nord}}, \binits{B.}},
\oauthor{\bsnm{{Norman}}, \binits{D.}},
\oauthor{\bsnm{{Nugent}}, \binits{P.}},
\oauthor{\bsnm{{O'Brien}}, \binits{T.}},
\oauthor{\bsnm{{Oh}}, \binits{M.}},
\oauthor{\bsnm{{Olsen}}, \binits{K.A.G.}},
\oauthor{\bsnm{{Padilla}}, \binits{C.}},
\oauthor{\bsnm{{Padmanabhan}}, \binits{H.}},
\oauthor{\bsnm{{Padmanabhan}}, \binits{N.}},
\oauthor{\bsnm{{Palanque-Delabrouille}}, \binits{N.}},
\oauthor{\bsnm{{Palmese}}, \binits{A.}},
\oauthor{\bsnm{{Pappalardo}}, \binits{D.}},
\oauthor{\bsnm{{P{\^a}ris}}, \binits{I.}},
\oauthor{\bsnm{{Park}}, \binits{C.}},
\oauthor{\bsnm{{Patej}}, \binits{A.}},
\oauthor{\bsnm{{Peacock}}, \binits{J.A.}},
\oauthor{\bsnm{{Peiris}}, \binits{H.V.}},
\oauthor{\bsnm{{Peng}}, \binits{X.}},
\oauthor{\bsnm{{Percival}}, \binits{W.J.}},
\oauthor{\bsnm{{Perruchot}}, \binits{S.}},
\oauthor{\bsnm{{Pieri}}, \binits{M.M.}},
\oauthor{\bsnm{{Pogge}}, \binits{R.}},
\oauthor{\bsnm{{Pollack}}, \binits{J.E.}},
\oauthor{\bsnm{{Poppett}}, \binits{C.}},
\oauthor{\bsnm{{Prada}}, \binits{F.}},
\oauthor{\bsnm{{Prakash}}, \binits{A.}},
\oauthor{\bsnm{{Probst}}, \binits{R.G.}},
\oauthor{\bsnm{{Rabinowitz}}, \binits{D.}},
\oauthor{\bsnm{{Raichoor}}, \binits{A.}},
\oauthor{\bsnm{{Ree}}, \binits{C.H.}},
\oauthor{\bsnm{{Refregier}}, \binits{A.}},
\oauthor{\bsnm{{Regal}}, \binits{X.}},
\oauthor{\bsnm{{Reid}}, \binits{B.}},
\oauthor{\bsnm{{Reil}}, \binits{K.}},
\oauthor{\bsnm{{Rezaie}}, \binits{M.}},
\oauthor{\bsnm{{Rockosi}}, \binits{C.M.}},
\oauthor{\bsnm{{Roe}}, \binits{N.}},
\oauthor{\bsnm{{Ronayette}}, \binits{S.}},
\oauthor{\bsnm{{Roodman}}, \binits{A.}},
\oauthor{\bsnm{{Ross}}, \binits{A.J.}},
\oauthor{\bsnm{{Ross}}, \binits{N.P.}},
\oauthor{\bsnm{{Rossi}}, \binits{G.}},
\oauthor{\bsnm{{Rozo}}, \binits{E.}},
\oauthor{\bsnm{{Ruhlmann-Kleider}}, \binits{V.}},
\oauthor{\bsnm{{Rykoff}}, \binits{E.S.}},
\oauthor{\bsnm{{Sabiu}}, \binits{C.}},
\oauthor{\bsnm{{Samushia}}, \binits{L.}},
\oauthor{\bsnm{{Sanchez}}, \binits{E.}},
\oauthor{\bsnm{{Sanchez}}, \binits{J.}},
\oauthor{\bsnm{{Schlegel}}, \binits{D.J.}},
\oauthor{\bsnm{{Schneider}}, \binits{M.}},
\oauthor{\bsnm{{Schubnell}}, \binits{M.}},
\oauthor{\bsnm{{Secroun}}, \binits{A.}},
\oauthor{\bsnm{{Seljak}}, \binits{U.}},
\oauthor{\bsnm{{Seo}}, \binits{H.-J.}},
\oauthor{\bsnm{{Serrano}}, \binits{S.}},
\oauthor{\bsnm{{Shafieloo}}, \binits{A.}},
\oauthor{\bsnm{{Shan}}, \binits{H.}},
\oauthor{\bsnm{{Sharples}}, \binits{R.}},
\oauthor{\bsnm{{Sholl}}, \binits{M.J.}},
\oauthor{\bsnm{{Shourt}}, \binits{W.V.}},
\oauthor{\bsnm{{Silber}}, \binits{J.H.}},
\oauthor{\bsnm{{Silva}}, \binits{D.R.}},
\oauthor{\bsnm{{Sirk}}, \binits{M.M.}},
\oauthor{\bsnm{{Slosar}}, \binits{A.}},
\oauthor{\bsnm{{Smith}}, \binits{A.}},
\oauthor{\bsnm{{Smoot}}, \binits{G.F.}},
\oauthor{\bsnm{{Som}}, \binits{D.}},
\oauthor{\bsnm{{Song}}, \binits{Y.-S.}},
\oauthor{\bsnm{{Sprayberry}}, \binits{D.}},
\oauthor{\bsnm{{Staten}}, \binits{R.}},
\oauthor{\bsnm{{Stefanik}}, \binits{A.}},
\oauthor{\bsnm{{Tarle}}, \binits{G.}},
\oauthor{\bsnm{{Sien Tie}}, \binits{S.}},
\oauthor{\bsnm{{Tinker}}, \binits{J.L.}},
\oauthor{\bsnm{{Tojeiro}}, \binits{R.}},
\oauthor{\bsnm{{Valdes}}, \binits{F.}},
\oauthor{\bsnm{{Valenzuela}}, \binits{O.}},
\oauthor{\bsnm{{Valluri}}, \binits{M.}},
\oauthor{\bsnm{{Vargas-Magana}}, \binits{M.}},
\oauthor{\bsnm{{Verde}}, \binits{L.}},
\oauthor{\bsnm{{Walker}}, \binits{A.R.}},
\oauthor{\bsnm{{Wang}}, \binits{J.}},
\oauthor{\bsnm{{Wang}}, \binits{Y.}},
\oauthor{\bsnm{{Weaver}}, \binits{B.A.}},
\oauthor{\bsnm{{Weaverdyck}}, \binits{C.}},
\oauthor{\bsnm{{Wechsler}}, \binits{R.H.}},
\oauthor{\bsnm{{Weinberg}}, \binits{D.H.}},
\oauthor{\bsnm{{White}}, \binits{M.}},
\oauthor{\bsnm{{Yang}}, \binits{Q.}},
\oauthor{\bsnm{{Yeche}}, \binits{C.}},
\oauthor{\bsnm{{Zhang}}, \binits{T.}},
\oauthor{\bsnm{{Zhao}}, \binits{G.-B.}},
\oauthor{\bsnm{{Zheng}}, \binits{Y.}},
\oauthor{\bsnm{{Zhou}}, \binits{X.}},
\oauthor{\bsnm{{Zhou}}, \binits{Z.}},
\oauthor{\bsnm{{Zhu}}, \binits{Y.}},
\oauthor{\bsnm{{Zou}}, \binits{H.}},
\oauthor{\bsnm{{Zu}}, \binits{Y.}}:
{The DESI Experiment Part II: Instrument Design}.
arXiv e-prints,
1611--00037
(2016)
{\href{https://arxiv.org/abs/1611.00037}{{arXiv:1611.00037}}}
{[astro-ph.IM]}
\end{botherref}
\endbibitem

\bibitem[\protect\citeauthoryear{{DESI Collaboration} et~al.}{2022}]{DESI_Collaboration2022}
\begin{barticle}
\bauthor{\bsnm{{DESI Collaboration}}},
\bauthor{\bsnm{{Abareshi}}, \binits{B.}},
\bauthor{\bsnm{{Aguilar}}, \binits{J.}},
\bauthor{\bsnm{{Ahlen}}, \binits{S.}},
\bauthor{\bsnm{{Alam}}, \binits{S.}},
\bauthor{\bsnm{{Alexander}}, \binits{D.M.}},
\bauthor{\bsnm{{Alfarsy}}, \binits{R.}},
\bauthor{\bsnm{{Allen}}, \binits{L.}},
\bauthor{\bsnm{{Allende Prieto}}, \binits{C.}},
\bauthor{\bsnm{{Alves}}, \binits{O.}},
\bauthor{\bsnm{{Ameel}}, \binits{J.}},
\bauthor{\bsnm{{Armengaud}}, \binits{E.}},
\bauthor{\bsnm{{Asorey}}, \binits{J.}},
\bauthor{\bsnm{{Aviles}}, \binits{A.}},
\bauthor{\bsnm{{Bailey}}, \binits{S.}},
\bauthor{\bsnm{{Balaguera-Antol{\'\i}nez}}, \binits{A.}},
\bauthor{\bsnm{{Ballester}}, \binits{O.}},
\bauthor{\bsnm{{Baltay}}, \binits{C.}},
\bauthor{\bsnm{{Bault}}, \binits{A.}},
\bauthor{\bsnm{{Beltran}}, \binits{S.F.}},
\bauthor{\bsnm{{Benavides}}, \binits{B.}},
\bauthor{\bsnm{{BenZvi}}, \binits{S.}},
\bauthor{\bsnm{{Berti}}, \binits{A.}},
\bauthor{\bsnm{{Besuner}}, \binits{R.}},
\bauthor{\bsnm{{Beutler}}, \binits{F.}},
\bauthor{\bsnm{{Bianchi}}, \binits{D.}},
\bauthor{\bsnm{{Blake}}, \binits{C.}},
\bauthor{\bsnm{{Blanc}}, \binits{P.}},
\bauthor{\bsnm{{Blum}}, \binits{R.}},
\bauthor{\bsnm{{Bolton}}, \binits{A.}},
\bauthor{\bsnm{{Bose}}, \binits{S.}},
\bauthor{\bsnm{{Bramall}}, \binits{D.}},
\bauthor{\bsnm{{Brieden}}, \binits{S.}},
\bauthor{\bsnm{{Brodzeller}}, \binits{A.}},
\bauthor{\bsnm{{Brooks}}, \binits{D.}},
\bauthor{\bsnm{{Brownewell}}, \binits{C.}},
\bauthor{\bsnm{{Buckley-Geer}}, \binits{E.}},
\bauthor{\bsnm{{Cahn}}, \binits{R.N.}},
\bauthor{\bsnm{{Cai}}, \binits{Z.}},
\bauthor{\bsnm{{Canning}}, \binits{R.}},
\bauthor{\bsnm{{Capasso}}, \binits{R.}},
\bauthor{\bsnm{{Carnero Rosell}}, \binits{A.}},
\bauthor{\bsnm{{Carton}}, \binits{P.}},
\bauthor{\bsnm{{Casas}}, \binits{R.}},
\bauthor{\bsnm{{Castander}}, \binits{F.J.}},
\bauthor{\bsnm{{Cervantes-Cota}}, \binits{J.L.}},
\bauthor{\bsnm{{Chabanier}}, \binits{S.}},
\bauthor{\bsnm{{Chaussidon}}, \binits{E.}},
\bauthor{\bsnm{{Chuang}}, \binits{C.}},
\bauthor{\bsnm{{Circosta}}, \binits{C.}},
\bauthor{\bsnm{{Cole}}, \binits{S.}},
\bauthor{\bsnm{{Cooper}}, \binits{A.P.}},
\bauthor{\bsnm{{da Costa}}, \binits{L.}},
\bauthor{\bsnm{{Cousinou}}, \binits{M.-C.}},
\bauthor{\bsnm{{Cuceu}}, \binits{A.}},
\bauthor{\bsnm{{Davis}}, \binits{T.M.}},
\bauthor{\bsnm{{Dawson}}, \binits{K.}},
\bauthor{\bsnm{{de la Cruz-Noriega}}, \binits{R.}},
\bauthor{\bsnm{{de la Macorra}}, \binits{A.}},
\bauthor{\bsnm{{de Mattia}}, \binits{A.}},
\bauthor{\bsnm{{Della Costa}}, \binits{J.}},
\bauthor{\bsnm{{Demmer}}, \binits{P.}},
\bauthor{\bsnm{{Derwent}}, \binits{M.}},
\bauthor{\bsnm{{Dey}}, \binits{A.}},
\bauthor{\bsnm{{Dey}}, \binits{B.}},
\bauthor{\bsnm{{Dhungana}}, \binits{G.}},
\bauthor{\bsnm{{Ding}}, \binits{Z.}},
\bauthor{\bsnm{{Dobson}}, \binits{C.}},
\bauthor{\bsnm{{Doel}}, \binits{P.}},
\bauthor{\bsnm{{Donald-McCann}}, \binits{J.}},
\bauthor{\bsnm{{Donaldson}}, \binits{J.}},
\bauthor{\bsnm{{Douglass}}, \binits{K.}},
\bauthor{\bsnm{{Duan}}, \binits{Y.}},
\bauthor{\bsnm{{Dunlop}}, \binits{P.}},
\bauthor{\bsnm{{Edelstein}}, \binits{J.}},
\bauthor{\bsnm{{Eftekharzadeh}}, \binits{S.}},
\bauthor{\bsnm{{Eisenstein}}, \binits{D.J.}},
\bauthor{\bsnm{{Enriquez-Vargas}}, \binits{M.}},
\bauthor{\bsnm{{Escoffier}}, \binits{S.}},
\bauthor{\bsnm{{Evatt}}, \binits{M.}},
\bauthor{\bsnm{{Fagrelius}}, \binits{P.}},
\bauthor{\bsnm{{Fan}}, \binits{X.}},
\bauthor{\bsnm{{Fanning}}, \binits{K.}},
\bauthor{\bsnm{{Fawcett}}, \binits{V.A.}},
\bauthor{\bsnm{{Ferraro}}, \binits{S.}},
\bauthor{\bsnm{{Ereza}}, \binits{J.}},
\bauthor{\bsnm{{Flaugher}}, \binits{B.}},
\bauthor{\bsnm{{Font-Ribera}}, \binits{A.}},
\bauthor{\bsnm{{Forero-Romero}}, \binits{J.E.}},
\bauthor{\bsnm{{Frenk}}, \binits{C.S.}},
\bauthor{\bsnm{{Fromenteau}}, \binits{S.}},
\bauthor{\bsnm{{G{\"a}nsicke}}, \binits{B.T.}},
\bauthor{\bsnm{{Garcia-Quintero}}, \binits{C.}},
\bauthor{\bsnm{{Garrison}}, \binits{L.}},
\bauthor{\bsnm{{Gazta{\~n}aga}}, \binits{E.}},
\bauthor{\bsnm{{Gerardi}}, \binits{F.}},
\bauthor{\bsnm{{Gil-Mar{\'\i}n}}, \binits{H.}},
\bauthor{\bsnm{{Gontcho a Gontcho}}, \binits{S.}},
\bauthor{\bsnm{{Gonzalez-Morales}}, \binits{A.X.}},
\bauthor{\bsnm{{Gonzalez-de-Rivera}}, \binits{G.}},
\bauthor{\bsnm{{Gonzalez-Perez}}, \binits{V.}},
\bauthor{\bsnm{{Gordon}}, \binits{C.}},
\bauthor{\bsnm{{Graur}}, \binits{O.}},
\bauthor{\bsnm{{Green}}, \binits{D.}},
\bauthor{\bsnm{{Grove}}, \binits{C.}},
\bauthor{\bsnm{{Gruen}}, \binits{D.}},
\bauthor{\bsnm{{Gutierrez}}, \binits{G.}},
\bauthor{\bsnm{{Guy}}, \binits{J.}},
\bauthor{\bsnm{{Hahn}}, \binits{C.}},
\bauthor{\bsnm{{Harris}}, \binits{S.}},
\bauthor{\bsnm{{Herrera}}, \binits{D.}},
\bauthor{\bsnm{{Herrera-Alcantar}}, \binits{H.K.}},
\bauthor{\bsnm{{Honscheid}}, \binits{K.}},
\bauthor{\bsnm{{Howlett}}, \binits{C.}},
\bauthor{\bsnm{{Huterer}}, \binits{D.}},
\bauthor{\bsnm{{Ir{\v{s}}i{\v{c}}}}, \binits{V.}},
\bauthor{\bsnm{{Ishak}}, \binits{M.}},
\bauthor{\bsnm{{Jelinsky}}, \binits{P.}},
\bauthor{\bsnm{{Jiang}}, \binits{L.}},
\bauthor{\bsnm{{Jimenez}}, \binits{J.}},
\bauthor{\bsnm{{Jing}}, \binits{Y.P.}},
\bauthor{\bsnm{{Joyce}}, \binits{R.}},
\bauthor{\bsnm{{Jullo}}, \binits{E.}},
\bauthor{\bsnm{{Juneau}}, \binits{S.}},
\bauthor{\bsnm{{Kara{\c{c}}ayl{\i}}}, \binits{N.G.}},
\bauthor{\bsnm{{Karamanis}}, \binits{M.}},
\bauthor{\bsnm{{Karcher}}, \binits{A.}},
\bauthor{\bsnm{{Karim}}, \binits{T.}},
\bauthor{\bsnm{{Kehoe}}, \binits{R.}},
\bauthor{\bsnm{{Kent}}, \binits{S.}},
\bauthor{\bsnm{{Kirkby}}, \binits{D.}},
\bauthor{\bsnm{{Kisner}}, \binits{T.}},
\bauthor{\bsnm{{Kitaura}}, \binits{F.}},
\bauthor{\bsnm{{Koposov}}, \binits{S.E.}},
\bauthor{\bsnm{{Kov{\'a}cs}}, \binits{A.}},
\bauthor{\bsnm{{Kremin}}, \binits{A.}},
\bauthor{\bsnm{{Krolewski}}, \binits{A.}},
\bauthor{\bsnm{{L'Huillier}}, \binits{B.}},
\bauthor{\bsnm{{Lahav}}, \binits{O.}},
\bauthor{\bsnm{{Lambert}}, \binits{A.}},
\bauthor{\bsnm{{Lamman}}, \binits{C.}},
\bauthor{\bsnm{{Lan}}, \binits{T.-W.}},
\bauthor{\bsnm{{Landriau}}, \binits{M.}},
\bauthor{\bsnm{{Lane}}, \binits{S.}},
\bauthor{\bsnm{{Lang}}, \binits{D.}},
\bauthor{\bsnm{{Lange}}, \binits{J.U.}},
\bauthor{\bsnm{{Lasker}}, \binits{J.}},
\bauthor{\bsnm{{Le Guillou}}, \binits{L.}},
\bauthor{\bsnm{{Leauthaud}}, \binits{A.}},
\bauthor{\bsnm{{Le Van Suu}}, \binits{A.}},
\bauthor{\bsnm{{Levi}}, \binits{M.E.}},
\bauthor{\bsnm{{Li}}, \binits{T.S.}},
\bauthor{\bsnm{{Magneville}}, \binits{C.}},
\bauthor{\bsnm{{Manera}}, \binits{M.}},
\bauthor{\bsnm{{Manser}}, \binits{C.J.}},
\bauthor{\bsnm{{Marshall}}, \binits{B.}},
\bauthor{\bsnm{{Martini}}, \binits{P.}},
\bauthor{\bsnm{{McCollam}}, \binits{W.}},
\bauthor{\bsnm{{McDonald}}, \binits{P.}},
\bauthor{\bsnm{{Meisner}}, \binits{A.M.}},
\bauthor{\bsnm{{Mena-Fern{\'a}ndez}}, \binits{J.}},
\bauthor{\bsnm{{Meneses-Rizo}}, \binits{J.}},
\bauthor{\bsnm{{Mezcua}}, \binits{M.}},
\bauthor{\bsnm{{Miller}}, \binits{T.}},
\bauthor{\bsnm{{Miquel}}, \binits{R.}},
\bauthor{\bsnm{{Montero-Camacho}}, \binits{P.}},
\bauthor{\bsnm{{Moon}}, \binits{J.}},
\bauthor{\bsnm{{Moustakas}}, \binits{J.}},
\bauthor{\bsnm{{Mueller}}, \binits{E.}},
\bauthor{\bsnm{{Mu{\~n}oz-Guti{\'e}rrez}}, \binits{A.}},
\bauthor{\bsnm{{Myers}}, \binits{A.D.}},
\bauthor{\bsnm{{Nadathur}}, \binits{S.}},
\bauthor{\bsnm{{Najita}}, \binits{J.}},
\bauthor{\bsnm{{Napolitano}}, \binits{L.}},
\bauthor{\bsnm{{Neilsen}}, \binits{E.}},
\bauthor{\bsnm{{Newman}}, \binits{J.A.}},
\bauthor{\bsnm{{Nie}}, \binits{J.D.}},
\bauthor{\bsnm{{Ning}}, \binits{Y.}},
\bauthor{\bsnm{{Niz}}, \binits{G.}},
\bauthor{\bsnm{{Norberg}}, \binits{P.}},
\bauthor{\bsnm{{Noriega}}, \binits{H.E.}},
\bauthor{\bsnm{{O'Brien}}, \binits{T.}},
\bauthor{\bsnm{{Obuljen}}, \binits{A.}},
\bauthor{\bsnm{{Palanque-Delabrouille}}, \binits{N.}},
\bauthor{\bsnm{{Palmese}}, \binits{A.}},
\bauthor{\bsnm{{Zhiwei}}, \binits{P.}},
\bauthor{\bsnm{{Pappalardo}}, \binits{D.}},
\bauthor{\bsnm{{PENG}}, \binits{X.}},
\bauthor{\bsnm{{Percival}}, \binits{W.J.}},
\bauthor{\bsnm{{Perruchot}}, \binits{S.}},
\bauthor{\bsnm{{Pogge}}, \binits{R.}},
\bauthor{\bsnm{{Poppett}}, \binits{C.}},
\bauthor{\bsnm{{Porredon}}, \binits{A.}},
\bauthor{\bsnm{{Prada}}, \binits{F.}},
\bauthor{\bsnm{{Prochaska}}, \binits{J.}},
\bauthor{\bsnm{{Pucha}}, \binits{R.}},
\bauthor{\bsnm{{P{\'e}rez-Fern{\'a}ndez}}, \binits{A.}},
\bauthor{\bsnm{{P{\'e}rez-R{\`a}fols}}, \binits{I.}},
\bauthor{\bsnm{{Rabinowitz}}, \binits{D.}},
\bauthor{\bsnm{{Raichoor}}, \binits{A.}},
\bauthor{\bsnm{{Ramirez-Solano}}, \binits{S.}},
\bauthor{\bsnm{{Ram{\'\i}rez-P{\'e}rez}}, \binits{C.}},
\bauthor{\bsnm{{Ravoux}}, \binits{C.}},
\bauthor{\bsnm{{Reil}}, \binits{K.}},
\bauthor{\bsnm{{Rezaie}}, \binits{M.}},
\bauthor{\bsnm{{Rocher}}, \binits{A.}},
\bauthor{\bsnm{{Rockosi}}, \binits{C.}},
\bauthor{\bsnm{{Roe}}, \binits{N.A.}},
\bauthor{\bsnm{{Roodman}}, \binits{A.}},
\bauthor{\bsnm{{Ross}}, \binits{A.J.}},
\bauthor{\bsnm{{Rossi}}, \binits{G.}},
\bauthor{\bsnm{{Ruggeri}}, \binits{R.}},
\bauthor{\bsnm{{Ruhlmann-Kleider}}, \binits{V.}},
\bauthor{\bsnm{{Sabiu}}, \binits{C.G.}},
\bauthor{\bsnm{{Safonova}}, \binits{S.}},
\bauthor{\bsnm{{Said}}, \binits{K.}},
\bauthor{\bsnm{{Saintonge}}, \binits{A.}},
\bauthor{\bsnm{{Salas Catonga}}, \binits{J.}},
\bauthor{\bsnm{{Samushia}}, \binits{L.}},
\bauthor{\bsnm{{Sanchez}}, \binits{E.}},
\bauthor{\bsnm{{Saulder}}, \binits{C.}},
\bauthor{\bsnm{{Schaan}}, \binits{E.}},
\bauthor{\bsnm{{Schlafly}}, \binits{E.}},
\bauthor{\bsnm{{Schlegel}}, \binits{D.}},
\bauthor{\bsnm{{Schmoll}}, \binits{J.}},
\bauthor{\bsnm{{Scholte}}, \binits{D.}},
\bauthor{\bsnm{{Schubnell}}, \binits{M.}},
\bauthor{\bsnm{{Secroun}}, \binits{A.}},
\bauthor{\bsnm{{Seo}}, \binits{H.}},
\bauthor{\bsnm{{Serrano}}, \binits{S.}},
\bauthor{\bsnm{{Sharples}}, \binits{R.M.}},
\bauthor{\bsnm{{Sholl}}, \binits{M.J.}},
\bauthor{\bsnm{{Silber}}, \binits{J.H.}},
\bauthor{\bsnm{{Silva}}, \binits{D.R.}},
\bauthor{\bsnm{{Sirk}}, \binits{M.}},
\bauthor{\bsnm{{Siudek}}, \binits{M.}},
\bauthor{\bsnm{{Smith}}, \binits{A.}},
\bauthor{\bsnm{{Sprayberry}}, \binits{D.}},
\bauthor{\bsnm{{Staten}}, \binits{R.}},
\bauthor{\bsnm{{Stupak}}, \binits{B.}},
\bauthor{\bsnm{{Tan}}, \binits{T.}},
\bauthor{\bsnm{{Tarl{\'e}}}, \binits{G.}},
\bauthor{\bsnm{{Tie}}, \binits{S.S.}},
\bauthor{\bsnm{{Tojeiro}}, \binits{R.}},
\bauthor{\bsnm{{Ure{\~n}a-L{\'o}pez}}, \binits{L.A.}},
\bauthor{\bsnm{{Valdes}}, \binits{F.}},
\bauthor{\bsnm{{Valenzuela}}, \binits{O.}},
\bauthor{\bsnm{{Valluri}}, \binits{M.}},
\bauthor{\bsnm{{Vargas-Maga{\~n}a}}, \binits{M.}},
\bauthor{\bsnm{{Verde}}, \binits{L.}},
\bauthor{\bsnm{{Walther}}, \binits{M.}},
\bauthor{\bsnm{{Wang}}, \binits{B.}},
\bauthor{\bsnm{{Wang}}, \binits{M.S.}},
\bauthor{\bsnm{{Weaver}}, \binits{B.A.}},
\bauthor{\bsnm{{Weaverdyck}}, \binits{C.}},
\bauthor{\bsnm{{Wechsler}}, \binits{R.}},
\bauthor{\bsnm{{Wilson}}, \binits{M.J.}},
\bauthor{\bsnm{{Yang}}, \binits{J.}},
\bauthor{\bsnm{{Yu}}, \binits{Y.}},
\bauthor{\bsnm{{Yuan}}, \binits{S.}},
\bauthor{\bsnm{{Y{\`e}che}}, \binits{C.}},
\bauthor{\bsnm{{Zhang}}, \binits{H.}},
\bauthor{\bsnm{{Zhang}}, \binits{K.}},
\bauthor{\bsnm{{Zhao}}, \binits{C.}},
\bauthor{\bsnm{{Zhou}}, \binits{R.}},
\bauthor{\bsnm{{Zhou}}, \binits{Z.}},
\bauthor{\bsnm{{Zou}}, \binits{H.}},
\bauthor{\bsnm{{Zou}}, \binits{J.}},
\bauthor{\bsnm{{Zou}}, \binits{S.}},
\bauthor{\bsnm{{Zu}}, \binits{Y.}},
\bauthor{\bsnm{{DESI Collaboration}}}:
\batitle{{Overview of the Instrumentation for the Dark Energy Spectroscopic Instrument}}.
\bjtitle{AJ}
\bvolume{164}(\bissue{5}),
\bfpage{207}
(\byear{2022})
\doiurl{10.3847/1538-3881/ac882b}
{\href{https://arxiv.org/abs/2205.10939}{{arXiv:2205.10939}}}
{[astro-ph.IM]}
\end{barticle}
\endbibitem

\bibitem[\protect\citeauthoryear{{Levi} et~al.}{2013}]{Levi2013}
\begin{botherref}
\oauthor{\bsnm{{Levi}}, \binits{M.}},
\oauthor{\bsnm{{Bebek}}, \binits{C.}},
\oauthor{\bsnm{{Beers}}, \binits{T.}},
\oauthor{\bsnm{{Blum}}, \binits{R.}},
\oauthor{\bsnm{{Cahn}}, \binits{R.}},
\oauthor{\bsnm{{Eisenstein}}, \binits{D.}},
\oauthor{\bsnm{{Flaugher}}, \binits{B.}},
\oauthor{\bsnm{{Honscheid}}, \binits{K.}},
\oauthor{\bsnm{{Kron}}, \binits{R.}},
\oauthor{\bsnm{{Lahav}}, \binits{O.}},
\oauthor{\bsnm{{McDonald}}, \binits{P.}},
\oauthor{\bsnm{{Roe}}, \binits{N.}},
\oauthor{\bsnm{{Schlegel}}, \binits{D.}},
\oauthor{\bsnm{{representing the DESI collaboration}}}:
{The DESI Experiment, a whitepaper for Snowmass 2013}.
arXiv e-prints,
1308--0847
(2013)
{\href{https://arxiv.org/abs/1308.0847}{{arXiv:1308.0847}}}
{[astro-ph.CO]}
\end{botherref}
\endbibitem

\bibitem[\protect\citeauthoryear{{Silber} et~al.}{2023}]{Silber2023}
\begin{barticle}
\bauthor{\bsnm{{Silber}}, \binits{J.H.}},
\bauthor{\bsnm{{Fagrelius}}, \binits{P.}},
\bauthor{\bsnm{{Fanning}}, \binits{K.}},
\bauthor{\bsnm{{Schubnell}}, \binits{M.}},
\bauthor{\bsnm{{Aguilar}}, \binits{J.N.}},
\bauthor{\bsnm{{Ahlen}}, \binits{S.}},
\bauthor{\bsnm{{Ameel}}, \binits{J.}},
\bauthor{\bsnm{{Ballester}}, \binits{O.}},
\bauthor{\bsnm{{Baltay}}, \binits{C.}},
\bauthor{\bsnm{{Bebek}}, \binits{C.}},
\bauthor{\bsnm{{Benton Beard}}, \binits{D.}},
\bauthor{\bsnm{{Besuner}}, \binits{R.}},
\bauthor{\bsnm{{Cardiel-Sas}}, \binits{L.}},
\bauthor{\bsnm{{Casas}}, \binits{R.}},
\bauthor{\bsnm{{Castander}}, \binits{F.J.}},
\bauthor{\bsnm{{Claybaugh}}, \binits{T.}},
\bauthor{\bsnm{{Dobson}}, \binits{C.}},
\bauthor{\bsnm{{Duan}}, \binits{Y.}},
\bauthor{\bsnm{{Dunlop}}, \binits{P.}},
\bauthor{\bsnm{{Edelstein}}, \binits{J.}},
\bauthor{\bsnm{{Emmet}}, \binits{W.T.}},
\bauthor{\bsnm{{Elliott}}, \binits{A.}},
\bauthor{\bsnm{{Evatt}}, \binits{M.}},
\bauthor{\bsnm{{Gershkovich}}, \binits{I.}},
\bauthor{\bsnm{{Guy}}, \binits{J.}},
\bauthor{\bsnm{{Harris}}, \binits{S.}},
\bauthor{\bsnm{{Heetderks}}, \binits{H.}},
\bauthor{\bsnm{{Heetderks}}, \binits{I.}},
\bauthor{\bsnm{{Honscheid}}, \binits{K.}},
\bauthor{\bsnm{{Illa}}, \binits{J.M.}},
\bauthor{\bsnm{{Jelinsky}}, \binits{P.}},
\bauthor{\bsnm{{Jelinsky}}, \binits{S.R.}},
\bauthor{\bsnm{{Jimenez}}, \binits{J.}},
\bauthor{\bsnm{{Karcher}}, \binits{A.}},
\bauthor{\bsnm{{Kent}}, \binits{S.}},
\bauthor{\bsnm{{Kirkby}}, \binits{D.}},
\bauthor{\bsnm{{Kneib}}, \binits{J.-P.}},
\bauthor{\bsnm{{Lambert}}, \binits{A.}},
\bauthor{\bsnm{{Lampton}}, \binits{M.}},
\bauthor{\bsnm{{Leitner}}, \binits{D.}},
\bauthor{\bsnm{{Levi}}, \binits{M.}},
\bauthor{\bsnm{{McCauley}}, \binits{J.}},
\bauthor{\bsnm{{Meisner}}, \binits{A.}},
\bauthor{\bsnm{{Miller}}, \binits{T.N.}},
\bauthor{\bsnm{{Miquel}}, \binits{R.}},
\bauthor{\bsnm{{Mundet}}, \binits{J.}},
\bauthor{\bsnm{{Poppett}}, \binits{C.}},
\bauthor{\bsnm{{Rabinowitz}}, \binits{D.}},
\bauthor{\bsnm{{Reil}}, \binits{K.}},
\bauthor{\bsnm{{Roman}}, \binits{D.}},
\bauthor{\bsnm{{Schlegel}}, \binits{D.}},
\bauthor{\bsnm{{Serrano}}, \binits{S.}},
\bauthor{\bsnm{{Van Shourt}}, \binits{W.}},
\bauthor{\bsnm{{Sprayberry}}, \binits{D.}},
\bauthor{\bsnm{{Tarl{\'e}}}, \binits{G.}},
\bauthor{\bsnm{{Tie}}, \binits{S.S.}},
\bauthor{\bsnm{{Weaverdyck}}, \binits{C.}},
\bauthor{\bsnm{{Zhang}}, \binits{K.}},
\bauthor{\bsnm{{Azzaro}}, \binits{M.}},
\bauthor{\bsnm{{Bailey}}, \binits{S.}},
\bauthor{\bsnm{{Becerril}}, \binits{S.}},
\bauthor{\bsnm{{Blackwell}}, \binits{T.}},
\bauthor{\bsnm{{Bouri}}, \binits{M.}},
\bauthor{\bsnm{{Brooks}}, \binits{D.}},
\bauthor{\bsnm{{Buckley-Geer}}, \binits{E.}},
\bauthor{\bsnm{{Castro}}, \binits{J.P.}},
\bauthor{\bsnm{{Derwent}}, \binits{M.}},
\bauthor{\bsnm{{Dey}}, \binits{A.}},
\bauthor{\bsnm{{Dhungana}}, \binits{G.}},
\bauthor{\bsnm{{Doel}}, \binits{P.}},
\bauthor{\bsnm{{Eisenstein}}, \binits{D.J.}},
\bauthor{\bsnm{{Fahim}}, \binits{N.}},
\bauthor{\bsnm{{Garcia-Bellido}}, \binits{J.}},
\bauthor{\bsnm{{Gazta{\~n}aga}}, \binits{E.}},
\bauthor{\bsnm{{A Gontcho}}, \binits{S.G.}},
\bauthor{\bsnm{{Gutierrez}}, \binits{G.}},
\bauthor{\bsnm{{H{\"o}rler}}, \binits{P.}},
\bauthor{\bsnm{{Kehoe}}, \binits{R.}},
\bauthor{\bsnm{{Kisner}}, \binits{T.}},
\bauthor{\bsnm{{Kremin}}, \binits{A.}},
\bauthor{\bsnm{{Kronig}}, \binits{L.}},
\bauthor{\bsnm{{Landriau}}, \binits{M.}},
\bauthor{\bsnm{{Le Guillou}}, \binits{L.}},
\bauthor{\bsnm{{Martini}}, \binits{P.}},
\bauthor{\bsnm{{Moustakas}}, \binits{J.}},
\bauthor{\bsnm{{Palanque-Delabrouille}}, \binits{N.}},
\bauthor{\bsnm{{Peng}}, \binits{X.}},
\bauthor{\bsnm{{Percival}}, \binits{W.}},
\bauthor{\bsnm{{Prada}}, \binits{F.}},
\bauthor{\bsnm{{Allende Prieto}}, \binits{C.}},
\bauthor{\bsnm{{de Rivera}}, \binits{G.G.}},
\bauthor{\bsnm{{Sanchez}}, \binits{E.}},
\bauthor{\bsnm{{Sanchez}}, \binits{J.}},
\bauthor{\bsnm{{Sharples}}, \binits{R.}},
\bauthor{\bsnm{{Soares-Santos}}, \binits{M.}},
\bauthor{\bsnm{{Schlafly}}, \binits{E.}},
\bauthor{\bsnm{{Weaver}}, \binits{B.A.}},
\bauthor{\bsnm{{Zhou}}, \binits{Z.}},
\bauthor{\bsnm{{Zhu}}, \binits{Y.}},
\bauthor{\bsnm{{Zou}}, \binits{H.}},
\bauthor{\bsnm{{DESI Collaboration}}}:
\batitle{{The Robotic Multiobject Focal Plane System of the Dark Energy Spectroscopic Instrument (DESI)}}.
\bjtitle{AJ}
\bvolume{165}(\bissue{1}),
\bfpage{9}
(\byear{2023})
\doiurl{10.3847/1538-3881/ac9ab1}
{\href{https://arxiv.org/abs/2205.09014}{{arXiv:2205.09014}}}
{[astro-ph.IM]}
\end{barticle}
\endbibitem

\bibitem[\protect\citeauthoryear{{Miller} et~al.}{2023}]{Miller2023}
\begin{botherref}
\oauthor{\bsnm{{Miller}}, \binits{T.N.}},
\oauthor{\bsnm{{Doel}}, \binits{P.}},
\oauthor{\bsnm{{Gutierrez}}, \binits{G.}},
\oauthor{\bsnm{{Besuner}}, \binits{R.}},
\oauthor{\bsnm{{Brooks}}, \binits{D.}},
\oauthor{\bsnm{{Gallo}}, \binits{G.}},
\oauthor{\bsnm{{Heetderks}}, \binits{H.}},
\oauthor{\bsnm{{Jelinsky}}, \binits{P.}},
\oauthor{\bsnm{{Kent}}, \binits{S.M.}},
\oauthor{\bsnm{{Lampton}}, \binits{M.}},
\oauthor{\bsnm{{Levi}}, \binits{M.}},
\oauthor{\bsnm{{Liang}}, \binits{M.}},
\oauthor{\bsnm{{Meisner}}, \binits{A.}},
\oauthor{\bsnm{{Sholl}}, \binits{M.J.}},
\oauthor{\bsnm{{Silber}}, \binits{J.H.}},
\oauthor{\bsnm{{Sprayberry}}, \binits{D.}},
\oauthor{\bsnm{{Aguilar}}, \binits{J.N.}},
\oauthor{\bsnm{{de la Macorra}}, \binits{A.}},
\oauthor{\bsnm{{Eisenstein}}, \binits{D.}},
\oauthor{\bsnm{{Fanning}}, \binits{K.}},
\oauthor{\bsnm{{Font-Ribera}}, \binits{A.}},
\oauthor{\bsnm{{Gaztanaga}}, \binits{E.}},
\oauthor{\bsnm{{Gontcho}}, \binits{S.G.A.}},
\oauthor{\bsnm{{Honscheid}}, \binits{K.}},
\oauthor{\bsnm{{Jimenez}}, \binits{J.}},
\oauthor{\bsnm{{Joyce}}, \binits{D.}},
\oauthor{\bsnm{{Kehoe}}, \binits{R.}},
\oauthor{\bsnm{{Kisner}}, \binits{T.}},
\oauthor{\bsnm{{Kremin}}, \binits{A.}},
\oauthor{\bsnm{{Landriau}}, \binits{M.}},
\oauthor{\bsnm{{Le Guillou}}, \binits{L.}},
\oauthor{\bsnm{{Magneville}}, \binits{C.}},
\oauthor{\bsnm{{Martini}}, \binits{P.}},
\oauthor{\bsnm{{Miquel}}, \binits{R.}},
\oauthor{\bsnm{{Moustakas}}, \binits{J.}},
\oauthor{\bsnm{{Nie}}, \binits{J.}},
\oauthor{\bsnm{{Percival}}, \binits{W.}},
\oauthor{\bsnm{{Poppett}}, \binits{C.}},
\oauthor{\bsnm{{Prada}}, \binits{F.}},
\oauthor{\bsnm{{Rossi}}, \binits{G.}},
\oauthor{\bsnm{{Schlegel}}, \binits{D.}},
\oauthor{\bsnm{{Schubnell}}, \binits{M.}},
\oauthor{\bsnm{{Seo}}, \binits{H.-J.}},
\oauthor{\bsnm{{Sharples}}, \binits{R.}},
\oauthor{\bsnm{{Tarle}}, \binits{G.}},
\oauthor{\bsnm{{Vargas-Magana}}, \binits{M.}},
\oauthor{\bsnm{{Zhou}}, \binits{Z.}}:
{The Optical Corrector for the Dark Energy Spectroscopic Instrument}.
arXiv e-prints,
2306--06310
(2023)
\doiurl{10.48550/arXiv.2306.06310}
{\href{https://arxiv.org/abs/2306.06310}{{arXiv:2306.06310}}}
{[astro-ph.IM]}
\end{botherref}
\endbibitem

\bibitem[\protect\citeauthoryear{{Zou} et~al.}{2017}]{Zou2017}
\begin{barticle}
\bauthor{\bsnm{{Zou}}, \binits{H.}},
\bauthor{\bsnm{{Zhou}}, \binits{X.}},
\bauthor{\bsnm{{Fan}}, \binits{X.}},
\bauthor{\bsnm{{Zhang}}, \binits{T.}},
\bauthor{\bsnm{{Zhou}}, \binits{Z.}},
\bauthor{\bsnm{{Nie}}, \binits{J.}},
\bauthor{\bsnm{{Peng}}, \binits{X.}},
\bauthor{\bsnm{{McGreer}}, \binits{I.}},
\bauthor{\bsnm{{Jiang}}, \binits{L.}},
\bauthor{\bsnm{{Dey}}, \binits{A.}},
\bauthor{\bsnm{{Fan}}, \binits{D.}},
\bauthor{\bsnm{{He}}, \binits{B.}},
\bauthor{\bsnm{{Jiang}}, \binits{Z.}},
\bauthor{\bsnm{{Lang}}, \binits{D.}},
\bauthor{\bsnm{{Lesser}}, \binits{M.}},
\bauthor{\bsnm{{Ma}}, \binits{J.}},
\bauthor{\bsnm{{Mao}}, \binits{S.}},
\bauthor{\bsnm{{Schlegel}}, \binits{D.}},
\bauthor{\bsnm{{Wang}}, \binits{J.}}:
\batitle{{Project Overview of the Beijing-Arizona Sky Survey}}.
\bjtitle{PASP}
\bvolume{129}(\bissue{976}),
\bfpage{064101}
(\byear{2017})
\doiurl{10.1088/1538-3873/aa65ba}
{\href{https://arxiv.org/abs/1702.03653}{{arXiv:1702.03653}}}
{[astro-ph.GA]}
\end{barticle}
\endbibitem

\bibitem[\protect\citeauthoryear{{Dey} et~al.}{2019}]{Dey2019}
\begin{barticle}
\bauthor{\bsnm{{Dey}}, \binits{A.}},
\bauthor{\bsnm{{Schlegel}}, \binits{D.J.}},
\bauthor{\bsnm{{Lang}}, \binits{D.}},
\bauthor{\bsnm{{Blum}}, \binits{R.}},
\bauthor{\bsnm{{Burleigh}}, \binits{K.}},
\bauthor{\bsnm{{Fan}}, \binits{X.}},
\bauthor{\bsnm{{Findlay}}, \binits{J.R.}},
\bauthor{\bsnm{{Finkbeiner}}, \binits{D.}},
\bauthor{\bsnm{{Herrera}}, \binits{D.}},
\bauthor{\bsnm{{Juneau}}, \binits{S.}},
\bauthor{\bsnm{{Landriau}}, \binits{M.}},
\bauthor{\bsnm{{Levi}}, \binits{M.}},
\bauthor{\bsnm{{McGreer}}, \binits{I.}},
\bauthor{\bsnm{{Meisner}}, \binits{A.}},
\bauthor{\bsnm{{Myers}}, \binits{A.D.}},
\bauthor{\bsnm{{Moustakas}}, \binits{J.}},
\bauthor{\bsnm{{Nugent}}, \binits{P.}},
\bauthor{\bsnm{{Patej}}, \binits{A.}},
\bauthor{\bsnm{{Schlafly}}, \binits{E.F.}},
\bauthor{\bsnm{{Walker}}, \binits{A.R.}},
\bauthor{\bsnm{{Valdes}}, \binits{F.}},
\bauthor{\bsnm{{Weaver}}, \binits{B.A.}},
\bauthor{\bsnm{{Y{\`e}che}}, \binits{C.}},
\bauthor{\bsnm{{Zou}}, \binits{H.}},
\bauthor{\bsnm{{Zhou}}, \binits{X.}},
\bauthor{\bsnm{{Abareshi}}, \binits{B.}},
\bauthor{\bsnm{{Abbott}}, \binits{T.M.C.}},
\bauthor{\bsnm{{Abolfathi}}, \binits{B.}},
\bauthor{\bsnm{{Aguilera}}, \binits{C.}},
\bauthor{\bsnm{{Alam}}, \binits{S.}},
\bauthor{\bsnm{{Allen}}, \binits{L.}},
\bauthor{\bsnm{{Alvarez}}, \binits{A.}},
\bauthor{\bsnm{{Annis}}, \binits{J.}},
\bauthor{\bsnm{{Ansarinejad}}, \binits{B.}},
\bauthor{\bsnm{{Aubert}}, \binits{M.}},
\bauthor{\bsnm{{Beechert}}, \binits{J.}},
\bauthor{\bsnm{{Bell}}, \binits{E.F.}},
\bauthor{\bsnm{{BenZvi}}, \binits{S.Y.}},
\bauthor{\bsnm{{Beutler}}, \binits{F.}},
\bauthor{\bsnm{{Bielby}}, \binits{R.M.}},
\bauthor{\bsnm{{Bolton}}, \binits{A.S.}},
\bauthor{\bsnm{{Brice{\~n}o}}, \binits{C.}},
\bauthor{\bsnm{{Buckley-Geer}}, \binits{E.J.}},
\bauthor{\bsnm{{Butler}}, \binits{K.}},
\bauthor{\bsnm{{Calamida}}, \binits{A.}},
\bauthor{\bsnm{{Carlberg}}, \binits{R.G.}},
\bauthor{\bsnm{{Carter}}, \binits{P.}},
\bauthor{\bsnm{{Casas}}, \binits{R.}},
\bauthor{\bsnm{{Castander}}, \binits{F.J.}},
\bauthor{\bsnm{{Choi}}, \binits{Y.}},
\bauthor{\bsnm{{Comparat}}, \binits{J.}},
\bauthor{\bsnm{{Cukanovaite}}, \binits{E.}},
\bauthor{\bsnm{{Delubac}}, \binits{T.}},
\bauthor{\bsnm{{DeVries}}, \binits{K.}},
\bauthor{\bsnm{{Dey}}, \binits{S.}},
\bauthor{\bsnm{{Dhungana}}, \binits{G.}},
\bauthor{\bsnm{{Dickinson}}, \binits{M.}},
\bauthor{\bsnm{{Ding}}, \binits{Z.}},
\bauthor{\bsnm{{Donaldson}}, \binits{J.B.}},
\bauthor{\bsnm{{Duan}}, \binits{Y.}},
\bauthor{\bsnm{{Duckworth}}, \binits{C.J.}},
\bauthor{\bsnm{{Eftekharzadeh}}, \binits{S.}},
\bauthor{\bsnm{{Eisenstein}}, \binits{D.J.}},
\bauthor{\bsnm{{Etourneau}}, \binits{T.}},
\bauthor{\bsnm{{Fagrelius}}, \binits{P.A.}},
\bauthor{\bsnm{{Farihi}}, \binits{J.}},
\bauthor{\bsnm{{Fitzpatrick}}, \binits{M.}},
\bauthor{\bsnm{{Font-Ribera}}, \binits{A.}},
\bauthor{\bsnm{{Fulmer}}, \binits{L.}},
\bauthor{\bsnm{{G{\"a}nsicke}}, \binits{B.T.}},
\bauthor{\bsnm{{Gaztanaga}}, \binits{E.}},
\bauthor{\bsnm{{George}}, \binits{K.}},
\bauthor{\bsnm{{Gerdes}}, \binits{D.W.}},
\bauthor{\bsnm{{Gontcho}}, \binits{S.G.A.}},
\bauthor{\bsnm{{Gorgoni}}, \binits{C.}},
\bauthor{\bsnm{{Green}}, \binits{G.}},
\bauthor{\bsnm{{Guy}}, \binits{J.}},
\bauthor{\bsnm{{Harmer}}, \binits{D.}},
\bauthor{\bsnm{{Hernandez}}, \binits{M.}},
\bauthor{\bsnm{{Honscheid}}, \binits{K.}},
\bauthor{\bsnm{{Huang}}, \binits{L.W.}},
\bauthor{\bsnm{{James}}, \binits{D.J.}},
\bauthor{\bsnm{{Jannuzi}}, \binits{B.T.}},
\bauthor{\bsnm{{Jiang}}, \binits{L.}},
\bauthor{\bsnm{{Joyce}}, \binits{R.}},
\bauthor{\bsnm{{Karcher}}, \binits{A.}},
\bauthor{\bsnm{{Karkar}}, \binits{S.}},
\bauthor{\bsnm{{Kehoe}}, \binits{R.}},
\bauthor{\bsnm{{Kneib}}, \binits{J.-P.}},
\bauthor{\bsnm{{Kueter-Young}}, \binits{A.}},
\bauthor{\bsnm{{Lan}}, \binits{T.-W.}},
\bauthor{\bsnm{{Lauer}}, \binits{T.R.}},
\bauthor{\bsnm{{Le Guillou}}, \binits{L.}},
\bauthor{\bsnm{{Le Van Suu}}, \binits{A.}},
\bauthor{\bsnm{{Lee}}, \binits{J.H.}},
\bauthor{\bsnm{{Lesser}}, \binits{M.}},
\bauthor{\bsnm{{Perreault Levasseur}}, \binits{L.}},
\bauthor{\bsnm{{Li}}, \binits{T.S.}},
\bauthor{\bsnm{{Mann}}, \binits{J.L.}},
\bauthor{\bsnm{{Marshall}}, \binits{R.}},
\bauthor{\bsnm{{Mart{\'\i}nez-V{\'a}zquez}}, \binits{C.E.}},
\bauthor{\bsnm{{Martini}}, \binits{P.}},
\bauthor{\bsnm{{du Mas des Bourboux}}, \binits{H.}},
\bauthor{\bsnm{{McManus}}, \binits{S.}},
\bauthor{\bsnm{{Meier}}, \binits{T.G.}},
\bauthor{\bsnm{{M{\'e}nard}}, \binits{B.}},
\bauthor{\bsnm{{Metcalfe}}, \binits{N.}},
\bauthor{\bsnm{{Mu{\~n}oz-Guti{\'e}rrez}}, \binits{A.}},
\bauthor{\bsnm{{Najita}}, \binits{J.}},
\bauthor{\bsnm{{Napier}}, \binits{K.}},
\bauthor{\bsnm{{Narayan}}, \binits{G.}},
\bauthor{\bsnm{{Newman}}, \binits{J.A.}},
\bauthor{\bsnm{{Nie}}, \binits{J.}},
\bauthor{\bsnm{{Nord}}, \binits{B.}},
\bauthor{\bsnm{{Norman}}, \binits{D.J.}},
\bauthor{\bsnm{{Olsen}}, \binits{K.A.G.}},
\bauthor{\bsnm{{Paat}}, \binits{A.}},
\bauthor{\bsnm{{Palanque-Delabrouille}}, \binits{N.}},
\bauthor{\bsnm{{Peng}}, \binits{X.}},
\bauthor{\bsnm{{Poppett}}, \binits{C.L.}},
\bauthor{\bsnm{{Poremba}}, \binits{M.R.}},
\bauthor{\bsnm{{Prakash}}, \binits{A.}},
\bauthor{\bsnm{{Rabinowitz}}, \binits{D.}},
\bauthor{\bsnm{{Raichoor}}, \binits{A.}},
\bauthor{\bsnm{{Rezaie}}, \binits{M.}},
\bauthor{\bsnm{{Robertson}}, \binits{A.N.}},
\bauthor{\bsnm{{Roe}}, \binits{N.A.}},
\bauthor{\bsnm{{Ross}}, \binits{A.J.}},
\bauthor{\bsnm{{Ross}}, \binits{N.P.}},
\bauthor{\bsnm{{Rudnick}}, \binits{G.}},
\bauthor{\bsnm{{Safonova}}, \binits{S.}},
\bauthor{\bsnm{{Saha}}, \binits{A.}},
\bauthor{\bsnm{{S{\'a}nchez}}, \binits{F.J.}},
\bauthor{\bsnm{{Savary}}, \binits{E.}},
\bauthor{\bsnm{{Schweiker}}, \binits{H.}},
\bauthor{\bsnm{{Scott}}, \binits{A.}},
\bauthor{\bsnm{{Seo}}, \binits{H.-J.}},
\bauthor{\bsnm{{Shan}}, \binits{H.}},
\bauthor{\bsnm{{Silva}}, \binits{D.R.}},
\bauthor{\bsnm{{Slepian}}, \binits{Z.}},
\bauthor{\bsnm{{Soto}}, \binits{C.}},
\bauthor{\bsnm{{Sprayberry}}, \binits{D.}},
\bauthor{\bsnm{{Staten}}, \binits{R.}},
\bauthor{\bsnm{{Stillman}}, \binits{C.M.}},
\bauthor{\bsnm{{Stupak}}, \binits{R.J.}},
\bauthor{\bsnm{{Summers}}, \binits{D.L.}},
\bauthor{\bsnm{{Sien Tie}}, \binits{S.}},
\bauthor{\bsnm{{Tirado}}, \binits{H.}},
\bauthor{\bsnm{{Vargas-Maga{\~n}a}}, \binits{M.}},
\bauthor{\bsnm{{Vivas}}, \binits{A.K.}},
\bauthor{\bsnm{{Wechsler}}, \binits{R.H.}},
\bauthor{\bsnm{{Williams}}, \binits{D.}},
\bauthor{\bsnm{{Yang}}, \binits{J.}},
\bauthor{\bsnm{{Yang}}, \binits{Q.}},
\bauthor{\bsnm{{Yapici}}, \binits{T.}},
\bauthor{\bsnm{{Zaritsky}}, \binits{D.}},
\bauthor{\bsnm{{Zenteno}}, \binits{A.}},
\bauthor{\bsnm{{Zhang}}, \binits{K.}},
\bauthor{\bsnm{{Zhang}}, \binits{T.}},
\bauthor{\bsnm{{Zhou}}, \binits{R.}},
\bauthor{\bsnm{{Zhou}}, \binits{Z.}}:
\batitle{{Overview of the DESI Legacy Imaging Surveys}}.
\bjtitle{AJ}
\bvolume{157}(\bissue{5}),
\bfpage{168}
(\byear{2019})
\doiurl{10.3847/1538-3881/ab089d}
{\href{https://arxiv.org/abs/1804.08657}{{arXiv:1804.08657}}}
{[astro-ph.IM]}
\end{barticle}
\endbibitem

\bibitem[\protect\citeauthoryear{{Zhou} et~al.}{2021}]{Zhou2021}
\begin{barticle}
\bauthor{\bsnm{{Zhou}}, \binits{R.}},
\bauthor{\bsnm{{Newman}}, \binits{J.A.}},
\bauthor{\bsnm{{Mao}}, \binits{Y.-Y.}},
\bauthor{\bsnm{{Meisner}}, \binits{A.}},
\bauthor{\bsnm{{Moustakas}}, \binits{J.}},
\bauthor{\bsnm{{Myers}}, \binits{A.D.}},
\bauthor{\bsnm{{Prakash}}, \binits{A.}},
\bauthor{\bsnm{{Zentner}}, \binits{A.R.}},
\bauthor{\bsnm{{Brooks}}, \binits{D.}},
\bauthor{\bsnm{{Duan}}, \binits{Y.}},
\bauthor{\bsnm{{Landriau}}, \binits{M.}},
\bauthor{\bsnm{{Levi}}, \binits{M.E.}},
\bauthor{\bsnm{{Prada}}, \binits{F.}},
\bauthor{\bsnm{{Tarle}}, \binits{G.}}:
\batitle{{The clustering of DESI-like luminous red galaxies using photometric redshifts}}.
\bjtitle{MNRAS}
\bvolume{501}(\bissue{3}),
\bfpage{3309}--\blpage{3331}
(\byear{2021})
\doiurl{10.1093/mnras/staa3764}
{\href{https://arxiv.org/abs/2001.06018}{{arXiv:2001.06018}}}
{[astro-ph.CO]}
\end{barticle}
\endbibitem

\bibitem[\protect\citeauthoryear{{Darragh-Ford} et~al.}{2023}]{Darragh-Ford2023}
\begin{barticle}
\bauthor{\bsnm{{Darragh-Ford}}, \binits{E.}},
\bauthor{\bsnm{{Wu}}, \binits{J.F.}},
\bauthor{\bsnm{{Mao}}, \binits{Y.-Y.}},
\bauthor{\bsnm{{Wechsler}}, \binits{R.H.}},
\bauthor{\bsnm{{Geha}}, \binits{M.}},
\bauthor{\bsnm{{Forero-Romero}}, \binits{J.E.}},
\bauthor{\bsnm{{Hahn}}, \binits{C.}},
\bauthor{\bsnm{{Kallivayalil}}, \binits{N.}},
\bauthor{\bsnm{{Moustakas}}, \binits{J.}},
\bauthor{\bsnm{{Nadler}}, \binits{E.O.}},
\bauthor{\bsnm{{Nowotka}}, \binits{M.}},
\bauthor{\bsnm{{Peek}}, \binits{J.E.G.}},
\bauthor{\bsnm{{Tollerud}}, \binits{E.J.}},
\bauthor{\bsnm{{Weiner}}, \binits{B.}},
\bauthor{\bsnm{{Aguilar}}, \binits{J.}},
\bauthor{\bsnm{{Ahlen}}, \binits{S.}},
\bauthor{\bsnm{{Brooks}}, \binits{D.}},
\bauthor{\bsnm{{Cooper}}, \binits{A.P.}},
\bauthor{\bsnm{{de la Macorra}}, \binits{A.}},
\bauthor{\bsnm{{Dey}}, \binits{A.}},
\bauthor{\bsnm{{Fanning}}, \binits{K.}},
\bauthor{\bsnm{{Font-Ribera}}, \binits{A.}},
\bauthor{\bsnm{{Gontcho A Gontcho}}, \binits{S.}},
\bauthor{\bsnm{{Honscheid}}, \binits{K.}},
\bauthor{\bsnm{{Kisner}}, \binits{T.}},
\bauthor{\bsnm{{Kremin}}, \binits{A.}},
\bauthor{\bsnm{{Landriau}}, \binits{M.}},
\bauthor{\bsnm{{Levi}}, \binits{M.E.}},
\bauthor{\bsnm{{Martini}}, \binits{P.}},
\bauthor{\bsnm{{Meisner}}, \binits{A.M.}},
\bauthor{\bsnm{{Miquel}}, \binits{R.}},
\bauthor{\bsnm{{Myers}}, \binits{A.D.}},
\bauthor{\bsnm{{Nie}}, \binits{J.}},
\bauthor{\bsnm{{Palanque-Delabrouille}}, \binits{N.}},
\bauthor{\bsnm{{Percival}}, \binits{W.J.}},
\bauthor{\bsnm{{Prada}}, \binits{F.}},
\bauthor{\bsnm{{Schlegel}}, \binits{D.}},
\bauthor{\bsnm{{Schubnell}}, \binits{M.}},
\bauthor{\bsnm{{Tarl{\'e}}}, \binits{G.}},
\bauthor{\bsnm{{Vargas-Maga{\~n}a}}, \binits{M.}},
\bauthor{\bsnm{{Zhou}}, \binits{Z.}},
\bauthor{\bsnm{{Zou}}, \binits{H.}}:
\batitle{{Target Selection and Sample Characterization for the DESI LOW-Z Secondary Target Program}}.
\bjtitle{\apj}
\bvolume{954}(\bissue{2}),
\bfpage{149}
(\byear{2023})
\doiurl{10.3847/1538-4357/ace902}
{\href{https://arxiv.org/abs/2212.07433}{{arXiv:2212.07433}}}
{[astro-ph.GA]}
\end{barticle}
\endbibitem

\bibitem[\protect\citeauthoryear{{Scholte} et~al.}{2024}]{Dirk2024}
\begin{barticle}
\bauthor{\bsnm{{Scholte}}, \binits{D.}},
\bauthor{\bsnm{{Saintonge}}, \binits{A.}},
\bauthor{\bsnm{{Moustakas}}, \binits{J.}},
\bauthor{\bsnm{{Catinella}}, \binits{B.}},
\bauthor{\bsnm{{Zou}}, \binits{H.}},
\bauthor{\bsnm{{Dey}}, \binits{B.}},
\bauthor{\bsnm{{Aguilar}}, \binits{J.}},
\bauthor{\bsnm{{Ahlen}}, \binits{S.}},
\bauthor{\bsnm{{Anand}}, \binits{A.}},
\bauthor{\bsnm{{Blum}}, \binits{R.}},
\bauthor{\bsnm{{Brooks}}, \binits{D.}},
\bauthor{\bsnm{{Circosta}}, \binits{C.}},
\bauthor{\bsnm{{Claybaugh}}, \binits{T.}},
\bauthor{\bsnm{{de la Macorra}}, \binits{A.}},
\bauthor{\bsnm{{Doel}}, \binits{P.}},
\bauthor{\bsnm{{Font-Ribera}}, \binits{A.}},
\bauthor{\bsnm{{F{\"o}rster}}, \binits{P.U.}},
\bauthor{\bsnm{{Forero-Romero}}, \binits{J.E.}},
\bauthor{\bsnm{{Gazta{\~n}aga}}, \binits{E.}},
\bauthor{\bsnm{{Gontcho A Gontcho}}, \binits{S.}},
\bauthor{\bsnm{{Juneau}}, \binits{S.}},
\bauthor{\bsnm{{Kehoe}}, \binits{R.}},
\bauthor{\bsnm{{Kisner}}, \binits{T.}},
\bauthor{\bsnm{{Koposov}}, \binits{S.E.}},
\bauthor{\bsnm{{Kremin}}, \binits{A.}},
\bauthor{\bsnm{{Lambert}}, \binits{A.}},
\bauthor{\bsnm{{Landriau}}, \binits{M.}},
\bauthor{\bsnm{{Maraston}}, \binits{C.}},
\bauthor{\bsnm{{Martini}}, \binits{P.}},
\bauthor{\bsnm{{Meisner}}, \binits{A.}},
\bauthor{\bsnm{{Mighty}}, \binits{A.S.}},
\bauthor{\bsnm{{Miquel}}, \binits{R.}},
\bauthor{\bsnm{{Myers}}, \binits{A.D.}},
\bauthor{\bsnm{{Nie}}, \binits{J.}},
\bauthor{\bsnm{{Poppett}}, \binits{C.}},
\bauthor{\bsnm{{Prada}}, \binits{F.}},
\bauthor{\bsnm{{Rezaie}}, \binits{M.}},
\bauthor{\bsnm{{Rossi}}, \binits{G.}},
\bauthor{\bsnm{{Sanchez}}, \binits{E.}},
\bauthor{\bsnm{{Schubnell}}, \binits{M.}},
\bauthor{\bsnm{{Silber}}, \binits{J.}},
\bauthor{\bsnm{{Sprayberry}}, \binits{D.}},
\bauthor{\bsnm{{Siudek}}, \binits{M.}},
\bauthor{\bsnm{{Speranza}}, \binits{F.}},
\bauthor{\bsnm{{Tarl{\'e}}}, \binits{G.}},
\bauthor{\bsnm{{Tojeiro}}, \binits{R.}},
\bauthor{\bsnm{{Weaver}}, \binits{B.A.}}:
\batitle{{The atomic gas sequence and mass-metallicity relation from dwarfs to massive galaxies}}.
\bjtitle{\mnras}
\bvolume{535}(\bissue{3}),
\bfpage{2341}--\blpage{2356}
(\byear{2024})
\doiurl{10.1093/mnras/stae2477}
{\href{https://arxiv.org/abs/2408.03996}{{arXiv:2408.03996}}}
{[astro-ph.GA]}
\end{barticle}
\endbibitem

\bibitem[\protect\citeauthoryear{{Pucha} et~al.}{2025}]{Pucha2025}
\begin{barticle}
\bauthor{\bsnm{{Pucha}}, \binits{R.}},
\bauthor{\bsnm{{Juneau}}, \binits{S.}},
\bauthor{\bsnm{{Dey}}, \binits{A.}},
\bauthor{\bsnm{{Siudek}}, \binits{M.}},
\bauthor{\bsnm{{Mezcua}}, \binits{M.}},
\bauthor{\bsnm{{Moustakas}}, \binits{J.}},
\bauthor{\bsnm{{BenZvi}}, \binits{S.}},
\bauthor{\bsnm{{Hainline}}, \binits{K.}},
\bauthor{\bsnm{{Hviding}}, \binits{R.}},
\bauthor{\bsnm{{Mao}}, \binits{Y.-Y.}},
\bauthor{\bsnm{{Alexander}}, \binits{D.M.}},
\bauthor{\bsnm{{Alfarsy}}, \binits{R.}},
\bauthor{\bsnm{{Circosta}}, \binits{C.}},
\bauthor{\bsnm{{Guo}}, \binits{W.-J.}},
\bauthor{\bsnm{{Manwadkar}}, \binits{V.}},
\bauthor{\bsnm{{Martini}}, \binits{P.}},
\bauthor{\bsnm{{Weaver}}, \binits{B.A.}},
\bauthor{\bsnm{{Aguilar}}, \binits{J.}},
\bauthor{\bsnm{{Ahlen}}, \binits{S.}},
\bauthor{\bsnm{{Bianchi}}, \binits{D.}},
\bauthor{\bsnm{{Brooks}}, \binits{D.}},
\bauthor{\bsnm{{Canning}}, \binits{R.}},
\bauthor{\bsnm{{Claybaugh}}, \binits{T.}},
\bauthor{\bsnm{{Dawson}}, \binits{K.}},
\bauthor{\bsnm{{de la Macorra}}, \binits{A.}},
\bauthor{\bsnm{{Dey}}, \binits{B.}},
\bauthor{\bsnm{{Doel}}, \binits{P.}},
\bauthor{\bsnm{{Font-Ribera}}, \binits{A.}},
\bauthor{\bsnm{{Forero-Romero}}, \binits{J.E.}},
\bauthor{\bsnm{{Gazta{\~n}aga}}, \binits{E.}},
\bauthor{\bsnm{{Gontcho A Gontcho}}, \binits{S.}},
\bauthor{\bsnm{{Gutierrez}}, \binits{G.}},
\bauthor{\bsnm{{Honscheid}}, \binits{K.}},
\bauthor{\bsnm{{Kehoe}}, \binits{R.}},
\bauthor{\bsnm{{Koposov}}, \binits{S.E.}},
\bauthor{\bsnm{{Lambert}}, \binits{A.}},
\bauthor{\bsnm{{Landriau}}, \binits{M.}},
\bauthor{\bsnm{{Le Guillou}}, \binits{L.}},
\bauthor{\bsnm{{Meisner}}, \binits{A.}},
\bauthor{\bsnm{{Miquel}}, \binits{R.}},
\bauthor{\bsnm{{Prada}}, \binits{F.}},
\bauthor{\bsnm{{Rossi}}, \binits{G.}},
\bauthor{\bsnm{{Sanchez}}, \binits{E.}},
\bauthor{\bsnm{{Schlegel}}, \binits{D.}},
\bauthor{\bsnm{{Schubnell}}, \binits{M.}},
\bauthor{\bsnm{{Seo}}, \binits{H.}},
\bauthor{\bsnm{{Sprayberry}}, \binits{D.}},
\bauthor{\bsnm{{Tarl{\'e}}}, \binits{G.}},
\bauthor{\bsnm{{Zou}}, \binits{H.}}:
\batitle{{Tripling the Census of Dwarf AGN Candidates Using DESI Early Data}}.
\bjtitle{\apj}
\bvolume{982}(\bissue{1}),
\bfpage{10}
(\byear{2025})
\doiurl{10.3847/1538-4357/adb1dd}
{\href{https://arxiv.org/abs/2411.00091}{{arXiv:2411.00091}}}
{[astro-ph.GA]}
\end{barticle}
\endbibitem

\bibitem[\protect\citeauthoryear{{G{\'o}rski} et~al.}{2005}]{Gorski2005}
\begin{barticle}
\bauthor{\bsnm{{G{\'o}rski}}, \binits{K.M.}},
\bauthor{\bsnm{{Hivon}}, \binits{E.}},
\bauthor{\bsnm{{Banday}}, \binits{A.J.}},
\bauthor{\bsnm{{Wandelt}}, \binits{B.D.}},
\bauthor{\bsnm{{Hansen}}, \binits{F.K.}},
\bauthor{\bsnm{{Reinecke}}, \binits{M.}},
\bauthor{\bsnm{{Bartelmann}}, \binits{M.}}:
\batitle{{HEALPix: A Framework for High-Resolution Discretization and Fast Analysis of Data Distributed on the Sphere}}.
\bjtitle{ApJ}
\bvolume{622}(\bissue{2}),
\bfpage{759}--\blpage{771}
(\byear{2005})
\doiurl{10.1086/427976}
{\href{https://arxiv.org/abs/astro-ph/0409513}{{arXiv:astro-ph/0409513}}}
{[astro-ph]}
\end{barticle}
\endbibitem

\bibitem[\protect\citeauthoryear{{Blanton} and {Roweis}}{2007}]{Blanton2007}
\begin{barticle}
\bauthor{\bsnm{{Blanton}}, \binits{M.R.}},
\bauthor{\bsnm{{Roweis}}, \binits{S.}}:
\batitle{{K-Corrections and Filter Transformations in the Ultraviolet, Optical, and Near-Infrared}}.
\bjtitle{AJ}
\bvolume{133}(\bissue{2}),
\bfpage{734}--\blpage{754}
(\byear{2007})
\doiurl{10.1086/510127}
{\href{https://arxiv.org/abs/astro-ph/0606170}{{arXiv:astro-ph/0606170}}}
{[astro-ph]}
\end{barticle}
\endbibitem

\bibitem[\protect\citeauthoryear{{Siudek} et~al.}{2024}]{Siudek2024}
\begin{barticle}
\bauthor{\bsnm{{Siudek}}, \binits{M.}},
\bauthor{\bsnm{{Pucha}}, \binits{R.}},
\bauthor{\bsnm{{Mezcua}}, \binits{M.}},
\bauthor{\bsnm{{Juneau}}, \binits{S.}},
\bauthor{\bsnm{{Aguilar}}, \binits{J.}},
\bauthor{\bsnm{{Ahlen}}, \binits{S.}},
\bauthor{\bsnm{{Brooks}}, \binits{D.}},
\bauthor{\bsnm{{Circosta}}, \binits{C.}},
\bauthor{\bsnm{{Claybaugh}}, \binits{T.}},
\bauthor{\bsnm{{Cole}}, \binits{S.}},
\bauthor{\bsnm{{Dawson}}, \binits{K.}},
\bauthor{\bsnm{{de la Macorra}}, \binits{A.}},
\bauthor{\bsnm{{Dey}}, \binits{A.}},
\bauthor{\bsnm{{Dey}}, \binits{B.}},
\bauthor{\bsnm{{Doel}}, \binits{P.}},
\bauthor{\bsnm{{Font-Ribera}}, \binits{A.}},
\bauthor{\bsnm{{Forero-Romero}}, \binits{J.E.}},
\bauthor{\bsnm{{Gazta{\~n}aga}}, \binits{E.}},
\bauthor{\bsnm{{Gontcho A Gontcho}}, \binits{S.}},
\bauthor{\bsnm{{Gutierrez}}, \binits{G.}},
\bauthor{\bsnm{{Honscheid}}, \binits{K.}},
\bauthor{\bsnm{{Howlett}}, \binits{C.}},
\bauthor{\bsnm{{Ishak}}, \binits{M.}},
\bauthor{\bsnm{{Kehoe}}, \binits{R.}},
\bauthor{\bsnm{{Kirkby}}, \binits{D.}},
\bauthor{\bsnm{{Kisner}}, \binits{T.}},
\bauthor{\bsnm{{Kremin}}, \binits{A.}},
\bauthor{\bsnm{{Lambert}}, \binits{A.}},
\bauthor{\bsnm{{Landriau}}, \binits{M.}},
\bauthor{\bsnm{{Le Guillou}}, \binits{L.}},
\bauthor{\bsnm{{Manera}}, \binits{M.}},
\bauthor{\bsnm{{Martini}}, \binits{P.}},
\bauthor{\bsnm{{Meisner}}, \binits{A.}},
\bauthor{\bsnm{{Miquel}}, \binits{R.}},
\bauthor{\bsnm{{Moustakas}}, \binits{J.}},
\bauthor{\bsnm{{Newman}}, \binits{J.A.}},
\bauthor{\bsnm{{Niz}}, \binits{G.}},
\bauthor{\bsnm{{Pan}}, \binits{Z.}},
\bauthor{\bsnm{{Percival}}, \binits{W.J.}},
\bauthor{\bsnm{{Poppett}}, \binits{C.}},
\bauthor{\bsnm{{Prada}}, \binits{F.}},
\bauthor{\bsnm{{Rossi}}, \binits{G.}},
\bauthor{\bsnm{{Saintonge}}, \binits{A.}},
\bauthor{\bsnm{{Sanchez}}, \binits{E.}},
\bauthor{\bsnm{{Schlegel}}, \binits{D.}},
\bauthor{\bsnm{{Scholte}}, \binits{D.}},
\bauthor{\bsnm{{Schubnell}}, \binits{M.}},
\bauthor{\bsnm{{Seo}}, \binits{H.}},
\bauthor{\bsnm{{Speranza}}, \binits{F.}},
\bauthor{\bsnm{{Sprayberry}}, \binits{D.}},
\bauthor{\bsnm{{Tarl{\'e}}}, \binits{G.}},
\bauthor{\bsnm{{Weaver}}, \binits{B.A.}},
\bauthor{\bsnm{{Zou}}, \binits{H.}}:
\batitle{{Value-added catalog of physical properties for more than 1.3 million galaxies from the DESI survey}}.
\bjtitle{A\&A}
\bvolume{691},
\bfpage{308}
(\byear{2024})
\doiurl{10.1051/0004-6361/202451761}
{\href{https://arxiv.org/abs/2409.19066}{{arXiv:2409.19066}}}
{[astro-ph.GA]}
\end{barticle}
\endbibitem

\bibitem[\protect\citeauthoryear{{Moustakas} et~al.}{2023}]{2023ascl.soft08005M}
\begin{botherref}
\oauthor{\bsnm{{Moustakas}}, \binits{J.}},
\oauthor{\bsnm{{Buhler}}, \binits{J.}},
\oauthor{\bsnm{{Scholte}}, \binits{D.}},
\oauthor{\bsnm{{Dey}}, \binits{B.}},
\oauthor{\bsnm{{Khederlarian}}, \binits{A.}}:
{FastSpecFit: Fast Spectral Synthesis and Emission-line Fitting of DESI Spectra}
\end{botherref}
\endbibitem

\bibitem[\protect\citeauthoryear{{Ciesla} et~al.}{2015}]{Ciesla2015}
\begin{barticle}
\bauthor{\bsnm{{Ciesla}}, \binits{L.}},
\bauthor{\bsnm{{Charmandaris}}, \binits{V.}},
\bauthor{\bsnm{{Georgakakis}}, \binits{A.}},
\bauthor{\bsnm{{Bernhard}}, \binits{E.}},
\bauthor{\bsnm{{Mitchell}}, \binits{P.D.}},
\bauthor{\bsnm{{Buat}}, \binits{V.}},
\bauthor{\bsnm{{Elbaz}}, \binits{D.}},
\bauthor{\bsnm{{LeFloc'h}}, \binits{E.}},
\bauthor{\bsnm{{Lacey}}, \binits{C.G.}},
\bauthor{\bsnm{{Magdis}}, \binits{G.E.}},
\bauthor{\bsnm{{Xilouris}}, \binits{M.}}:
\batitle{{Constraining the properties of AGN host galaxies with spectral energy distribution modelling}}.
\bjtitle{A\&A}
\bvolume{576},
\bfpage{10}
(\byear{2015})
\doiurl{10.1051/0004-6361/201425252}
{\href{https://arxiv.org/abs/1501.03672}{{arXiv:1501.03672}}}
{[astro-ph.GA]}
\end{barticle}
\endbibitem

\bibitem[\protect\citeauthoryear{{Yang} et~al.}{2023}]{Yang2023}
\begin{barticle}
\bauthor{\bsnm{{Yang}}, \binits{G.}},
\bauthor{\bsnm{{Caputi}}, \binits{K.I.}},
\bauthor{\bsnm{{Papovich}}, \binits{C.}},
\bauthor{\bsnm{{Arrabal Haro}}, \binits{P.}},
\bauthor{\bsnm{{Bagley}}, \binits{M.B.}},
\bauthor{\bsnm{{Behroozi}}, \binits{P.}},
\bauthor{\bsnm{{Bell}}, \binits{E.F.}},
\bauthor{\bsnm{{Bisigello}}, \binits{L.}},
\bauthor{\bsnm{{Buat}}, \binits{V.}},
\bauthor{\bsnm{{Burgarella}}, \binits{D.}},
\bauthor{\bsnm{{Cheng}}, \binits{Y.}},
\bauthor{\bsnm{{Cleri}}, \binits{N.J.}},
\bauthor{\bsnm{{Dav{\'e}}}, \binits{R.}},
\bauthor{\bsnm{{Dickinson}}, \binits{M.}},
\bauthor{\bsnm{{Elbaz}}, \binits{D.}},
\bauthor{\bsnm{{Ferguson}}, \binits{H.C.}},
\bauthor{\bsnm{{Finkelstein}}, \binits{S.L.}},
\bauthor{\bsnm{{Grogin}}, \binits{N.A.}},
\bauthor{\bsnm{{Hathi}}, \binits{N.P.}},
\bauthor{\bsnm{{Hirschmann}}, \binits{M.}},
\bauthor{\bsnm{{Holwerda}}, \binits{B.W.}},
\bauthor{\bsnm{{Huertas-Company}}, \binits{M.}},
\bauthor{\bsnm{{Hutchison}}, \binits{T.A.}},
\bauthor{\bsnm{{Iani}}, \binits{E.}},
\bauthor{\bsnm{{Kartaltepe}}, \binits{J.S.}},
\bauthor{\bsnm{{Kirkpatrick}}, \binits{A.}},
\bauthor{\bsnm{{Kocevski}}, \binits{D.D.}},
\bauthor{\bsnm{{Koekemoer}}, \binits{A.M.}},
\bauthor{\bsnm{{Kokorev}}, \binits{V.}},
\bauthor{\bsnm{{Larson}}, \binits{R.L.}},
\bauthor{\bsnm{{Lucas}}, \binits{R.A.}},
\bauthor{\bsnm{{P{\'e}rez-Gonz{\'a}lez}}, \binits{P.G.}},
\bauthor{\bsnm{{Rinaldi}}, \binits{P.}},
\bauthor{\bsnm{{Shen}}, \binits{L.}},
\bauthor{\bsnm{{Trump}}, \binits{J.R.}},
\bauthor{\bsnm{{de la Vega}}, \binits{A.}},
\bauthor{\bsnm{{Yung}}, \binits{L.Y.A.}},
\bauthor{\bsnm{{Zavala}}, \binits{J.A.}}:
\batitle{{CEERS Key Paper. VI. JWST/MIRI Uncovers a Large Population of Obscured AGN at High Redshifts}}.
\bjtitle{ApJL}
\bvolume{950}(\bissue{1}),
\bfpage{5}
(\byear{2023})
\doiurl{10.3847/2041-8213/acd639}
{\href{https://arxiv.org/abs/2303.11736}{{arXiv:2303.11736}}}
{[astro-ph.GA]}
\end{barticle}
\endbibitem

\bibitem[\protect\citeauthoryear{{Brinchmann} et~al.}{2004}]{Brinchmann2004}
\begin{barticle}
\bauthor{\bsnm{{Brinchmann}}, \binits{J.}},
\bauthor{\bsnm{{Charlot}}, \binits{S.}},
\bauthor{\bsnm{{White}}, \binits{S.D.M.}},
\bauthor{\bsnm{{Tremonti}}, \binits{C.}},
\bauthor{\bsnm{{Kauffmann}}, \binits{G.}},
\bauthor{\bsnm{{Heckman}}, \binits{T.}},
\bauthor{\bsnm{{Brinkmann}}, \binits{J.}}:
\batitle{{The physical properties of star-forming galaxies in the low-redshift Universe}}.
\bjtitle{MNRAS}
\bvolume{351}(\bissue{4}),
\bfpage{1151}--\blpage{1179}
(\byear{2004})
\doiurl{10.1111/j.1365-2966.2004.07881.x}
{\href{https://arxiv.org/abs/astro-ph/0311060}{{arXiv:astro-ph/0311060}}}
{[astro-ph]}
\end{barticle}
\endbibitem

\bibitem[\protect\citeauthoryear{{Salim} et~al.}{2007}]{Salim2007}
\begin{barticle}
\bauthor{\bsnm{{Salim}}, \binits{S.}},
\bauthor{\bsnm{{Rich}}, \binits{R.M.}},
\bauthor{\bsnm{{Charlot}}, \binits{S.}},
\bauthor{\bsnm{{Brinchmann}}, \binits{J.}},
\bauthor{\bsnm{{Johnson}}, \binits{B.D.}},
\bauthor{\bsnm{{Schiminovich}}, \binits{D.}},
\bauthor{\bsnm{{Seibert}}, \binits{M.}},
\bauthor{\bsnm{{Mallery}}, \binits{R.}},
\bauthor{\bsnm{{Heckman}}, \binits{T.M.}},
\bauthor{\bsnm{{Forster}}, \binits{K.}},
\bauthor{\bsnm{{Friedman}}, \binits{P.G.}},
\bauthor{\bsnm{{Martin}}, \binits{D.C.}},
\bauthor{\bsnm{{Morrissey}}, \binits{P.}},
\bauthor{\bsnm{{Neff}}, \binits{S.G.}},
\bauthor{\bsnm{{Small}}, \binits{T.}},
\bauthor{\bsnm{{Wyder}}, \binits{T.K.}},
\bauthor{\bsnm{{Bianchi}}, \binits{L.}},
\bauthor{\bsnm{{Donas}}, \binits{J.}},
\bauthor{\bsnm{{Lee}}, \binits{Y.-W.}},
\bauthor{\bsnm{{Madore}}, \binits{B.F.}},
\bauthor{\bsnm{{Milliard}}, \binits{B.}},
\bauthor{\bsnm{{Szalay}}, \binits{A.S.}},
\bauthor{\bsnm{{Welsh}}, \binits{B.Y.}},
\bauthor{\bsnm{{Yi}}, \binits{S.K.}}:
\batitle{{UV Star Formation Rates in the Local Universe}}.
\bjtitle{ApJS}
\bvolume{173}(\bissue{2}),
\bfpage{267}--\blpage{292}
(\byear{2007})
\doiurl{10.1086/519218}
{\href{https://arxiv.org/abs/0704.3611}{{arXiv:0704.3611}}}
{[astro-ph]}
\end{barticle}
\endbibitem

\bibitem[\protect\citeauthoryear{{Tamburri} et~al.}{2014}]{Tamburri2014}
\begin{barticle}
\bauthor{\bsnm{{Tamburri}}, \binits{S.}},
\bauthor{\bsnm{{Saracco}}, \binits{P.}},
\bauthor{\bsnm{{Longhetti}}, \binits{M.}},
\bauthor{\bsnm{{Gargiulo}}, \binits{A.}},
\bauthor{\bsnm{{Lonoce}}, \binits{I.}},
\bauthor{\bsnm{{Ciocca}}, \binits{F.}}:
\batitle{{The population of early-type galaxies: how it evolves with time and how it differs from passive and late-type galaxies}}.
\bjtitle{A\&A}
\bvolume{570},
\bfpage{102}
(\byear{2014})
\doiurl{10.1051/0004-6361/201424040}
{\href{https://arxiv.org/abs/1409.1028}{{arXiv:1409.1028}}}
{[astro-ph.GA]}
\end{barticle}
\endbibitem

\bibitem[\protect\citeauthoryear{{Salim} et~al.}{2018}]{Salim2018}
\begin{barticle}
\bauthor{\bsnm{{Salim}}, \binits{S.}},
\bauthor{\bsnm{{Boquien}}, \binits{M.}},
\bauthor{\bsnm{{Lee}}, \binits{J.C.}}:
\batitle{{Dust Attenuation Curves in the Local Universe: Demographics and New Laws for Star-forming Galaxies and High-redshift Analogs}}.
\bjtitle{ApJ}
\bvolume{859}(\bissue{1}),
\bfpage{11}
(\byear{2018})
\doiurl{10.3847/1538-4357/aabf3c}
{\href{https://arxiv.org/abs/1804.05850}{{arXiv:1804.05850}}}
{[astro-ph.GA]}
\end{barticle}
\endbibitem

\bibitem[\protect\citeauthoryear{{Pozzetti} et~al.}{2010}]{Pozzetti2010}
\begin{barticle}
\bauthor{\bsnm{{Pozzetti}}, \binits{L.}},
\bauthor{\bsnm{{Bolzonella}}, \binits{M.}},
\bauthor{\bsnm{{Zucca}}, \binits{E.}},
\bauthor{\bsnm{{Zamorani}}, \binits{G.}},
\bauthor{\bsnm{{Lilly}}, \binits{S.}},
\bauthor{\bsnm{{Renzini}}, \binits{A.}},
\bauthor{\bsnm{{Moresco}}, \binits{M.}},
\bauthor{\bsnm{{Mignoli}}, \binits{M.}},
\bauthor{\bsnm{{Cassata}}, \binits{P.}},
\bauthor{\bsnm{{Tasca}}, \binits{L.}},
\bauthor{\bsnm{{Lamareille}}, \binits{F.}},
\bauthor{\bsnm{{Maier}}, \binits{C.}},
\bauthor{\bsnm{{Meneux}}, \binits{B.}},
\bauthor{\bsnm{{Halliday}}, \binits{C.}},
\bauthor{\bsnm{{Oesch}}, \binits{P.}},
\bauthor{\bsnm{{Vergani}}, \binits{D.}},
\bauthor{\bsnm{{Caputi}}, \binits{K.}},
\bauthor{\bsnm{{Kova{\v{c}}}}, \binits{K.}},
\bauthor{\bsnm{{Cimatti}}, \binits{A.}},
\bauthor{\bsnm{{Cucciati}}, \binits{O.}},
\bauthor{\bsnm{{Iovino}}, \binits{A.}},
\bauthor{\bsnm{{Peng}}, \binits{Y.}},
\bauthor{\bsnm{{Carollo}}, \binits{M.}},
\bauthor{\bsnm{{Contini}}, \binits{T.}},
\bauthor{\bsnm{{Kneib}}, \binits{J.-P.}},
\bauthor{\bsnm{{Le F{\'e}vre}}, \binits{O.}},
\bauthor{\bsnm{{Mainieri}}, \binits{V.}},
\bauthor{\bsnm{{Scodeggio}}, \binits{M.}},
\bauthor{\bsnm{{Bardelli}}, \binits{S.}},
\bauthor{\bsnm{{Bongiorno}}, \binits{A.}},
\bauthor{\bsnm{{Coppa}}, \binits{G.}},
\bauthor{\bsnm{{de la Torre}}, \binits{S.}},
\bauthor{\bsnm{{de Ravel}}, \binits{L.}},
\bauthor{\bsnm{{Franzetti}}, \binits{P.}},
\bauthor{\bsnm{{Garilli}}, \binits{B.}},
\bauthor{\bsnm{{Kampczyk}}, \binits{P.}},
\bauthor{\bsnm{{Knobel}}, \binits{C.}},
\bauthor{\bsnm{{Le Borgne}}, \binits{J.-F.}},
\bauthor{\bsnm{{Le Brun}}, \binits{V.}},
\bauthor{\bsnm{{Pell{\`o}}}, \binits{R.}},
\bauthor{\bsnm{{Perez Montero}}, \binits{E.}},
\bauthor{\bsnm{{Ricciardelli}}, \binits{E.}},
\bauthor{\bsnm{{Silverman}}, \binits{J.D.}},
\bauthor{\bsnm{{Tanaka}}, \binits{M.}},
\bauthor{\bsnm{{Tresse}}, \binits{L.}},
\bauthor{\bsnm{{Abbas}}, \binits{U.}},
\bauthor{\bsnm{{Bottini}}, \binits{D.}},
\bauthor{\bsnm{{Cappi}}, \binits{A.}},
\bauthor{\bsnm{{Guzzo}}, \binits{L.}},
\bauthor{\bsnm{{Koekemoer}}, \binits{A.M.}},
\bauthor{\bsnm{{Leauthaud}}, \binits{A.}},
\bauthor{\bsnm{{Maccagni}}, \binits{D.}},
\bauthor{\bsnm{{Marinoni}}, \binits{C.}},
\bauthor{\bsnm{{McCracken}}, \binits{H.J.}},
\bauthor{\bsnm{{Memeo}}, \binits{P.}},
\bauthor{\bsnm{{Porciani}}, \binits{C.}},
\bauthor{\bsnm{{Scaramella}}, \binits{R.}},
\bauthor{\bsnm{{Scarlata}}, \binits{C.}},
\bauthor{\bsnm{{Scoville}}, \binits{N.}}:
\batitle{{zCOSMOS - 10k-bright spectroscopic sample. The bimodality in the galaxy stellar mass function: exploring its evolution with redshift}}.
\bjtitle{\aap}
\bvolume{523},
\bfpage{13}
(\byear{2010})
\doiurl{10.1051/0004-6361/200913020}
{\href{https://arxiv.org/abs/0907.5416}{{arXiv:0907.5416}}}
{[astro-ph.CO]}
\end{barticle}
\endbibitem

\bibitem[\protect\citeauthoryear{{Loveday} et~al.}{2015}]{Loveday2015}
\begin{barticle}
\bauthor{\bsnm{{Loveday}}, \binits{J.}},
\bauthor{\bsnm{{Norberg}}, \binits{P.}},
\bauthor{\bsnm{{Baldry}}, \binits{I.K.}},
\bauthor{\bsnm{{Bland-Hawthorn}}, \binits{J.}},
\bauthor{\bsnm{{Brough}}, \binits{S.}},
\bauthor{\bsnm{{Brown}}, \binits{M.J.I.}},
\bauthor{\bsnm{{Driver}}, \binits{S.P.}},
\bauthor{\bsnm{{Kelvin}}, \binits{L.S.}},
\bauthor{\bsnm{{Phillipps}}, \binits{S.}}:
\batitle{{Galaxy and Mass Assembly (GAMA): maximum-likelihood determination of the luminosity function and its evolution}}.
\bjtitle{MNRAS}
\bvolume{451}(\bissue{2}),
\bfpage{1540}--\blpage{1552}
(\byear{2015})
\doiurl{10.1093/mnras/stv1013}
{\href{https://arxiv.org/abs/1505.01003}{{arXiv:1505.01003}}}
{[astro-ph.GA]}
\end{barticle}
\endbibitem

\bibitem[\protect\citeauthoryear{{Xu} et~al.}{2025}]{Xu2025}
\begin{barticle}
\bauthor{\bsnm{{Xu}}, \binits{K.}},
\bauthor{\bsnm{{Jing}}, \binits{Y.P.}},
\bauthor{\bsnm{{Cole}}, \binits{S.}},
\bauthor{\bsnm{{Frenk}}, \binits{C.S.}},
\bauthor{\bsnm{{Bose}}, \binits{S.}},
\bauthor{\bsnm{{Elbers}}, \binits{W.}},
\bauthor{\bsnm{{Wang}}, \binits{W.}},
\bauthor{\bsnm{{Wang}}, \binits{Y.}},
\bauthor{\bsnm{{Moore}}, \binits{S.}},
\bauthor{\bsnm{{Aguilar}}, \binits{J.}},
\bauthor{\bsnm{{Ahlen}}, \binits{S.}},
\bauthor{\bsnm{{Bianchi}}, \binits{D.}},
\bauthor{\bsnm{{Brooks}}, \binits{D.}},
\bauthor{\bsnm{{Claybaugh}}, \binits{T.}},
\bauthor{\bsnm{{de la Macorra}}, \binits{A.}},
\bauthor{\bsnm{{Dey}}, \binits{A.}},
\bauthor{\bsnm{{Forero-Romero}}, \binits{J.E.}},
\bauthor{\bsnm{{Gazta{\~n}aga}}, \binits{E.}},
\bauthor{\bsnm{{Gontcho}}, \binits{S.G.A.}},
\bauthor{\bsnm{{Gutierrez}}, \binits{G.}},
\bauthor{\bsnm{{Honscheid}}, \binits{K.}},
\bauthor{\bsnm{{Ishak}}, \binits{M.}},
\bauthor{\bsnm{{Kisner}}, \binits{T.}},
\bauthor{\bsnm{{Koposov}}, \binits{S.E.}},
\bauthor{\bsnm{{Landriau}}, \binits{M.}},
\bauthor{\bsnm{{Le Guillou}}, \binits{L.}},
\bauthor{\bsnm{{Miquel}}, \binits{R.}},
\bauthor{\bsnm{{Moustakas}}, \binits{J.}},
\bauthor{\bsnm{{Poppett}}, \binits{C.}},
\bauthor{\bsnm{{Prada}}, \binits{F.}},
\bauthor{\bsnm{{P{\'e}rez-R{\`a}fols}}, \binits{I.}},
\bauthor{\bsnm{{Rossi}}, \binits{G.}},
\bauthor{\bsnm{{Sanchez}}, \binits{E.}},
\bauthor{\bsnm{{Sprayberry}}, \binits{D.}},
\bauthor{\bsnm{{Tarl{\'e}}}, \binits{G.}},
\bauthor{\bsnm{{Weaver}}, \binits{B.A.}},
\bauthor{\bsnm{{Zou}}, \binits{H.}}:
\batitle{{PAC in DESI. I. Galaxy stellar mass function into the {}10$^{6}$ M$_{{\ensuremath{\odot}}}$ frontier}}.
\bjtitle{MNRAS}
\bvolume{540}(\bissue{2}),
\bfpage{1635}--\blpage{1667}
(\byear{2025})
\doiurl{10.1093/mnras/staf782}
{\href{https://arxiv.org/abs/2503.01948}{{arXiv:2503.01948}}}
{[astro-ph.GA]}
\end{barticle}
\endbibitem

\bibitem[\protect\citeauthoryear{{Moore} et~al.}{2025}]{Moore2025}
\begin{botherref}
\oauthor{\bsnm{{Moore}}, \binits{S.G.}},
\oauthor{\bsnm{{Cole}}, \binits{S.}},
\oauthor{\bsnm{{Wilson}}, \binits{M.}},
\oauthor{\bsnm{{Norberg}}, \binits{P.}},
\oauthor{\bsnm{{Moustakas}}, \binits{J.}},
\oauthor{\bsnm{{Aguilar}}, \binits{J.}},
\oauthor{\bsnm{{Ahlen}}, \binits{S.}},
\oauthor{\bsnm{{Anand}}, \binits{A.}},
\oauthor{\bsnm{{Bianchi}}, \binits{D.}},
\oauthor{\bsnm{{Brooks}}, \binits{D.}},
\oauthor{\bsnm{{Castander}}, \binits{F.J.}},
\oauthor{\bsnm{{Claybaugh}}, \binits{T.}},
\oauthor{\bsnm{{Cuceu}}, \binits{A.}},
\oauthor{\bsnm{{de la Macorra}}, \binits{A.}},
\oauthor{\bsnm{{Dey}}, \binits{A.}},
\oauthor{\bsnm{{Dey}}, \binits{B.}},
\oauthor{\bsnm{{Ferraro}}, \binits{S.}},
\oauthor{\bsnm{{Font-Ribera}}, \binits{A.}},
\oauthor{\bsnm{{Forero-Romero}}, \binits{J.E.}},
\oauthor{\bsnm{{Gaztanaga}}, \binits{E.}},
\oauthor{\bsnm{{Gontcho}}, \binits{S.G.A.}},
\oauthor{\bsnm{{Gutierrez}}, \binits{G.}},
\oauthor{\bsnm{{Herrera-Alcantar}}, \binits{H.K.}},
\oauthor{\bsnm{{Honscheid}}, \binits{K.}},
\oauthor{\bsnm{{Ishak}}, \binits{M.}},
\oauthor{\bsnm{{Joyce}}, \binits{R.}},
\oauthor{\bsnm{{Juneau}}, \binits{S.}},
\oauthor{\bsnm{{Kehoe}}, \binits{R.}},
\oauthor{\bsnm{{Kisner}}, \binits{T.}},
\oauthor{\bsnm{{Koposov}}, \binits{S.E.}},
\oauthor{\bsnm{{Kremin}}, \binits{A.}},
\oauthor{\bsnm{{Lahav}}, \binits{O.}},
\oauthor{\bsnm{{Lamman}}, \binits{C.}},
\oauthor{\bsnm{{Landriau}}, \binits{M.}},
\oauthor{\bsnm{{Le Guillou}}, \binits{L.}},
\oauthor{\bsnm{{Levi}}, \binits{M.E.}},
\oauthor{\bsnm{{Manera}}, \binits{M.}},
\oauthor{\bsnm{{Meisner}}, \binits{A.}},
\oauthor{\bsnm{{Miquel}}, \binits{R.}},
\oauthor{\bsnm{{Nadathur}}, \binits{S.}},
\oauthor{\bsnm{{Percival}}, \binits{W.J.}},
\oauthor{\bsnm{{Poppett}}, \binits{C.}},
\oauthor{\bsnm{{Prada}}, \binits{F.}},
\oauthor{\bsnm{{Ross}}, \binits{A.J.}},
\oauthor{\bsnm{{Rossi}}, \binits{G.}},
\oauthor{\bsnm{{Sanchez}}, \binits{E.}},
\oauthor{\bsnm{{Schlegel}}, \binits{D.}},
\oauthor{\bsnm{{Schubnell}}, \binits{M.}},
\oauthor{\bsnm{{Seo}}, \binits{H.}},
\oauthor{\bsnm{{Silber}}, \binits{J.}},
\oauthor{\bsnm{{Sprayberry}}, \binits{D.}},
\oauthor{\bsnm{{Tarl{\'e}}}, \binits{G.}},
\oauthor{\bsnm{{Weaver}}, \binits{B.A.}},
\oauthor{\bsnm{{Wechsler}}, \binits{R.H.}},
\oauthor{\bsnm{{Zhou}}, \binits{R.}},
\oauthor{\bsnm{{Zou}}, \binits{H.}}:
{DESI DR2 Galaxy Luminosity Functions}.
arXiv e-prints,
2511--01803
(2025)
\doiurl{10.48550/arXiv.2511.01803}
{\href{https://arxiv.org/abs/2511.01803}{{arXiv:2511.01803}}}
{[astro-ph.GA]}
\end{botherref}
\endbibitem

\bibitem[\protect\citeauthoryear{{Han} et~al.}{2025}]{Han2025}
\begin{barticle}
\bauthor{\bsnm{{Han}}, \binits{J.}},
\bauthor{\bsnm{{Li}}, \binits{M.}},
\bauthor{\bsnm{{Jiang}}, \binits{W.}},
\bauthor{\bsnm{{Chen}}, \binits{Z.}},
\bauthor{\bsnm{{Wang}}, \binits{H.}},
\bauthor{\bsnm{{Wei}}, \binits{C.}},
\bauthor{\bsnm{{He}}, \binits{F.}},
\bauthor{\bsnm{{He}}, \binits{J.}},
\bauthor{\bsnm{{Zhang}}, \binits{J.}},
\bauthor{\bsnm{{Liu}}, \binits{Y.}},
\bauthor{\bsnm{{Cui}}, \binits{W.}},
\bauthor{\bsnm{{Gu}}, \binits{Y.}},
\bauthor{\bsnm{{Guo}}, \binits{Q.}},
\bauthor{\bsnm{{Jing}}, \binits{Y.}},
\bauthor{\bsnm{{Kang}}, \binits{X.}},
\bauthor{\bsnm{{Li}}, \binits{G.}},
\bauthor{\bsnm{{Luo}}, \binits{X.}},
\bauthor{\bsnm{{Luo}}, \binits{Y.}},
\bauthor{\bsnm{{Pei}}, \binits{W.}},
\bauthor{\bsnm{{Qiu}}, \binits{Y.}},
\bauthor{\bsnm{{Tan}}, \binits{Z.}},
\bauthor{\bsnm{{Xie}}, \binits{L.}},
\bauthor{\bsnm{{Yang}}, \binits{X.}},
\bauthor{\bsnm{{Yu}}, \binits{H.}},
\bauthor{\bsnm{{Yu}}, \binits{Y.}},
\bauthor{\bsnm{{Zhou}}, \binits{J.}}:
\batitle{{The Jiutian simulations for the CSST extra-galactic surveys}}.
\bjtitle{Science China Physics, Mechanics, and Astronomy}
\bvolume{68}(\bissue{10}),
\bfpage{109511}
(\byear{2025})
\doiurl{10.1007/s11433-025-2712-1}
{\href{https://arxiv.org/abs/2503.21368}{{arXiv:2503.21368}}}
{[astro-ph.CO]}
\end{barticle}
\endbibitem

\bibitem[\protect\citeauthoryear{{Planck Collaboration} et~al.}{2020}]{Planck2020}
\begin{barticle}
\bauthor{\bsnm{{Planck Collaboration}}},
\bauthor{\bsnm{{Aghanim}}, \binits{N.}},
\bauthor{\bsnm{{Akrami}}, \binits{Y.}},
\bauthor{\bsnm{{Ashdown}}, \binits{M.}},
\bauthor{\bsnm{{Aumont}}, \binits{J.}},
\bauthor{\bsnm{{Baccigalupi}}, \binits{C.}},
\bauthor{\bsnm{{Ballardini}}, \binits{M.}},
\bauthor{\bsnm{{Banday}}, \binits{A.J.}},
\bauthor{\bsnm{{Barreiro}}, \binits{R.B.}},
\bauthor{\bsnm{{Bartolo}}, \binits{N.}},
\bauthor{\bsnm{{Basak}}, \binits{S.}},
\bauthor{\bsnm{{Battye}}, \binits{R.}},
\bauthor{\bsnm{{Benabed}}, \binits{K.}},
\bauthor{\bsnm{{Bernard}}, \binits{J.-P.}},
\bauthor{\bsnm{{Bersanelli}}, \binits{M.}},
\bauthor{\bsnm{{Bielewicz}}, \binits{P.}},
\bauthor{\bsnm{{Bock}}, \binits{J.J.}},
\bauthor{\bsnm{{Bond}}, \binits{J.R.}},
\bauthor{\bsnm{{Borrill}}, \binits{J.}},
\bauthor{\bsnm{{Bouchet}}, \binits{F.R.}},
\bauthor{\bsnm{{Boulanger}}, \binits{F.}},
\bauthor{\bsnm{{Bucher}}, \binits{M.}},
\bauthor{\bsnm{{Burigana}}, \binits{C.}},
\bauthor{\bsnm{{Butler}}, \binits{R.C.}},
\bauthor{\bsnm{{Calabrese}}, \binits{E.}},
\bauthor{\bsnm{{Cardoso}}, \binits{J.-F.}},
\bauthor{\bsnm{{Carron}}, \binits{J.}},
\bauthor{\bsnm{{Challinor}}, \binits{A.}},
\bauthor{\bsnm{{Chiang}}, \binits{H.C.}},
\bauthor{\bsnm{{Chluba}}, \binits{J.}},
\bauthor{\bsnm{{Colombo}}, \binits{L.P.L.}},
\bauthor{\bsnm{{Combet}}, \binits{C.}},
\bauthor{\bsnm{{Contreras}}, \binits{D.}},
\bauthor{\bsnm{{Crill}}, \binits{B.P.}},
\bauthor{\bsnm{{Cuttaia}}, \binits{F.}},
\bauthor{\bsnm{{de Bernardis}}, \binits{P.}},
\bauthor{\bsnm{{de Zotti}}, \binits{G.}},
\bauthor{\bsnm{{Delabrouille}}, \binits{J.}},
\bauthor{\bsnm{{Delouis}}, \binits{J.-M.}},
\bauthor{\bsnm{{Di Valentino}}, \binits{E.}},
\bauthor{\bsnm{{Diego}}, \binits{J.M.}},
\bauthor{\bsnm{{Dor{\'e}}}, \binits{O.}},
\bauthor{\bsnm{{Douspis}}, \binits{M.}},
\bauthor{\bsnm{{Ducout}}, \binits{A.}},
\bauthor{\bsnm{{Dupac}}, \binits{X.}},
\bauthor{\bsnm{{Dusini}}, \binits{S.}},
\bauthor{\bsnm{{Efstathiou}}, \binits{G.}},
\bauthor{\bsnm{{Elsner}}, \binits{F.}},
\bauthor{\bsnm{{En{\ss}lin}}, \binits{T.A.}},
\bauthor{\bsnm{{Eriksen}}, \binits{H.K.}},
\bauthor{\bsnm{{Fantaye}}, \binits{Y.}},
\bauthor{\bsnm{{Farhang}}, \binits{M.}},
\bauthor{\bsnm{{Fergusson}}, \binits{J.}},
\bauthor{\bsnm{{Fernandez-Cobos}}, \binits{R.}},
\bauthor{\bsnm{{Finelli}}, \binits{F.}},
\bauthor{\bsnm{{Forastieri}}, \binits{F.}},
\bauthor{\bsnm{{Frailis}}, \binits{M.}},
\bauthor{\bsnm{{Fraisse}}, \binits{A.A.}},
\bauthor{\bsnm{{Franceschi}}, \binits{E.}},
\bauthor{\bsnm{{Frolov}}, \binits{A.}},
\bauthor{\bsnm{{Galeotta}}, \binits{S.}},
\bauthor{\bsnm{{Galli}}, \binits{S.}},
\bauthor{\bsnm{{Ganga}}, \binits{K.}},
\bauthor{\bsnm{{G{\'e}nova-Santos}}, \binits{R.T.}},
\bauthor{\bsnm{{Gerbino}}, \binits{M.}},
\bauthor{\bsnm{{Ghosh}}, \binits{T.}},
\bauthor{\bsnm{{Gonz{\'a}lez-Nuevo}}, \binits{J.}},
\bauthor{\bsnm{{G{\'o}rski}}, \binits{K.M.}},
\bauthor{\bsnm{{Gratton}}, \binits{S.}},
\bauthor{\bsnm{{Gruppuso}}, \binits{A.}},
\bauthor{\bsnm{{Gudmundsson}}, \binits{J.E.}},
\bauthor{\bsnm{{Hamann}}, \binits{J.}},
\bauthor{\bsnm{{Handley}}, \binits{W.}},
\bauthor{\bsnm{{Hansen}}, \binits{F.K.}},
\bauthor{\bsnm{{Herranz}}, \binits{D.}},
\bauthor{\bsnm{{Hildebrandt}}, \binits{S.R.}},
\bauthor{\bsnm{{Hivon}}, \binits{E.}},
\bauthor{\bsnm{{Huang}}, \binits{Z.}},
\bauthor{\bsnm{{Jaffe}}, \binits{A.H.}},
\bauthor{\bsnm{{Jones}}, \binits{W.C.}},
\bauthor{\bsnm{{Karakci}}, \binits{A.}},
\bauthor{\bsnm{{Keih{\"a}nen}}, \binits{E.}},
\bauthor{\bsnm{{Keskitalo}}, \binits{R.}},
\bauthor{\bsnm{{Kiiveri}}, \binits{K.}},
\bauthor{\bsnm{{Kim}}, \binits{J.}},
\bauthor{\bsnm{{Kisner}}, \binits{T.S.}},
\bauthor{\bsnm{{Knox}}, \binits{L.}},
\bauthor{\bsnm{{Krachmalnicoff}}, \binits{N.}},
\bauthor{\bsnm{{Kunz}}, \binits{M.}},
\bauthor{\bsnm{{Kurki-Suonio}}, \binits{H.}},
\bauthor{\bsnm{{Lagache}}, \binits{G.}},
\bauthor{\bsnm{{Lamarre}}, \binits{J.-M.}},
\bauthor{\bsnm{{Lasenby}}, \binits{A.}},
\bauthor{\bsnm{{Lattanzi}}, \binits{M.}},
\bauthor{\bsnm{{Lawrence}}, \binits{C.R.}},
\bauthor{\bsnm{{Le Jeune}}, \binits{M.}},
\bauthor{\bsnm{{Lemos}}, \binits{P.}},
\bauthor{\bsnm{{Lesgourgues}}, \binits{J.}},
\bauthor{\bsnm{{Levrier}}, \binits{F.}},
\bauthor{\bsnm{{Lewis}}, \binits{A.}},
\bauthor{\bsnm{{Liguori}}, \binits{M.}},
\bauthor{\bsnm{{Lilje}}, \binits{P.B.}},
\bauthor{\bsnm{{Lilley}}, \binits{M.}},
\bauthor{\bsnm{{Lindholm}}, \binits{V.}},
\bauthor{\bsnm{{L{\'o}pez-Caniego}}, \binits{M.}},
\bauthor{\bsnm{{Lubin}}, \binits{P.M.}},
\bauthor{\bsnm{{Ma}}, \binits{Y.-Z.}},
\bauthor{\bsnm{{Mac{\'\i}as-P{\'e}rez}}, \binits{J.F.}},
\bauthor{\bsnm{{Maggio}}, \binits{G.}},
\bauthor{\bsnm{{Maino}}, \binits{D.}},
\bauthor{\bsnm{{Mandolesi}}, \binits{N.}},
\bauthor{\bsnm{{Mangilli}}, \binits{A.}},
\bauthor{\bsnm{{Marcos-Caballero}}, \binits{A.}},
\bauthor{\bsnm{{Maris}}, \binits{M.}},
\bauthor{\bsnm{{Martin}}, \binits{P.G.}},
\bauthor{\bsnm{{Martinelli}}, \binits{M.}},
\bauthor{\bsnm{{Mart{\'\i}nez-Gonz{\'a}lez}}, \binits{E.}},
\bauthor{\bsnm{{Matarrese}}, \binits{S.}},
\bauthor{\bsnm{{Mauri}}, \binits{N.}},
\bauthor{\bsnm{{McEwen}}, \binits{J.D.}},
\bauthor{\bsnm{{Meinhold}}, \binits{P.R.}},
\bauthor{\bsnm{{Melchiorri}}, \binits{A.}},
\bauthor{\bsnm{{Mennella}}, \binits{A.}},
\bauthor{\bsnm{{Migliaccio}}, \binits{M.}},
\bauthor{\bsnm{{Millea}}, \binits{M.}},
\bauthor{\bsnm{{Mitra}}, \binits{S.}},
\bauthor{\bsnm{{Miville-Desch{\^e}nes}}, \binits{M.-A.}},
\bauthor{\bsnm{{Molinari}}, \binits{D.}},
\bauthor{\bsnm{{Montier}}, \binits{L.}},
\bauthor{\bsnm{{Morgante}}, \binits{G.}},
\bauthor{\bsnm{{Moss}}, \binits{A.}},
\bauthor{\bsnm{{Natoli}}, \binits{P.}},
\bauthor{\bsnm{{N{\o}rgaard-Nielsen}}, \binits{H.U.}},
\bauthor{\bsnm{{Pagano}}, \binits{L.}},
\bauthor{\bsnm{{Paoletti}}, \binits{D.}},
\bauthor{\bsnm{{Partridge}}, \binits{B.}},
\bauthor{\bsnm{{Patanchon}}, \binits{G.}},
\bauthor{\bsnm{{Peiris}}, \binits{H.V.}},
\bauthor{\bsnm{{Perrotta}}, \binits{F.}},
\bauthor{\bsnm{{Pettorino}}, \binits{V.}},
\bauthor{\bsnm{{Piacentini}}, \binits{F.}},
\bauthor{\bsnm{{Polastri}}, \binits{L.}},
\bauthor{\bsnm{{Polenta}}, \binits{G.}},
\bauthor{\bsnm{{Puget}}, \binits{J.-L.}},
\bauthor{\bsnm{{Rachen}}, \binits{J.P.}},
\bauthor{\bsnm{{Reinecke}}, \binits{M.}},
\bauthor{\bsnm{{Remazeilles}}, \binits{M.}},
\bauthor{\bsnm{{Renzi}}, \binits{A.}},
\bauthor{\bsnm{{Rocha}}, \binits{G.}},
\bauthor{\bsnm{{Rosset}}, \binits{C.}},
\bauthor{\bsnm{{Roudier}}, \binits{G.}},
\bauthor{\bsnm{{Rubi{\~n}o-Mart{\'\i}n}}, \binits{J.A.}},
\bauthor{\bsnm{{Ruiz-Granados}}, \binits{B.}},
\bauthor{\bsnm{{Salvati}}, \binits{L.}},
\bauthor{\bsnm{{Sandri}}, \binits{M.}},
\bauthor{\bsnm{{Savelainen}}, \binits{M.}},
\bauthor{\bsnm{{Scott}}, \binits{D.}},
\bauthor{\bsnm{{Shellard}}, \binits{E.P.S.}},
\bauthor{\bsnm{{Sirignano}}, \binits{C.}},
\bauthor{\bsnm{{Sirri}}, \binits{G.}},
\bauthor{\bsnm{{Spencer}}, \binits{L.D.}},
\bauthor{\bsnm{{Sunyaev}}, \binits{R.}},
\bauthor{\bsnm{{Suur-Uski}}, \binits{A.-S.}},
\bauthor{\bsnm{{Tauber}}, \binits{J.A.}},
\bauthor{\bsnm{{Tavagnacco}}, \binits{D.}},
\bauthor{\bsnm{{Tenti}}, \binits{M.}},
\bauthor{\bsnm{{Toffolatti}}, \binits{L.}},
\bauthor{\bsnm{{Tomasi}}, \binits{M.}},
\bauthor{\bsnm{{Trombetti}}, \binits{T.}},
\bauthor{\bsnm{{Valenziano}}, \binits{L.}},
\bauthor{\bsnm{{Valiviita}}, \binits{J.}},
\bauthor{\bsnm{{Van Tent}}, \binits{B.}},
\bauthor{\bsnm{{Vibert}}, \binits{L.}},
\bauthor{\bsnm{{Vielva}}, \binits{P.}},
\bauthor{\bsnm{{Villa}}, \binits{F.}},
\bauthor{\bsnm{{Vittorio}}, \binits{N.}},
\bauthor{\bsnm{{Wandelt}}, \binits{B.D.}},
\bauthor{\bsnm{{Wehus}}, \binits{I.K.}},
\bauthor{\bsnm{{White}}, \binits{M.}},
\bauthor{\bsnm{{White}}, \binits{S.D.M.}},
\bauthor{\bsnm{{Zacchei}}, \binits{A.}},
\bauthor{\bsnm{{Zonca}}, \binits{A.}}:
\batitle{{Planck 2018 results. VI. Cosmological parameters}}.
\bjtitle{\aap}
\bvolume{641},
\bfpage{6}
(\byear{2020})
\doiurl{10.1051/0004-6361/201833910}
{\href{https://arxiv.org/abs/1807.06209}{{arXiv:1807.06209}}}
{[astro-ph.CO]}
\end{barticle}
\endbibitem

\bibitem[\protect\citeauthoryear{{Davis} et~al.}{1985}]{Davis1985}
\begin{barticle}
\bauthor{\bsnm{{Davis}}, \binits{M.}},
\bauthor{\bsnm{{Efstathiou}}, \binits{G.}},
\bauthor{\bsnm{{Frenk}}, \binits{C.S.}},
\bauthor{\bsnm{{White}}, \binits{S.D.M.}}:
\batitle{{The evolution of large-scale structure in a universe dominated by cold dark matter}}.
\bjtitle{\apj}
\bvolume{292},
\bfpage{371}--\blpage{394}
(\byear{1985})
\doiurl{10.1086/163168}
\end{barticle}
\endbibitem

\bibitem[\protect\citeauthoryear{{Han} et~al.}{2012}]{Han2012}
\begin{barticle}
\bauthor{\bsnm{{Han}}, \binits{J.}},
\bauthor{\bsnm{{Jing}}, \binits{Y.P.}},
\bauthor{\bsnm{{Wang}}, \binits{H.}},
\bauthor{\bsnm{{Wang}}, \binits{W.}}:
\batitle{{Resolving subhaloes' lives with the Hierarchical Bound-Tracing algorithm}}.
\bjtitle{\mnras}
\bvolume{427}(\bissue{3}),
\bfpage{2437}--\blpage{2449}
(\byear{2012})
\doiurl{10.1111/j.1365-2966.2012.22111.x}
{\href{https://arxiv.org/abs/1103.2099}{{arXiv:1103.2099}}}
{[astro-ph.CO]}
\end{barticle}
\endbibitem

\bibitem[\protect\citeauthoryear{{Han} et~al.}{2018}]{Han2018}
\begin{barticle}
\bauthor{\bsnm{{Han}}, \binits{J.}},
\bauthor{\bsnm{{Cole}}, \binits{S.}},
\bauthor{\bsnm{{Frenk}}, \binits{C.S.}},
\bauthor{\bsnm{{Benitez-Llambay}}, \binits{A.}},
\bauthor{\bsnm{{Helly}}, \binits{J.}}:
\batitle{{HBT+: an improved code for finding subhaloes and building merger trees in cosmological simulations}}.
\bjtitle{\mnras}
\bvolume{474}(\bissue{1}),
\bfpage{604}--\blpage{617}
(\byear{2018})
\doiurl{10.1093/mnras/stx2792}
{\href{https://arxiv.org/abs/1708.03646}{{arXiv:1708.03646}}}
{[astro-ph.CO]}
\end{barticle}
\endbibitem

\bibitem[\protect\citeauthoryear{{Xu} et~al.}{2026}]{Xu2026}
\begin{botherref}
\oauthor{\bsnm{{Xu}}, \binits{X.}},
\oauthor{\bsnm{{Yang}}, \binits{X.}},
\oauthor{\bsnm{{Zhai}}, \binits{Z.}},
\oauthor{\bsnm{{Guo}}, \binits{Y.}},
\oauthor{\bsnm{{Gu}}, \binits{Y.}},
\oauthor{\bsnm{{Wang}}, \binits{Y.}},
\oauthor{\bsnm{{Han}}, \binits{J.}},
\oauthor{\bsnm{{Tan}}, \binits{Z.}},
\oauthor{\bsnm{{Li}}, \binits{J.}}:
{What can galaxy clustering really tell us about the galaxy-halo connections?}
arXiv e-prints,
2604--05561
(2026)
\doiurl{10.48550/arXiv.2604.05561}
{\href{https://arxiv.org/abs/2604.05561}{{arXiv:2604.05561}}}
{[astro-ph.GA]}
\end{botherref}
\endbibitem

\bibitem[\protect\citeauthoryear{{Gu} et~al.}{2024}]{Gu2024}
\begin{barticle}
\bauthor{\bsnm{{Gu}}, \binits{Y.}},
\bauthor{\bsnm{{Yang}}, \binits{X.}},
\bauthor{\bsnm{{Han}}, \binits{J.}},
\bauthor{\bsnm{{Wang}}, \binits{Y.}},
\bauthor{\bsnm{{Li}}, \binits{Q.}},
\bauthor{\bsnm{{Tan}}, \binits{Z.}},
\bauthor{\bsnm{{Jiang}}, \binits{W.}},
\bauthor{\bsnm{{Wang}}, \binits{Y.}},
\bauthor{\bsnm{{Wang}}, \binits{J.}},
\bauthor{\bsnm{{Katsianis}}, \binits{A.}},
\bauthor{\bsnm{{Xu}}, \binits{X.}},
\bauthor{\bsnm{{Xu}}, \binits{H.}},
\bauthor{\bsnm{{Hong}}, \binits{W.}},
\bauthor{\bsnm{{Mo}}, \binits{H.}},
\bauthor{\bsnm{{Wen}}, \binits{R.}},
\bauthor{\bsnm{{Zheng}}, \binits{X.}},
\bauthor{\bsnm{{Shi}}, \binits{F.}},
\bauthor{\bsnm{{Zhang}}, \binits{P.}},
\bauthor{\bsnm{{Zhai}}, \binits{Z.}},
\bauthor{\bsnm{{Liu}}, \binits{C.}},
\bauthor{\bsnm{{Wang}}, \binits{W.}},
\bauthor{\bsnm{{Zu}}, \binits{Y.}},
\bauthor{\bsnm{{Guo}}, \binits{H.}},
\bauthor{\bsnm{{Zhang}}, \binits{Y.}},
\bauthor{\bsnm{{Lu}}, \binits{Y.}},
\bauthor{\bsnm{{Zheng}}, \binits{Y.}},
\bauthor{\bsnm{{Han}}, \binits{Y.}},
\bauthor{\bsnm{{Zou}}, \binits{H.}},
\bauthor{\bsnm{{Wang}}, \binits{X.}},
\bauthor{\bsnm{{Wei}}, \binits{C.}},
\bauthor{\bsnm{{Li}}, \binits{M.}},
\bauthor{\bsnm{{Luo}}, \binits{Y.}}:
\batitle{{CSST large-scale structure analysis pipeline: I. Constructing reference mock galaxy redshift surveys}}.
\bjtitle{\mnras}
\bvolume{529}(\bissue{4}),
\bfpage{4015}--\blpage{4027}
(\byear{2024})
\doiurl{10.1093/mnras/stae762}
{\href{https://arxiv.org/abs/2403.10754}{{arXiv:2403.10754}}}
{[astro-ph.GA]}
\end{barticle}
\endbibitem

\bibitem[\protect\citeauthoryear{Yang et~al.}{2004}]{Yang2004}
\begin{barticle}
\bauthor{\bsnm{Yang}, \binits{X.}},
\bauthor{\bsnm{Mo}, \binits{H.J.}},
\bauthor{\bsnm{Jing}, \binits{Y.P.}},
\bauthor{\bsnm{Van Den~Bosch}, \binits{F.C.}},
\bauthor{\bsnm{Chu}, \binits{Y.}}:
\batitle{Populating dark matter haloes with galaxies: comparing the 2dfgrs with mock galaxy redshift surveys}.
\bjtitle{Monthly Notices of the Royal Astronomical Society}
\bvolume{350}(\bissue{4}),
\bfpage{1153}--\blpage{1173}
(\byear{2004})
\doiurl{10.1111/j.1365-2966.2004.07744.x}
{\href{https://arxiv.org/abs/https://academic.oup.com/mnras/article-pdf/350/4/1153/3083337/350-4-1153.pdf}{{https://academic.oup.com/mnras/article-pdf/350/4/1153/3083337/350-4-1153.pdf}}}
\end{barticle}
\endbibitem

\end{thebibliography}

\end{document}